%% file: mainARXIV.tex
\documentclass[10pt,journal]{IEEEtran}

\usepackage[utf8]{inputenc}

\usepackage{subcaption}
\usepackage{multicol}
\usepackage{color,soul}
\usepackage{amsmath,amssymb,amsfonts,mathrsfs,amsthm}
\usepackage{pgfplots}
\usepackage{array}
\usepackage{mathbbol}
\usepackage{tikz}
\usepackage{svg}
\usepackage{enumitem}
\usepackage{blindtext}
\usepackage{scrextend}
\addtokomafont{labelinglabel}{\sffamily}
\usepackage{xfrac}   
\usepackage[author={noi}]{pdfcomment}

\usepackage{bm}
\usepackage{algorithmicx}
\usepackage[ruled]{algorithm}
\usepackage[noend]{algpseudocode}
\algrenewcommand\algorithmicindent{1.80em}%
\usepackage{lmodern}
\usepackage{hyperref}
\usepackage{cleveref}

\addtokomafont{labelinglabel}{\sffamily}

\DeclareMathAlphabet{\mathbbmsl}{U}{bbm}{m}{sl}

\newtheorem{definition}{Definition}[section]
\newtheorem{theorem}{Theorem}[section]
\newtheorem{observation}{Observation}[section]
\newtheorem{proposition}{Proposition}[section]

\newcommand{\Paths}{\mathcal{M}}
\newcommand{\FPaths}{\mathcal{F}}
\newcommand{\RPaths}{\mathcal{R}}
\newcommand{\TPaths}{\mathcal{T}}
\newcommand{\Obst}{O_{\mathcal{T}}}
\newcommand{\NList}{\boldsymbol{V}_f}
\newcommand{\Fone}{\mathcal{F}_{1}^{(m_i,\mathcal{T})}}
\newcommand{\Fonem}{\mathcal{F}_{1}^{(m,\mathcal{T})}}
\newcommand{\Pc}{\mathcal{P}_v}

\newcommand{\pathft}{m^{(\mathcal{T})}}
\newcommand{\hpathft}{\hat{m}^{(\mathcal{T})}}
\allowdisplaybreaks

\graphicspath{{./imgs/}}
\DeclareGraphicsExtensions{.pdf,.jpeg,.png}

\title{Static and Dynamic Failure Localization through \\Progressive Network Tomography}

\author{        Viviana Arrigoni,~\IEEEmembership{Student Member,~IEEE,}        Novella Bartolini,~\IEEEmembership{Senior Member, ~IEEE,}\\
Annalisa Massini, ~\IEEEmembership{Member,~IEEE,}  
Federico Trombetti,~\IEEEmembership{Student Member,~IEEE,}
 }       

\begin{document}
\maketitle

\begin{abstract}
We aim at assessing the states of the nodes in a network by means of end-to-end monitoring paths. 
    The contribution of this paper is twofold. First, we consider a \emph{static failure scenario}. In this context, we aim at minimizing the 
    number of probes to obtain failure identification.
    To face this problem 
    we propose a progressive approach to failure localization based on 
    stochastic optimization, whose solution is the optimal sequence of monitoring paths to probe. 
    We address the complexity of the problem by proposing a greedy strategy in two variants: one considers exact calculation of posterior probabilities of node failures, given the observation, whereas the other approximates these values by means of a novel failure centrality metric. Secondly, we adapt these two strategies to a \emph{dynamic failure scenario} where nodes states can change throughout a monitoring period. 
By means of numerical experiments conducted on real network topologies, we demonstrate the practical applicability of our approach. Our performance evaluation evidences the superiority of our algorithms with respect to  state of the art solutions based on classic Boolean Network Tomography as well as approaches based on sequential group testing.
\end{abstract}

\section{Introduction}
Boolean Network Tomography (BNT) provides a series of powerful tools to localize network failures by using  end-to-end monitoring paths connecting deployed monitors.
BNT approaches characterize network components in terms of their state identifiability (working or failed) under any failure scenario.
The major challenge  of this approach comes from  the fact that observations of the outcome of monitoring paths (working/failed) induce a system of Boolean equations that is commonly under-determined, hence allowing multiple solutions \cite{nguyen07infocom}. 
Moreover, exact assessment of the status of each network component is not always achievable if monitors can only be deployed on a given subset of nodes, and routing of probing paths  is not controllable.
In addition, when the number of potentially concurrent failures is unbounded, 
maximum identification of failed components may require an enormous number of monitoring paths and related probes \cite{bartolini2020fundamental,bth_inf17}, which severely limits the applicability of the approach.

However, we notice that executing the probing activity in a progressive manner, according to which the next probing path is selected on the basis of the information obtained from the previous probes, is particularly helpful in reducing the number of required probes to assess the status of the network under a specific failure scenario.
According to this approach, hereby referred to  as  {\em progressive BNT}, the outcome of any new network measurement is used to simplify the problem instance.
In fact, we observe that if a monitoring path is traversed successfully, we can  ascertain the  status of all the traversed components as working. In contrast, if a monitoring path fails, it certainly contains at least a failed component.
It follows that, depending on the current observation scenario, monitoring a path may contribute valuable knowledge to different degrees. 
To measure the incremental value of monitoring paths in a progressive probing activity, we introduce a new notion of path utility which takes account of the added failure localization information with respect to the previously obtained network assessment.
By using the information obtained by monitoring a given subset of the available paths, we can calculate the posterior expectation of the utility of monitoring any of the paths which have not been probed yet.
By applying a Bayesian approach we are able to design a stochastic optimization problem which maximizes the expected utility over a progressive monitoring activity.
We formulate a dynamic programming approach to derive the optimal progressive policy to maximize failure identification. However, we point out that the aforementioned optimization is computationally intractable for two reasons. The first reason is the large size of the state space representation, where each path may contribute different pieces of information, depending on its outcome. The second reason is because the computation of the posterior probabilities of a path to work properly, is exponential in the number of paths composing the network.
In order to cope with the described complexity we propose a simplified approach based on two fundamental pillars.

On the one hand, rather than resorting to dynamic programming which would require the exploration of exponentially-many intermediate states in the progressive execution of the probing activity, we propose a greedy approach, called Posterior Probability Greedy (PoPGreedy), that selects the path that more likely contributes disambiguation of the state of a large number of network components.

On the other hand we approximate the posterior probability of a path failure by means of a 
polynomially computable approximation metric, to which we refer with the name of {\em failure centrality}.
Failure centrality of a node reflects the probability that a node is broken, based on the currently available observation.
We call this approach {\em Failure Centrality Greedy algorithm (FaCeGreedy)}

In order to measure the failure localization capability of the proposed approaches and be able to provide a quantitative  evaluation, we introduce four novel metrics to measure the accuracy in properly localizing working as well as failed nodes.
We compare FaCeGreedy and PoPGreedy both in terms of failure detection performance and related execution time.
The experiments show that FaCeGreedy provides an excellent approximation of the stochastic optimization approach, in a negligible time. 
Instances that require an execution time of a week for the exact optimization, are solved in a matter of minutes by FaCeGreedy.

We also compare the performance of FaCeGreedy with  algorithms for failure assessment 
based on classic BNT approaches. 
Simulations show that, as expected, FaCeGreedy has superior performance as it localizes more failures with fewer probing paths than BNT approaches.

To complete our analysis, we compare FaCeGreedy with AdaptiveFinder \cite{karbasi2012sequential}, a state of the art solution based on sequential group testing, and with Adaptive Path Construction (APC), \cite{mukamoto2015adaptive}, a routing-constraint algorithm, also based on sequential group testing, for link failure detection. 
We highlight that, being graph-constraint rather than routing-constraint, AdaptiveFinder has much more freedom than FaCeGreedy in selecting the composition of the probing sets in terms of network components. 
Despite the higher flexibility in selecting testing sets, AdaptiveFinder performs worse than FaCeGreedy when the number of paths is given, and during its progressive execution.
APC instead investigates on the state of the network by means of end-to-end paths that are given as an input, similarly to our scenario. We translate the link failure problem into a problem of node failure localization. 
The experiments show that in all the experimented settings, setting the number of tests to the minimum required by FaCeGreedy to localize all the failed nodes, AdaptiveFinder only localizes about half of the failed components. AdaptiveFinder requires many more probing sets than FaCeGreedy to correctly localize all the failures. In addition, although APC works well on small networks where only few nodes fail, it performs worse than FaCeGreedy when large networks are involved and many multiple failures occur, when it requires up to three times more path probes than FaCeGreedy. 
Finally, we also show that the approach introduced in this paper may be easily extended to deal with dynamic changes within the network. This evolution to our original approach comes with low computational cost.  

\noindent Our original contributions are the following:
\vspace{-.1cm} \begin{itemize}[leftmargin=*]
    \item We formulate the problem of progressive network tomography in terms of stochastic optimization and Bayesian analysis.
    \item We give an exact solution approach and discuss its complexity, motivating the need to resort to polynomial heuristic approaches.
    \item We formulate a novel failure centrality metric to approximate the failure probability 
    of a node, given the observation of the outcome of a given set of probing paths.
    \item We formulate four novel metrics to quantitatively measure the capability of a monitoring algorithm to properly localize network failure and reduce the localization uncertainty.
    \item We propose two greedy approaches, called PoPGreedy and FaCeGreedy, based on Bayesian utility maximization.
    \item We prove optimality approximation for PoPGreedy. 
    \item By means of simulations conducted on real network topologies, we compare FaCeGreedy and PoPGreedy against classic Boolean Tomography approaches, as well as approaches based on sequential group testing, showing that the our solutions outperform the others in all the performance metrics, and in all the considered scenarios.
    \item We show computational inexpensive altered versions of PoPGreedy and FaceGreey to deal with dynamic changes.
\end{itemize}

\section{Related Work}

{\color{black} Network tomography employs path probing to localize network failures. 
Network tomography techniques are broadly categorized in two families depending on the metric of interest for the inspection, additive or non-additive. An additive metric establishes a linear relationship between the measurement of a path and  measurements of individual links and nodes composing the path. 
 Along this line of research, Tati et al. \cite{Tati14ICDCS} proposed a path selection algorithm to improve link metric identifiability, by maximizing the rank of successful measurements subject to random link failures. The work of Ren et al. \cite{Ren:Infocom16}  proposed algorithms to determine which link metrics can be identified and where to place monitors to maximize the number of identifiable links, subject to a bounded number of link failures.
Additive metric tomography was also studied in   \cite{ting1,ting2}, to  identify of additive link metrics under topology changes. 


In contrast, non-additive tomography refers to non linear relationships between the path and its component metrics. The most relevant examples are those related to congestion or failure localization, where the dominant factor of a path state is the state of its worst performing component.
In this paper, we focus on the second of the aforementioned families, namely on the case of non-additive tomography, and more specifically  Boolean Network Tomography.

 The early works on this topic focused on best-effort inference. For example, Duffield et al. \cite{Duffield03,Duffield06TInfo} and Kompella et al. \cite{Kompella07infocom} aimed at finding the minimum set of failures that can explain the observed measurements, and Nguyen et al. \cite{nguyen07infocom} aimed at finding the most likely failure set that explains the observations.
Later, the identifiability problem attracted attention. Ma et al. characterized in \cite{Ma&etal:14IMC} the maximum number of simultaneous failures that can be uniquely localized, and then extended the results in \cite{Ma16TON} to characterize the maximum number of failures under which the states of specified nodes can be uniquely identified as well as the number of nodes whose states can be identified under a given number of failures.
In contrast to \cite{Ma16TON}, the work in \cite{bartolini2020fundamental, bth_inf17} 
provide fundamental bounds that 
are topology agnostic, i.e.,  only based on 
 the number of monitoring paths and high level routing consistency properties.
The related optimization problems have also been studied under different formulations. For instance, the work by Bejerano et al. in \cite{Bejerano03INFOCOM}
formulates the problem of optimally placing monitors to detect failed nodes via round-trip probing and demonstrate its  NP-hardness.
The work by Cheraghchi et al. \cite{Cheraghchi:TOI2012} formulates the identifiability problem for a graph-based group-testing framework, where the test sets are constrained by the topology. Nevertheless, in the addressed framework the test sets are not end-to-end paths, but just connected components determined by random walks on the monitored network graph.

 Ma et al. \cite{Ma15PE} proposed polynomial time heuristics to deploy a minimum number of monitors to uniquely localize a given number of failures under various routing constraints.
When monitoring is performed at the service layer, He et al. \cite{usICDCS16} proposed service placement algorithms to maximize the number of identifiable nodes by monitoring the paths connecting clients and servers.

Boolean Network Tomography  suffers from two problems which severely limit its practical applicability to real settings. 
A first limitation is in the usual assumption of knowing  an upper bound on the number of congested or failed links.
Such a limitation is mostly due to the size explosion of the candidate failure set scenarios. 
Most of the proposed works are not designed to work under an unbounded number of failing components.
Second,  the cited
works aim at designing monitoring paths so as to ensure node failure identifiability, according to the definition given in \cite{Ma&etal:14IMC}, under any failure scenario that meet the above mentioned constraint on the number of failed elements. 

Unlike these works  we do not assume any  bound on the number of failures and we focus on actual node state identification rather than on identifiability. 
In fact, we observe that by monitoring paths in a real failure scenario, it is possible to identify the state of network components that are not theoretically identifiable in all failure scenarios as prescribed in \cite{Ma&etal:14IMC}, but are so in the considered real case setting.


With a similar goal, the authors of \cite{Qlearn} model the tomography problem as a Markov Decision Process, and solve it with a $Q$-learning technique.
The  actions of the decision process are related to the diagnosis of the congestion status of individual links. 
The work in \cite{LiangArkiv} also utilizes machine learning techniques based on neural networks to infer a network topology from incrementally selected paths, with the purpose to predict the performance of  paths that are not directly probed.

We  adopt incremental path selection, with the purpose of identifying the state of individual network components on known topology networks. 
Based on the information progressively gathered, the instance of the failure identification problem is updated and reduced.

One approach towards this the same goal is the algorithm AdaptiveFinder \cite{karbasi2012sequential}, which considers progressive monitoring of graph-based test groups.
We consider this proposal as a benchmark for performance comparisons against our own approach.
AdaptiveFinder considers a network graph, and creates arbitrary sets of connected network components to determine the 
next paths to test according to a progressive approach.
Unlike this work
we consider testing sets which are end-to-end monitoring paths, where pairs of monitors are connected by a series of nodes that strictly follows the routing protocol in use by the considered network.
Similarly, Adaptive Path Construction (APC), \cite{mukamoto2015adaptive}, is a group-testing routing-constraint algorithm for detecting link failures by means of BNT techniques, that aims at minimizing the number of path probes. Differently from our Bayesian approach, in APC the choice of the next path to probe follows a binary search based idea. 
As we will see in Section \ref{sec:results} thanks to the incremental knowledge constructed through our Bayesian decision support, our approach overcomes the limitation imposed by the routing algorithm and provides superior performance than the AdaptiveFinder and APC approaches. 
}
In this paper, we also take into account dynamic changes in the nodes of the network, and we show how our two proposed algorithms can be adapted to an online setting where a network outsider wants to infer the state of the nodes continuously. 
The problem of detecting failures occurring dynamically within a network attracted attention in recent years. A large portions of the available literature focuses on specific networks (e.g., data centres~\cite{arzani2018007} and Wireless Sensor Networks (WSNs)~\cite{kamal2014failure,muhammed2017analysis,swain2018heterogeneous,roy2017d3}).
In~\cite{huang2008practical}, Huang et. al. highlight practical issues when tomography techniques are used to infer link degradation within a network. Their approach is divided into an initial offline phase (a set of paths covering the network is selected), followed by an online phase (where monitor nodes periodically probe measurements along the defined paths in order to track possible changes in the performance of the links). 
In~\cite{johnsson2014online}, Johnsson et. al. propose a two-step algorithm to interpret and analyze the outcome of path probes in order to detect and localize failures. Differently from these works, we consider path selection to be the key part of the online phase: we not only provide a way to interpret data, but also we show how to obtain the most informative data. 

\section{Problem Formulation}


We consider a network modeled as a graph  $G=(V,E)$, and a set of monitor nodes $V_M \subseteq V$ (shortly called {\em monitors}). For  each ordered pair $(i,j)$ of monitors in $V_M$ we consider a  unique monitoring path whose sequence of nodes is only determined by the routing algorithm in use.  We consider uncontrollable routing, (see \cite{Ma&etal:14IMC}), i.e., monitoring paths are determined by the routing protocol
used by the network, not controllable by the monitors. Routing between the monitors $i$ and $j$ is not necessarily symmetric, but is assumed to be  deterministic, and known.
We shortly denote with $\hat{m}$ the set of nodes traversed by the monitoring path $m$. We refer to $\mathcal{M}$ as to the set of monitoring paths available for probing the network.

By probing the paths of $\mathcal{M}$ it is possible to obtain indirect information on the state of the traversed nodes. 
Both working and failed paths 
provide helpful information for the state assessment of the network components.
In particular, all the nodes of a working paths must be properly functioning, whereas a non working path must contain at least a broken component.
By probing paths in a sequence, it is possible to determine the most suitable choice for the next path to probe, on the basis of the information gathered so far. 
We address the problem of designing a {\em Progressive Monitoring Policy (PMP)}, i.e., a sequence of paths to be probed one by one, such that we can identify the status of the largest number of nodes, in the minimum number of steps (number of paths).
We refer to this problem as the {\em PMP problem}. 

In the following, we denote with $S_v$ the event ‘node $v$ works', and $\bar{S}_v$ the event ‘node $v$ is broken'. 
If no information is available concerning the distribution of failures in the network, it is reasonable to assume uniform probability
of node failures, namely that all  nodes have equal {\em prior probability} $p$ to be damaged, that is $P(\bar{S}_i)=p$ for all $i \in V$, while we denote with $P(S_i)=(1-p)$ the prior probability that  $v_i$ is a working node. 
A path fails if at least one of its nodes fails, while a path works if all of its node work.  
Unlike classic tomography approaches, we do not assume any prior knowledge of the exact number of failed nodes.

{\color{black}
\begin{table}[t!]
\caption{Summary of notations}\label{tab:notation}
\renewcommand{\arraystretch}{.9}
\centering
{\footnotesize{
\hspace{-3mm}
\begin{tabular}{|p{0.075\textwidth}|p{0.38\textwidth}|}
\hline
\bfseries Notation & \bfseries Description \\
\hline\hline
$\hat{m}$ & set of nodes traversed by path $m\in \mathcal{M}$\\
\hline
$S_i$  &  state of node $v_i$ (failed if $S_i=0$, working otherwise)\\
\hline
$Z_j$  &  state of path  $m_j$ (failed if $Z_j=0$, working otherwise)\\
\hline
$Z^o_j$  &  observed (tested) state of path  $m_j$\\
\hline
$\mathcal{A}$ & $\mathcal{A} \triangleq \{a_1, a_2, \ldots, a_{|\mathcal{M}|} \}$: set of monitoring decisions\\
\hline
$\mathcal{T} \subseteq{\mathcal{M}}$  & set of tested monitoring paths\\
\hline
$\mathcal{R} = \mathcal{M} \setminus \mathcal{T}$  & set of not yet tested monitoring paths\\
\hline
$\FPaths\subseteq \TPaths$ & set of failed tested monitoring paths \\ \hline
$O_\mathcal{T}$  &  set of observed outcomes of paths in $\mathcal{T}$\\
\hline
      \end{tabular}
}}
\vspace{-0.6cm}
\end{table}
%
%
Classic approaches to Boolean Network Tomography adopt the concept of 
$k$-identifiability \cite{bartolini2020fundamental,bth_inf17,Ma&etal:14IMC}, which 
refers to the capability of inferring the state of individual nodes from the state of the monitoring paths.  A node $v$ is $k$-identifiable in $\Paths$ if any two sets of failing nodes $F_1$ and $F_2$ of size at most $k$, which differ at least in  $v$ (i.e., one contains $v$ and the other does not), cause the failures of different  subsets of paths in $\mathcal{M}$. 
The concept of $k$-identifiability  assumes knowledge of an upper bound $k$ to the number of occurring failures, and characterizes nodes regardless of their status, failed or working, but only in terms of whether their status can be uniquely inferred by observing the outcome of the monitoring paths of $\mathcal{M}$.
However, our setting is characterized by (1) 
absence of  a bound on the number of simultaneous failures,
(2) uncontrollable position of  monitor nodes and (3) given routing algorithm. In such a setting, 
the PMP problem is particularly challenging 
as no node is guaranteed to be identifiable according to classic tomography.
 We note that, especially for large values of $k$, $k$-identifiability of a node is an unlikely condition that is hardly verified in real networks.

\subsection{Bayesian utility of path probing}\label{sec:bayes}

Let  $Z_j$ be the event that  path
$m_j \in \mathcal{M}$ properly works,  and $\bar{Z}_j$ the event that path $m_j$ fails.
Under the assumption of uniform probability
of node failures a {\em prior estimate} of the state probability of path  $m_i \in \mathcal{M}$ is $P(Z_i)= (1-p)^{|\hat{m}_i|}
$, and $P(\bar{Z}_i)= 1-P(Z_i)$.

In our problem setting, the state of the network can only be observed by probing monitoring paths in a sequence of monitoring interventions.
We denote by $\mathcal{A}\triangleq \{a_1, a_2, \ldots, a_{|\mathcal{M}|} \}$ the set of possible monitoring decisions, where decision $a_i$  implies monitoring the network through path $m_i$.

We denote by $\mathcal{T} \subseteq \mathcal{M}$ the   already  monitored paths. 
We denote with $Z_i^o$ the outcome of the probing activity along path $m_i$, for any $m_i \in \mathcal{T}$.
Knowledge of the outcome of the paths in $\mathcal{T}$ constitutes a source of additional information $O_\mathcal{T} \triangleq\{Z^o_j | m_j \in \mathcal{T} \}$ that can be used to produce a better --- a posteriori --- estimate, of the network status. 
We denote with $\mathcal{R}$  the current residual set of paths, i.e., paths which have not been monitored yet, namely $\mathcal{R} \triangleq \mathcal{M} \setminus \mathcal{T}$.
By knowing the values of the random variables in $O_\mathcal{T}$ we can update the} {\em posterior estimate} of the state probability of any path $m_i \in \mathcal{R}$, which is
$P(Z_i | O_\mathcal{T})$. 

We note that the outcome of any monitoring path is as informative as it contributes identification of the status of individual network components, or decreases the size of the identification problem instance.
More specifically, we observe that monitoring working or non-working paths contributes useful information for  failure identification in different manners.

\noindent
{\em{Information obtained by monitoring working paths:}}

\begin{figure}[]	
	\centering
	    \begin{subfigure}{.48\columnwidth}
        \centering
\begin{tikzpicture}
\tikzset{mylab/.style={black, minimum size=0.5cm}}
\tikzset{pallino/.style={draw,circle,fill=white,minimum size=4pt,inner sep=0pt}}
\tikzset{working/.style={draw,circle,fill=green,minimum size=4pt,inner sep=0pt}}
    \draw (0,0) node (1) [pallino,label=left:\footnotesize{$v_1$}] {}
        -- ++(0:0.6cm) node (3) [working] {}
        -- ++(0:0.6cm) node (2) [pallino,label=right:\footnotesize{$v_2  $}] {};
 \node [mylab,below of = 3, node distance=0.1in] (blank) {\vspace{-1cm}\footnotesize{$w $} };
\end{tikzpicture}
        \caption{Original topology}
        \label{fig:ot1}
    \end{subfigure}%
	\begin{subfigure}{.48\columnwidth}
        \centering
       \begin{tikzpicture}
\tikzset{mylab/.style={black, minimum size=0.5cm}}
\tikzset{working/.style={draw,circle,fill=green,minimum size=4pt,inner sep=0pt}}
\tikzset{pallino/.style={draw,circle,fill=white,minimum size=4pt,inner sep=0pt}}
    \draw (0,0) node (1) [pallino,label=left:\footnotesize{$v_1$}] {}
               -- ++(0:1.0cm) node (2) [pallino,label=right:\footnotesize{$v_2  $}] {};
 \node [mylab,below of = 3, node distance=0.1in] (blank) {\vspace{-1cm}\footnotesize{} };
\end{tikzpicture}
        \caption{Pruned topology}
        \label{fig:l18_bis_a}
    \end{subfigure}%
	 \caption{Removal of a working node (in green)} \label{fig:shortcircuit}
\end{figure}
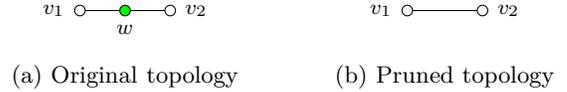

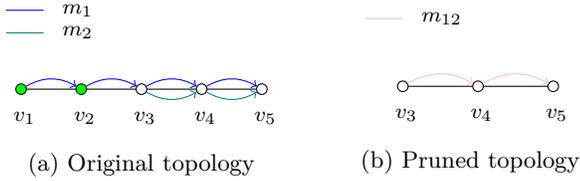
\begin{figure}[h!]		
\begin{center}
	    \begin{subfigure}{0.48\columnwidth}
        \centering

\begin{tikzpicture}

\tikzset{pallino/.style={draw,circle,fill=white,minimum size=4pt,
                            inner sep=0pt}}
\tikzset{mylab/.style={black, minimum size=0.5cm}}
\tikzset{working/.style={draw,circle,fill=green,minimum size=4pt,inner sep=0pt}}

  \draw (.7,0) node (1) [working] {}        
  --++(0:.8cm) node (2) [working]{}
           --++ (0:.8cm) node (3) [pallino]{}
            --++ (0:.8cm) node (4) [pallino]{}
            --++ (0:.8cm) node (5) [pallino]{};

        \node [mylab,below of = 1, node distance=0.15in] (blank) {\footnotesize{$\ v_1 $} };
        \node [mylab,below of = 2, node distance=0.15in] (blank) {\footnotesize{$\ v_2  $} };
        \node [mylab,below of = 3, node distance=0.15in] (blank) {\footnotesize{$\ v_3$} };
         \node [mylab,below of = 4, node distance=0.15in] (blank) {\footnotesize{$\ v_4 $} };
         \node [mylab,below of = 5, node distance=0.15in] (blank) {\footnotesize{$\ v_5 $} };

\draw [->,blue] (1) to [out=30,in=150] (2);
\draw [->,blue] (2) to [out=30,in=150] (3);
\draw [->,blue] (3) to [out=30,in=150] (4);
\draw [->,blue] (4) to [out=30,in=150] (5);
\draw [->,teal] (3) to [out=-30,in=-150] (4);
\draw [->,teal] (4) to [out=-30,in=-150] (5);

\draw[color=teal] (0.5cm, .75cm) --  (1cm, .75cm) node[draw=none,fill=none] (pippo) [label=right:{{\color{black}\footnotesize{$m_2$}}}]{};
\draw[blue] (0.5cm, 1.05cm) -- (1cm, 1.05cm) node[draw=none,fill=none] (pippo) [label=right:{{\color{black}\footnotesize{$m_1$}}}]{};

	\end{tikzpicture}
         \caption{Original topology}
        \label{fig:l18_bis_b}
    \end{subfigure}
     \begin{subfigure}{0.48\columnwidth}
        \centering

\begin{tikzpicture}

\tikzset{pallino/.style={draw,circle,fill=white,minimum size=4pt,
                            inner sep=0pt}}
\tikzset{mylab/.style={black, minimum size=0.5cm}}
\tikzset{working/.style={draw,circle,fill=green,minimum size=4pt,inner sep=0pt}}

  \draw (0:0) node (3) [pallino]{}
            --++ (0:1cm) node (4) [pallino]{}
            --++ (0:1cm) node (5) [pallino]{};

        
        \node [mylab,below of = 3, node distance=0.15in] (blank) {\footnotesize{$\ v_3$} };
         \node [mylab,below of = 4, node distance=0.15in] (blank) {\footnotesize{$\ v_4 $} };
         \node [mylab,below of = 5, node distance=0.15in] (blank) {\footnotesize{$\ v_5 $} };

\draw [->,pink] (3) to [out=30,in=150] (4);
\draw [->,pink] (4) to [out=30,in=150] (5);

\draw[color=white] (-.5cm, 1.05cm) --  (0cm, 1.05cm) node[draw=none,fill=none] (pippo) []{};
\draw[color=pink] (-.5cm, 0.9cm) --  (0cm, 0.9cm) node[draw=none,fill=none] (pippo) [label=right:{{\color{black}\footnotesize{$m_{12}$}}}]{};
	\end{tikzpicture}
        \caption{Pruned topology}
        \label{fig:l18_bis_c}
    \end{subfigure}\\
    \end{center}
	 \caption{Merge of  paths after  node pruning} \label{fig:two_paths_become_one}
\end{figure}

If a probed path works then all its traversed nodes work as well. After the observation of a working path, the current instance of the failure identification problem  can be reduced by considering a logical representation of the network graph constructed by  {\em pruning the working nodes} and short circuiting the incident edges, as in Figure \ref{fig:shortcircuit}. We call $\pathft$ and $\hpathft$ the path $m$ in the pruned logical graph,  after testing the paths of $\TPaths$, and the set of nodes that the pruned path traverses, respectively.
       Similarly, the residual set of paths to be monitored $\mathcal{R}$ may be reduced as well when any two paths $m_i$ and $m_j$ traverse the same set of nodes after pruning working nodes, namely when $\hpathft_i= \hpathft_j$,
    after pruning, as in Figure \ref{fig:two_paths_become_one}.
    Likewise, paths consisting only of nodes which have been found to be working, will be removed as well, as a consequence of the pruning of all their nodes.

\begin{figure}[h!]
\centering
\begin{tikzpicture}
\centering
\tikzset{pallino/.style={draw,circle,fill=green,minimum size=4pt,     inner sep=0pt}}
\tikzset{mylab/.style={black, minimum size=0.5cm}}
\tikzset{working/.style={draw,circle,fill=green,minimum size=4pt,inner sep=0pt}}
\tikzset{failed/.style={draw,circle,fill=red,minimum size=4pt,inner sep=0pt}}
  \draw (.7,0) node (1) [pallino] {}       
  --++(0:.8cm) node (2) [pallino]{}
           --++ (0:.8cm) node (3) [pallino]{}
            --++ (0:.8cm) node (4) [pallino]{}
            --++ (0:.8cm) node (5) [pallino]{};
            
\draw (3.1,.8) node (6) [failed]{};

    \node [mylab,above of = 6, node distance=0.15in] (blank) {\footnotesize{$\ v_6 $} };
        \node [mylab,below of = 1, node distance=0.15in] (blank) {\footnotesize{$\ v_1 $} };
        \node [mylab,below of = 2, node distance=0.15in] (blank) {\footnotesize{$\ v_2  $} };
        \node [mylab,below of = 3, node distance=0.15in] (blank) {\footnotesize{$\ v_3$} };
         \node [mylab,below of = 4, node distance=0.15in] (blank) {\footnotesize{$\ v_4 $} };
         \node [mylab,below of = 5, node distance=0.15in] (blank) {\footnotesize{$\ v_5 $} };
    \draw [->,blue] (1) to [out=30,in=150] (2);
\draw [->,blue] (2) to [out=30,in=150] (3);
\draw [->,blue] (3) to [out=30,in=150] (4);
\draw [->,blue] (4) to [out=30,in=150] (5);
\draw [->,red] (4) to [out=210,in=-30] (3);
\draw (6) --++ (4);
\draw [->,red] (6) to [out=-60,in=60] (4);

\draw[color=red] (0.5cm, 0.75cm) --  (1cm, 0.75cm) node[draw=none,fill=none] (pippo) [label=right:{{\color{black}\footnotesize{$m_2$}}}]{};
\draw[blue] (0.5cm, 1.05cm) -- (1cm, 1.05cm) node[draw=none,fill=none] (pippo) [label=right:{{\color{black}\footnotesize{$m_1$}}}]{};
	\end{tikzpicture}
	\caption{Identification by exclusion}
	\label{fig:identification_by_exclusion}
	\end{figure}
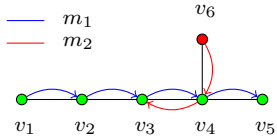

    For this reason, the utility deriving from probing a working path is set proportional to its number of nodes. More precisely,  the utility of probing a  path $\pathft_i$ (obtained as a logical representation of $m_i$ after pruning all the working nodes) is proportional to the amount of newly found working nodes $|\hpathft_i|$.
    In addition to this we notice that  the longer a working path is, the smaller the search area for locating failed nodes will be.  
   In particular, by pruning certainly working nodes, it may also happen that some non-working nodes are identified {\em by exclusion}, namely because they belong to non-working paths, already monitored, which identify subsets of candidate non-working nodes, reduced to size one after  pruning the newly found working nodes. 
    We define as $\mathcal{F}^{(\mathcal{T})}_1$ the set of failed paths $m_i$ which have already been monitored (i.e. $m_i \in \mathcal{T}$) and have length equal to 1 after node pruning.
    Hence, we  consider another additive term, to the utility of monitoring working paths, i.e. $|\mathcal{F}^{(\mathcal{T})}_1|$.
        Consider the example of Figure \ref{fig:identification_by_exclusion}.
    Assume that the monitoring activity starts by probing path $m_2$ first, which is found to be non-working, hence $m_2$ is inserted in $\mathcal{T}$.
    Then the monitoring activity proceeds by considering path $m_1$ which is properly working. Knowledge of the outcome of $m_1$ allows us to assess the status of the nodes $v_i$, with $i=1, \ldots, 5$, as working. As a consequence these nodes are all pruned, and can be removed by all the non working paths included in $\mathcal{T}$.
    Due to the pruning of $v_4$ and $v_3$ the length of the already monitored path $m_2$ reduces to 1  in the logical representation of the network graph with pruned components, which implies 
     $|\hpathft_2 |= 1$. Hence,   $\pathft_2$ turns to be a failing path of one only node,  $v_6$, whose state must be failed, by exclusion. 
    
\vspace{-0.3cm}

\begin{figure}[h!]		\hspace{.5cm}	
\begin{subfigure}{0.48\columnwidth} 
\begin{tikzpicture}
\centering
\tikzset{pallino/.style={draw,circle,fill=white,minimum size=4pt,     inner sep=0pt}}
\tikzset{mylab/.style={black, minimum size=0.5cm}}
\tikzset{working/.style={draw,circle,fill=green,minimum size=4pt,inner sep=0pt}}
\tikzset{failed/.style={draw,circle,fill=red,minimum size=4pt,inner sep=0pt}}
  \draw (.7,0) node (1) [pallino] {}       
  --++(0:.8cm) node (2) [pallino]{}
           --++ (0:.8cm) node (3) [pallino]{}
            --++ (0:.8cm) node (4) [failed]{};
            
\draw (3.1,.8) node (5) [working]{};

    \node [mylab,above of = 5, node distance=0.15in] (blank) {\footnotesize{$\ v_5 $} };
        \node [mylab,below of = 1, node distance=0.15in] (blank) {\footnotesize{$\ v_1 $} };
        \node [mylab,below of = 2, node distance=0.15in] (blank) {\footnotesize{$\ v_2  $} };
        \node [mylab,below of = 3, node distance=0.15in] (blank) {\footnotesize{$\ v_3$} };
         \node [mylab,below of = 4, node distance=0.15in] (blank) {\footnotesize{$\ v_4 $} };
    \draw [->,blue] (1) to [out=30,in=150] (2);
\draw [->,blue] (2) to [out=30,in=150] (3);
\draw [->,blue] (3) to [out=30,in=150] (4);
\draw (5) --++ (4);
\draw [->,red] (5) to [out=-60,in=60] (4);
\draw [->,teal] (1) to [out=-30,in=-150] (2);
\draw [->,teal] (2) to [out=-30,in=-150] (3);

\draw[color=blue] (0.5cm, .75cm) --  (1cm, .75cm) node[draw=none,fill=none] (pippo) [label=right:{{\color{black}\footnotesize{$m_3$}}}]{};
\draw[teal] (0.5cm, 1.05cm) -- (1cm, 1.05cm) node[draw=none,fill=none] (pippo) [label=right:{{\color{black}\footnotesize{$m_2$}}}]{};
\draw[red] (0.5cm, 1.35cm) -- (1cm, 1.35cm) node[draw=none,fill=none] (pippo) [label=right:{{\color{black}\footnotesize{$m_1$}}}]{};
	\end{tikzpicture}
	\caption{Original topology}
	\end{subfigure}
\hspace{-.5cm}
	\begin{subfigure}{0.48\columnwidth}
	\centering
	
	\begin{tikzpicture}
\centering
\tikzset{pallino/.style={draw,circle,fill=white,minimum size=4pt,
                            inner sep=0pt}}
\tikzset{mylab/.style={black, minimum size=0.5cm}}
\tikzset{working/.style={draw,circle,fill=green,minimum size=4pt,inner sep=0pt}}
\tikzset{failed/.style={draw,circle,fill=red,minimum size=4pt,inner sep=0pt}}

  \draw (.7,0) node (1) [pallino] {}        
  --++(0:.8cm) node (2) [pallino]{}
           --++ (0:.8cm) node (3) [pallino]{}
           --++ (0:.8cm) node (4) [failed]{};
            
   
        \node [mylab,below of = 1, node distance=0.15in] (blank) {\footnotesize{$\ v_1 $} };
        \node [mylab,below of = 2, node distance=0.15in] (blank) {\footnotesize{$\ v_2  $} };
        \node [mylab,below of = 3, node distance=0.15in] (blank) {\footnotesize{$\ v_3$} }; 
        \node [mylab,below of = 4, node distance=0.15in] (blank) {\footnotesize{$\ v_4$} }; 

\draw [->,teal] (1) to [out=-30,in=-150] (2);
\draw [->,teal] (2) to [out=-30,in=-150] (3);

\draw[color=white] (0.5cm, .75cm) --  (1cm, .75cm) node[draw=none,fill=none] (pippo) [label=right:]{};
\draw[white] (0.5cm, 1.05cm) -- (1cm, 1.05cm) node[draw=none,fill=none] (pippo) [label=right:]{};
\draw[white] (0.5cm, 1.35cm) -- (1cm, 1.35cm) node[draw=none,fill=none] (pippo) [label=right:]{};
	
	\end{tikzpicture}
	\caption{Pruned paths}
	\end{subfigure}
	\caption{Pruning  of monitoring paths}
	\label{fig:path_removal}
	
    \vspace{-0.3cm}
	\end{figure}
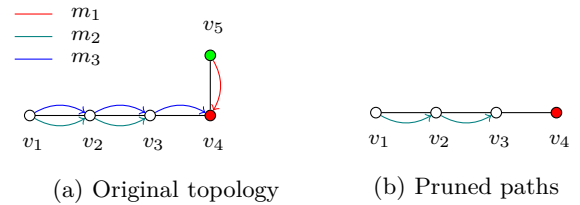

    Finally, we notice also that it never happens that working nodes are discovered by exclusion as they must belong to working paths, in which case they would have already been pruned\footnote{Discovery of working node by exclusion may instead happen if there is knowledge of the number of failed nodes, which is not considered here.}.
 \noindent
{\em{Information obtained by monitoring non-working paths:}   }
 When probing paths fail, we also have relevant information on the network status.
    A failing path corresponds to a subset of nodes containing at least a failed node.
     When a path failure occurs, the nodes of the path must undergo additional monitoring, i.e. probing intersecting paths, to obtain precise failure identification. Indeed, short failed paths allow to localize node failures more precisely than long ones. 
    Moreover, finding failed nodes, or set of nodes containing at least a failure, suggests  not to probe paths that, containing at least a failed node, will certainly fail if probed.
    In the example of Figure \ref{fig:path_removal} we consider a scenario in which, after some probing activity (not shown in the figure), $v_5$ is found to be properly functioning.
    Nevertheless, by observing the failure of monitoring along path $m_1$ we assess the failure of node $v_4$ 
    This suggests the removal of path $m_3$ from the set of monitoring paths, as its failure can be deduced from the failure of the included nodes $v_4$ which is known to be broken.
    As a consequence, whenever a monitoring path $m_i$ fails, the monitoring problem can be simplified by removing all the paths $m$ including the entire set of nodes of the failed path $m_i$, i.e. the set $\{ m_j | \hat{m_j} \supseteq \hat{m_i}\}$. 
    Additionally, we note that after pruning  nodes, we may end up with some degenerate paths with cardinality one. If this occurs, the probing of these paths gives direct information on the node states, both if the path works and if it does not. While this situation is already considered in the utility of working paths, to take it into account also for the case of a non working path $m_i$, we  consider a further  information utility component, in the form of an additive term $\lfloor 1/|\hpathft_i| \rfloor$
    which is equal to one only if  path $\pathft_i$ traverses one only node, and is zero otherwise. 
   
    \vspace{-0.3cm}
In conclusion, every time we have certainty of the state of a node (either working or broken), we prune the node (if it works) or the paths including it (if it does not work). When a certain path $m$ is probed, if $\pathft$ works, then all of its sub-paths certainly works as well (i.e., paths $m'\,s.t.\, \hat{m}'^{(\TPaths)}\subseteq \hpathft$). If $\pathft$ fails, then all of its super-paths are failing, too (i.e., paths $m'\,s.t.\,  \hpathft \subseteq \hat{m}'^{(\TPaths)}$). When the status of  non tested paths can be assessed with certainty due to the described pruning actions on the logical graph, we do not consider them for successive probes, and the set of available actions $\mathcal{A}$ is updated consequently.
 
In summary, if we make decision $a_i$ corresponding to monitoring path $m_i$, the information utility is proportional to $|\hpathft_i| + |\mathcal{F}_1|$ if the path $m_i$ works, and to $\lfloor 1/|\hpathft_i| \rfloor$ otherwise.
We can then formulate the information utility function, for each decision $a_i \in \mathcal{A}$ as follows:
\begin{equation}
\lambda(a_i | Z_i) = 
\left\{
\begin{array}{cc}
  \color{black}{|\hpathft_i|} + \color{black}{ |\mathcal{F}_1|}   & \textrm{if }Z_i=1
  \\
\color{black}{\lfloor 1/|\hpathft_i| \rfloor}   & \textrm{if }Z_i=0
\end{array}
\right.
    \label{eq:conditional_utility_given_path_state}
\end{equation}
%

%


Correspondingly, we calculate the following {\em expectation of conditional utility given the observation}:

\begin{equation}
\mathcal{U}(a_i | O_\mathcal{T}) = 
\lambda(a_i | Z_i) P(Z_i|O_{\mathcal{T}})+\lambda(a_i | \bar{Z}_i) P(\bar{Z}_i|O_{\mathcal{T}})
    \label{eq:exp_conditional_utility_given_path_state}
\end{equation}
As the available paths may give a different contribution to the identification task, some of them may become redundant, depending on the probing order, which brings our attention to determine an efficient progressive monitoring policy, i.e. to solve the PMP problem.
Formally, a PMP policy is a sequence of monitoring actions of $\mathcal{A}$.
In the following we aim at defining a PMP policy which 
maximizes the number of nodes whose state is identified, i.e. the utility defined above.

\section{Stochastic optimization of PMP}
We consider a decision process, in the discrete time, which may end when one of the following conditions occurs:
\begin{itemize}[leftmargin=*]
    \item Every node status is known
    \item There are no more paths to monitor (each of the remaining path cannot add any information on the node states)
    \item The maximum number of probing steps has been reached.
\end{itemize}

At each step, the process may make one of the decisions in $\mathcal{A}$, whose utility depends on the outcome of the related monitoring path. 
The number of steps before termination is uncertain. An upper bound is given by the number of monitoring paths $N$. We recall that we do not assume symmetric routing, i.e. the upper bound on the number of monitoring paths is given by the number of ordered pairs of monitoring nodes
$N= V_M\cdot(V_M-1)$.

Considering the discussion made in Section \ref{sec:bayes}, we formulate the failure identification problem in terms of {\em stochastic optimization}.
At the $n$-th step, the state of the decision process is given by the set $O^{(n)}_\mathcal{T}$, which reflects the observations made until step $n$, provides the current knowledge of the status of network components at step $n$ and determines the future action utility values, according to Equation \ref{eq:exp_conditional_utility_given_path_state}.

As actions cannot be repeated in consecutive monitoring steps,
we denote the actions available at step $n$, $\mathcal{A}(O^{(n)}_\mathcal{T})$, shortly as follows: $\mathcal{A}^{(n)} \triangleq 
\mathcal{A}(O^{(n)}_\mathcal{T})=
\{ a_i:  m_i \in \mathcal{R}^{(n)}\}$,
where $\mathcal{R}^{(n)}$ is the set of monitoring paths which have not been tested yet at the $n$-th step.

We seek a decision policy that maximizes the expected sum of the utilities incurred by its decisions.
The optimal decision policy depends on the utilities, on the number of steps taken to assess the state of the network, and our confidence that obtaining such information through monitoring is actually possible.

Let $V(O^{(n)}_\mathcal{T},n)$ denote the expected  information (utility) that will  be obtained by the optimal decision policy (e.g. nodes still to be assessed), starting from the observation $O^{(n)}_\mathcal{T}$ at step $n$.
If we choose action $a_i$ the expected gain at the following step is given by Equation \ref{eq:exp_conditional_utility_given_path_state}.
Now, let $E_U(O^{(n)}_\mathcal{T},n)$ 
be the optimal remaining utility, after step $n+1$, given a monitoring decision in state $O^{(n)}_\mathcal{T}$.
$E_U(O^{(n)}_\mathcal{T},n)$ describes the optimal decision policy utility after step $n+1$ and so it is stated in terms of $V(O^{(n+1)}_\mathcal{T},n+1)$.
In particular, given that the monitoring path selected at step $n$ is the one pointed by action $a^{(n)}=a_i^*$, corresponding to path $m_{i^*}$, we have
$$E_U(O^{(n)}_\mathcal{T},n|a_i^*)=P(Z_{i^*}) \cdot 
V(O^{(n)}_\mathcal{T} \cup \{Z_{i^*}=1\},n+1)+$$
$$+P(\bar{Z}_{i^*}) \cdot 
V(O^{(n)}_\mathcal{T} \cup \{Z_{i^*}=0\},n+1).
$$
By the principle of optimality (Bellman equation) we have the following:

{\footnotesize{
$$V(O^{(n)}_\mathcal{T} ,n)=$$
$$=\max_{a_i \in \mathcal{A}^{(n)}
}
\left\{P(Z_{i^*}) \cdot 
\left(|\hat{m}_i^{(\TPaths)}|+|\FPaths_1^{(m_i,\TPaths)(n)}|+
V(O^{(n)}_\mathcal{T} \cup \{Z_{i^*}\},n+1)
)
\right)+\right.$$
$$+\left.P(\bar{Z}_{i^*}) \cdot 
\left(
\left\lfloor \frac{1}{
|\hat{m}_i^{(\TPaths)}|} \right\rfloor+
V(O^{(n)}_\mathcal{T} \cup \{\bar{Z}_{i^*}\},n+1)
)
\right)
\right\}
.$$}}
While the equation suggests the use of a dynamic programming approach, over a finite horizon, to solve the PMP problem, we underline  the following challenges.
(1) The computational complexity in the calculation of the posterior probability $P(Z_i | O_\mathcal{T})$ is exponential in the number of paths. 
(2) There is the well known curse of dimensionality in the representation of the state space of the process,  which needs to take account of the outcome of each monitored path. In fact, we note that it is not sufficient to represent the state with the vector of the identified node status, because of the possibility to have delayed assessment of broken nodes, that must be considered to properly calculate the utility terms expressed by Equation \ref{eq:exp_conditional_utility_given_path_state} for the case of working paths. Therefore non working monitored paths must be part of the state representation. Working monitored paths must also be stored in the state representation, to determine the available decisions. Hence the state space of the process is exponential in the number of paths.

We will devote the next sections to polynomial approaches to the design of efficient PMP policies and to metrics to quantitatively measure such efficiency.

\subsection{The PoPGreedy approach}
\label{sec:algProb}
A {\em Bayesian greedy strategy}  to monitoring path selection and probing is one that progressively selects the next path based on the current {\em utility maximization rule} and updates the overall observation for the next step.
 Initially (at step 0) $O_\mathcal{T}=\emptyset$, therefore the calculation of the initial action utility is based on prior probabilities as follows:
 \begin{equation}
\mathcal{U}^{(0)}(a_i) = 
\lambda(a_i) P(Z_i)+\lambda(a_i) P(\bar{Z}_i).
    \label{eq:util_step_zero}
\end{equation}
Hence, at step 0, the Bayesian strategy consists in  selecting 
the action that maximizes the utility based on prior knowledge: $$a^{(1)}=\arg \max_{a_i \in \mathcal{A}}\ \mathcal{U}^{(0)}(a_i). $$
Anytime a new path is monitored, it produces an outcome which requires the update of the current estimate of path failure probabilities.

At step $n+1$ the Bayesian strategy selects the action $a^{(n+1)}$ that maximizes the expectation of the  utility given the current observation:
\begin{align*} a^{(n+1)}=\arg \max_{a_i \in \mathcal{A}^{(n)}} \mathcal{U}^{(n)}(a_i|O^{(n)}_\mathcal{T})
\end{align*}




The testing procedure is described in Algorithm~\ref{alg:minpaths}. Given a graph $G$ representing the network topology, a set of paths $\Paths$ and a prior probability of node failure $p$, the algorithm returns the posterior probability of failure of all nodes as is obtained after probing at most $K$ paths in $\Paths$ and a related ranking.

At each iteration the algorithm selects the path $m$ with  maximum expected utility (ties are broken by considering a priority based on the path index, i.e. if $m_i$ and $m_j$, with $i<j$ have the same utility, the algorithm selects $m_i$). 
Depending on the outcome of the test, either the set of failed paths (if $m$ failed) or the set of working nodes (otherwise) is updated, together with actions corresponding to testing non visited super-paths and sub-paths of a failing/working paths, respectively (lines~\ref{lst:line_super} and \ref{lst:line_sub}). 
We refer to this approach as to PoPGreedy (Posterior Probability Greedy), and detail it in Algorithm~\ref{alg:minpaths}.
\begin{algorithm}
\caption{PoPGreedy}
\label{alg:minpaths}
\textbf{Input:} $G=(V,E)$: graph representing a network topology.\\ $\Paths$: set of walkable paths in the network.\\$p$: initial probability of node failures.\\ $K$: maximum number of path probes.
\textbf{Output:} sorted sequence of nodes depending on their probability of failure: $\NList$. 
\begin{algorithmic}[1]
\State $W = \emptyset$ (set of working nodes)
\State $\TPaths=\emptyset$ (set of tested paths)
\State $\FPaths = \emptyset, \FPaths\subseteq \TPaths$ (set of failed paths)
\State $\RPaths = \Paths\setminus \TPaths$  (set of non visited paths)
\State $\mathcal{A}^{(0)}=\{a_1,\ldots,a_{|\Paths|}\}$ 
\State $\Obst = \emptyset$ 
\For{ $i = 1,\ldots,K$}
    \State $a^{(i)} = \arg \max\limits_{a \in \mathcal{A}^{(i-1)}} \mathcal{U}(a|\Obst)$ \label{lst:line:l15}
    \If {$\mathcal{U}(a^{(i)}|\Obst) = 0$}\label{lst:line:l16}
        \State return list of nodes sorted by $P(\bar{S}_v|\Obst)$, $\NList$\label{lst:line:l18}
    \EndIf
    \State $\TPaths \leftarrow \TPaths \cup \left \{m_i\right \}$
    \State $\RPaths \leftarrow \RPaths \setminus \left \{m_i\right \}$
    \State $\blacktriangleright$ \emph{\textbf{test}} \bm{$m_i$}
    \If {$m_i$ fails}
        \State $\FPaths \leftarrow \FPaths \cup \left \{m_i\right \}$
        \State $\Obst = \Obst \cup (\bar{Z}_i)$
        \State $\mathcal{A}^{(i)}\leftarrow \mathcal{A}^{(i - 1)}\setminus (\{a_i\}\cup\{a_j\,:\,\hpathft_i\subseteq \hpathft_j\})$\label{lst:line_super}
    \Else 
        \State $W \leftarrow W \cup \left \{\hat{m}_i\right \}$
        \State $\Obst = \Obst \cup (Z_i)$
        \State $\mathcal{A}^{(i)}\leftarrow \mathcal{A}^{(i - 1)}\setminus (\{a_i\}\cup\{a_j\,:\,\hpathft_j\subseteq \hpathft_i\})$\label{lst:line_sub}
    \EndIf
\EndFor

\end{algorithmic}
\end{algorithm}

{\em An example of execution of PoPGreedy: }
We show an example of execution on the network represented  in Figure~\ref{fig:example}. We assume priori probability $p = 0.1$ and let node $v_9$ be the only failed node in the network. Nodes $v_1, v_2, v_3$ and $v_4$ are monitors, and consider undirected paths. We consider the 6 monitoring paths shown in the figure. 
\begin{itemize}[leftmargin=*]
    \item[-] {\em Step 1:} The paths that maximize utility at the first iteration are $m_3$, $m_5$ and $m_6$. For the tie breaking rule we choose path $m_3$. The path works. Hence $ \mathcal{R} = \Paths \setminus \{m_3\}$, $O^{\TPaths} = \{Z_3\}$ and the set of working nodes $W = \{1,4,7,8\}$. 
    It results that $a_4 = \arg \max_{a\in\mathcal{A}}u(a|\Obst)$, $u(a_4|\Obst) = 2.187$. 
    
    \item[-] {\em Step 2: } Test $m_4$, $\RPaths=\RPaths\setminus \{m_4\}$. The path fails, therefore $\FPaths = \{m_4\}$ and $\Obst = \Obst \cup \bar{Z}_4$.  
    It holds that $a_1 = \arg \max_{a\in\mathcal{A}}$, with $u(a_1|\Obst)= 1.1358$.
    \item[-] {\em Step 3: }  Test $m_1$, $\RPaths=\RPaths\setminus \{m_1\}$. The path works, hence: $W = W\ \cup \  \hpathft_1$ and $\Obst = \Obst \cup Z_1$. At this point, it results that $u(a_2|\Obst) =  1.278$, while $u(a_6|\Obst) =0$. By knowing that $m_4$ failed while node $v_2$ works, we can claim with certainty that $m_6$ will also fail, as it steps onto all the remaining nodes of $m_4$. 
    \item[-] {\em Step 4: } Test $m_2$, $\RPaths=\RPaths\setminus \{m_2\}$. The path works, hence: $W = \{1,2,3,4,5,6,7,8\}$ and $\Obst = \Obst  \cup Z_2 $. The utility of the two non visited paths, $m_5$ and $m_6$ is zero, therefore the execution is over. The algorithm  returns the failure probabilities: $P(\bar{S}_9| \Obst) = 1$, $P(\bar{S}_{10}|\Obst)=0.1$ and $P(\bar{S}_v|\Obst)=0$ for all the other nodes.
\end{itemize}
It must be noted that, although the algorithm leaves some uncertainty on the state assessment of node $v_{10}$ this is due to the impossibility of obtaining certain status identification for $v_{10}$ with the available paths. None of the existing paths can disambiguate the status of such a node as the only paths traversing it fail because of the failure of node $v_9$.

{\color{black}
\begin{figure}[h!]
\centering
\begin{tikzpicture}
\centering
\tikzset{quadrato/.style={draw,rectangle,fill=white,minimum size=7pt,     inner sep=0pt}}
\tikzset{pallino/.style={draw,circle,fill=white,minimum size=7pt,     inner sep=0pt}}
\tikzset{mylab/.style={black, minimum size=0.5cm}}
\tikzset{working/.style={draw,circle,fill=green,minimum size=4pt,inner sep=0pt}}
\tikzset{failed/.style={draw,circle,fill=red,minimum size=7pt,inner sep=0pt}}

\draw (0.2,0) node (1) [quadrato] {}       
  --++(0:1.5cm) node (5) [pallino]{}
  --++ (0:1.5cm) node (2) [quadrato]{};
            
\draw (1.4,-.8) node (6) [pallino]{}
    --++(-.8,-.7) node (7) [pallino]{}
    --++(-.6,-.6) node (8) [pallino]{}
    --++(1,-.6) node(4) [quadrato]{}
    --++(1.2,.2) node(10) [pallino]{}
    --++(1,1) node(9) [failed]{}
    --++(-1,.3) node(3) [quadrato]{};
\draw (6) --++(3);
\draw (6) --++(2);
\draw (7)--++(10);
\draw (1) --++(6);
\draw (2) --++(9);
\draw (1) --++(7);

    \node [mylab,below of = 6, node distance=0.15in] (blank) {\footnotesize{$\ v_6 $} };
        \node [mylab,left of = 1, node distance=0.17in] (blank) {\footnotesize{$\ v_1 $} };
       \node [mylab,right of = 2, node distance=0.15in] (blank) {\footnotesize{$\ v_2  $} };
       \node [mylab,below of = 3, node distance=0.15in] (blank) {\footnotesize{$\ v_3$} };
         \node [mylab,below of = 4, node distance=0.15in] (blank) {\footnotesize{$\ v_4 $} };
         \node [mylab,below of = 5, node distance=0.12in] (blank) {\footnotesize{$\ v_5 $} };
        \node [mylab,below of = 7, node distance=0.15in] (blank) {\footnotesize{$\ v_7 $} };
        \node [mylab,below of = 8, node distance=0.15in] (blank) {\footnotesize{$\ v_8 $} };
        \node [mylab,below of = 9, node distance=0.15in] (blank) {\footnotesize{$\ v_9 $} };
        \node [mylab,below of = 10, node distance=0.15in] (blank) {\footnotesize{$\ v_{10} $} };
    \draw [->,blue] (1) to [out=30,in=150] (5);
    \draw [->,blue] (5) to [out=30,in=150] (2);
    \draw [->,red] (1) to[bend left](6);
    \draw [->,red] (6) to [bend left] (3);
     \draw [->,green] (1) to [bend right] (7);
     \draw [->,green] (7) to [bend right] (8);
    \draw [->,green] (8) to [bend right] (4);
    \draw [->,pink] (2) to [bend right] (9);
     \draw [->,pink] (9) to [bend right] (3);
     \draw [->,yellow] (2) to [bend left] (9);
     \draw [->,yellow] (9) to [bend left] (10);
     \draw [->,yellow] (10) to [bend left] (4);
     Periwinkle
     \draw [->,cyan] (3) to [bend right] (9);
     \draw [->,cyan] (9) to [bend right] (10);
     \draw [->,cyan] (10) to [bend right] (4);

\draw[blue] (-.7cm, 0cm) -- (-1.2cm, 0cm) node[draw=none,fill=none] (pippo) [label=left:{{\color{black}\footnotesize{$m_1$}}}]{};
\draw[red] (-.7cm, -.4cm) -- (-1.2cm, -.4cm) node[draw=none,fill=none] (pippo) [label=left:{{\color{black}\footnotesize{$m_2$}}}]{};
\draw[green] (-.7cm, -.8cm) -- (-1.2cm, -.8cm) node[draw=none,fill=none] (pippo) [label=left:{{\color{black}\footnotesize{$m_3$}}}]{};
\draw[pink] (-.7cm, -1.2cm) -- (-1.2cm, -1.2cm) node[draw=none,fill=none] (pippo) [label=left:{{\color{black}\footnotesize{$m_4$}}}]{};
\draw[yellow] (-.7cm, -1.6cm) -- (-1.2cm, -1.6cm) node[draw=none,fill=none] (pippo) [label=left:{{\color{black}\footnotesize{$m_5$}}}]{};
\draw[cyan] (-.7cm, -2cm) -- (-1.2cm, -2cm) node[draw=none,fill=none] (pippo) [label=left:{{\color{black}\footnotesize{$m_6$}}}]{};
	\end{tikzpicture}
	\caption{Example topology with a failure in $v_9$}
	\label{fig:example}
	\end{figure}
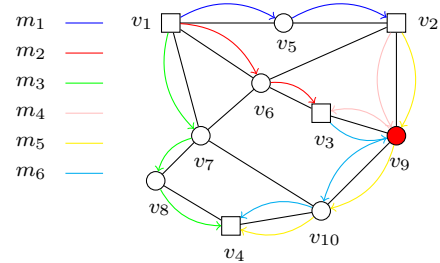}

\input{OptimalityApproximation}

\subsection{Computational Complexity}
\begin{theorem}
The computational complexity of PoPGreedy (Algorithm~\ref{alg:minpaths}) is  $O(K\cdot|\Paths|\cdot2^{|\FPaths|} )$, where $K$ is the maximum number of path probes.
\end{theorem}
\begin{proof}
 At each of the $O(K)$ steps of the algorithm, expected utilities are updated (line~\ref{lst:line:l15}). This operations requires computing $P(\bar{Z}|\Obst)$ for all paths that are not tested yet:
\begin{equation}
\label{eq:condProb}
    P(\bar{Z}|O_{\mathcal{T}}) = {P(\bar{Z}\land O_{\mathcal{T}})}/{P(O_{\mathcal{T}})}.
\end{equation}
Observe that, when computing the joint probability of the outcomes of previously tested paths, the contribution given by working paths simply results in pruning working nodes from non working paths. 
Therefore, the joint probabilities in equation~(\ref{eq:condProb}) may be computed as follows:
\begin{equation}
    \label{eq:jointPaths}
    P(\Obst) = P\left(\bigwedge\nolimits_{m\in \FPaths}\bar{Z}^o_{\pathft} \right ).
\end{equation}
The expression in the previous equation requires a number of addends that is exponential in the number of failed paths ($2^{|\FPaths|}$). Computing node failure probabilities (line~\ref{lst:line:l118}) requires the same number of operations.  The final cost is $O(K\cdot |\Paths|\cdot 2^{|\FPaths|})$.
\end{proof} 
\noindent Even considering sporadic failures, it is hard to predict how much the exponential factor may grow. Even within the same topology, the time required for computing the joint probability $P(O_{\mathcal{T}})$ is highly dependent on where failures occur: if highly connected nodes fail, the number of failed paths may be big, which makes the computation of the failure probabilities extremely time consuming.\\
The above reasoning motivates the use of polynomially computable metrics to approximate the nodes' conditioned failure probabilities.
In the next section, we define a polynomially computable centrality metric that captures the trend of how node failure probabilities are influenced when conditioned by iterative observations on test outcomes.

\section{Failure centrality}
   We hereby define the {\em failure centrality} of a node $v$ given the observation $O_\mathcal{T}$. 
    \begin{definition}
    \label{def:centrality}
        The {\em failure centrality} of a node $v$ given the observation $O_\mathcal{T}$ is $c(v|\Obst)=0$ if $v$ is traversed by some working paths in $\TPaths$, it is equal to the prior probability of failure if $v$ is not traversed by any path in 
 $\TPaths$, otherwise  
     $  c(v|O_\mathcal{T} ) =\max\{T_1;T_2\}$,
   where:
    \begin{equation}
    \label{eq:T1}
        T_1 = \left\lceil 
                {\sum\limits_{m\in \Paths_v^{(\TPaths)}\cap \FPaths} \left \lfloor \frac{1}{|\hpathft|}    \right \rfloor}/{|\Paths_v^{(\TPaths)}\cap \FPaths|}
        \right\rceil,
    \end{equation}
    
\begin{align}
        T_2&= \Pc + H(\left \lfloor \Pc\right \rfloor-1)\cdot \left(  1-\frac{\epsilon}{\Pc}-\Pc\right), \label{eq:T2}\\
        \Pc &= {|\Paths_v^{(\TPaths)}\cap \FPaths|}/{|\bigcup\limits_{m \in \Paths_v^{(\TPaths)}\cap \FPaths }\hpathft|},\label{eq:Pc}
    \end{align}
    where $\Paths_v^{(\TPaths)}$ is the set of monitoring paths $m^{(\TPaths)}$ crossing  node $v$. 
    $H(x)$ is the Heaviside function ($H(x)=0$, if $x<0$, and $H(x)=1$ when $ x\ge0$).  
    $\epsilon>0$ is a small constant.
 \end{definition}
Node centrality is used to approximate the value of $P(\bar{S}_v|\Obst)$ in 
the calculation of the posterior estimate of the state probability of any path $m_i \in \mathcal{R}$, which is
$P(Z_i | O_\mathcal{T})$,  that may be time consuming. 
The possible values of $c(v|O_\mathcal{T})$ span in the interval $[0,1]$ and, in analogy with probabilities, $c(v|O_\mathcal{T})=0$ means that the failure probability of node $v$ is 0, that is, $v \in W$, whereas if $c(v|O_\mathcal{T}) = 1$ implies that the node is broken. 

In the following we give some observations and proposition to characterize the values of the node failure centrality given the observation.

\begin{observation}
Firstly, observe that $T_1\in\{0,1\}$. Indeed, for any failed path it holds that $|\hpathft|\ge 1$, therefore the maximum value of the numerator in equation~(\ref{eq:T1}) is $|\Paths_v\cap \FPaths|$, proving that $T_1$ can not be greater than 1. When there is at least one path $m$ $s.t.\, |\hpathft|= 1$,  $T_1=1$, otherwise $T_1=0$.
\end{observation}
\begin{proposition}
\label{prop:2nd<1}
For all nodes $v$ and observations $\Obst$ it holds that $0 \leq T_2< 1$. 
\end{proposition}
\begin{proof}
While it is trivially true that $T_2\ge 0$, we prove that $T_2$ cannot be greater than or equal to $1$. 
We observe that if $\Pc < 1$, then $T_2=\Pc$. When $\Pc=1$, $T_2$ becomes $1-\epsilon$, while if $\Pc > 1$, $T_2 = 1-\frac{\epsilon}{\Pc}$, that is a monotonically growing function with a horizontal asymptote in $1$. 
\end{proof}

\begin{proposition}
    Let $v\in V$ be a node and $\Obst$ the outcome of some path probes. If $c(v|\Obst)=1\implies v$ is broken.
\end{proposition}
\begin{proof}
In Proposition~\ref{prop:2nd<1} we prove that $T_2<1$, hence $c(v)=1\iff T_1 = 1$. When there is at least a failed path $m$ traversing $v$ such that $|\hpathft|=1$, the numerator $num$ of $T_1$ is $0<num\leq |\Paths_v\cap \FPaths|$ and therefore $T_1=c(v|\Obst) =1$. When this situation occurs, the  probability of failure of node $v$ is indeed 1, as this means that the failure of path $m$ is only due to the failure of node $v$.
\end{proof}
\begin{proposition}
\label{prop:firstTerm1}
   Let $v\in V$ be a $k$-identifiable node with respect to the set of paths $\TPaths$, where $k$ is the number of failures in the network,  and let $\Obst$ be the outcomes of path probes on $\TPaths$. If $v$ is broken $\implies c(v|\Obst)=1$. 
\end{proposition}
\begin{proof}
 Since $v$ is $k$-identifiable, this means that the set of paths crossing $v$ is different from the sets of paths crossing any other set of nodes of size at most $k$. In particular, it is different from the set of paths crossing the other $k-1$ broken nodes. Hence there must be at least one path $m$ that passes through $v$ and not through any other failed node, and therefore $\hpathft=\{v\}$. What is left to prove is that there is some set of observations $\Obst$ that allows to disambiguate $v$ by finding out that indeed $\hpathft=\{v\}$. If $|\hpathft|=1$, then this is trivially true. Also if $k=1$, by definition, node $v$ must be traversed by a set of paths different than the set of paths traversing any other node laying in $m$; therefore there exist some working path that passes through nodes in $\hat{m}\setminus\{v\}$. Otherwise, again from the definition of $k$-identifiability, the failure of node $v$ must produce different sets of failed paths than the ones resulting from simultaneous failures of $v$ and any other node in $\hat{m}\setminus\{v\}$. As a consequence, there must be some working path passing through the nodes in $\hat{m}\setminus\{v\}$ and not through $v$, making it possible to verify through end-to-end monitoring measurements that $\hpathft=\{v\}$, which results in $T_1=c(v|\Obst)=1$. 
\end{proof}

To conclude the discussion on the formulation of the centrality, we comment on the choice of term $T_2$ in equation~(\ref{eq:T2}). This formulation is motivated by the observation that node failure probabilities are directly proportional to the number of failed paths traversing a node, and  inversely proportional to the number of nodes $w\not\in W$ being traversed by such paths. This property is satisfied by both $\Pc$ and $1-\frac{\epsilon}{\Pc}$. Furthermore, by experimental observations, we noticed that $P(\bar{S}_v|\Obst)$ grows steeply with the number of terms $\bar{Z}_i$ (where $v\in \hat{m}_i^{(\TPaths)}$) in $\Obst$ when $P(\bar{S}_i|\Obst)\ll1$, while it slowly converges to 1 for $P(\bar{S}_v|\Obst)\lesssim 1$ for increasing numbers of negative tests on paths passing through $v$. Similarly, $T_2 = \Pc$ when $\Pc<1$, while $T_2=1-\frac{\epsilon}{\Pc}$ for $\Pc \ge 1$. \\
In order to tune the value of  $\epsilon$ we observe that if $q^* = \max \Pc$ $s.t. \,\Pc < 1$, then $q^*\le \frac{d-1}{d}$, where  $d=|\bigcup_{m\in \Paths_v}\hpathft|$. Therefore, for $\epsilon < 1-q^*$, the growing trend of $T_2$ would be still satisfied when $\Pc$ exceeds 1.
\subsection{Centrality-based Utility}
Because of the dependencies among path failures, computing the joint probability $P(\Obst)$ requires teh computation of $2^{|\FPaths|}$ addends. In order to reduce computational costs, we approximate the probability that a path works, conditioned on the observation,  as follows:
\begin{equation}
    \label{eq:pathcent}
    \Tilde{P}_c(Z_i|\Obst) = \prod\limits_{v \in \hat{m}_i^{(\TPaths)}}(1-c(v|\Obst)).
\end{equation}
\begin{definition}
\label{def:Cutility}
    The expected conditional utility based on failure node centrality is given by the formula:
    \begin{equation}
    \label{eq:utilityC}
        \mathcal{U}_c(a_i|\Obst) = \lambda(a_i|Z_i)\Tilde{P}_c(Z_i|\Obst)+\lambda(a_i|\bar{Z}_i)\Tilde{P}_c(\bar{Z}_i|\Obst)
    \end{equation}
    if $\nexists \; m' \in \FPaths: \hat{m}'^{(\TPaths)}\subseteq \hat{m}_i^{(\TPaths)}$. Otherwise $\mathcal{U}_c(a_i|\Obst) = 0$. Here, $\lambda(a_i|Z_i)$ is defined as in equation~(\ref{eq:conditional_utility_given_path_state}) and $\Tilde{P}_c(\bar{Z}|\Obst) = 1-\Tilde{P}_c(Z|\Obst)$. 
\end{definition}
The condition that equation~(\ref{eq:utilityC}) is valid  if $\nexists \; m' \in \FPaths: \hat{m}'^{(\TPaths)}\subseteq \hpathft_i$ serves to recognize situations as the one described in Figure~\ref{fig:path_removal}, where we observed that if a path fails, every of its super-path is going to be failing, too. Thanks to prior observation we can assess the state of such paths and therefore there is no need to probe them. 

\subsection{Probing Algorithm with Centrality: FaCeGreedy}
Algorithm~\ref{alg:minpaths} may be adapted to use this metric instead of the exact conditional probability by applying the following modifications:
\begin{itemize}[leftmargin=*]
    \item Input: change $p$ for $c$ as initial node centrality.  
    \item Lines~\ref{lst:line:l15} and \ref{lst:line:l16}: substitute $\mathcal{U}(a|\Obst)$ with $\mathcal{U}_c(a|\Obst)$.
    \item Line~\ref{lst:line:l18}: replace $P(\bar{S}_v|\Obst)$ with $c(v|\Obst)$.
\end{itemize}

We hereby call FaCeGreedy (Failure Centrality Greedy algorithm) the Algorithm~\ref{alg:minpaths} with the modifications described above. 

{\em An example of execution of FaCeGreedy: }
By running FaCeGreey on the example in Figure~\ref{fig:example}, with initial node centrality $c=0.1$ and $\epsilon = 0.05$, the path probe sequence is the same as the one resulting by PoPGreedy, and final node centralities are $c(v_9|\Obst) = 1$, $c(v_{10}|\Obst) = 0.1$ while $c(v|\Obst) = 0$ $\forall v \in W$.

\subsubsection{Computational Complexity} The computational complexity of Algorithm~\ref{alg:minpaths} changes when centrality and centrality-based utility (Definitions~\ref{def:centrality} and \ref{def:Cutility}) are implemented instead of probability and utility (equation~\ref{eq:exp_conditional_utility_given_path_state}). 
\begin{theorem}
The computational complexity of FaCeGreedy (Algorithm~\ref{alg:minpaths} with the changes described above) is $O(K\cdot (|V|\cdot|\FPaths|^2+|\hat{m}_{max}|))$, where $K$ is the maximum number of path probes, $V$ is the set of nodes of the network, $\FPaths$ is the set of failing tested paths and $|\hat{m}_{max}|$ is the maximum path length. 
\end{theorem}
\begin{proof}
The total number of tests is $O(K)$. computing the centrality of a node $v$ requires scrolling the failed paths and searching for possible sub-paths in order to compute $\Pc$ (equation~(\ref{eq:Pc})). This is comprehensive of computing $|\Paths_v\cap \FPaths|$ in equation~(\ref{eq:T1}) and requires $O(|\FPaths|^2)$ operations. This is done for all nodes $v$ at each iteration. Computing the centrality-based utility of a path requires a number of operations that is linear in the number of nodes paths pass through. The overall cost of the algorithm is $O(K\cdot (|V|\cdot|\FPaths|^2+|\hat{m}_{max}|))$, where $|\hat{m}_{max}|$ is the maximum path length. 
\end{proof}
\subsection{Dynamic Failures}
In this section we show how our algorithms for node state classification can be exploited to develop an online, state-change aware monitoring system, where we consider that the failure scenario may change throughout the monitoring activity. When we consider this scenario, past observations do not guarantee certain information, in contrast with the static model that we adopted in the previous sections. Furthermore, while in PoPGreedy and FaCeGreedy the path probe activity would naturally stop when the expected utility function of non-tested path results to be 0, the dynamic failure scenario that we are introducing can rather be classified as an infinite horizon problem. We adapt PoPGreedy and FaCeGreedy to take into account the newly introduced dependency on time by considering the following facts: \emph{i.}~we do not suppose to have any knowledge about prior node failure probabilities nor on the maximum number of failures; \emph{ii.}~we do not assume knowledge on the time required for a node to be fixed, nor on a node's life time. We call these dynamic-aware algorithms \emph{Dynamic PoPGreedy (DPoPGreedy)} and \emph{Dynamic FaCeGreedy (DFaCeGreedy)}. To model this scenario, we discretize time into the intervals between path probes, and we assign to each node $v$ a probability to transition from working to failed ($p_{W\rightarrow F}$) and a probability to transition from failed to working (($p_{F\rightarrow W}$) at each time step. We assume that it is more likely for a node to be fixed, rather than for a node to fail (i.e., $p_{W\rightarrow F}< p_{F\rightarrow W}$). We base our procedure on the observation that information gained in the past progressively expires by the passing of the time. Because of the computational complexity that would result in a Bayesian analysis where probabilities are explicitly time-dependent, we consider the following simplified and easily computable approach: we define a \emph{window} that is the set of the last probed paths. We assume that the width of the window $\ell_w$ is big enough to ensure at least network coverage. The window slides progressively: at each time step the least recently probed path in the window is removed from it, and a new path is probed and brought inside the window. We consider valid the information obtained by the last $\ell_w$ path probes, whereas we consider previous observations expired. DPoPGreedy and DFaCeGreedy work as their static counterparts inside the window, unless a contradiction is detected. A contradiction inside a window occurs when the joint probability of the last $\ell_w$ path probe outcomes is 0. This could happen either because a path traversing working nodes fails, or because a super-path of a failed path works. When this occurs, we locate the most recent path that causes a contradiction, and we remove it together with all the older paths from the window, as the information they provided is corrupted. 
\section{Experimental Results}
\label{sec:results}
In the following we provide a performance evaluation of both the variants PoPGreedy and FaCeGreedy of our approach, against state of the art solutions for classic boolean network tomography and sequential graph-based group testing. In the experiments we assume cycle-free routing between monitor nodes. 
Our evaluation considers the following metrics: Section~\ref{sec:metrics}. 
If not explicitly stated otherwise, initial failure probability and centrality are set to 0.1. 
\subsection{Metrics}
\label{sec:metrics}
We consider the output of any of the probing algorithms in terms of the probability associated to each node failure.
We compare the performance of the heuristics with respect to the results that would be obtained by using all the monitoring paths.

In the following, we call $W_{\Paths}$ and $B_{\Paths}$ the set of nodes correctly classified as working (failure probability 0) and broken (failure probability 1), respectively when all paths of $\Paths$ are probed. 
Similarly, we denote with  $W_h$ and $B_h$ the same sets according to the classification made by any of the heuristics  $h$, which selects progressive monitoring policy, probing only  a subset of the paths in $\Paths$.

We denote with $a_W \triangleq {|W_{h}|}/{|W_{\Paths}|}$ the  \emph{accuracy of detection of working nodes}, namely the fraction of nodes classified as  working by the heuristics,  over the number of nodes recognized as working when all available paths are probed.

Similarly we denote with 
$a_B \triangleq {|B_{h}|}/{|B_{\Paths}|}$ the  \emph{accuracy of detection of broken nodes}.

The next two  metrics measure the correctness of the ranking $\NList$ produced by the heuristics to sort the nodes in terms of failure probability:
$R_1 \triangleq   \frac{\big|\NList[1:k]\cap F\big|}{k}$, where $F$ is the set of failed nodes, $k = |F|$, and with $\NList[1:k]$ we denote the nodes in the first $k$ positions in the rank $\NList$, i.e. with highest failure probability;
$R_2 \triangleq \frac{k}{n-i +1}$, where $i$ is the index of a failed node appearing in $\NList$.
If $k$ is the number of truly failed nodes, $R_1$ counts how many of those appear in the highest $k$ positions in the ranking, while $R_2$ says how many nodes' state we should verify before finding all the failed nodes. 

It holds that $R_1\leq 1$ and $R_2\le 1$, and $R_1=1$ when the top $k$ positions are indeed occupied by the truly failed nodes $F$; $R_2 = 1$ when the nodes whose failing probability is  1 appear in the top $k$ positions of the rank $\NList$. Therefore, $R_1=1\iff R_2 = 1$. $R_1$ metric is similar to the recall metric of ML \cite{powers2011evaluation}, but it evaluates probabilistic outcomes instead of binary classifications.
In addition to the metrics $a_W$, $a_B$, $R_1$ and $R_2$, we also consider the number of probes required to reach convergence and the execution time, when comparing our approaches to the previous solutions.\\

For evaluating DPoPGreedy and DFaCeGreedy, we use metrics that capture the ability to detect node state changes, and metrics that measure the reliability of the classification results step by step. For the first category, we compute the percentage of detected node state changes in both ways ($W\rightarrow F$ and $F\rightarrow W$), and the time for detection in terms of time stamps. For assessing the classification reliability in each sliding window, we use the classical definitions of precision and recall:
\begin{equation*}
    P = \frac{tp}{tp+fp},\qquad R = \frac{tp}{tp+fn}
\end{equation*}
where $tp$ (true positive) is the number of correctly classified nodes; $fp$ (false positive) is the number of nodes erroneously classified either as working or as failed; $fn$ (false negative) is the sum of the number of real working nodes that are not classified as working, and of the number of real failed nodes that are not classified as failed. Notice that the recall is similar to $R_1$, except that $R_1$ evaluates probabilistic outcomes instead of binary classifications.

\vspace{-0.2cm}
\subsection{Benchmark solutions}
To validate our approach we compare it with previous solutions based on classical Boolean Network Tomography as well as an approach based on progressive graph-constrained group testing. For the first set of benchmarks we consider the \emph{greedy for coverage}, \emph{greedy for identifiability} and \emph{greedy for distinguishability} (GC, GI, GD) heuristics defined in~\cite{usICDCS16}. At each iteration, the next path to probe among the available input paths is chosen as the one that maximizes network coverage/identifiability/distinguishability, respectively. When the greedy procedures meet some stopping criteria, node failure probabilities $P(\bar{S}_v|\Obst)$ are computed and the outcome is evaluated in terms of the metrics described in Section~\ref{sec:metrics}. 

Together with this, we compare our method to the adaptive, graph-constrained group testing algorithm introduced in~\cite{karbasi2012sequential}, to which we  refer to as AdaptiveFinder, (AF). The goal of AdaptiveFinder is to detect the set of defective items (nodes) in a graph with the least number of probes. The main differences between our setting and the one adopted by AdaptiveFinder are that, although graph-constrained, AdaptiveFinder is not routing-constrained, meaning that monitoring probes are not limited to move along end-to-end paths that are determined by the routing scheme implemented in the network, and given as input (i.e., they can be trees or contain cycles); in addition, direct node inspection is allowed through degenerate paths composed of only one node, meaning that all nodes are monitoring nodes. These two facts result in a major flexibility of AdaptiveFinder, i.e. an advantageous  degree of freedom that is not available to our approach. Nevertheless, we note that this constitute an unrealistic capability in a general network tomography scenario and it is more expensive to implement on an actual network because it  requires all nodes of the network to be provided with a monitoring system software and also assumes fully controllable routing.
The set of paths available to our algorithm is limited to  a small subset of the possible paths that AdaptiveFinder is allowed to walk across. For these reasons, accuracy metrics for AdaptiveFinder are taken with respect to the ground truth. 

Notice that, because of the possibility of direct node inspection, there is no uncertainty in the sets of nodes classified as failed by AdaptiveFinder, hence this algorithm is not susceptible to lack of identifiability, that instead is an ascertained issue in network tomography. As a consequence, when the algorithm is run until convergence and a maximum number of recursive steps is not fixed, it manages to assess with certainty the state of all nodes, even when they are not $k-$identifiable by means of end-to-end given monitoring paths, being $k$ the number of broken nodes.\\  

Finally, we also compare our results with the Adaptive Path Construction (APC) algorithm, \cite{mukamoto2015adaptive}. Similarly to PoPGreedy and FaCeGreedy, APC investigates on the state of the network by means of end-to-end monitoring paths that are given and determined by uncontrollable routing. 
APC may be divided in two phases. In the first phase and differently from us, a greedy for coverage is applied. The outcomes of the path probes used in this phase are then analysed, and if they are not sufficient for identifying the status of all nodes, the adaptive group testing phase is executed: the decision on the action to take at a certain step (i.e., the next path to probe) follows the binary search idea: the path whose number of nodes is closer to the half of the number of still unclassified nodes in the network is tested. The original output of APC is the set of the failed nodes and of the candidate nodes (nodes that are not classified as working and that might be failed). In PoPGreedy and FaceCeGreedy, these sets correspond to the set of nodes whose failing probability/centrality is 1, and to the set of nodes whose failing probability/centrality lies in $(0,1)$, respectively.  In order to compare APC in terms of the metrics introduced in Section \ref{sec:metrics}, we compute the failure probability of the nodes in the candidate set, we set to 1 the failure probability of the nodes in the identified set, and as 0 the failure probability of all remaining nodes.\\
FINALLY: we highlight two errors in the algorithmic procedure of APC as it is decribed in \cite{mukamoto2015adaptive}: in Algorithm 1 (CBP), line 10, it must also hold that $w_m\cap \mathcal{E}_I = \emptyset$. The third condition that a path $w$ must not satisfy in order to belong to the set $\mathcal{W}_{eff}$ is actually $w\cap \mathcal{E}_C \supseteq w_i \cap (\mathcal{E}_C \cup \mathcal{E}_I), \exists w_i \in \{w_j \in \mathcal{W}:y_j = 1\}$. 

\subsection{Tests}
We perform experiments by considering different  settings. In particular, we show experiments conducted on two different networks, an internet network in Europe, BICS \cite{knight2011internet}, and a fiber network topology of Minnesota \cite{auroranet}. We use the first network for understanding thoroughly the behaviour of our algorithms and benchmark methods, and we see that such considerations hold on the bigger network. 

\noindent In Table~\ref{table:tabPath} features of the two topologies  are detailed (left) as well as networks' features taking into account monitor-to-monitor path choices.
 Table~\ref{table:tabPath} details features of the two topologies  as well as networks' features taking into account monitor-to-monitor path choices.
We use the smaller network, BICS,  for running a thorough study of   the behaviour of our algorithms and benchmarks, before extending our conclusions to the case of the larger Minnesota network.
In the experiments, the set of candidate monitors is  chosen randomly, with several paths between the same monitor pairs, to ensure broad network coverage. 

\footnotesize

\begin{table}[ht]
\centering
\renewcommand{\arraystretch}{1.185}
\begin{tabular}{|c|c|c|}
\hline

      & BICS & MN  \\ \hline \hline
     $|V|$ & 33 & 681\\
     $|E|$ & 48 & 921\\
     $\delta_{min}$ & 1 & 1\\
     $\delta_{max}$ & 8 & 13\\
     $\delta_{avg}$ & 3 & 2.7\\
     diameter & 9 & 29\\
     $n_{\delta = 1}$ & 5 &134\\
     \hline
\end{tabular}
\renewcommand{\arraystretch}{1}
\begin{tabular}{|c|c|c|}
\hline
      & BICS &  MN  \\ \hline \hline
     $|V^{C}|$ & 33 & 450\\
     $|E^{C}|$ & 43 & 610\\
     $\delta^{\Paths}_{min}$ & 1 & 1\\
     $\delta^{\Paths}_{max}$ & 29 &  631\\
     $\delta^{\Paths}_{avg}$ & 9.9  & 60.7\\
     longest path & 9 & 27\\
     $|V_M|$ & 10 & 62\\
     $|\Paths|$ & 55 & 1996\\
     \hline
\end{tabular}

\caption{On the left: experimental settings. $\delta$ = node degree, $n_{\delta = 1}$ = number of dangling nodes (degree 1). On the right: path characteristics. $V^{C}$ is the set of covered nodes; $E^{C}$  is the set of covered links;  $\delta^{\Paths}_i$ = number of paths in $\Paths$ traversing node $v_i$. 
}
\label{table:tabPath}
\end{table}
\normalsize


 \begin{figure}[t!]
    \hspace{-3mm}\begin{subfigure}{.49\columnwidth}
        \subcaptionbox{\label{fig:AW1it}}{\input{imgs/BICS1_it/AW1it}}
    \end{subfigure}
    \begin{subfigure}{.49\columnwidth}
        \subcaptionbox{\label{fig:AB1it}}{\input{imgs/BICS1_it/AB1it}}
    \end{subfigure}
    \hspace{-3mm}\begin{subfigure}{.49\columnwidth}
        \subcaptionbox{\label{fig:R11it}}{\input{imgs/BICS1_it/R11it}}
    \end{subfigure}
    \begin{subfigure}{.49\columnwidth}
        \subcaptionbox{\label{fig:R21it}}{\input{imgs/BICS1_it/R21it}}
    \end{subfigure}
    
    \caption{Metrics evolution,
    iteration-wise (BICS network with 4 failed nodes). Bounded. 
    }
    \label{fig:random1it}
\end{figure}

 \begin{figure}[h!]
    \hspace{-3mm}\begin{subfigure}{.49\columnwidth}
        \subcaptionbox{\label{fig:AWbics1}}{\input{imgs/BICS1/AW1}}
    \end{subfigure}
    \begin{subfigure}{.49\columnwidth}
        \subcaptionbox{\label{fig:ABbics1}}{\input{imgs/BICS1/AB1}}
    \end{subfigure}
    
     \hspace{-3mm}\begin{subfigure}{.49\columnwidth}
        \subcaptionbox{\label{fig:R1bics1}}{\input{imgs/BICS1/R11}}
    \end{subfigure}
    \begin{subfigure}{.49\columnwidth}
         \subcaptionbox{\label{fig:R2bics1}}{\input{imgs/BICS1/R21}}
    \end{subfigure}

    \hspace{-3mm}\begin{subfigure}{.49\columnwidth}
       \subcaptionbox{ \label{fig:ET1}}{\input{imgs/BICS1/BICS_ETs}} 
    \end{subfigure}
    \begin{subfigure}{.49\columnwidth}
        \subcaptionbox{ \label{fig:NT1it}}{\input{imgs/BICS1/BICS_NTs}}
    \end{subfigure}
    \caption{Tests on BICS network. Bounded
    }
    \label{fig:random1}
\end{figure}

\subsubsection{Experiments on BICS network}
Figures \ref{fig:random1it} to \ref{fig:random2} are related to the BICS network. All curves are averaged on 20 experiments and show the value of the metrics defined in Section~\ref{sec:metrics} on PoPGreedy, FaCeGreedy and all benchmarks. Shades/bars depict standard deviation. In the experimental configurations shown in Figures~\ref{fig:random1it} and \ref{fig:random1} all the approaches stop either when they reach convergence or when they reach a maximum number of path probes.   Such bound is the number of path probes needed by PoPGreedy to converge (i.e., expected utility is 0) for each experiment.  In particular, in Figure~\ref{fig:random1it}, we show the evolution of the metrics iterative-wise when four failures occur in the network. In Figure~\ref{fig:random1} instead,  we show how the aforementioned metrics, as well as the elapsed times and the average number of tested paths, change for a growing number of failed nodes (from 1 to 5 failures). Notice that FaCeGreedy and GC always reach convergence before PoPGreedy (Figure~\ref{fig:NT1it}), but, while GC has poor, non-improvable performance in terms of node classifications, FaCeGreedy, together with PoPGreedy, always reach the same performances achieved by probing all paths (see Figures~\ref{fig:AWbics1} to \ref{fig:R2bics1}), that is the upper-bound to the ability of node states assessment by means of end-to-end monitoring paths.  Greedy identifiability and greedy distinguishability instead stop before convergence for all tests. AdaptiveFinder manages to converge with a very small number of paths only when a single failure occurs in the network. In contrast, APC converges with a a similar number of paths as FaCeGreedy. This is because the number of failures considered in this experimental scenario is small, and the initial coverage phase implemented by APC helps with the detection of many working nodes. Observing  Figure~\ref{fig:random1it} we can notice that 
since the number of tests changes depending on where the 4 failed nodes are located in the network, curves may be subject to oscillations at the end, as fewer tests reach the highest numbers of tested paths. Within one single test, $a_W$ and $a_B$ have a monotone growing trend, while $R_1$ and $R_2$ may oscillate: as a matter of facts, during intermediate probes working nodes may gain a high failure probability (hence moving to the top positions of the sorted node failure probability chart) and then their failure probability goes abruptly to 0 when a working path traverse them. An observable phenomenon is that $a_W$ curves are concave and they grow steeply with the very first experiments, and become less steep when they approach the maximum value (i.e., $a_W=1$). This is because of the sporadic failures scenario that we are considering: failed nodes are a small percentage of the set of all nodes, and therefore working paths are more likely to exist with respect to failing ones. Consequently, correct working node classification is easier and faster to achieve within the first tests. On the contrary, $a_B$ curves follow a convex function trend and in the first tests they may be 0. This is because it takes a number of tests before a node can be classified as failed (i.e., failure probability equal to 1).

\begin{figure}[t!]
    \hspace{-3mm}\begin{subfigure}{.49\columnwidth}
        \subcaptionbox{\label{fig:my_label}}{\input{imgs/BICS2_it/AW2it}}
    \end{subfigure}
    \begin{subfigure}{.49\columnwidth}
        \subcaptionbox{\label{fig:my_label}}{\input{imgs/BICS2_it/AB2it}}
    \end{subfigure}
    \hspace{-3mm}\begin{subfigure}{.49\columnwidth}
        \subcaptionbox{\label{fig:my_label}}{\input{imgs/BICS2_it/R12it}}
    \end{subfigure}
    \begin{subfigure}{.49\columnwidth}
        \subcaptionbox{\label{fig:my_label}}{\input{imgs/BICS2_it/R22it}}
    \end{subfigure}
    
    \caption{Metrics evolution iteration-wise (BICS network with 4 failed nodes). Unbounded. }
    \label{fig:random2it}
\end{figure}
 \begin{figure}[h!]
    \hspace{-3mm}\begin{subfigure}{.49\columnwidth}
        \subcaptionbox{\label{fig:AW2}}{\input{imgs/BICS2/AW2}}
    \end{subfigure}
    \begin{subfigure}{.49\columnwidth}
        \subcaptionbox{\label{fig:AB2}}{\input{imgs/BICS2/AB2}}
    \end{subfigure}
    \hspace{-3mm}\begin{subfigure}{.49\columnwidth}
        \subcaptionbox{\label{fig:R12}}{\input{imgs/BICS2/R12}}
    \end{subfigure}
    \begin{subfigure}{.49\columnwidth}
        \subcaptionbox{\label{fig:R22}}{\input{imgs/BICS2/R22}}
    \end{subfigure}
    \hspace{-3mm}\begin{subfigure}[b]{.49\columnwidth}
        \subcaptionbox{\label{fig:ET2}}{\input{imgs/BICS2/ET2}}
    \end{subfigure}
    \begin{subfigure}[b]{.49\columnwidth}
        \subcaptionbox{\label{fig:NT2}}{\input{imgs/BICS2/NT2}}
    \end{subfigure}
    
    \caption{Tests on BICS network. Unbounded}
    \label{fig:random2}
\end{figure}

Similar considerations on the evolutionary curves hold for the experiments shown in Figure~\ref{fig:random2it}. Figures~\ref{fig:random2it} and \ref{fig:random2} are again related to the BICS network. In this experimental configuration, all methods stop either because the reach convergence, or because they probe all available monitoring paths. The latter condition does not hold for AdaptiveFinder, since it is not limited to move along given paths between monitors. Except for greedy coverage, consistently with what we observed for the previous set of experiments, AdaptiveFinder requires a greater number of tests than those used by PopGreedy and FaCeGreedy to converge (see Figure~\ref{fig:NT2}), while greedy identifiability and greedy distinguishability always probe all available paths. When we do not give constraints on the maximum number of paths to probe, AdaptiveFinder converges to the ground truth: it correctly classifies all nodes. We stress that this is due to its possibility to monitor single nodes directly and to its freedom to walk on the network without the restriction of moving along given paths. Once again, PoPGreedy and FaCeGreedy achieve the same performance as probing all paths would do, but testing with little portions of available monitoring paths. This holds also for APC in this failure scenario.
As expected, the average elapsed time required by PoPGreedy considerably increases with the number of failed nodes, even on a small network (see Figures~\ref{fig:ET1} and \ref{fig:ET2}). High variance is due to its exponentially dependence on the number of failed paths, amplifying the discrepancy of when central or non central nodes fail. For this reason, in the next set of experiments, we are not going to consider such method. 
\subsubsection{Minnesota}
Figures~\ref{fig:minnesotaNet} and \ref{fig:minnesotaNetUnbounded} show our experiments on the  Minnesota network. In Figure~\ref{fig:minnesotaNet},  tests are run until convergence or until a maximum number of tests $K$ has been reached, whichever occurs earlier. In this case, the bound $K$ is given by the number of path probes needed by FaCeGreedy to converge. In contrast, experiments in Figure~\ref{fig:minnesotaNetUnbounded} are run until convergence or until all available paths are probed. Again, we observe that GI and GD need to test all available paths and are still unable to converge because of their inability to take account of the progressively available information which can be obtained by probing the paths in a sequence. In fact, in Figure~\ref{fig:minnesotaNet}, GI and GD use the same number of paths as FaCeGreedy  but with much inferior classification performance, whereas for the unbounded tests in Figure~\ref{fig:minnesotaNet}, they reach the same performances of FaCeGreedy by probing all available paths. On the other hand, FaCeGreedy is able to obtain full network information by converging with less than 9\% of all the available paths. As in the previous experiments, CG  is faster in covering the network, but performs poorly in terms of failure detection. Once again, in this configuration, in the unbounded case of Figure~\ref{fig:minnesotaNet}, AF is able to correctly detect all the failures within the maximum number of tests $K$ only when the failure set is very small. In contrast, in the unbounded scenario, AdaptiveFinder reaches convergence with a higher number of tests than the ones required by FaCeGreedy (Figure~\ref{fig:NTMin}). Despite the good performance of APC in the previous network, when APC is applied to a bigger network and when many failures occur, it reaches convergence with many more paths than the ones used by FaCeGreedy (Figure~\ref{fig:NTMin}), and performs poorly in the bounded tests (Figure~\ref{fig:minnesotaNet}). This can be explained with the following two factors: ensuring network coverage may be convenient for small networks with a little number of failed nodes, but it is not as effective in large networks with many failed nodes. Similarly, using a binary search approach is not as convenient when many multiple failures occur.
 \begin{figure}[t!]
    \begin{subfigure}{.49\columnwidth}
      \hspace{-3mm}\subcaptionbox{\label{fig:AW1MN}}{\input{imgs/MINNESOTA1/AW1MN}}
    \end{subfigure}
    \begin{subfigure}{.49\columnwidth}
     \subcaptionbox{\label{fig:AB1MN}}{\input{imgs/MINNESOTA1/AB1MN}} 
    \end{subfigure}
    \hspace{-3mm}\begin{subfigure}{.49\columnwidth}
        \subcaptionbox{\label{fig:R11MN}}{\input{imgs/MINNESOTA1/R11MN}}
    \end{subfigure}
    \begin{subfigure}{.49\columnwidth}
        \subcaptionbox{\label{fig:R21MN}}{\input{imgs/MINNESOTA1/R21MN}}
    \end{subfigure}
    \caption{Tests on the Minnesota topology. Bounded. }
    \label{fig:minnesotaNet}
\end{figure}

\begin{figure}[t!]
    \hspace{-3mm}\begin{subfigure}{.49\columnwidth}
        \subcaptionbox{\label{fig:AW2MN}}{ \input{imgs/MINNESOTA2/AW2MN}}
    \end{subfigure}
    \begin{subfigure}{.49\columnwidth}
        \subcaptionbox{\label{fig:AB2MN}}{\input{imgs/MINNESOTA2/AB2MN}}
    \end{subfigure}
    
   \hspace{-3mm}  
    \begin{subfigure}{.49\columnwidth}        
        \subcaptionbox{\label{fig:R12MN}}{ \input{imgs/MINNESOTA2/R12MN}} 
    \end{subfigure}
    \begin{subfigure}{.49\columnwidth}        
        \subcaptionbox{\label{fig:R22MN}}{\input{imgs/MINNESOTA2/R22MN}}
    \end{subfigure}
    
    \hspace{-3mm}\begin{subfigure}{.49\columnwidth}        
        \subcaptionbox{\label{fig:NTMin}}{\input{imgs/MINNESOTA2/NT2MN}}
    \end{subfigure}
    \begin{subfigure}{.49\columnwidth}
        \subcaptionbox{\label{fig:varP}}{ \input{imgs/MINNESOTA/ABcen}}
    \end{subfigure}
    \caption{Tests on the Minnesota topology. Unbounded}
    \label{fig:minnesotaNetUnbounded}
\end{figure}

Together with the aforementioned metrics, we also study how different choices of prior centrality values ($c$) may affect the performance of FaCeGreedy in terms of $a_B$. Figure~\ref{fig:varP} depicts how the accuracy of detection of broken nodes changes at each iteration of FaCeGreedy for $c=0.05,\,0.08,\,0.1$. For each experiment, 35 failed nodes (~8\% of the total number of covered nodes) are generated. Despite curves vary throughout intermediate iterations, and despite small differences in the final number of tested paths FaCeGreedy is able to reach maximal accuracy (i.e., $a_B=1$) also for under and over estimated choices of $c$ (that is, $c=0.05$ and $c= 0.1$) for all choices of $c$, proving its consistency,
and robustness against potentially wrong settings of the prior probability or centrality of a node.

\subsubsection{Experiments on Dynamic Failures}
\begin{figure}[t!]
    \hspace{-3mm}\begin{subfigure}{.49\columnwidth}
        \subcaptionbox{\label{fig:precision}}{\input{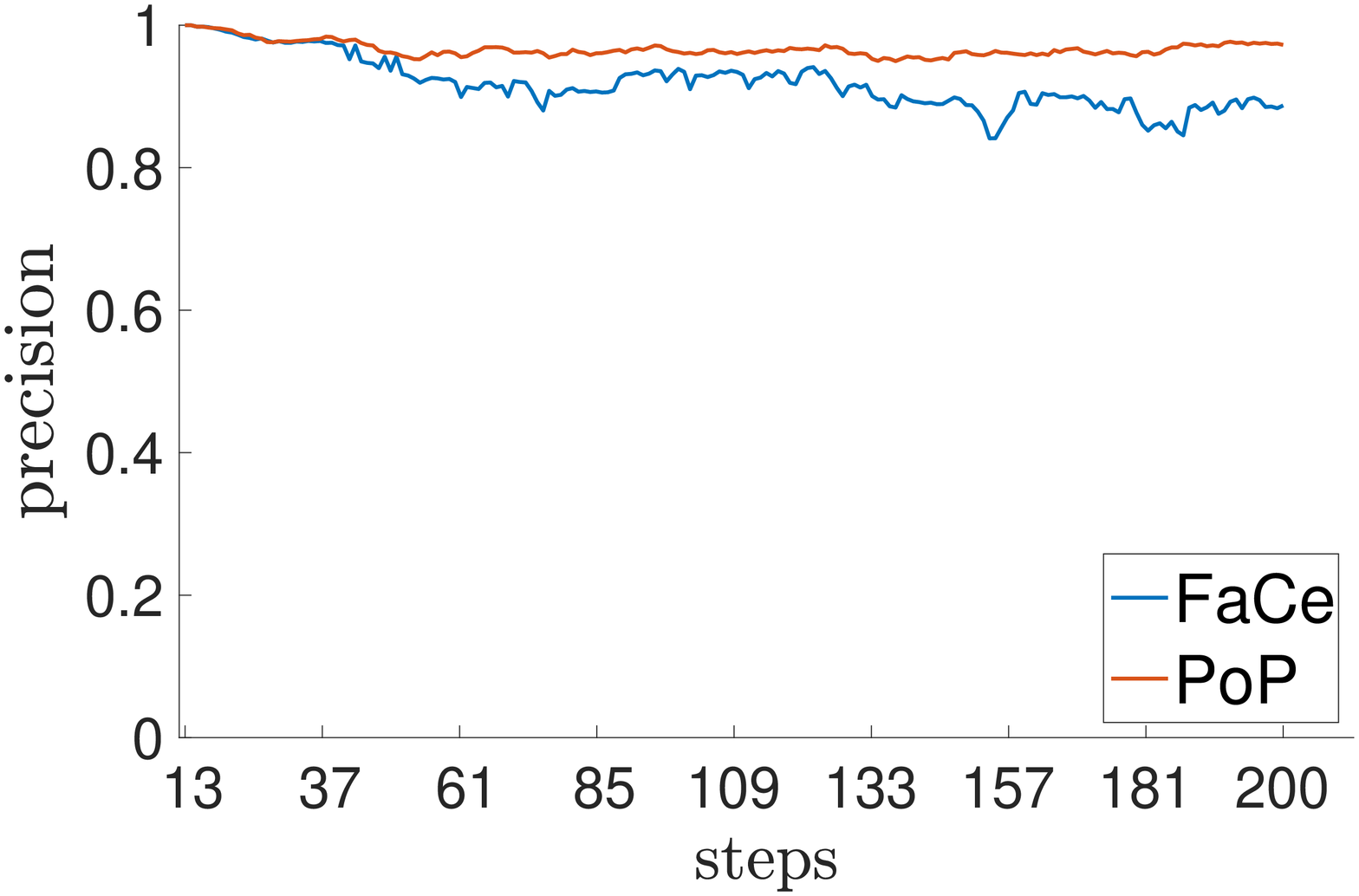}}
    \end{subfigure}
    \begin{subfigure}{.49\columnwidth}
        \subcaptionbox{\label{fig:recall}}{\input{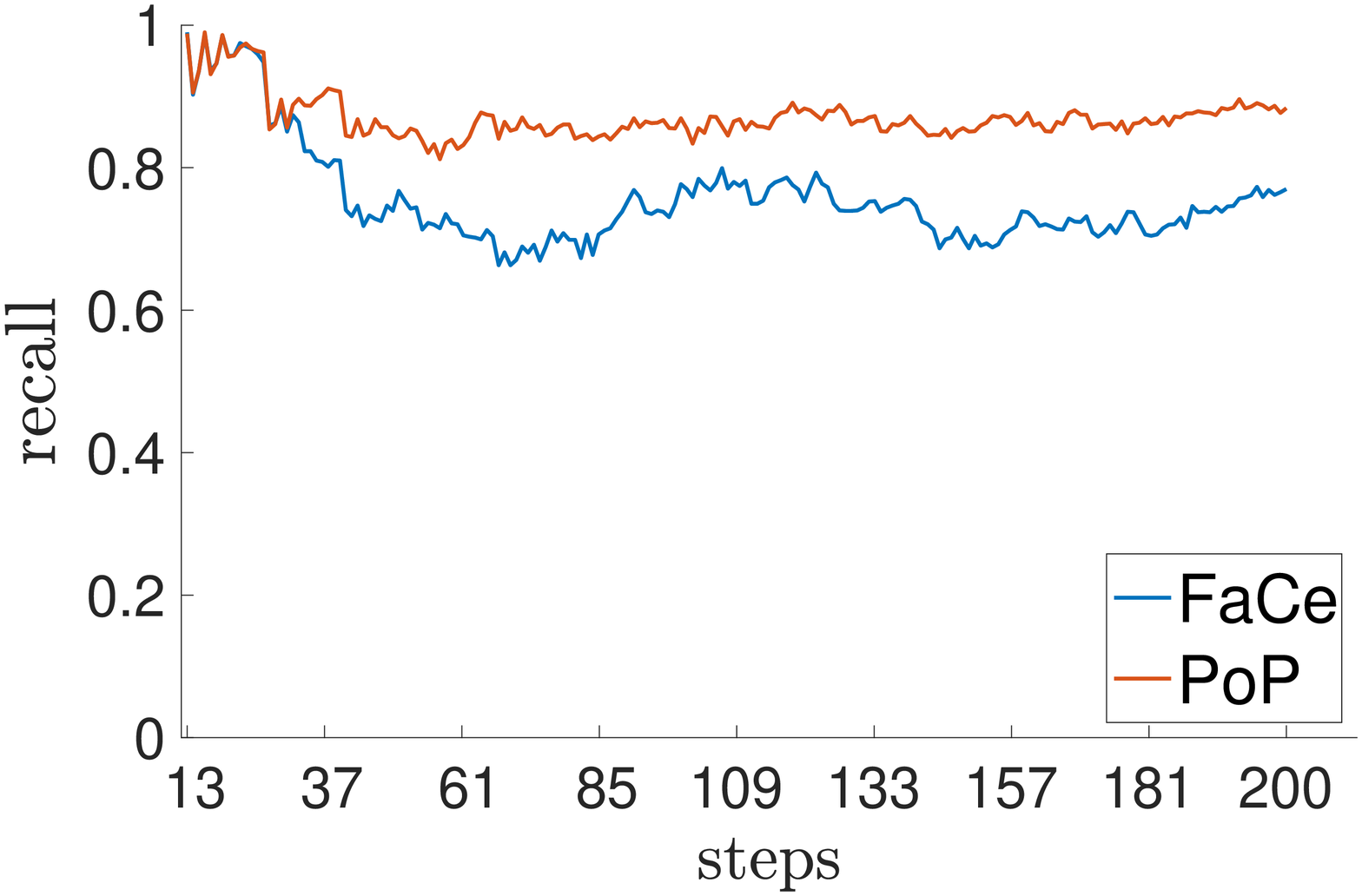}}
    \end{subfigure}
    \caption{Precision and Recall of DFaCeGreedy and DPoPGreedy.}
    \label{fig:dynamicFig}
   \end{figure}

Figure~\ref{fig:dynamicFig} shows the average precision and recall of DPoPGreedy and DFaCeGreedy on the BICS network. We run the algorithms for 200 steps, and we consider the window size $\ell_w$ to be 12. We compute the evaluation metrics at each time step, from 13 to 200. In Table ..., we show the percentage of detected node state changes and the average time for change detection. The experiments are leveraged on 200 experiments, and standard deviations are provided between parenthesis.

   \begin{table}[]
  \renewcommand{\arraystretch}{1}
       \centering
       \begin{tabular}{p{0.06\columnwidth}|c|c|c|c}
            & \%$_{F\rightarrow W}$ & \%$_{W\rightarrow F}$ & $t_{F\rightarrow W}$ & $t_{W\rightarrow F }$\\ \hline
            PoP  & 82.3 ($\pm$15.6)  & 89.7 ($\pm$9.1)  & 6.6 ($\pm$2.4) &  12.7 ($\pm$6.3) \\
         FaCe & 56.7 ($\pm$21.5)  & 89.7 ($\pm$13.6) & 14.0 ($\pm$8.7) & 14.5 ($\pm$7.6)
       \end{tabular}
       \caption{Average percentages of detected state changes and average time for change detection, and their standard deviation. }
       \label{tab:dynamictable}
   \end{table}
\section{Conclusions}

Boolean Network Tomography (BNT) provides the design of end-to-end monitoring paths to ensure network failure localization. However,  when the number of concurrent failures is unknown, BNT techniques hit the snag of the huge dimension and intractability of the solution space. With this paper we  propose a progressive approach to failure localization in the challenging scenario where failures may occur in an unknown and unbounded number. A set of monitoring paths is probed in a progressive manner, and decisions on which path to probe are made on the basis of a Bayesian approach which optimizes the expected value of the failure related information that can be obtained by incrementally monitoring new paths. To face the complexity of calculating posterior failure probabilities at each monitoring step, we propose a {\em failure centrality metric}, computable in polynomial time, which reflects the likelihood of a node to be the site of a failure. We use such a metric to guide decision making and provide a conclusive assessment of the state of network components. We extend these metrics to an online scenario where node states change dynamically throughout the experimental period. By means of numerical experiments conducted on synthetic as well as real network topologies, we demonstrate the practical applicability of our approach. The experiments show that our approach outperforms state of the art solutions based on classic Boolean Network Tomography as well as approaches based on progressive group testing.

\clearpage
\bibliographystyle{IEEEtran}
\bibliography{IEEEabrv,mybib}
\clearpage
\input{Appendix}

\end{document}

%% file: OptimalityApproximation.tex
\subsection{Optimality approximation}
\label{sec:optApp}
The definition of utility of an action given prior observations (Equation~\ref{eq:conditional_utility_given_path_state}) can be extended in order to characterize the utility of a set of actions (corresponding to the utility of probing a set of distinct paths, $\TPaths$) as follows:
\footnotesize
\begin{align*}
    \lambda&(\{a^{(1)},\ldots,a^{(|\TPaths|)}\},\Obst) =\\& \Bigg | \bigcup\limits_{m_i\in \TPaths\setminus\FPaths}\hat{m}_i\Bigg| + \Bigg | \bigcup\limits_{m_i\in \TPaths\setminus\FPaths}\Fonem\Bigg| + F_{\TPaths},
\end{align*}
\normalsize
where $F_{\TPaths}:=\{w\in V:P(\bar{S}_w|\Obst) = 1\}$ is the set of detected failed nodes by probing paths $\TPaths$. 
This definition allows us to formulate the problem of assessing the maximum number of node states with $K$ path probes as a formal maximization problem:
\begin{equation}
    \begin{split}
    \label{eq:cumUtility}
        \max_{\mathcal{A}^{(\TPaths)}\subseteq \mathcal{A}}\lambda(\TPaths|\Obst)\\
        s.t.\quad |\TPaths|\leq K
    \end{split}
\end{equation}
where $\mathcal{A}^{(\TPaths)} = \{a^{(1)},\ldots,a^{(|\TPaths|)}\}$. 
Constant approximations for deterministic optimization problems where the objective function has properties of monotonicity and submodularity were proved in \cite{nemhauser1978best}. More recently, the concept of \emph{adaptive} monotonicity and submodularity, originally introduced in \cite{golovin2011adaptive} and lately revised in \cite{esfandiari2019adaptivity}, extended such properties to the context of stochastic optimization problems, that is where our scenario belongs. In a stochastic maximization problem, the function to be maximized depends on a set of observations $\Obst$ on the state of the elements of the ground set, in our case, the state of paths in $\Paths$. In this context, greedy policies choose at each step the action that maximizes the expected value of the utility, that is known with certainty only after tests take place. Notice that PoPGreedy in Algorithm~\ref{alg:minpaths} follows the Adaptive Greedy Algorithm structure shown in \cite{golovin2011adaptive}. Before reporting the definitions of adaptive monotonocity and submodularity, we give definition of conditional expected marginal benefit (\cite{golovin2011adaptive}). In the following definitions, $X$ is a finite set of elements. 
\begin{definition}
\label{def:benefit}
    Let $Y\subset X$ and let $x\in X\setminus Y$. The \emph{conditional expected marginal benefit} of $x$ with respect of a function $f$ having observed $O_Y$ is:
    \begin{equation}
        \Delta(x|O_{Y}):=\mathbb{E}[f(Y\cup \{x\},O_Y)-f(Y,O_Y)],
    \end{equation}
    where by $O_Y$ we mean the restriction of the observations $O_X$ to the subset $Y$. 
\end{definition}
\begin{definition}
    A function $f : 2^X\times O_X\to\mathbb{R}$ is \emph{adaptive monotone} if $\forall x \in X$ and $\forall Y\subseteq X$ it holds that $\Delta(x|O_Y)\ge 0$.
\end{definition}
\begin{definition}
    A function $f:2^X\times O_X\rightarrow\mathbb{R}$ is \emph{adaptive submodular} if $\forall Z\subseteq Y\subset X$ and $\forall x \in X\setminus Y$ we have $\Delta(x|O_Z) \ge \Delta(x|O_Y)$. 
\end{definition}
In our scenario, the ground set $X$ is the set of all possible actions $\mathcal{A}$ on paths $\Paths$, and the state of a path is either normal or defective (or equivalently, 1 or 0). Paths' states can be assessed only through observations on path probes. Observe that the definition of  conditional  expected marginal benefit corresponds to the definition of expected utility given in Equation~\ref{eq:exp_conditional_utility_given_path_state}:

\begin{footnotesize}
\begin{align*}
    &\Delta(a|\Obst) =\\ 
    &\Bigg (\Big |\hat{m}\cup \bigcup\limits_{m_i\in \TPaths\setminus\FPaths}\hat{m}_i\Bigg| + \Big | \Fonem \cup\bigcup\limits_{m_i\in \TPaths\setminus\FPaths}\Fone\Bigg|+ |F_{\TPaths}|\Bigg ) P(Z|\Obst)\\
    &+\Bigg (\Big | \bigcup\limits_{m_i\in \TPaths\setminus\FPaths}\hat{m}_i\Big| + \left | \bigcup\limits_{m_i\in \TPaths\setminus\FPaths}\Fonem\right|+|F_{\TPaths}|+\left\lfloor\frac{1}{|\hat{m}^{(\TPaths)}|} \right \rfloor\Bigg ) P(\bar{Z}|\Obst)\\
    &-\Big | \bigcup\limits_{m_i\in \TPaths\setminus\FPaths}\hat{m}_i\Big| - \Big | \bigcup\limits_{m_i\in \TPaths\setminus\FPaths}\Fonem\Big|-|F_{\TPaths}|=\\
    &=(|\hat{m}^{(\TPaths)}|+|\Fonem|)P(Z|\Obst)+ \left\lfloor\frac{1}{|\hat{m}^{(\TPaths)}|} \right \rfloor P(\bar{Z}|\Obst)=\\&=\mathcal{U}(a_i|\Obst). 
\end{align*}
\end{footnotesize}
For adaptive monotone and submodular functions, solutions achieved by greedy policies are constant approximations to the optimal solutions (\cite{golovin2011adaptive},\cite{esfandiari2019adaptivity}). While it is trivial to prove adaptive monotonicity for our utility function, we can exhibit an example (depicted in Figure~\ref{fig:ex_no_sub}) showing that $\lambda$ is not adaptive submdular. Assume $p = 0.1$ is the a priori node failure probability and let $\TPaths = \{m_1\}$ and $\TPaths'=\{m_1,m_2\}$, with $\FPaths = \{m_1,m_2\}$. Then $P(Z_3|\Obst)= 1-\frac{1-(1-p)^2-(1-p)^4+(1-p)^5}{1-(1-p)^4}=0.638$, which leads to $\mathcal{U}(a_3|\Obst) = |\hat{m}^{(\TPaths)}_3|P(Z_3|\Obst) = 1.2766$, whereas $P(Z_3|O_{\TPaths'}) = 1-\frac{1-(1-p)^2-2(1-p)^4 + 2(1-p)^5}{1-2(1-p)^4+(1-p)^5}=0.789$, and hence $\mathcal{U}(a_3|O_{\TPaths'}) =|\hat{m}^{(\TPaths')}_3|P(Z_3|O_{\TPaths'})=1.578 $, which is greater than $\mathcal{U}(a_3|\Obst)$. 
\begin{figure}[h!]
\centering
\begin{tikzpicture}
\centering
\tikzset{pallino/.style={draw,circle,fill=white,minimum size=4pt,     inner sep=0pt}}
\tikzset{mylab/.style={black, minimum size=0.5cm}}
\tikzset{working/.style={draw,circle,fill=green,minimum size=4pt,inner sep=0pt}}
\tikzset{failed/.style={draw,circle,fill=red,minimum size=4pt,inner sep=0pt}}
  \draw (.7,0) node (1) [pallino] {}       
  --++(0:.8cm) node (2) [pallino]{}
           --++ (0:.8cm) node (3) [pallino]{}
            --++ (0:.8cm) node (4) [pallino]{};
          
   \draw (2.3,.8) node (5) [pallino]{};         
\draw (3.1,.8) node (6) [pallino]{};
\draw (5) --++(3);

    \node [mylab,above of = 6, node distance=0.15in] (blank) {\footnotesize{$\ v_6 $} };
        \node [mylab,below of = 1, node distance=0.15in] (blank) {\footnotesize{$\ v_1 $} };
        \node [mylab,below of = 2, node distance=0.15in] (blank) {\footnotesize{$\ v_2  $} };
        \node [mylab,below of = 3, node distance=0.15in] (blank) {\footnotesize{$\ v_3$} };
         \node [mylab,below of = 4, node distance=0.15in] (blank) {\footnotesize{$\ v_4 $} };
         \node [mylab,above of = 5, node distance=0.15in] (blank) {\footnotesize{$\ v_5 $} };
    \draw [->,blue] (1) to [out=30,in=150] (2);
\draw [->,blue] (2) to [out=30,in=150] (3);
\draw [->,blue] (3) to [out=30,in=150] (4);
\draw [->,green](1) to [out=-30,in=-150] (2);
\draw [->,green](2) to [out=-30,in=-150] (3);
\draw [->,green](3) to [out=60,in=-60] (5);
\draw (6) --++ (4);
\draw [->,red] (6) to [out=-60,in=60] (4);

\draw[color=red] (0.5cm, 0.55cm) --  (1cm, 0.55cm) node[draw=none,fill=none] (pippo) [label=right:{{\color{black}\footnotesize{$m_3$}}}]{};
\draw[color=green] (0.5cm, 0.85cm) --  (1cm, 0.85cm) node[draw=none,fill=none] (pippo) [label=right:{{\color{black}\footnotesize{$m_2$}}}]{};
\draw[blue] (0.5cm, 1.15cm) -- (1cm, 1.15cm) node[draw=none,fill=none] (pippo) [label=right:{{\color{black}\footnotesize{$m_1$}}}]{};
	\end{tikzpicture}
	\caption{Example showing non adaptive submodularity of $f$.}
	\label{fig:ex_no_sub}
	\end{figure}
\\When the objective function of a maximization problem is not adaptive submodular, as for our definition of utility, it is still possible to study an approximation of the solution obtained by a greedy policy with respect to the optimal one by bounding the \emph{adaptive submodularity ratio} $\gamma_{\Obst, k}(\lambda, p)$, with a scalar $\alpha\in(0,1]$. The resulting approximation is:
\begin{equation}
\label{eq:thm1fujiii}
    \lambda_{avg}(\pi^G)\geq \left ( 1-\exp\left (-\frac{\alpha K}{h}\right)\right )\lambda_{avg}(\pi^*)
\end{equation}
where $\lambda_{avg}(\pi^G/\pi^*)$ are the average quantity of information gained by the greedy and the optimal policies $\pi^G$ and $\pi^*$, respectively. Parameters $K$ and $h$ are the constraint to the maximum number of tests and the height of the decision tree of policy $\pi^*$, respectively. This result, together with the definition of adaptive submodular ratio, was recently proposed in \cite{fujii2019beyond}. 
\subsubsection{Upper-bound to adaptive submodularity ratio}
The goal of this section is to exhibit a scalar $\alpha > 0$  such that
\begin{equation}
\label{eq:asratio}
    \frac{\sum\limits_{m\in \Paths}{P(m\in \TPaths^{\pi})\Delta(a|O_{\TPaths}})}{\Delta(\pi|O_{\TPaths})}\ge \alpha.
\end{equation}
The adaptive submodularity ratio is upperbounded by 1 and it is equal to 1 if and only if $\lambda$ is adaptive submodular. 
Here $\TPaths^{\pi}$ is the set of paths chosen by a policy $\pi$, whereas $\Obst$ is a set of partial observations over a set of path $\TPaths$. It holds that:
\begin{align}
\label{eq:inequality}
    &\frac{\sum\limits_{m\in \Paths}{P(m\in \TPaths^{\pi})\Delta(a|O_{\TPaths}})}{\Delta(\pi|O_{\TPaths})}\ge \alpha \Rightarrow\\
    &\frac{\sum\limits_{m\in \Paths}P(m\in \TPaths^{\pi})\Delta(a|O_{\TPaths})}{\sum\limits_{m\in \Paths}P(m\in \TPaths^{\pi})\Delta(a|O_{\TPaths'})}\ge \alpha
\end{align}
where $O_{\TPaths'}$ is the set of observations such that the next path chosen by policy $\pi$ is $m$, and $\TPaths\subset \TPaths'$. From the discussion in \cite{fujii2019beyond}, it holds that the inequality~\ref{eq:inequality} can be equivalently expressed as follows:
\begin{align}
&\sum\limits_{m\in \Paths}{P(m\in \TPaths^{\pi})\Delta(a|O_{\TPaths}})\ge \alpha \sum\limits_{m\in \Paths}P(m\in \TPaths^{\pi})\Delta(a|O_{\TPaths'}) \nonumber\\
&\sum\limits_{m\in \Paths}P(m\in \TPaths^{\pi})d_a(\alpha)\ge 0, \label{eq:alpha}
\end{align}
where $d_a(\alpha) = \Delta(a|O_{\TPaths})-\alpha \Delta(a|O_{\TPaths'})\equiv \mathcal{U}(a|O_{\TPaths})-\alpha \mathcal{U}(a|O_{\TPaths'})$.
To prove the previous relation, we want to show that for every $a\in \mathcal{A}$ (action corresponding to probing path $m$ such that $m\not \in \TPaths'$) it holds that $d_{a}(\alpha)\ge 0$.
Among all path choices, we just need to study the contribution of those such that $d_{a}(1)< 0$. Therefore, we exclude all actions corresponding to the following sets of paths from our analysis: \emph{i)} all paths $m\in \TPaths'$ that were already tested; \emph{ii)} all paths for which it holds that $P(Z|O_{\TPaths'}) = 0$ or 1 (as in such case, $d_{a}(\alpha) = \mathcal{U}(a|\Obst)\ge 0$); \emph{iii)} all paths such that $P(Z|\Obst) = 0$ or 1, as this implies $P(Z|O_{\TPaths'}) = 0$ or 1 respectively, and therefore $d_{a}(1) = 0$; \emph{iv)} paths $m$ such that $\mathcal{U}(a|O_{\TPaths}) \ge \mathcal{U}(a|O_{\TPaths'})$. For all the listed cases, $d_{a}(1)\ge 0$. We study the maximum difference (i.e., the maximum value of  $|d_{a}(\alpha)|$ with $d_{a}(\alpha)<0$) that may occur between $\mathcal{U}(a|O_{\TPaths})$ and $\mathcal{U}(a|O_{\TPaths}')$ for all other paths. 

In order to accomplish to this task, we need to study the smallest non-zero value of $\mathcal{U}(a|\Obst)$, $\Delta_{min}$, and the greatest value of $\mathcal{U}(a|O_{\TPaths'})$, $\Delta'_{max}$. By choosing $\alpha = \frac{\Delta_{min}}{\Delta'_{max}}$, the relation in Equation~\ref{eq:alpha} holds always. 

\paragraph{Smallest value of $\mathcal{U}(a|\Obst)$}
  $P(Z|\Obst)$ is positive and minimum when every node $v$ of $m$ is traversed by failing paths of length 2. This is because the probability of failure of a node $v$ is directly proportional to the number of failing paths traversing it and inversely proportional to the number of nodes traversed by such paths. Nevertheless, observe that if even just one of such paths had length 1 (i.e., it would only pass through a node of $m$), then the probability of failure of $m$ would be 1 and its utility would be 0; this situation would fall into the set \emph{iii.} of paths that we exclude from this analysis. Hence, when a node is traversed only by failing paths of length 2, its failure probability is maximal and therefore its working probability is minimal,  excluding the case where $P(Z|\Obst)=0$.  We call $P_{min}$ such value of $P(Z|\Obst)$. In this case, $\mathcal{U}(a|\Obst) = (|\hat{m}^{(\TPaths)}|+|\Fonem|)P_{min}$. Notice that this expression exhibits explicit growing dependency of $\mathcal{U}(a|\Obst)$ on $|\Fonem|$. In Appendix 
 \ref{app:appA} we show that $ (|\hat{m}^{(\TPaths)}|+|\Fonem|)P_{min}$ is indeed the smallest value of $\mathcal{U}(a|\Obst)$ subject to $P(Z|\Obst)\in (0,1)$ despite the presence of the term $|\Fone|$. 
 Let us study now how $\mathcal{U}(a|\Obst)$ changes with respect to $|\hat{m}|$. Note that $\mathcal{U}(a|\Obst)$ grows linearly with $|\hat{m}|$ if $P(Z|\Obst)$ is fixed. Nevertheless, if we consider the case where every node of $m^{(\TPaths)}$ is traversed by some failing paths of length 2, then the contribution given by each node to the decrease of $P(Z|\Obst)$ is greater than adding 1 to $|\hat{m}|$, i.e., one such node would contribute to the exponential decrease of $P(Z|\Obst)$, while it would make $|\hat{m}|$ increase just by 1. This consideration emerges explicitly in the equations that follow. The path working probability $P(Z|\Obst)=P_{min}$ is equal to:
\begingroup
\allowdisplaybreaks
\begin{align*}
        P_{min} &= \prod\limits_{v\in \hat{m}}P(S_v|\Obst)=\\
        &=\prod\limits_{v\in\hat{m}} 1 - \frac{p}{1 - \sum\limits_{i = 2}^{\partial_v + 1}(-1)^i(1-p)^i \binom{\partial_v}{i-1}}\ge\\
        &\ge \Big [1 - \frac{p}{1 - \sum\limits_{i = 2}^{\partial_{max} + 1}(-1)^i(1-p)^i \binom{\partial_{max}}{i-1}}\Big ]^{|\hat{m}|}\ge\\
        &\ge \Big [1 - \frac{p}{1 - \sum\limits_{i = 2}^{\partial_{max} + 1}(-1)^i(1-p)^i \binom{\partial_{max}}{i-1}}\Big ]^{|\hat{m}_{max}|}
\end{align*}
\endgroup
where $\partial_v$ is the number of failing paths traversing $v$,  $\partial_{max}$ is the biggest among them and $|\hat{m}_{max}|$ is the length of the longest path. Notice that $\partial_{max}\leq |\Fonem|$. The denominator appearing in the last expression can be written as follows:
\begingroup
\allowdisplaybreaks
\begin{equation}
    \begin{split}
        &1 - \sum\limits_{i = 2}^{\partial_{max} + 1}(-1)^i(1-p)^i \binom{\partial_{max}}{i-1}=\\
        &1+(1-p)\cdot \sum\limits_{i = 1}^{\partial_{max}}(-1)^i(1-p)^i\binom{\partial_{max}}{i} = \\
        &1+(1-p)\cdot \sum\limits_{i = 1}^{\partial_{max}}(p-1)^i\binom{\partial_{max}}{i} = \quad \text{ \footnotesize{[Newton's binomial]}}\\
        &1 +(1+p)\cdot [(p-1 + 1)^{\partial_{max}}-1] =\\
        &1+ (1-p)(p^{\partial_{max}}-1).
    \end{split}
\end{equation}
\endgroup

\noindent Therefore the minimum value of $\mathcal{U}(a|\Obst)$, $\Delta_{min}$ is:

\begin{footnotesize}
\begin{equation}
\begin{split}
\label{eq:DeltaMin}
    \Delta_{min} &= (|\hat{m}^{(\TPaths)}|+|\Fonem|) \prod\limits_{v\in\hat{m}} 1 - \frac{p}{1+(1-p)(p^{\partial_{v}}-1)}\\ &\ge (|\hat{m}^{(\TPaths)}|+|\Fonem|)\left [1 - \frac{p}{1+(1-p)(p^{\partial_{max}}-1)}\right ]^{|\hat{m}_{max}|}.
    \end{split}
\end{equation}
\end{footnotesize}

Notice that we are excluding from our analysis the case $|\hat{m}^{(\TPaths)}|=1$, that is the only situation in which the second term of the utility $\left \lfloor \frac{1}{|\hat{m}^{(\TPaths)}|}\right \rfloor$ is non zero: as a matter of fact, Equation \ref{eq:DeltaMin} proves that the larger the path length, appearing as the exponent of $P_{min}$, the smaller is $\Delta_{min}$. To recap, the smallest value of $\mathcal{U}(a|\Obst)$ is achieved when path $m^{(\TPaths)}$ is long and all of its nodes are traversed by failing paths of length 2.  
\paragraph{Greatest value of $\mathcal{U}(a|O_{\TPaths'})$}
From the situation described above, we can analyse what is the greatest value of $\mathcal{U}(a|O_{\TPaths'})$. First of all, observe that $\forall \TPaths,\,\TPaths'$ such that $\TPaths\subset\TPaths'$ it holds that $|\hat{m}^{(\TPaths)}|\ge|\hat{m}^{(\TPaths')}|$. As a consequence, it holds that $\mathcal{U}(a|O_{\TPaths'})>\mathcal{U}(a|\Obst)$ if and only if $P(Z|O_{\TPaths'})$ is sufficiently larger than $P(Z|\Obst)$.  In general, $P(Z|O_{\TPaths'})>P(Z|\Obst)$ with $\Obst\subset O_{\TPaths'}$ in two occasions: either because of the presence of some functioning paths in $\TPaths'\setminus\TPaths$ traversing $m$, or because it was possible to localize failures of paths traversing nodes of $m$ on some other nodes. In the first case it results that  $|\hat{m}^{(O_{\TPaths'})}|<|\hat{m}^{(\Obst)}|$, therefore the increase  of $P(Z|O_{\TPaths'})$ could be contrasted by the decrease of $|\hat{m}^{(O_{\TPaths'})}|$, possibly resulting in $\mathcal{U}(a|\Obst)>\mathcal{U}(a|O_{\TPaths'})$. In the second case instead, it holds that $|\hat{m}^{(\TPaths)}|=|\hat{m}^{(\TPaths')}|$. Hybrid situations may occur, too. We shall first consider the second case: assume it was possible to assess as "failed" all nodes not in $\hat{m}$ appearing in the two-length paths traversing $m$. Therefore, $\mathcal{U}(a|O_{\TPaths'})$ becomes equal to $|\hat{m}^{(\TPaths')}|(1-p)^{|\hat{m}^{(\TPaths')}|}$. 
Notice that this then this the initial expected value of the utility function of $a$, when no observations were made. As a matter of fact, the working probability of a node only grows  when a working path traverses it (and in such case it becomes 1). Now we can also consider the case in which $|m^{(\TPaths')}|$ is reduced if some working path $m'$ partially covering nodes of $m$ was tested. Assuming $m^{(\TPaths)}$ is long enough, we want to analyse the growing trend of $|\hat{m}^{(\TPaths')}|(1-p)^{|\hat{m}^{(\TPaths')}|}$. Indeed, the first term of this expression $|\hat{m}^{(\TPaths')}|$ trivially grows linearly, whereas $(1-p)^{|\hat{m}^{(\TPaths')}|}$ decreases with $|\hat{m}^{(\TPaths')}|$. The trend of their products depends on the value of $p$. Excluding the trivial cases where $p = 0$ or 1, it is easy to prove analytically 
that the maximum value of $|\hat{m}^{(\TPaths')}|(1-p)^{|\hat{m}^{(\TPaths')}|}$ is for $|\hat{m}^{(\TPaths')}| = -\left [\frac{1}{\ln(1-p)}\right ]$. Here $n = [x]$ is the rounded natural value of $x$. Therefore the maximum value of $\mathcal{U}(a|O_{\TPaths'})$ is  $\Delta'_{max}=- \left [\frac{1}{\ln(1-p)} \right ](1-p)^{-\left [ \frac{1}{\ln(1-p)} \right ]}$. Notice that the maximum value of $\mathcal{U}(a|O_{\TPaths'})$ is reached when $\Fonem=\emptyset$. Indeed, in case $\Fonem\neq \emptyset$, the working probability of $m$ would decrease exponentially, at the face of a linear growth of the deterministic multiplier $(|\hat{m}^{(\TPaths)}|+|\Fonem|)$, as explained in Appendix A.

\paragraph{Solution approximation}
By choosing $\alpha = \frac{\Delta_{min}}{\Delta'_{max}}$ as discussed in the previous sections, it holds that for all paths $m$ and observations $\Obst$ and $O_{\TPaths'}$, $d_m^{\alpha}\ge 0$, implying soundness of Equation~\ref{eq:asratio}. Given a lower bound to the adaptive submodularity ratio, we may use the result shown in~\cite{fujii2019beyond} to claim the following:
\begin{proposition}
If $\pi^G$ is the  policy representing the adaptive greedy algorithm using $K$ steps, and $\lambda:2^{\Paths}\times O_{\Paths}\rightarrow\mathbb{R}_{\ge 0}$ is the utility function defined in Equation~\ref{eq:cumUtility}, then:
\[
\lambda_{avg}(\pi^G) \ge \left ( 1-\exp\left (-\frac{\alpha K}{k}\right ) \right )\lambda_{avg}(\pi^*)
\]
where $\pi^*$ is the optimal policy, $k$ is the number of steps that $\pi^*$ takes to reach convergence and $\alpha = \frac{\Delta_{min}}{\Delta'_{max}}$.
\end{proposition}
\begin{proof}
    The statement is a direct consequence of the following facts: \emph{i.} $\lambda$ is adaptive monotone. \emph{ii.} Theorem 1 in \cite{fujii2019beyond}, reported in Equation~\ref{eq:thm1fujiii}. \emph{iii.} PoPGreedy is an Adaptive Greedy Algorithm. 
\end{proof}
Notice that $\alpha$ is dependent on controllable parameters $\partial_{max}$ and $|\hat{m}_{max}|$ that do not depend on the network topology but only on the routing paths choice.\\
When PoPGreedy is run on the simple example shown in Figure \ref{fig:example}, only in two occasions it happens that $\mathcal{U}(a|\Obst)<\mathcal{U}(a|O_{\TPaths'})$. The one marking maximum difference holds for $a = a_5$, $\TPaths = \{m_3,m_4\}$ and $\TPaths'=\{m_1,m_3,m_4\}$, where $\mathcal{U}(a_5|\Obst) = 1.076$ and $\mathcal{U}(a^{(5)}|O_{\TPaths'})=1.278$. For this example, $\alpha = 0.842$. 

%% file: imgs/BICS1_it/AW1it.tex
%
%
\definecolor{mycolor1}{rgb}{0.00000,0.44700,0.74100}%
\definecolor{mycolor2}{rgb}{0.85000,0.32500,0.09800}%
\definecolor{mycolor3}{rgb}{0.92900,0.69400,0.12500}%
\definecolor{mycolor4}{rgb}{0.49400,0.18400,0.55600}%
\definecolor{mycolor5}{rgb}{0.46600,0.67400,0.18800}%
\definecolor{mycolor6}{rgb}{0.30100,0.74500,0.93300}%
\definecolor{mycolor7}{rgb}{0.63500,0.07800,0.18400}%
\pgfplotsset{
compat=1.11,
legend image code/.code={
\draw[mark repeat=2,mark phase=2]
plot coordinates {
(0cm,0cm)
(0.15cm,0cm)        
(0.3cm,0cm)         
};%
}
}
\begin{tikzpicture}

\begin{axis}[%
width=.85\columnwidth,
height=.55\columnwidth,
scale only axis,
xmin=0,
xmax=21,
xtick={1,3,5,7,9,11,13,15,17,19},
xticklabels={{1},{3},{5},{7},{9},{11},{13},{15},{17},{19}},
xticklabel style={font=\footnotesize},
xlabel style={font=\color{white!15!black}\footnotesize,yshift=3pt},
xlabel={\# tested paths},
ymin=0,
ymax=1.05,
ytick={0,0.2,0.4,0.6,0.8,1},
yticklabels={{0},{.2},{.4},{.6},{.8},{1}},
yticklabel style={font=\footnotesize},
ylabel style={font=\color{white!15!black}\footnotesize,yshift=-5pt},
ylabel={$a_W$},
axis background/.style={fill=white},
legend style={at={(0.42,0.0)}, legend columns=2, anchor=south west, legend cell align=left, align=left, fill=none, draw=none, font=\scriptsize}
]

\addplot [color=mycolor1, line width=.6pt, mark size=1.5pt, mark=text,text mark={\large $\star$}, mark options={solid, fill=mycolor1, mycolor1}]
  table[row sep=crcr]{%
1	0\\
2	0.137436813186813\\
3	0.235676536426536\\
4	0.355680606430606\\
5	0.542003154253154\\
6	0.604083943833944\\
7	0.681199633699634\\
8	0.773842999593\\
9	0.837605921855922\\
10	0.880877594627595\\
11	0.909016585266585\\
12	0.943041107041107\\
13	0.973729194781826\\
14	0.982880658436214\\
15	1\\
};
\addlegendentry{FaCe}

\addplot [color=mycolor2, line width=.6pt, mark size=1.5pt, mark=o, mark options={solid, mycolor2}]
  table[row sep=crcr]{%
1	0\\
2	0.137436813186813\\
3	0.312486263736264\\
4	0.378369454619455\\
5	0.507534086284086\\
6	0.554400183150183\\
7	0.659995014245014\\
8	0.728708587708588\\
9	0.783965201465202\\
10	0.821858770858771\\
11	0.865554029304029\\
12	0.902148962148962\\
13	0.925062442430863\\
14	0.930832472370934\\
15	0.966603355492244\\
16	0.984126984126984\\
17	0.981481481481482\\
18	1\\
19	1\\
20	1\\
};
\addlegendentry{PoP}

\addplot [color=mycolor3, line width=.6pt, mark size=1.5pt, mark=asterisk, mark options={solid, mycolor3}]
  table[row sep=crcr]{%
1	0\\
2	0.254195054945055\\
3	0.345658221408221\\
4	0.507914021164021\\
5	0.603723544973545\\
6	0.660553520553521\\
7	0.696417175417176\\
8	0.787204110704111\\
9	0.821663614163614\\
10	0.81163381591953\\
};
\addlegendentry{GC}

\addplot [color=mycolor4, line width=.6pt, mark size=1.5pt, mark=triangle, mark options={solid, mycolor4}]
  table[row sep=crcr]{%
1	0\\
2	0.132632275132275\\
3	0.240416564916565\\
4	0.336469881969882\\
5	0.387592287342287\\
6	0.459886955636956\\
7	0.508800671550672\\
8	0.556916971916972\\
9	0.604436100936101\\
10	0.649780321530322\\
11	0.711806471306471\\
12	0.762425824175824\\
13	0.787280915965127\\
14	0.810232459847845\\
15	0.838918509474065\\
16	0.855825629397058\\
17	0.88260582010582\\
18	0.839285714285714\\
19	0.821428571428571\\
20	0.892857142857143\\
};
\addlegendentry{GI}

\addplot [color=mycolor5, line width=.6pt, mark size=1.2pt, mark=square, mark options={solid, mycolor5}]
  table[row sep=crcr]{%
1	0\\
2	0.187979242979243\\
3	0.392328856328856\\
4	0.477079568579569\\
5	0.549382376882377\\
6	0.612642857142857\\
7	0.677602564102564\\
8	0.745974053724054\\
9	0.789086182336182\\
10	0.834805657305657\\
11	0.872578449328449\\
12	0.88748544973545\\
13	0.907146391620076\\
14	0.938647662878432\\
15	0.973951973951974\\
16	0.982600732600733\\
17	1\\
18	1\\
19	1\\
20	1\\
};
\addlegendentry{GD}

\addplot [color=mycolor6, line width=.6pt, mark size=1.5pt, mark=+, mark options={solid, mycolor6}]
  table[row sep=crcr]{%
1	0\\
2	0.0285714285714286\\
3	0.128571428571429\\
4	0.15\\
5	0.164285714285714\\
6	0.175\\
7	0.175\\
8	0.289285714285714\\
9	0.358928571428571\\
10	0.4125\\
11	0.417857142857143\\
12	0.426785714285714\\
13	0.419172932330827\\
14	0.535714285714286\\
15	0.523809523809524\\
16	0.540816326530612\\
17	0.544642857142857\\
18	0.607142857142857\\
19	0.607142857142857\\
20	0.607142857142857\\
};
\addlegendentry{AF}

\addplot [color=mycolor7, line width=.6pt, mark size=1.5pt, mark=x, mark options={solid, mycolor7}]
  table[row sep=crcr]{%
1	0.226236263736264\\
2	0.270324933687002\\
3	0.373191655088207\\
4	0.50093732895457\\
5	0.556109742747674\\
6	0.669630752389373\\
7	0.713036819502337\\
8	0.741944023409541\\
9	0.821522777988295\\
10	0.890980906067113\\
11	0.933714896214896\\
12	0.958333333333333\\
13	0.962962962962963\\
};
\addlegendentry{APC}

\end{axis}
\end{tikzpicture}%

%% file: imgs/BICS1_it/AB1it.tex
%
%
\definecolor{mycolor1}{rgb}{0.00000,0.44700,0.74100}%
\definecolor{mycolor2}{rgb}{0.85000,0.32500,0.09800}%
\definecolor{mycolor3}{rgb}{0.92900,0.69400,0.12500}%
\definecolor{mycolor4}{rgb}{0.49400,0.18400,0.55600}%
\definecolor{mycolor5}{rgb}{0.46600,0.67400,0.18800}%
\definecolor{mycolor6}{rgb}{0.30100,0.74500,0.93300}%
\definecolor{mycolor7}{rgb}{0.63500,0.07800,0.18400}%
\pgfplotsset{
compat=1.11,
legend image code/.code={
\draw[mark repeat=2,mark phase=2]
plot coordinates {
(0cm,0cm)
(0.15cm,0cm)        
(0.3cm,0cm)         
};%
}
}
\begin{tikzpicture}

\begin{axis}[%
width=.85\columnwidth,
height=.55\columnwidth,
scale only axis,
xmin=0,
xmax=21,
xtick={1,3,5,7,9,11,13,15,17,19},
xticklabels={{1},{3},{5},{7},{9},{11},{13},{15},{17},{19}},
xticklabel style={font=\footnotesize},
xlabel style={font=\color{white!15!black}\footnotesize,yshift=3pt},
xlabel={\# tested paths},
ymin=0,
ymax=1.05,
ytick={0,0.2,0.4,0.6,0.8,1},
yticklabels={{0},{.2},{.4},{.6},{.8},{1}},
yticklabel style={font=\footnotesize},
ylabel style={font=\color{white!15!black}\footnotesize,yshift=-5pt},
ylabel={$a_B$},
axis background/.style={fill=white},
legend style={at={(0.0,0.25)}, legend columns=2, anchor=south west, legend cell align=left, align=left, fill=none, draw=none, font=\scriptsize}
]

\addplot [color=mycolor1, line width=.6pt, mark size=1.5pt, mark=text,text mark={\large $\star$}, mark options={solid, fill=mycolor1, mycolor1}]
  table[row sep=crcr]{%
1	0\\
2	0\\
3	0\\
4	0\\
5	0\\
6	0\\
7	0.0964912280701754\\
8	0.0964912280701754\\
9	0.109649122807018\\
10	0.201754385964912\\
11	0.293859649122807\\
12	0.447368421052632\\
13	0.767543859649123\\
14	0.962962962962963\\
15	1\\
};
\addlegendentry{FaCe}

\addplot [color=mycolor2, line width=.6pt, mark size=1.5pt, mark=o, mark options={solid, mycolor2}]
  table[row sep=crcr]{%
1	0\\
2	0\\
3	0\\
4	0\\
5	0\\
6	0\\
7	0\\
8	0\\
9	0.0138888888888889\\
10	0.0277777777777778\\
11	0.134259259259259\\
12	0.384259259259259\\
13	0.651960784313725\\
14	0.673611111111111\\
15	0.6875\\
16	0.785714285714286\\
17	1\\
18	1\\
19	1\\
20	1\\
};
\addlegendentry{PoP}

\addplot [color=mycolor3, line width=.6pt, mark size=1.5pt, mark=asterisk, mark options={solid, mycolor3}]
  table[row sep=crcr]{%
1	0\\
2	0\\
3	0\\
4	0\\
5	0\\
6	0\\
7	0.0357142857142857\\
8	0.297619047619048\\
9	0.452380952380952\\
10	0.604166666666667\\
};
\addlegendentry{GC}

\addplot [color=mycolor4, line width=.6pt, mark size=1.5pt, mark=triangle, mark options={solid, mycolor4}]
  table[row sep=crcr]{%
1	0\\
2	0\\
3	0\\
4	0\\
5	0.05\\
6	0.1\\
7	0.172222222222222\\
8	0.244444444444444\\
9	0.3\\
10	0.344444444444444\\
11	0.344444444444444\\
12	0.344444444444444\\
13	0.522222222222222\\
14	0.759259259259259\\
15	0.69047619047619\\
16	0.783333333333333\\
17	0.854166666666667\\
18	0.875\\
19	0.75\\
20	0.75\\
};
\addlegendentry{GI}

\addplot [color=mycolor5, line width=.6pt, mark size=1.2pt, mark=square, mark options={solid, mycolor5}]
  table[row sep=crcr]{%
1	0\\
2	0\\
3	0\\
4	0\\
5	0\\
6	0\\
7	0.0294117647058824\\
8	0.0294117647058824\\
9	0.0637254901960784\\
10	0.102941176470588\\
11	0.161764705882353\\
12	0.480392156862745\\
13	0.645833333333333\\
14	0.757575757575757\\
15	0.75\\
16	0.791666666666667\\
17	0.916666666666667\\
18	0.875\\
19	0.75\\
20	0.75\\
};
\addlegendentry{GD}

\addplot [color=mycolor6, line width=.6pt, mark size=1.5pt, mark=+, mark options={solid, mycolor6}]
  table[row sep=crcr]{%
1	0\\
2	0\\
3	0\\
4	0\\
5	0\\
6	0.25\\
7	0.25\\
8	0.25\\
9	0.25\\
10	0.25\\
11	0.275\\
12	0.4875\\
13	0.5\\
14	0.5\\
15	0.5\\
16	0.5\\
17	0.5\\
18	0.75\\
19	0.75\\
20	0.75\\
};
\addlegendentry{AF}

\addplot [color=mycolor7, line width=.6pt, mark size=1.5pt, mark=x, mark options={solid, mycolor7}]
  table[row sep=crcr]{%
1	0\\
2	0\\
3	0\\
4	0\\
5	0\\
6	0\\
7	0\\
8	0.0964912280701754\\
9	0.201754385964912\\
10	0.368421052631579\\
11	0.657894736842105\\
12	1\\
13	1\\
};
\addlegendentry{APC}

\end{axis}
\end{tikzpicture}%

%% file: imgs/BICS1_it/R11it.tex
%
%
\definecolor{mycolor1}{rgb}{0.00000,0.44700,0.74100}%
\definecolor{mycolor2}{rgb}{0.85000,0.32500,0.09800}%
\definecolor{mycolor3}{rgb}{0.92900,0.69400,0.12500}%
\definecolor{mycolor4}{rgb}{0.49400,0.18400,0.55600}%
\definecolor{mycolor5}{rgb}{0.46600,0.67400,0.18800}%
\definecolor{mycolor6}{rgb}{0.30100,0.74500,0.93300}%
\definecolor{mycolor7}{rgb}{0.63500,0.07800,0.18400}%
\pgfplotsset{
compat=1.11,
legend image code/.code={
\draw[mark repeat=2,mark phase=2]
plot coordinates {
(0cm,0cm)
(0.15cm,0cm)        
(0.3cm,0cm)         
};%
}
}
\begin{tikzpicture}

\begin{axis}[%
width=.85\columnwidth,
height=.55\columnwidth,
scale only axis,
xmin=0,
xmax=21,
xtick={1,3,5,7,9,11,13,15,17,19},
xticklabels={{1},{3},{5},{7},{9},{11},{13},{15},{17},{19}},
xticklabel style={font=\footnotesize},
xlabel style={font=\color{white!15!black}\footnotesize,yshift=3pt},
xlabel={\# tested paths},
ymin=0,
ymax=1.05,
ytick={0,0.2,0.4,0.6,0.8,1},
yticklabels={{0},{.2},{.4},{.6},{.8},{1}},
yticklabel style={font=\footnotesize},
ylabel style={font=\color{white!15!black}\footnotesize,yshift=-5pt},
ylabel={$R_1$},
axis background/.style={fill=white},
legend style={at={(0.45,0.0)}, nodes={scale=0.8, transform shape}, legend columns=2,, anchor=south west, legend cell align=left, align=left, fill=none, draw=none, font=\scriptsize}
]

\addplot [color=mycolor1, line width=.6pt, mark size=1.5pt, mark=text,text mark={\large $\star$}, mark options={solid, fill=mycolor1, mycolor1}]
  table[row sep=crcr]{%
1	0.138888888888889\\
2	0.25\\
3	0.0694444444444444\\
4	0.125\\
5	0.236111111111111\\
6	0.291666666666667\\
7	0.472222222222222\\
8	0.513888888888889\\
9	0.513888888888889\\
10	0.569444444444444\\
11	0.666666666666667\\
12	0.763888888888889\\
13	0.838235294117647\\
14	0.8125\\
15	1\\
};
\addlegendentry{FaCe}

\addplot [color=mycolor2, line width=.6pt, mark size=1.5pt, mark=o, mark options={solid, mycolor2}]
  table[row sep=crcr]{%
1	0.138888888888889\\
2	0.25\\
3	0.166666666666667\\
4	0.180555555555556\\
5	0.208333333333333\\
6	0.222222222222222\\
7	0.291666666666667\\
8	0.319444444444444\\
9	0.402777777777778\\
10	0.486111111111111\\
11	0.583333333333333\\
12	0.708333333333333\\
13	0.779411764705882\\
14	0.791666666666667\\
15	0.90625\\
16	0.916666666666667\\
17	1\\
18	1\\
19	1\\
20	1\\
};
\addlegendentry{PoP}

\addplot [color=mycolor3, line width=.6pt, mark size=1.5pt, mark=asterisk, mark options={solid, mycolor3}]
  table[row sep=crcr]{%
1	0.138888888888889\\
2	0.166666666666667\\
3	0.0833333333333333\\
4	0.111111111111111\\
5	0.194444444444444\\
6	0.236111111111111\\
7	0.347222222222222\\
8	0.458333333333333\\
9	0.513888888888889\\
10	0.5\\
};
\addlegendentry{GC}

\addplot [color=mycolor4, line width=.6pt, mark size=1.5pt, mark=triangle, mark options={solid, mycolor4}]
  table[row sep=crcr]{%
1	0.138888888888889\\
2	0.152777777777778\\
3	0.194444444444444\\
4	0.222222222222222\\
5	0.236111111111111\\
6	0.319444444444444\\
7	0.430555555555556\\
8	0.472222222222222\\
9	0.513888888888889\\
10	0.541666666666667\\
11	0.569444444444444\\
12	0.583333333333333\\
13	0.617647058823529\\
14	0.604166666666667\\
15	0.65625\\
16	0.708333333333333\\
17	0.833333333333333\\
18	0.875\\
19	0.75\\
20	0.75\\
};
\addlegendentry{GI}

\addplot [color=mycolor5, line width=.6pt, mark size=1.2pt, mark=square, mark options={solid, mycolor5}]
  table[row sep=crcr]{%
1	0.138888888888889\\
2	0.194444444444444\\
3	0.263888888888889\\
4	0.263888888888889\\
5	0.305555555555556\\
6	0.319444444444444\\
7	0.361111111111111\\
8	0.416666666666667\\
9	0.472222222222222\\
10	0.527777777777778\\
11	0.555555555555556\\
12	0.625\\
13	0.676470588235294\\
14	0.770833333333333\\
15	0.8125\\
16	0.791666666666667\\
17	0.916666666666667\\
18	1\\
19	1\\
20	1\\
};
\addlegendentry{GD}

\addplot [color=mycolor6, line width=.6pt, mark size=1.5pt, mark=+, mark options={solid, mycolor6}]
  table[row sep=crcr]{%
1	0.138888888888889\\
2	0.125\\
3	0.208333333333333\\
4	0.277777777777778\\
5	0.319444444444444\\
6	0.361111111111111\\
7	0.361111111111111\\
8	0.416666666666667\\
9	0.458333333333333\\
10	0.416666666666667\\
11	0.555555555555556\\
12	0.597222222222222\\
13	0.602941176470588\\
14	0.604166666666667\\
15	0.625\\
16	0.666666666666667\\
17	0.75\\
18	0.875\\
19	0.75\\
20	0.75\\
};
\addlegendentry{AF}

\addplot [color=mycolor7, line width=.6pt, mark size=1.5pt, mark=x, mark options={solid, mycolor7}]
  table[row sep=crcr]{%
1	0\\
2	0\\
3	0\\
4	0\\
5	0\\
6	0\\
7	0\\
8	0.0964912280701754\\
9	0.201754385964912\\
10	0.368421052631579\\
11	0.657894736842105\\
12	1\\
13	1\\
};
\addlegendentry{APC}

\end{axis}
\end{tikzpicture}%

%% file: imgs/BICS1_it/R21it.tex
%
%
\definecolor{mycolor1}{rgb}{0.00000,0.44700,0.74100}%
\definecolor{mycolor2}{rgb}{0.85000,0.32500,0.09800}%
\definecolor{mycolor3}{rgb}{0.92900,0.69400,0.12500}%
\definecolor{mycolor4}{rgb}{0.49400,0.18400,0.55600}%
\definecolor{mycolor5}{rgb}{0.46600,0.67400,0.18800}%
\definecolor{mycolor6}{rgb}{0.30100,0.74500,0.93300}%
\definecolor{mycolor7}{rgb}{0.63500,0.07800,0.18400}%
\pgfplotsset{
compat=1.11,
legend image code/.code={
\draw[mark repeat=2,mark phase=2]
plot coordinates {
(0cm,0cm)
(0.15cm,0cm)        
(0.3cm,0cm)         
};%
}
}
\begin{tikzpicture}

\begin{axis}[%
width=.85\columnwidth,
height=.55\columnwidth,
scale only axis,
xmin=0,
xmax=21,
xtick={1,3,5,7,9,11,13,15,17,19},
xticklabels={{1},{3},{5},{7},{9},{11},{13},{15},{17},{19}},
xticklabel style={font=\footnotesize},
xlabel style={font=\color{white!15!black}\footnotesize,yshift=3pt},
xlabel={\# tested paths},
ymin=0,
ymax=1.05,
ytick={0,0.2,0.4,0.6,0.8,1},
yticklabels={{0},{.2},{.4},{.6},{.8},{1}},
yticklabel style={font=\footnotesize},
ylabel style={font=\color{white!15!black}\footnotesize,yshift=-5pt},
ylabel={$R_2$},
axis background/.style={fill=white},
legend style={at={(0.00,0.27)}, legend columns=2, anchor=south west, legend cell align=left, align=left, fill=none, draw=none, font=\scriptsize}
]

\addplot [color=mycolor1, line width=.6pt, mark size=1.5pt, mark=text,text mark={\large $\star$}, mark options={solid, fill=mycolor1, mycolor1}]
  table[row sep=crcr]{%
1	0.156862745098039\\
2	0.158940397350993\\
3	0.177339901477832\\
4	0.196185286103542\\
5	0.25531914893617\\
6	0.289156626506024\\
7	0.344497607655502\\
8	0.411428571428571\\
9	0.47682119205298\\
10	0.545454545454546\\
11	0.626086956521739\\
12	0.705882352941176\\
13	0.829268292682927\\
14	0.842105263157895\\
15	1\\
};
\addlegendentry{FaCe}

\addplot [color=mycolor2, line width=.6pt, mark size=1.5pt, mark=o, mark options={solid, mycolor2}]
  table[row sep=crcr]{%
1	0.156862745098039\\
2	0.158940397350993\\
3	0.196185286103542\\
4	0.206303724928367\\
5	0.24\\
6	0.252631578947368\\
7	0.301255230125523\\
8	0.346153846153846\\
9	0.413793103448276\\
10	0.455696202531646\\
11	0.529411764705882\\
12	0.626086956521739\\
13	0.731182795698925\\
14	0.738461538461538\\
15	0.82051282051282\\
16	0.888888888888889\\
17	1\\
18	1\\
19	1\\
20	1\\
};
\addlegendentry{PoP}

\addplot [color=mycolor3, line width=.6pt, mark size=1.5pt, mark=asterisk, mark options={solid, mycolor3}]
  table[row sep=crcr]{%
1	0.156862745098039\\
2	0.183206106870229\\
3	0.198347107438017\\
4	0.240802675585284\\
5	0.282352941176471\\
6	0.306382978723404\\
7	0.328767123287671\\
8	0.413793103448276\\
9	0.464516129032258\\
10	0.472727272727273\\
};
\addlegendentry{GC}

\addplot [color=mycolor4, line width=.6pt, mark size=1.5pt, mark=triangle, mark options={solid, mycolor4}]
  table[row sep=crcr]{%
1	0.156862745098039\\
2	0.175182481751825\\
3	0.189473684210526\\
4	0.209912536443149\\
5	0.225705329153605\\
6	0.25\\
7	0.269662921348315\\
8	0.289156626506024\\
9	0.32579185520362\\
10	0.351219512195122\\
11	0.385026737967914\\
12	0.423529411764706\\
13	0.462585034013605\\
14	0.5\\
15	0.516129032258065\\
16	0.5\\
17	0.545454545454546\\
18	0.615384615384615\\
19	0.444444444444444\\
20	0.571428571428571\\
};
\addlegendentry{GI}

\addplot [color=mycolor5, line width=.6pt, mark size=1.2pt, mark=square, mark options={solid, mycolor5}]
  table[row sep=crcr]{%
1	0.156862745098039\\
2	0.184615384615385\\
3	0.215568862275449\\
4	0.232258064516129\\
5	0.250871080139373\\
6	0.275862068965517\\
7	0.322869955156951\\
8	0.352941176470588\\
9	0.389189189189189\\
10	0.444444444444444\\
11	0.51063829787234\\
12	0.541353383458647\\
13	0.62962962962963\\
14	0.705882352941176\\
15	0.761904761904762\\
16	0.75\\
17	0.923076923076923\\
18	1\\
19	1\\
20	1\\
};
\addlegendentry{GD}

\addplot [color=mycolor6, line width=.6pt, mark size=1.5pt, mark=+, mark options={solid, mycolor6}]
  table[row sep=crcr]{%
1	0.156862745098039\\
2	0.13926499032882\\
3	0.164009111617312\\
4	0.194070080862534\\
5	0.171428571428571\\
6	0.16551724137931\\
7	0.16551724137931\\
8	0.194070080862534\\
9	0.222222222222222\\
10	0.3\\
11	0.288\\
12	0.32579185520362\\
13	0.320754716981132\\
14	0.324324324324324\\
15	0.347826086956522\\
16	0.406779661016949\\
17	0.521739130434783\\
18	0.727272727272727\\
19	0.571428571428571\\
20	0.571428571428571\\
};
\addlegendentry{AF}

\addplot [color=mycolor7, line width=.6pt, mark size=1.5pt, mark=x, mark options={solid, mycolor7}]
  table[row sep=crcr]{%
1	0.136363636363636\\
2	0.163934426229508\\
3	0.18018018018018\\
4	0.204081632653061\\
5	0.237154150197628\\
6	0.277777777777778\\
7	0.310880829015544\\
8	0.338983050847458\\
9	0.444444444444444\\
10	0.530973451327434\\
11	0.638297872340426\\
12	0.631578947368421\\
13	0.692307692307692\\
};
\addlegendentry{APC}

\end{axis}
\end{tikzpicture}%

%% file: imgs/BICS1/AW1.tex
%
%
\definecolor{mycolor1}{rgb}{0.00000,0.44700,0.74100}%
\definecolor{mycolor2}{rgb}{0.85000,0.32500,0.09800}%
\definecolor{mycolor3}{rgb}{0.92900,0.69400,0.12500}%
\definecolor{mycolor4}{rgb}{0.49400,0.18400,0.55600}%
\definecolor{mycolor5}{rgb}{0.46600,0.67400,0.18800}%
\definecolor{mycolor6}{rgb}{0.30100,0.74500,0.93300}%
\definecolor{mycolor7}{rgb}{0.63500,0.07800,0.18400}%
\pgfplotsset{
compat=1.11,
legend image code/.code={
\draw[mark repeat=2,mark phase=2]
plot coordinates {
(0cm,0cm)
(0.15cm,0cm)        
(0.3cm,0cm)         
};%
}
}
\begin{tikzpicture}

\begin{axis}[%
width=.85\columnwidth,
height=.55\columnwidth,
scale only axis,
xmin=0.8,
xmax=5.2,
xtick={1,2,3,4,5},
xticklabel style={font=\footnotesize},
xlabel style={font=\color{white!15!black}\footnotesize,yshift=3pt},
xlabel={\# failed nodes},
ymin=0,
ymax=1.05,
ytick={0,0.2,0.4,0.6,0.8,1},
yticklabels={{0},{.2},{.4},{.6},{.8},{1}},
yticklabel style={font=\footnotesize},
ylabel style={font=\color{white!15!black}\footnotesize,yshift=-5pt},
ylabel={$a_W$},
axis background/.style={fill=white},
legend style={at={(0.1,0.03)}, legend columns=3, anchor=south west, legend cell align=left, align=left, fill=none, draw=none, font=\scriptsize}
]

\addplot[area legend, draw=none, fill=mycolor1, fill opacity=0.2, forget plot]
table[row sep=crcr] {%
x	y\\
1	1\\
2	1\\
3	1\\
4	1\\
5	1\\
5	1\\
4	1\\
3	1\\
2	1\\
1	1\\
}--cycle;
\addplot [color=mycolor1, line width=.6pt, mark size=2.5pt, mark=text,text mark={\LARGE $\star$}, mark options={solid, fill=mycolor1, mycolor1}]
  table[row sep=crcr]{%
1	1\\
2	1\\
3	1\\
4	1\\
5	1\\
};
\addlegendentry{FaCe}

\addplot [color=mycolor1, line width=.6pt, mark size=2.5pt,  mark=text,text mark={\LARGE $\star$}, mark options={solid, fill=mycolor1, mycolor1}, forget plot]
 plot [error bars/.cd, y dir = both, y explicit]
 table[row sep=crcr, y error plus index=2, y error minus index=3]{%
1	1	0	0\\
2	1	0	0\\
3	1	0	0\\
4	1	0	0\\
5	1	0	0\\
};

\addplot[area legend, draw=none, fill=mycolor2, fill opacity=0.2, forget plot]
table[row sep=crcr] {%
x	y\\
1	1\\
2	1\\
3	1\\
4	1\\
5	1\\
5	1\\
4	1\\
3	1\\
2	1\\
1	1\\
}--cycle;
\addplot [color=mycolor2, line width=.6pt, mark size=2.5pt, mark=o, mark options={solid, mycolor2}]
  table[row sep=crcr]{%
1	1\\
2	1\\
3	1\\
4	1\\
5	1\\
};
\addlegendentry{PoP}

\addplot [color=mycolor2, line width=.6pt, mark size=2.5pt, mark=o, mark options={solid, mycolor2}, forget plot]
 plot [error bars/.cd, y dir = both, y explicit]
 table[row sep=crcr, y error plus index=2, y error minus index=3]{%
1	1	0	0\\
2	1	0	0\\
3	1	0	0\\
4	1	0	0\\
5	1	0	0\\
};

\addplot[area legend, draw=none, fill=mycolor3, fill opacity=0.2, forget plot]
table[row sep=crcr] {%
x	y\\
1	0.963364550855206\\
2	0.923278900703582\\
3	0.905759941302508\\
4	0.866523178690978\\
5	0.841857722008468\\
5	0.796922904772158\\
4	0.816830504662705\\
3	0.871677499928037\\
2	0.893929637884267\\
1	0.941574270056919\\
}--cycle;
\addplot [color=mycolor3, line width=.6pt, mark size=2.5pt, mark=asterisk, mark options={solid, mycolor3}]
  table[row sep=crcr]{%
1	0.952469410456062\\
2	0.908604269293924\\
3	0.888718720615272\\
4	0.841676841676842\\
5	0.819390313390313\\
};
\addlegendentry{GC}

\addplot [color=mycolor3, line width=.6pt, mark size=2.5pt, mark=asterisk, mark options={solid, mycolor3}, forget plot]
 plot [error bars/.cd, y dir = both, y explicit]
 table[row sep=crcr, y error plus index=2, y error minus index=3]{%
1	0.952469410456062	0.0108951403991435	0.0108951403991435\\
2	0.908604269293924	0.0146746314096577	0.0146746314096577\\
3	0.888718720615272	0.0170412206872355	0.0170412206872355\\
4	0.841676841676842	0.0248463370141364	0.0248463370141364\\
5	0.819390313390313	0.022467408618155	0.022467408618155\\
};

\addplot[area legend, draw=none, fill=mycolor4, fill opacity=0.2, forget plot]
table[row sep=crcr] {%
x	y\\
1	0.724948748203951\\
2	0.824148341693946\\
3	0.842264524656748\\
4	0.922741586339301\\
5	0.97244208466466\\
5	0.929196091973517\\
4	0.879144860547146\\
3	0.809765916166796\\
2	0.785769556499814\\
1	0.684376427917666\\
}--cycle;
\addplot [color=mycolor4, line width=.6pt, mark size=2.5pt, mark=triangle, mark options={solid, mycolor4}]
  table[row sep=crcr]{%
1	0.704662588060808\\
2	0.80495894909688\\
3	0.826015220411772\\
4	0.900943223443223\\
5	0.950819088319088\\
};
\addlegendentry{GI}

\addplot [color=mycolor4, line width=.6pt, mark size=2.5pt, mark=triangle, mark options={solid, mycolor4}, forget plot]
 plot [error bars/.cd, y dir = both, y explicit]
 table[row sep=crcr, y error plus index=2, y error minus index=3]{%
1	0.704662588060808	0.0202861601431427	0.0202861601431427\\
2	0.80495894909688	0.0191893925970657	0.0191893925970657\\
3	0.826015220411772	0.016249304244976	0.016249304244976\\
4	0.900943223443223	0.0217983628960774	0.0217983628960774\\
5	0.950819088319088	0.0216229963455713	0.0216229963455713\\
};

\addplot[area legend, draw=none, fill=mycolor5, fill opacity=0.2, forget plot]
table[row sep=crcr] {%
x	y\\
1	0.940885361192639\\
2	0.958439829359782\\
3	0.969520038248905\\
4	0.992588843131904\\
5	0.984410089016689\\
5	0.96720672010012\\
4	0.981454705911645\\
3	0.948546920335296\\
2	0.9397046698192\\
1	0.918354534988302\\
}--cycle;
\addplot [color=mycolor5, line width=.6pt, mark size=2pt, mark=square, mark options={solid, mycolor5}]
  table[row sep=crcr]{%
1	0.929619948090471\\
2	0.949072249589491\\
3	0.9590334792921\\
4	0.987021774521774\\
5	0.975808404558405\\
};
\addlegendentry{GD}

\addplot [color=mycolor5, line width=.6pt, mark size=2pt, mark=square, mark options={solid, mycolor5}, forget plot]
 plot [error bars/.cd, y dir = both, y explicit]
 table[row sep=crcr, y error plus index=2, y error minus index=3]{%
1	0.929619948090471	0.0112654131021685	0.0112654131021685\\
2	0.949072249589491	0.00936757977029108	0.00936757977029108\\
3	0.9590334792921	0.0104865589568045	0.0104865589568045\\
4	0.987021774521774	0.00556706861012995	0.00556706861012995\\
5	0.975808404558405	0.00860168445828448	0.00860168445828448\\
};

\addplot[area legend, draw=none, fill=mycolor6, fill opacity=0.2, forget plot]
table[row sep=crcr] {%
x	y\\
1	1\\
2	0.915399835960422\\
3	0.644553801195755\\
4	0.688277849596719\\
5	0.627615907745347\\
5	0.568680388550949\\
4	0.626007864688995\\
3	0.596825509149072\\
2	0.831266830706245\\
1	1\\
}--cycle;
\addplot [color=mycolor6, line width=.6pt, mark size=2.5pt, mark=+, mark options={solid, mycolor6}]
  table[row sep=crcr]{%
1	1\\
2	0.873333333333333\\
3	0.620689655172414\\
4	0.657142857142857\\
5	0.598148148148148\\
};
\addlegendentry{AF}

\addplot [color=mycolor6, line width=.6pt, mark size=2.5pt, mark=+, mark options={solid, mycolor6}, forget plot]
 plot [error bars/.cd, y dir = both, y explicit]
 table[row sep=crcr, y error plus index=2, y error minus index=3]{%
 1	1	0	0\\
2	0.873333333333333	0.0420665026270886	0.0420665026270886\\
3	0.620689655172414	0.0238641460233416	0.0238641460233416\\
4	0.657142857142857	0.031134992453862	0.031134992453862\\
5	0.598148148148148	0.0294677595971986	0.0294677595971986\\
};
\addplot[area legend, draw=none, fill=mycolor7, fill opacity=0.2, forget plot]
table[row sep=crcr] {%
x	y\\
1	1\\
2	1\\
3	1\\
4	1\\
5	1\\
5	1\\
4	1\\
3	1\\
2	1\\
1	0.996774193548387\\
};--cycle;
\addplot [color=mycolor7, line width=.6pt, mark size=2.5pt, mark=x, mark options={solid, mycolor7}]
  table[row sep=crcr]{%
1	0.998387096774194\\
2	1\\
3	1\\
4	1\\
5	1\\
};
\addlegendentry{APC}

\addplot [color=mycolor7, line width=.6pt, mark size=2.5pt, mark=x, mark options={solid, mycolor7}, forget plot]
 plot [error bars/.cd, y dir = both, y explicit]
 table[row sep=crcr, y error plus index=2, y error minus index=3]{%
1	0.998387096774194	0.00161290322580645	0.00161290322580645\\
2	1	0	0\\
3	1	0	0\\
4	1	0	0\\
5	1	0	0\\
};
\end{axis}
\end{tikzpicture}%

%% file: imgs/BICS1/AB1.tex
%
%
\definecolor{mycolor1}{rgb}{0.00000,0.44700,0.74100}%
\definecolor{mycolor2}{rgb}{0.85000,0.32500,0.09800}%
\definecolor{mycolor3}{rgb}{0.92900,0.69400,0.12500}%
\definecolor{mycolor4}{rgb}{0.49400,0.18400,0.55600}%
\definecolor{mycolor5}{rgb}{0.46600,0.67400,0.18800}%
\definecolor{mycolor6}{rgb}{0.30100,0.74500,0.93300}%
\definecolor{mycolor7}{rgb}{0.63500,0.07800,0.18400}%
\pgfplotsset{
compat=1.11,
legend image code/.code={
\draw[mark repeat=2,mark phase=2]
plot coordinates {
(0cm,0cm)
(0.15cm,0cm)        
(0.3cm,0cm)         
};%
}
}
\begin{tikzpicture}

\begin{axis}[%
width=.85\columnwidth,
height=.55\columnwidth,
scale only axis,
xmin=0.8,
xmax=5.2,
xtick={1,2,3,4,5},
xticklabel style={font=\footnotesize},
xlabel style={font=\color{white!15!black}\footnotesize,yshift=3pt},
xlabel={\# failed nodes},
ymin=0,
ymax=1.05,
ytick={0,0.2,0.4,0.6,0.8,1},
yticklabels={{0},{.2},{.4},{.6},{.8},{1}},
yticklabel style={font=\footnotesize},
ylabel style={font=\color{white!15!black}\footnotesize,yshift=-5pt},
ylabel={$a_B$},
axis background/.style={fill=white},
legend style={at={(0.1,0.03)}, legend columns=3, anchor=south west, legend cell align=left, align=left, fill=none, draw=none, font=\scriptsize}
]

\addplot[area legend, draw=none, fill=mycolor1, fill opacity=0.2, forget plot]
table[row sep=crcr] {%
x	y\\
1	1\\
2	1\\
3	1\\
4	1\\
5	1\\
5	1\\
4	1\\
3	1\\
2	1\\
1	1\\
}--cycle;
\addplot [color=mycolor1, line width=.6pt, mark size=2.5pt, mark=text,text mark={\LARGE $\star$}, mark options={solid, fill=mycolor1, mycolor1}]
  table[row sep=crcr]{%
1	1\\
2	1\\
3	1\\
4	1\\
5	1\\
};
\addlegendentry{FaCe}

\addplot [color=mycolor1, line width=.6pt, mark size=2.5pt,  mark=text,text mark={\LARGE $\star$}, mark options={solid, fill=mycolor1, mycolor1}, forget plot]
 plot [error bars/.cd, y dir = both, y explicit]
 table[row sep=crcr, y error plus index=2, y error minus index=3]{%
1	1	0	0\\
2	1	0	0\\
3	1	0	0\\
4	1	0	0\\
5	1	0	0\\
};

\addplot[area legend, draw=none, fill=mycolor2, fill opacity=0.2, forget plot]
table[row sep=crcr] {%
x	y\\
1	1\\
2	1\\
3	1\\
4	1\\
5	1\\
5	1\\
4	1\\
3	1\\
2	1\\
1	1\\
}--cycle;
\addplot [color=mycolor2, line width=.6pt, mark size=2.5pt, mark=o, mark options={solid, mycolor2}]
  table[row sep=crcr]{%
1	1\\
2	1\\
3	1\\
4	1\\
5	1\\
};
\addlegendentry{PoP}

\addplot [color=mycolor2, line width=.6pt, mark size=2.5pt, mark=o, mark options={solid, mycolor2}, forget plot]
 plot [error bars/.cd, y dir = both, y explicit]
 table[row sep=crcr, y error plus index=2, y error minus index=3]{%
1	1	0	0\\
2	1	0	0\\
3	1	0	0\\
4	1	0	0\\
5	1	0	0\\
};

\addplot[area legend, draw=none, fill=mycolor3, fill opacity=0.2, forget plot]
table[row sep=crcr] {%
x	y\\
1	0.486803398874989\\
2	0.366895593556548\\
3	0.301087619359044\\
4	0.356376577376883\\
5	0.291184098964605\\
5	0.170482567702062\\
4	0.187483071745924\\
3	0.165579047307623\\
2	0.212051774864504\\
1	0.263196601125011\\
}--cycle;
\addplot [color=mycolor3, line width=.6pt, mark size=2.5pt, mark=asterisk, mark options={solid, mycolor3}]
  table[row sep=crcr]{%
1	0.375\\
2	0.289473684210526\\
3	0.233333333333333\\
4	0.271929824561404\\
5	0.230833333333333\\
};
\addlegendentry{GC}

\addplot [color=mycolor3, line width=.6pt, mark size=2.5pt, mark=asterisk, mark options={solid, mycolor3}, forget plot]
 plot [error bars/.cd, y dir = both, y explicit]
 table[row sep=crcr, y error plus index=2, y error minus index=3]{%
1	0.375	0.111803398874989	0.111803398874989\\
2	0.289473684210526	0.0774219093460219	0.0774219093460219\\
3	0.233333333333333	0.0677542860257106	0.0677542860257106\\
4	0.271929824561404	0.0844467528154791	0.0844467528154791\\
5	0.230833333333333	0.0603507656312718	0.0603507656312718\\
};

\addplot[area legend, draw=none, fill=mycolor4, fill opacity=0.2, forget plot]
table[row sep=crcr] {%
x	y\\
1	0.677064392373896\\
2	0.76928015723813\\
3	0.707475602137247\\
4	0.855149923439341\\
5	0.885479892230759\\
5	0.766186774435907\\
4	0.732569374806273\\
3	0.559191064529419\\
2	0.599140895393449\\
1	0.447935607626104\\
}--cycle;
\addplot [color=mycolor4, line width=.6pt, mark size=2.5pt, mark=triangle, mark options={solid, mycolor4}]
  table[row sep=crcr]{%
1	0.5625\\
2	0.684210526315789\\
3	0.633333333333333\\
4	0.793859649122807\\
5	0.825833333333333\\
};
\addlegendentry{GI}

\addplot [color=mycolor4, line width=.6pt, mark size=2.5pt, mark=triangle, mark options={solid, mycolor4}, forget plot]
 plot [error bars/.cd, y dir = both, y explicit]
 table[row sep=crcr, y error plus index=2, y error minus index=3]{%
1	0.5625	0.114564392373896	0.114564392373896\\
2	0.684210526315789	0.0850696309223401	0.0850696309223401\\
3	0.633333333333333	0.0741422688039138	0.0741422688039138\\
4	0.793859649122807	0.0612902743165339	0.0612902743165339\\
5	0.825833333333333	0.0596465588974259	0.0596465588974259\\
};

\addplot[area legend, draw=none, fill=mycolor5, fill opacity=0.2, forget plot]
table[row sep=crcr] {%
x	y\\
1	0.736803398874989\\
2	0.814734668849089\\
3	0.824978067289176\\
4	0.911830032275665\\
5	0.938044295512634\\
5	0.861955704487366\\
4	0.816240143162931\\
3	0.658355266044158\\
2	0.658949541677227\\
1	0.513196601125011\\
}--cycle;
\addplot [color=mycolor5, line width=.6pt, mark size=2pt, mark=square, mark options={solid, mycolor5}]
  table[row sep=crcr]{%
1	0.625\\
2	0.736842105263158\\
3	0.741666666666667\\
4	0.864035087719298\\
5	0.9\\
};
\addlegendentry{GD}

\addplot [color=mycolor5, line width=.6pt, mark size=2pt, mark=square, mark options={solid, mycolor5}, forget plot]
 plot [error bars/.cd, y dir = both, y explicit]
 table[row sep=crcr, y error plus index=2, y error minus index=3]{%
1	0.625	0.111803398874989	0.111803398874989\\
2	0.736842105263158	0.077892563585931	0.077892563585931\\
3	0.741666666666667	0.0833114006225091	0.0833114006225091\\
4	0.864035087719298	0.047794944556367	0.047794944556367\\
5	0.9	0.0380442955126341	0.0380442955126341\\
};

\addplot[area legend, draw=none, fill=mycolor6, fill opacity=0.2, forget plot]
table[row sep=crcr] {%
x	y\\
1	1\\
2	0.984412360080584\\
3	0.712702928145234\\
4	0.614856082745121\\
5	0.542478059477961\\
5	0.497521940522039\\
4	0.560143917254879\\
3	0.653963738521433\\
2	0.915587639919416\\
1	1\\
}--cycle;
\addplot [color=mycolor6, line width=.6pt, mark size=2.5pt, mark=+, mark options={solid, mycolor6}]
  table[row sep=crcr]{%
1	1\\
2	0.95\\
3	0.683333333333333\\
4	0.5875\\
5	0.52\\
};
\addlegendentry{AF}

\addplot [color=mycolor6, line width=.6pt, mark size=2.5pt, mark=+, mark options={solid, mycolor6}, forget plot]
 plot [error bars/.cd, y dir = both, y explicit]
 table[row sep=crcr, y error plus index=2, y error minus index=3]{%
1	1	0	0\\
2	0.95	0.0344123600805843	0.0344123600805843\\
3	0.683333333333333	0.0293695948119004	0.0293695948119004\\
4	0.5875	0.0273560827451208	0.0273560827451208\\
5	0.52	0.0224780594779606	0.0224780594779606\\
};
\addplot[area legend, draw=none, fill=mycolor7, fill opacity=0.2, forget plot]
table[row sep=crcr] {%
x	y\\
1	1\\
2	1\\
3	1\\
4	1\\
5	1\\
5	1\\
4	1\\
3	1\\
2	1\\
1	0.9\\
};--cycle;
\addplot [color=mycolor7, line width=.6pt, mark size=2.5pt, mark=x, mark options={solid, mycolor7}]
  table[row sep=crcr]{%
1	0.95\\
2	1\\
3	1\\
4	1\\
5	1\\
};
\addlegendentry{APC}

\addplot [color=mycolor7, line width=.6pt, mark size=2.5pt, mark=x, mark options={solid, mycolor7}, forget plot]
 plot [error bars/.cd, y dir = both, y explicit]
 table[row sep=crcr, y error plus index=2, y error minus index=3]{%
1	0.95	0.05	0.05\\
2	1	0	0\\
3	1	0	0\\
4	1	0	0\\
5	1	0	0\\
};
\end{axis}
\end{tikzpicture}%

%% file: imgs/BICS1/R11.tex
%
%
\definecolor{mycolor1}{rgb}{0.00000,0.44700,0.74100}%
\definecolor{mycolor2}{rgb}{0.85000,0.32500,0.09800}%
\definecolor{mycolor3}{rgb}{0.92900,0.69400,0.12500}%
\definecolor{mycolor4}{rgb}{0.49400,0.18400,0.55600}%
\definecolor{mycolor5}{rgb}{0.46600,0.67400,0.18800}%
\definecolor{mycolor6}{rgb}{0.30100,0.74500,0.93300}%
\definecolor{mycolor7}{rgb}{0.63500,0.07800,0.18400}%
\pgfplotsset{
compat=1.11,
legend image code/.code={
\draw[mark repeat=2,mark phase=2]
plot coordinates {
(0cm,0cm)
(0.15cm,0cm)        
(0.3cm,0cm)         
};%
}
}
\begin{tikzpicture}

\begin{axis}[%
width=.85\columnwidth,
height=.55\columnwidth,
scale only axis,
xmin=0.8,
xmax=5.2,
xtick={1,2,3,4,5},
xticklabel style={font=\footnotesize},
xlabel style={font=\color{white!15!black}\footnotesize,yshift=3pt},
xlabel={\# failed nodes},
ymin=0,
ymax=1.05,
ytick={0,0.2,0.4,0.6,0.8,1},
yticklabels={{0},{.2},{.4},{.6},{.8},{1}},
yticklabel style={font=\footnotesize},
ylabel style={font=\color{white!15!black}\footnotesize,yshift=-6pt},
ylabel={$R_1$},
axis background/.style={fill=white},
legend style={at={(-0.02,0.03)}, nodes={scale=0.9, transform shape}, legend columns=4, anchor=south west, legend cell align=left, align=left, fill=none, draw=none, font=\scriptsize}
]

\addplot[area legend, draw=none, fill=mycolor1, fill opacity=0.2, forget plot]
table[row sep=crcr] {%
x	y\\
1	1\\
2	1\\
3	0.989608240053723\\
4	0.951282874151892\\
5	0.922941573387056\\
5	0.877058426612944\\
4	0.898717125848108\\
3	0.943725093279611\\
2	0.95\\
1	0.9\\
}--cycle;
\addplot [color=mycolor1, line width=.6pt, mark size=2.5pt, mark=text,text mark={\LARGE $\star$}, mark options={solid, fill=mycolor1, mycolor1}]
  table[row sep=crcr]{%
1	0.95\\
2	0.975\\
3	0.966666666666667\\
4	0.925\\
5	0.9\\
};
\addlegendentry{FaCe}

\addplot [color=mycolor1, line width=.6pt, mark size=2.5pt,  mark=text,text mark={\LARGE $\star$}, mark options={solid, fill=mycolor1, mycolor1}, forget plot]
 plot [error bars/.cd, y dir = both, y explicit]
 table[row sep=crcr, y error plus index=2, y error minus index=3]{%
1	0.95	0.05	0.05\\
2	0.975	0.025	0.025\\
3	0.966666666666667	0.0229415733870562	0.0229415733870562\\
4	0.925	0.0262828741518923	0.0262828741518923\\
5	0.9	0.0229415733870562	0.0229415733870562\\
};

\addplot[area legend, draw=none, fill=mycolor2, fill opacity=0.2, forget plot]
table[row sep=crcr] {%
x	y\\
1	1\\
2	1\\
3	0.989608240053723\\
4	0.951282874151892\\
5	0.922941573387056\\
5	0.877058426612944\\
4	0.898717125848108\\
3	0.943725093279611\\
2	0.95\\
1	0.9\\
}--cycle;
\addplot [color=mycolor2, line width=.6pt, mark size=2.5pt, mark=o, mark options={solid, mycolor2}]
  table[row sep=crcr]{%
1	0.95\\
2	0.975\\
3	0.966666666666667\\
4	0.925\\
5	0.9\\
};
\addlegendentry{PoP}

\addplot [color=mycolor2, line width=.6pt, mark size=2.5pt, mark=o, mark options={solid, mycolor2}, forget plot]
 plot [error bars/.cd, y dir = both, y explicit]
 table[row sep=crcr, y error plus index=2, y error minus index=3]{%
1	0.95	0.05	0.05\\
2	0.975	0.025	0.025\\
3	0.966666666666667	0.0229415733870562	0.0229415733870562\\
4	0.925	0.0262828741518923	0.0262828741518923\\
5	0.9	0.0229415733870562	0.0229415733870562\\
};

\addplot[area legend, draw=none, fill=mycolor3, fill opacity=0.2, forget plot]
table[row sep=crcr] {%
x	y\\
1	0.614707866935281\\
2	0.658311400622509\\
3	0.578201613709809\\
4	0.595637201605567\\
5	0.564485361282348\\
5	0.475514638717652\\
4	0.479362798394433\\
3	0.455131719623524\\
2	0.491688599377491\\
1	0.385292133064719\\
}--cycle;
\addplot [color=mycolor3, line width=.6pt, mark size=2.5pt, mark=asterisk, mark options={solid, mycolor3}]
  table[row sep=crcr]{%
1	0.5\\
2	0.575\\
3	0.516666666666667\\
4	0.5375\\
5	0.52\\
};
\addlegendentry{GC}

\addplot [color=mycolor3, line width=.6pt, mark size=2.5pt, mark=asterisk, mark options={solid, mycolor3}, forget plot]
 plot [error bars/.cd, y dir = both, y explicit]
 table[row sep=crcr, y error plus index=2, y error minus index=3]{%
1	0.5	0.114707866935281	0.114707866935281\\
2	0.575	0.0833114006225091	0.0833114006225091\\
3	0.516666666666667	0.0615349470431424	0.0615349470431424\\
4	0.5375	0.0581372016055668	0.0581372016055668\\
5	0.52	0.0444853612823483	0.0444853612823483\\
};

\addplot[area legend, draw=none, fill=mycolor4, fill opacity=0.2, forget plot]
table[row sep=crcr] {%
x	y\\
1	0.664132886537902\\
2	0.83506963092234\\
3	0.681090026627425\\
4	0.856460957558653\\
5	0.852118202741879\\
5	0.787881797258121\\
4	0.768539042441347\\
3	0.585576640039241\\
2	0.66493036907766\\
1	0.435867113462098\\
}--cycle;
\addplot [color=mycolor4, line width=.6pt, mark size=2.5pt, mark=triangle, mark options={solid, mycolor4}]
  table[row sep=crcr]{%
1	0.55\\
2	0.75\\
3	0.633333333333333\\
4	0.8125\\
5	0.82\\
};
\addlegendentry{GI}

\addplot [color=mycolor4, line width=.6pt, mark size=2.5pt, mark=triangle, mark options={solid, mycolor4}, forget plot]
 plot [error bars/.cd, y dir = both, y explicit]
 table[row sep=crcr, y error plus index=2, y error minus index=3]{%
1	0.55	0.114132886537902	0.114132886537902\\
2	0.75	0.0850696309223401	0.0850696309223401\\
3	0.633333333333333	0.0477566932940919	0.0477566932940919\\
4	0.8125	0.0439609575586529	0.0439609575586529\\
5	0.82	0.0321182027418787	0.0321182027418787\\
};

\addplot[area legend, draw=none, fill=mycolor5, fill opacity=0.2, forget plot]
table[row sep=crcr] {%
x	y\\
1	0.891766293548225\\
2	0.924669963389939\\
3	0.901156222146749\\
4	0.908931044626915\\
5	0.874602096615832\\
5	0.825397903384168\\
4	0.841068955373085\\
3	0.798843777853251\\
2	0.825330036610061\\
1	0.708233706451775\\
}--cycle;
\addplot [color=mycolor5, line width=.6pt, mark size=2pt, mark=square, mark options={solid, mycolor5}]
  table[row sep=crcr]{%
1	0.8\\
2	0.875\\
3	0.85\\
4	0.875\\
5	0.85\\
};
\addlegendentry{GD}

\addplot [color=mycolor5, line width=.6pt, mark size=2pt, mark=square, mark options={solid, mycolor5}, forget plot]
 plot [error bars/.cd, y dir = both, y explicit]
 table[row sep=crcr, y error plus index=2, y error minus index=3]{%
1	0.8	0.0917662935482247	0.0917662935482247\\
2	0.875	0.0496699633899391	0.0496699633899391\\
3	0.85	0.0511562221467485	0.0511562221467485\\
4	0.875	0.0339310446269148	0.0339310446269148\\
5	0.85	0.0246020966158321	0.0246020966158321\\
};

\addplot[area legend, draw=none, fill=mycolor6, fill opacity=0.2, forget plot]
table[row sep=crcr] {%
x	y\\
1	1\\
2	1\\
3	0.783113308926626\\
4	0.755886038406067\\
5	0.720693733288049\\
5	0.659306266711951\\
4	0.694113961593933\\
3	0.716886691073374\\
2	0.95\\
1	1\\
}--cycle;
\addplot [color=mycolor6, line width=.6pt, mark size=2.5pt, mark=+, mark options={solid, mycolor6}]
  table[row sep=crcr]{%
1	1\\
2	0.975\\
3	0.75\\
4	0.725\\
5	0.69\\
};
\addlegendentry{AF}

\addplot [color=mycolor6, line width=.6pt, mark size=2.5pt, mark=+, mark options={solid, mycolor6}, forget plot]
 plot [error bars/.cd, y dir = both, y explicit]
 table[row sep=crcr, y error plus index=2, y error minus index=3]{%
1	1	0	0\\
2	0.975	0.025	0.025\\
3	0.75	0.0331133089266261	0.0331133089266261\\
4	0.725	0.0308860384060671	0.0308860384060671\\
5	0.69	0.0306937332880491	0.0306937332880491\\
};
\addplot[area legend, draw=none, fill=mycolor7, fill opacity=0.2, forget plot]
table[row sep=crcr] {%
x	y\\
1	1\\
2	1\\
3	0.989608240053723\\
4	0.951282874151892\\
5	0.922941573387056\\
5	0.877058426612944\\
4	0.898717125848108\\
3	0.943725093279611\\
2	0.95\\
1	0.9\\
};--cycle;
\addplot [color=mycolor7, line width=.6pt, mark size=2.5pt, mark=x, mark options={solid, mycolor7}]
  table[row sep=crcr]{%
1	0.95\\
2	0.975\\
3	0.966666666666667\\
4	0.925\\
5	0.9\\
};
\addlegendentry{APC}

\addplot [color=mycolor7, line width=.6pt, mark size=2.5pt, mark=x, mark options={solid, mycolor7}, forget plot]
 plot [error bars/.cd, y dir = both, y explicit]
 table[row sep=crcr, y error plus index=2, y error minus index=3]{%
1	0.95	0.05	0.05\\
2	0.975	0.025	0.025\\
3	0.966666666666667	0.0229415733870562	0.0229415733870562\\
4	0.925	0.0262828741518923	0.0262828741518923\\
5	0.9	0.0229415733870562	0.0229415733870562\\
};

\addplot[area legend, draw=none, fill=black, fill opacity=0.08, forget plot]
table[row sep=crcr] {%
x	y\\
1	1\\
2	1\\
3	0.989608240053723\\
4	0.951282874151892\\
5	0.922941573387056\\
5	0.877058426612944\\
4	0.898717125848108\\
3	0.943725093279611\\
2	0.95\\
1	0.9\\
}--cycle;
\addplot [color=black, dashed, line width=.8pt]
  table[row sep=crcr]{%
1	0.95\\
2	0.975\\
3	0.966666666666667\\
4	0.925\\
5	0.9\\
};
\addlegendentry{all}

\addplot [color=black, dashed, line width=.8pt, forget plot]
 plot [error bars/.cd, y dir = both, y explicit]
 table[row sep=crcr, y error plus index=2, y error minus index=3]{%
1	0.95	0.05	0.05\\
2	0.975	0.025	0.025\\
3	0.966666666666667	0.0229415733870562	0.0229415733870562\\
4	0.925	0.0262828741518923	0.0262828741518923\\
5	0.9	0.0229415733870562	0.0229415733870562\\
};
\end{axis}
\end{tikzpicture}%

%% file: imgs/BICS1/R21.tex
%
%
\definecolor{mycolor1}{rgb}{0.00000,0.44700,0.74100}%
\definecolor{mycolor2}{rgb}{0.85000,0.32500,0.09800}%
\definecolor{mycolor3}{rgb}{0.92900,0.69400,0.12500}%
\definecolor{mycolor4}{rgb}{0.49400,0.18400,0.55600}%
\definecolor{mycolor5}{rgb}{0.46600,0.67400,0.18800}%
\definecolor{mycolor6}{rgb}{0.30100,0.74500,0.93300}%
\definecolor{mycolor7}{rgb}{0.63500,0.07800,0.18400}%
\pgfplotsset{
compat=1.11,
legend image code/.code={
\draw[mark repeat=2,mark phase=2]
plot coordinates {
(0cm,0cm)
(0.15cm,0cm)        
(0.3cm,0cm)         
};%
}
}
\begin{tikzpicture}

\begin{axis}[%
width=.85\columnwidth,
height=.55\columnwidth,
scale only axis,
xmin=0.8,
xmax=5.2,
xtick={1,2,3,4,5},
xticklabel style={font=\footnotesize},
xlabel style={font=\color{white!15!black}\footnotesize,yshift=3pt},
xlabel={\# failed nodes},
ymin=0,
ymax=1.05,
ytick={0,0.2,0.4,0.6,0.8,1},
yticklabels={{0},{.2},{.4},{.6},{.8},{1}},
yticklabel style={font=\footnotesize},
ylabel style={font=\color{white!15!black}\footnotesize,yshift=-5pt},
ylabel={$R_2$},
axis background/.style={fill=white},
legend style={at={(-0.02,0.03)}, nodes={scale=0.9, transform shape}, legend columns=4, anchor=south west, legend cell align=left, align=left, fill=none, draw=none, font=\scriptsize}
]

\addplot[area legend, draw=none, fill=mycolor1, fill opacity=0.2, forget plot]
table[row sep=crcr] {%
x	y\\
1	1\\
2	1\\
3	0.992206180040292\\
4	0.957515875003667\\
5	0.932001412881402\\
5	0.889427158547169\\
4	0.909150791663\\
3	0.957793819959708\\
2	0.966666666666667\\
1	0.95\\
}--cycle;
\addplot [color=mycolor1, line width=.6pt, mark size=2.5pt, mark=text,text mark={\LARGE $\star$}, mark options={solid, fill=mycolor1, mycolor1}]
  table[row sep=crcr]{%
1	0.975\\
2	0.983333333333333\\
3	0.975\\
4	0.933333333333333\\
5	0.910714285714286\\
};
\addlegendentry{FaCe}

\addplot [color=mycolor1, line width=.6pt, mark size=2.5pt,  mark=text,text mark={\LARGE $\star$}, mark options={solid, fill=mycolor1, mycolor1}, forget plot]
 plot [error bars/.cd, y dir = both, y explicit]
 table[row sep=crcr, y error plus index=2, y error minus index=3]{%
1	0.975	0.025	0.025\\
2	0.983333333333333	0.0166666666666667	0.0166666666666667\\
3	0.975	0.0172061800402921	0.0172061800402921\\
4	0.933333333333333	0.0241825416703337	0.0241825416703337\\
5	0.910714285714286	0.0212871271671168	0.0212871271671168\\
};

\addplot[area legend, draw=none, fill=mycolor2, fill opacity=0.2, forget plot]
table[row sep=crcr] {%
x	y\\
1	1\\
2	1\\
3	0.992206180040292\\
4	0.957515875003667\\
5	0.932001412881402\\
5	0.889427158547169\\
4	0.909150791663\\
3	0.957793819959708\\
2	0.966666666666667\\
1	0.95\\
}--cycle;
\addplot [color=mycolor2, line width=.6pt, mark size=2.5pt, mark=o, mark options={solid, mycolor2}]
  table[row sep=crcr]{%
1	0.975\\
2	0.983333333333333\\
3	0.975\\
4	0.933333333333333\\
5	0.910714285714286\\
};
\addlegendentry{PoP}

\addplot [color=mycolor2, line width=.6pt, mark size=2.5pt, mark=o, mark options={solid, mycolor2}, forget plot]
 plot [error bars/.cd, y dir = both, y explicit]
 table[row sep=crcr, y error plus index=2, y error minus index=3]{%
1	0.975	0.025	0.025\\
2	0.983333333333333	0.0166666666666667	0.0166666666666667\\
3	0.975	0.0172061800402921	0.0172061800402921\\
4	0.933333333333333	0.0241825416703337	0.0241825416703337\\
5	0.910714285714286	0.0212871271671168	0.0212871271671168\\
};

\addplot[area legend, draw=none, fill=mycolor3, fill opacity=0.2, forget plot]
table[row sep=crcr] {%
x	y\\
1	0.786582224325248\\
2	0.744350440067388\\
3	0.65169556578766\\
4	0.610411864850951\\
5	0.575944093981392\\
5	0.501035176747879\\
4	0.508481464042378\\
3	0.56093430434221\\
2	0.630887655170708\\
1	0.655084442341419\\
}--cycle;
\addplot [color=mycolor3, line width=.6pt, mark size=2.5pt, mark=asterisk, mark options={solid, mycolor3}]
  table[row sep=crcr]{%
1	0.720833333333333\\
2	0.687619047619048\\
3	0.606314935064935\\
4	0.559446664446664\\
5	0.538489635364635\\
};
\addlegendentry{GC}

\addplot [color=mycolor3, line width=.6pt, mark size=2.5pt, mark=asterisk, mark options={solid, mycolor3}, forget plot]
 plot [error bars/.cd, y dir = both, y explicit]
 table[row sep=crcr, y error plus index=2, y error minus index=3]{%
1	0.720833333333333	0.0657488909919146	0.0657488909919146\\
2	0.687619047619048	0.0567313924483401	0.0567313924483401\\
3	0.606314935064935	0.0453806307227249	0.0453806307227249\\
4	0.559446664446664	0.0509652004042861	0.0509652004042861\\
5	0.538489635364635	0.0374544586167567	0.0374544586167567\\
};

\addplot[area legend, draw=none, fill=mycolor4, fill opacity=0.2, forget plot]
table[row sep=crcr] {%
x	y\\
1	0.753152277172665\\
2	0.854031549371936\\
3	0.604255168223657\\
4	0.851525518948331\\
5	0.868453644169504\\
5	0.794641593925734\\
4	0.764982417559606\\
3	0.5053876889192\\
2	0.714857339516953\\
1	0.578514389494001\\
}--cycle;
\addplot [color=mycolor4, line width=.6pt, mark size=2.5pt, mark=triangle, mark options={solid, mycolor4}]
  table[row sep=crcr]{%
1	0.665833333333333\\
2	0.784444444444444\\
3	0.554821428571429\\
4	0.808253968253968\\
5	0.831547619047619\\
};
\addlegendentry{GI}

\addplot [color=mycolor4, line width=.6pt, mark size=2.5pt, mark=triangle, mark options={solid, mycolor4}, forget plot]
 plot [error bars/.cd, y dir = both, y explicit]
 table[row sep=crcr, y error plus index=2, y error minus index=3]{%
1	0.665833333333333	0.0873189438393318	0.0873189438393318\\
2	0.784444444444444	0.0695871049274916	0.0695871049274916\\
3	0.554821428571429	0.0494337396522284	0.0494337396522284\\
4	0.808253968253968	0.0432715506943623	0.0432715506943623\\
5	0.831547619047619	0.0369060251218845	0.0369060251218845\\
};

\addplot[area legend, draw=none, fill=mycolor5, fill opacity=0.2, forget plot]
table[row sep=crcr] {%
x	y\\
1	0.934709804486337\\
2	0.935721058884517\\
3	0.925835883222209\\
4	0.924228399609409\\
5	0.867432210186518\\
5	0.805980488226181\\
4	0.869104933723925\\
3	0.854164116777791\\
2	0.847612274448816\\
1	0.818623528846997\\
}--cycle;
\addplot [color=mycolor5, line width=.6pt, mark size=2pt, mark=square, mark options={solid, mycolor5}]
  table[row sep=crcr]{%
1	0.876666666666667\\
2	0.891666666666667\\
3	0.89\\
4	0.896666666666667\\
5	0.836706349206349\\
};
\addlegendentry{GD}

\addplot [color=mycolor5, line width=.6pt, mark size=2pt, mark=square, mark options={solid, mycolor5}, forget plot]
 plot [error bars/.cd, y dir = both, y explicit]
 table[row sep=crcr, y error plus index=2, y error minus index=3]{%
1	0.876666666666667	0.0580431378196701	0.0580431378196701\\
2	0.891666666666667	0.0440543922178505	0.0440543922178505\\
3	0.89	0.0358358832222088	0.0358358832222088\\
4	0.896666666666667	0.0275617329427418	0.0275617329427418\\
5	0.836706349206349	0.0307258609801685	0.0307258609801685\\
};

\addplot[area legend, draw=none, fill=mycolor6, fill opacity=0.2, forget plot]
table[row sep=crcr] {%
x	y\\
1	1\\
2	1\\
3	0.644336832548645\\
4	0.534714594818682\\
5	0.522409391417282\\
5	0.439255536791764\\
4	0.437014231910144\\
3	0.520663167451355\\
2	0.95\\
1	1\\
}--cycle;
\addplot [color=mycolor6, line width=.6pt, mark size=2.5pt, mark=+, mark options={solid, mycolor6}]
  table[row sep=crcr]{%
1	1\\
2	0.975\\
3	0.5825\\
4	0.485864413364413\\
5	0.480832464104523\\
};
\addlegendentry{AF}

\addplot [color=mycolor6, line width=.6pt, mark size=2.5pt, mark=+, mark options={solid, mycolor6}, forget plot]
 plot [error bars/.cd, y dir = both, y explicit]
 table[row sep=crcr, y error plus index=2, y error minus index=3]{%
1	1	0	0\\
2	0.975	0.025	0.025\\
3	0.5825	0.0618368325486447	0.0618368325486447\\
4	0.485864413364413	0.048850181454269	0.048850181454269\\
5	0.480832464104523	0.0415769273127588	0.0415769273127588\\
};
\addplot[area legend, draw=none, fill=mycolor7, fill opacity=0.2, forget plot]
table[row sep=crcr] {%
x	y\\
1	1\\
2	1\\
3	0.992206180040292\\
4	0.957515875003667\\
5	0.932001412881402\\
5	0.889427158547169\\
4	0.909150791663\\
3	0.957793819959708\\
2	0.966666666666667\\
1	0.95\\
};--cycle;
\addplot [color=mycolor7, line width=.6pt, mark size=2.5pt, mark=x, mark options={solid, mycolor7}]
  table[row sep=crcr]{%
1	0.975\\
2	0.983333333333333\\
3	0.975\\
4	0.933333333333333\\
5	0.910714285714286\\
};
\addlegendentry{APC}

\addplot [color=mycolor7, line width=.6pt, mark size=2.5pt, mark=x, mark options={solid, mycolor7}, forget plot]
 plot [error bars/.cd, y dir = both, y explicit]
 table[row sep=crcr, y error plus index=2, y error minus index=3]{%
1	0.975	0.025	0.025\\
2	0.983333333333333	0.0166666666666667	0.0166666666666667\\
3	0.975	0.0172061800402921	0.0172061800402921\\
4	0.933333333333333	0.0241825416703337	0.0241825416703337\\
5	0.910714285714286	0.0212871271671168	0.0212871271671168\\
};

\addplot[area legend, draw=none, fill=black, fill opacity=0.08, forget plot]
table[row sep=crcr] {%
x	y\\
1	1\\
2	1\\
3	0.992206180040292\\
4	0.957515875003667\\
5	0.932001412881402\\
5	0.889427158547169\\
4	0.909150791663\\
3	0.957793819959708\\
2	0.966666666666667\\
1	0.95\\
}--cycle;
\addplot [color=black, dashed, line width=.8pt]
  table[row sep=crcr]{%
1	0.975\\
2	0.983333333333333\\
3	0.975\\
4	0.933333333333333\\
5	0.910714285714286\\
};
\addlegendentry{all}

\addplot [color=black, dashed, line width=.8pt, forget plot]
 plot [error bars/.cd, y dir = both, y explicit]
 table[row sep=crcr, y error plus index=2, y error minus index=3]{%
1	0.975	0.025	0.025\\
2	0.983333333333333	0.0166666666666667	0.0166666666666667\\
3	0.975	0.0172061800402921	0.0172061800402921\\
4	0.933333333333333	0.0241825416703337	0.0241825416703337\\
5	0.910714285714286	0.0212871271671168	0.0212871271671168\\
};
\end{axis}
\end{tikzpicture}%

%% file: imgs/BICS1/BICS_ETs.tex
%
%
\definecolor{mycolor1}{rgb}{0.00000,0.44700,0.74100}%
\definecolor{mycolor2}{rgb}{0.85000,0.32500,0.09800}%
\definecolor{mycolor3}{rgb}{0.92900,0.69400,0.12500}%
\definecolor{mycolor4}{rgb}{0.49400,0.18400,0.55600}%
\definecolor{mycolor5}{rgb}{0.46600,0.67400,0.18800}%
\definecolor{mycolor6}{rgb}{0.30100,0.74500,0.93300}%
\definecolor{mycolor7}{rgb}{0.63500,0.07800,0.18400}%
\pgfplotsset{
compat=1.11,
legend image code/.code={
\draw[mark repeat=2,mark phase=2]
plot coordinates {
(0cm,0cm)
(0.15cm,0cm)        
(0.3cm,0cm)         
};%
}
}
\begin{tikzpicture}

\begin{axis}[%
width=.85\columnwidth,
height=.55\columnwidth,
scale only axis,
xmin=0.8,
xmax=5.2,
xtick={1,2,3,4,5},
xticklabel style={font=\footnotesize},
xlabel style={font=\color{white!15!black}\footnotesize,yshift=3pt},
xlabel={\# failed nodes},
ymin=0,
ymax=6,
yticklabel style={font=\footnotesize},
ylabel style={font=\color{white!15!black}\footnotesize,yshift=-5pt},
ylabel={elapsed time (s)},
axis background/.style={fill=white},
legend style={at={(0.03,0.97)}, legend columns=2, anchor=north west, legend cell align=left, align=left, fill=none, draw=none, font=\scriptsize}
]

\addplot[area legend, draw=none, fill=mycolor1, fill opacity=0.2, forget plot]
table[row sep=crcr] {%
x	y\\
1	0.1420947541716\\
2	0.18700698405005\\
3	0.241078476330981\\
4	0.340703575426742\\
5	0.440364040098706\\
5	0.354948459901294\\
4	0.268671424573258\\
3	0.196421523669019\\
2	0.15361801594995\\
1	0.1047802458284\\
}--cycle;
\addplot [color=mycolor1, line width=.6pt, mark size=2.5pt, mark=text,text mark={\LARGE $\star$}, mark options={solid, fill=mycolor1, mycolor1}]
  table[row sep=crcr]{%
1	0.1234375\\
2	0.1703125\\
3	0.21875\\
4	0.3046875\\
5	0.39765625\\
};
\addlegendentry{FaCe}

\addplot [color=mycolor1, line width=.6pt, mark size=2.5pt,  mark=text,text mark={\LARGE $\star$}, mark options={solid, fill=mycolor1, mycolor1}, forget plot]
 plot [error bars/.cd, y dir = both, y explicit]
 table[row sep=crcr, y error plus index=2, y error minus index=3]{%
1	0.1234375	0.0186572541716	0.0186572541716\\
2	0.1703125	0.0166944840500498	0.0166944840500498\\
3	0.21875	0.0223284763309805	0.0223284763309805\\
4	0.3046875	0.0360160754267416	0.0360160754267416\\
5	0.39765625	0.042707790098706	0.042707790098706\\
};

\addplot[area legend, draw=none, fill=mycolor2, fill opacity=0.2, forget plot]
table[row sep=crcr] {%
x	y\\
1	0.0682230212846097\\
2	0.139155699681428\\
3	0.330424356810408\\
4	3.11079122915026\\
5	4.93702188264539\\
5	2.19891561735461\\
4	1.02983377084974\\
3	0.202388143189592\\
2	0.0811568003185723\\
1	0.0380269787153903\\
}--cycle;
\addplot [color=mycolor2, line width=.6pt, mark size=2.5pt, mark=o, mark options={solid, mycolor2}]
  table[row sep=crcr]{%
1	0.053125\\
2	0.11015625\\
3	0.26640625\\
4	2.0703125\\
5	3.56796875\\
};
\addlegendentry{PoP}

\addplot [color=mycolor2, line width=.6pt, mark size=2.5pt, mark=o, mark options={solid, mycolor2}, forget plot]
 plot [error bars/.cd, y dir = both, y explicit]
 table[row sep=crcr, y error plus index=2, y error minus index=3]{%
1	0.053125	0.0150980212846097	0.0150980212846097\\
2	0.11015625	0.0289994496814277	0.0289994496814277\\
3	0.26640625	0.0640181068104081	0.0640181068104081\\
4	2.0703125	1.04047872915026	1.04047872915026\\
5	3.56796875	1.36905313264539	1.36905313264539\\
};

\addplot[area legend, draw=none, fill=mycolor3, fill opacity=0.2, forget plot]
table[row sep=crcr] {%
x	y\\
1	0.032794201335274\\
2	0.0248329757229554\\
3	0.0168158995136157\\
4	0.0179429444385216\\
5	0.0274452279378326\\
5	0.0194297720621674\\
4	0.0117445555614784\\
3	0.0113091004863843\\
2	0.0142295242770446\\
1	0.018768298664726\\
}--cycle;
\addplot [color=mycolor3, line width=.6pt, mark size=2.5pt, mark=asterisk, mark options={solid, mycolor3}]
  table[row sep=crcr]{%
1	0.02578125\\
2	0.01953125\\
3	0.0140625\\
4	0.01484375\\
5	0.0234375\\
};
\addlegendentry{GC}

\addplot [color=mycolor3, line width=.6pt, mark size=2.5pt, mark=asterisk, mark options={solid, mycolor3}, forget plot]
 plot [error bars/.cd, y dir = both, y explicit]
 table[row sep=crcr, y error plus index=2, y error minus index=3]{%
1	0.02578125	0.00701295133527398	0.00701295133527398\\
2	0.01953125	0.00530172572295544	0.00530172572295544\\
3	0.0140625	0.00275339951361566	0.00275339951361566\\
4	0.01484375	0.00309919443852159	0.00309919443852159\\
5	0.0234375	0.00400772793783263	0.00400772793783263\\
};

\addplot[area legend, draw=none, fill=mycolor4, fill opacity=0.2, forget plot]
table[row sep=crcr] {%
x	y\\
1	0.0539379648044094\\
2	0.0609784396419293\\
3	0.0743834694089012\\
4	0.0818145879602458\\
5	0.0968242659955434\\
5	0.0766132340044566\\
4	0.0634979120397542\\
3	0.0537415305910988\\
2	0.0515215603580707\\
1	0.0366870351955906\\
}--cycle;
\addplot [color=mycolor4, line width=.6pt, mark size=2.5pt, mark=triangle, mark options={solid, mycolor4}]
  table[row sep=crcr]{%
1	0.0453125\\
2	0.05625\\
3	0.0640625\\
4	0.07265625\\
5	0.08671875\\
};
\addlegendentry{GI}

\addplot [color=mycolor4, line width=.6pt, mark size=2.5pt, mark=triangle, mark options={solid, mycolor4}, forget plot]
 plot [error bars/.cd, y dir = both, y explicit]
 table[row sep=crcr, y error plus index=2, y error minus index=3]{%
1	0.0453125	0.0086254648044094	0.0086254648044094\\
2	0.05625	0.00472843964192929	0.00472843964192929\\
3	0.0640625	0.0103209694089012	0.0103209694089012\\
4	0.07265625	0.00915833796024578	0.00915833796024578\\
5	0.08671875	0.0101055159955434	0.0101055159955434\\
};

\addplot[area legend, draw=none, fill=mycolor5, fill opacity=0.2, forget plot]
table[row sep=crcr] {%
x	y\\
1	0.118632095130455\\
2	0.111516107620899\\
3	0.130276660315387\\
4	0.152678101497164\\
5	0.163569607790723\\
5	0.137992892209277\\
4	0.119196898502836\\
3	0.107223339684613\\
2	0.0931713923791013\\
1	0.086055404869545\\
}--cycle;
\addplot [color=mycolor5, line width=.6pt, mark size=2pt, mark=square, mark options={solid, mycolor5}]
  table[row sep=crcr]{%
1	0.10234375\\
2	0.10234375\\
3	0.11875\\
4	0.1359375\\
5	0.15078125\\
};
\addlegendentry{GD}

\addplot [color=mycolor5, line width=.6pt, mark size=2pt, mark=square, mark options={solid, mycolor5}, forget plot]
 plot [error bars/.cd, y dir = both, y explicit]
 table[row sep=crcr, y error plus index=2, y error minus index=3]{%
1	0.10234375	0.016288345130455	0.016288345130455\\
2	0.10234375	0.00917235762089867	0.00917235762089867\\
3	0.11875	0.0115266603153869	0.0115266603153869\\
4	0.1359375	0.0167406014971638	0.0167406014971638\\
5	0.15078125	0.0127883577907233	0.0127883577907233\\
};

\addplot[area legend, draw=none, fill=mycolor6, fill opacity=0.2, forget plot]
table[row sep=crcr] {%
x	y\\
1	0.779677777119643\\
2	0.85137462176717\\
3	1.03040841847818\\
4	1.03808902134551\\
5	0.742475960489823\\
5	0.568461539510177\\
4	0.774410978654489\\
3	0.794591581521817\\
2	0.63612537823283\\
1	0.462509722880357\\
}--cycle;
\addplot [color=mycolor6, line width=.6pt, mark size=2.5pt, mark=+, mark options={solid, mycolor6}]
  table[row sep=crcr]{%
1	0.62109375\\
2	0.74375\\
3	0.9125\\
4	0.90625\\
5	0.65546875\\
};
\addlegendentry{AF}

\addplot [color=mycolor6, line width=.6pt, mark size=2.5pt, mark=+, mark options={solid, mycolor6}, forget plot]
 plot [error bars/.cd, y dir = both, y explicit]
 table[row sep=crcr, y error plus index=2, y error minus index=3]{%
1	0.62109375	0.158584027119643	0.158584027119643\\
2	0.74375	0.10762462176717	0.10762462176717\\
3	0.9125	0.117908418478183	0.117908418478183\\
4	0.90625	0.131839021345511	0.131839021345511\\
5	0.65546875	0.0870072104898232	0.0870072104898232\\
};

\addplot [color=mycolor7, line width=.6pt, mark size=2.5pt, mark=x, mark options={solid, mycolor7}, forget plot]
 plot [error bars/.cd, y dir = both, y explicit]
 table[row sep=crcr, y error plus index=2, y error minus index=3]{%
1	10.2	0.17167901505579	0.17167901505579\\
2	11.05	0.16975214129337	0.16975214129337\\
3	11.85	0.232548806422267	0.232548806422267\\
4	13	0.324442842261525	0.324442842261525\\
5	14.25	0.383028101209023	0.383028101209023\\
};
\addplot[area legend, draw=none, fill=mycolor7, fill opacity=0.2, forget plot]
table[row sep=crcr] {%
x	y\\
1	0.00888414759507351\\
2	0.02925618607642\\
3	0.15554855792375\\
4	0.78187632821906\\
5	1.99870127236706\\
5	0.940361227632938\\
4	0.41812367178094\\
3	0.0866389420762501\\
2	0.01449381392358\\
1	0.00361585240492649\\
};--cycle;
\addplot [color=mycolor7, line width=.6pt, mark size=2.5pt, mark=x, mark options={solid, mycolor7}]
  table[row sep=crcr]{%
1	0.00625\\
2	0.021875\\
3	0.12109375\\
4	0.6\\
5	1.46953125\\
};
\addlegendentry{APC}

\addplot [color=mycolor7, line width=.6pt, mark size=2.5pt, mark=x, mark options={solid, mycolor7}, forget plot]
 plot [error bars/.cd, y dir = both, y explicit]
 table[row sep=crcr, y error plus index=2, y error minus index=3]{%
1	0.00625	0.00263414759507351	0.00263414759507351\\
2	0.021875	0.00738118607642002	0.00738118607642002\\
3	0.12109375	0.0344548079237499	0.0344548079237499\\
4	0.6	0.18187632821906	0.18187632821906\\
5	1.46953125	0.529170022367062	0.529170022367062\\
};
\end{axis}
\end{tikzpicture}%

%% file: imgs/BICS1/BICS_NTs.tex
%
%
\definecolor{mycolor1}{rgb}{0.00000,0.44700,0.74100}%
\definecolor{mycolor2}{rgb}{0.85000,0.32500,0.09800}%
\definecolor{mycolor3}{rgb}{0.92900,0.69400,0.12500}%
\definecolor{mycolor4}{rgb}{0.49400,0.18400,0.55600}%
\definecolor{mycolor5}{rgb}{0.46600,0.67400,0.18800}%
\definecolor{mycolor6}{rgb}{0.30100,0.74500,0.93300}%
\definecolor{mycolor7}{rgb}{0.63500,0.07800,0.18400}%

\pgfplotsset{
compat=1.11,
legend image code/.code={
\draw[mark repeat=2,mark phase=2]
plot coordinates {
(0cm,0cm)
(0.15cm,0cm)        
(0.3cm,0cm)         
};%
}
}
\begin{tikzpicture}

\begin{axis}[%
width=.85\columnwidth,
height=.55\columnwidth,
scale only axis,
xmin=0.8,
xmax=5.2,
xtick={1,2,3,4,5},
xticklabel style={font=\footnotesize},
xlabel style={font=\color{white!15!black}\footnotesize,yshift=3pt},
xlabel={\# failed nodes},
ymin=0,
ymax=20,
yticklabel style={font=\footnotesize},
ylabel style={font=\color{white!15!black}\footnotesize,yshift=-5pt},
ylabel={\# tests},
axis background/.style={fill=white},
legend style={at={(-0.02,0.03)}, nodes={scale=0.85, transform shape}, legend columns=4, anchor=south west, legend cell align=left, align=left, fill=none, draw=none, font=\scriptsize}
]

\addplot[area legend, draw=none, fill=mycolor1, fill opacity=0.2, forget plot]
table[row sep=crcr] {%
x	y\\
1	10.0605910137094\\
2	11\\
3	11.6143263533071\\
4	12.9910959328216\\
5	14.0486817278958\\
5	13.3513182721042\\
4	12.4089040671784\\
3	11.0856736466929\\
2	10.5\\
1	9.73940898629061\\
}--cycle;
\addplot [color=mycolor1, line width=.6pt, mark size=2.5pt, mark=text,text mark={\LARGE $\star$}, mark options={solid, fill=mycolor1, mycolor1}]
  table[row sep=crcr]{%
1	9.9\\
2	10.75\\
3	11.35\\
4	12.7\\
5	13.7\\
};
\addlegendentry{FaCe}

\addplot [color=mycolor1, line width=.6pt, mark size=2.5pt,  mark=text,text mark={\LARGE $\star$}, mark options={solid, fill=mycolor1, mycolor1}, forget plot]
 plot [error bars/.cd, y dir = both, y explicit]
 table[row sep=crcr, y error plus index=2, y error minus index=3]{%
1	9.9	0.160591013709393	0.160591013709393\\
2	10.75	0.25	0.25\\
3	11.35	0.264326353307103	0.264326353307103\\
4	12.7	0.291095932821575	0.291095932821575\\
5	13.7	0.348681727895829	0.348681727895829\\
};

\addplot[area legend, draw=none, fill=mycolor2, fill opacity=0.2, forget plot]
table[row sep=crcr] {%
x	y\\
1	11.4153944898227\\
2	13.5292735225183\\
3	14.9697430291575\\
4	16.9544081718848\\
5	18.5525062514531\\
5	17.4474937485469\\
4	15.8455918281152\\
3	14.1302569708425\\
2	12.8707264774817\\
1	10.7846055101773\\
}--cycle;
\addplot [color=mycolor2, line width=.6pt, mark size=2.5pt, mark=o, mark options={solid, mycolor2}]
  table[row sep=crcr]{%
1	11.1\\
2	13.2\\
3	14.55\\
4	16.4\\
5	18\\
};
\addlegendentry{PoP}

\addplot [color=mycolor2, line width=.6pt, mark size=2.5pt, mark=o, mark options={solid, mycolor2}, forget plot]
 plot [error bars/.cd, y dir = both, y explicit]
 table[row sep=crcr, y error plus index=2, y error minus index=3]{%
1	11.1	0.315394489822708	0.315394489822708\\
2	13.2	0.329273522518254	0.329273522518254\\
3	14.55	0.419743029157502	0.419743029157502\\
4	16.4	0.554408171884787	0.554408171884787\\
5	18	0.552506251453083	0.552506251453083\\
};

\addplot[area legend, draw=none, fill=mycolor3, fill opacity=0.2, forget plot]
table[row sep=crcr] {%
x	y\\
1	8.80513149660757\\
2	8.89176629354823\\
3	8.6641328865379\\
4	8.61470786693528\\
5	8.61470786693528\\
5	8.38529213306472\\
4	8.38529213306472\\
3	8.4358671134621\\
2	8.70823370645178\\
1	8.59486850339243\\
}--cycle;
\addplot [color=mycolor3, line width=.6pt, mark size=2.5pt, mark=asterisk, mark options={solid, mycolor3}]
  table[row sep=crcr]{%
1	8.7\\
2	8.8\\
3	8.55\\
4	8.5\\
5	8.5\\
};
\addlegendentry{GC}

\addplot [color=mycolor3, line width=.6pt, mark size=2.5pt, mark=asterisk, mark options={solid, mycolor3}, forget plot]
 plot [error bars/.cd, y dir = both, y explicit]
 table[row sep=crcr, y error plus index=2, y error minus index=3]{%
1	8.7	0.105131496607569	0.105131496607569\\
2	8.8	0.0917662935482247	0.0917662935482247\\
3	8.55	0.114132886537902	0.114132886537902\\
4	8.5	0.114707866935281	0.114707866935281\\
5	8.5	0.114707866935281	0.114707866935281\\
};

\addplot[area legend, draw=none, fill=mycolor4, fill opacity=0.2, forget plot]
table[row sep=crcr] {%
x	y\\
1	11.4153944898227\\
2	13.5292735225183\\
3	14.9697430291575\\
4	16.9544081718848\\
5	18.5525062514531\\
5	17.4474937485469\\
4	15.8455918281152\\
3	14.1302569708425\\
2	12.8707264774817\\
1	10.7846055101773\\
}--cycle;
\addplot [color=mycolor4, line width=.6pt, mark size=2.5pt, mark=triangle, mark options={solid, mycolor4}]
  table[row sep=crcr]{%
1	11.1\\
2	13.2\\
3	14.55\\
4	16.4\\
5	18\\
};
\addlegendentry{GI}

\addplot [color=mycolor4, line width=.6pt, mark size=2.5pt, mark=triangle, mark options={solid, mycolor4}, forget plot]
 plot [error bars/.cd, y dir = both, y explicit]
 table[row sep=crcr, y error plus index=2, y error minus index=3]{%
1	11.1	0.315394489822708	0.315394489822708\\
2	13.2	0.329273522518254	0.329273522518254\\
3	14.55	0.419743029157502	0.419743029157502\\
4	16.4	0.554408171884787	0.554408171884787\\
5	18	0.552506251453083	0.552506251453083\\
};

\addplot[area legend, draw=none, fill=mycolor5, fill opacity=0.2, forget plot]
table[row sep=crcr] {%
x	y\\
1	11.4153944898227\\
2	13.5292735225183\\
3	14.9697430291575\\
4	16.9544081718848\\
5	18.5525062514531\\
5	17.4474937485469\\
4	15.8455918281152\\
3	14.1302569708425\\
2	12.8707264774817\\
1	10.7846055101773\\
}--cycle;
\addplot [color=mycolor5, line width=.6pt, mark size=2pt, mark=square, mark options={solid, mycolor5}]
  table[row sep=crcr]{%
1	11.1\\
2	13.2\\
3	14.55\\
4	16.4\\
5	18\\
};
\addlegendentry{GD}

\addplot [color=mycolor5, line width=.6pt, mark size=2pt, mark=square, mark options={solid, mycolor5}, forget plot]
 plot [error bars/.cd, y dir = both, y explicit]
 table[row sep=crcr, y error plus index=2, y error minus index=3]{%
1	11.1	0.315394489822708	0.315394489822708\\
2	13.2	0.329273522518254	0.329273522518254\\
3	14.55	0.419743029157502	0.419743029157502\\
4	16.4	0.554408171884787	0.554408171884787\\
5	18	0.552506251453083	0.552506251453083\\
};

\addplot[area legend, draw=none, fill=mycolor6, fill opacity=0.2, forget plot]
table[row sep=crcr] {%
x	y\\
1	7.23191780219091\\
2	12.7611995813393\\
3	14.9697430291575\\
4	16.9544081718848\\
5	18.5525062514531\\
5	17.4474937485469\\
4	15.8455918281152\\
3	14.1302569708425\\
2	12.3388004186607\\
1	7.06808219780909\\
}--cycle;
\addplot [color=mycolor6, line width=.6pt, mark size=2.5pt, mark=+, mark options={solid, mycolor6}]
  table[row sep=crcr]{%
1	7.15\\
2	12.55\\
3	14.55\\
4	16.4\\
5	18\\
};
\addlegendentry{AF}

\addplot [color=mycolor6, line width=.6pt, mark size=2.5pt, mark=+, mark options={solid, mycolor6}, forget plot]
 plot [error bars/.cd, y dir = both, y explicit]
 table[row sep=crcr, y error plus index=2, y error minus index=3]{%
1	7.15	0.0819178021909125	0.0819178021909125\\
2	12.55	0.211199581339298	0.211199581339298\\
3	14.55	0.419743029157502	0.419743029157502\\
4	16.4	0.554408171884787	0.554408171884787\\
5	18	0.552506251453083	0.552506251453083\\
};
\addplot[area legend, draw=none, fill=mycolor7, fill opacity=0.2, forget plot]
table[row sep=crcr] {%
x	y\\
1	10.145095250022\\
2	11.145095250022\\
3	12.0586926725569\\
4	12.6957534065727\\
5	13.7026201370835\\
5	13.0973798629165\\
4	12.2042465934273\\
3	11.6413073274431\\
2	10.854904749978\\
1	9.854904749978\\
};--cycle;
\addplot [color=mycolor7, line width=.6pt, mark size=2.5pt, mark=x, mark options={solid, mycolor7}]
  table[row sep=crcr]{%
1	10\\
2	11\\
3	11.85\\
4	12.45\\
5	13.4\\
};
\addlegendentry{APC}

\addplot [color=mycolor7, line width=.6pt, mark size=2.5pt, mark=x, mark options={solid, mycolor7}, forget plot]
 plot [error bars/.cd, y dir = both, y explicit]
 table[row sep=crcr, y error plus index=2, y error minus index=3]{%
1	10	0.145095250022002	0.145095250022002\\
2	11	0.145095250022002	0.145095250022002\\
3	11.85	0.208692672556914	0.208692672556914\\
4	12.45	0.245753406572738	0.245753406572738\\
5	13.4	0.302620137083475	0.302620137083475\\
};
\end{axis}
\end{tikzpicture}%

%% file: imgs/BICS2_it/AW2it.tex
%
%
\definecolor{mycolor1}{rgb}{0.00000,0.44700,0.74100}%
\definecolor{mycolor2}{rgb}{0.85000,0.32500,0.09800}%
\definecolor{mycolor3}{rgb}{0.92900,0.69400,0.12500}%
\definecolor{mycolor4}{rgb}{0.49400,0.18400,0.55600}%
\definecolor{mycolor5}{rgb}{0.46600,0.67400,0.18800}%
\definecolor{mycolor6}{rgb}{0.30100,0.74500,0.93300}%
\definecolor{mycolor7}{rgb}{0.63500,0.07800,0.18400}%
\pgfplotsset{
compat=1.11,
legend image code/.code={
\draw[mark repeat=2,mark phase=2]
plot coordinates {
(0cm,0cm)
(0.15cm,0cm)        
(0.3cm,0cm)         
};%
}
}
\begin{tikzpicture}

\begin{axis}[%
width=.85\columnwidth,
height=.55\columnwidth,
scale only axis,
xmin=0,
xmax=56,
xtick={1,5,10,15,20,25,30,35,40,45,50,55},
xticklabels={{1},{5},{},{15},{},{25},{},{35},{},{45},{},{55}},
xticklabel style={font=\footnotesize},
xlabel style={font=\color{white!15!black}\footnotesize,yshift=3pt},
xlabel={\# tested paths},
ymin=0,
ymax=1.05,
ytick={0,0.2,0.4,0.6,0.8,1},
yticklabels={{0},{.2},{.4},{.6},{.8},{1}},
yticklabel style={font=\footnotesize},
ylabel style={font=\color{white!15!black}\footnotesize,yshift=-5pt},
ylabel={$a_W$},
axis background/.style={fill=white},
legend style={at={(0.4,0.1)}, legend columns=2, anchor=south west, legend cell align=left, align=left, fill=none, draw=none, font=\scriptsize}
]

\addplot [color=mycolor1, line width=.6pt, mark size=1.5pt, mark=text,text mark={\large $\star$}, mark options={solid, fill=mycolor1, mycolor1}]
  table[row sep=crcr]{%
1	0\\
2	0.201373626373626\\
3	0.246016483516484\\
4	0.380413105413105\\
5	0.603072853072853\\
6	0.657432844932845\\
7	0.723600936100936\\
8	0.807961945461946\\
9	0.880718355718356\\
10	0.914417989417989\\
11	0.930962555962556\\
12	0.949562474562475\\
13	0.977587844254511\\
14	0.98989898989899\\
15	1\\
};
\addlegendentry{FaCe}

\addplot [color=mycolor2, line width=.6pt, mark size=1.5pt, mark=o, mark options={solid, mycolor2}]
  table[row sep=crcr]{%
1	0\\
2	0.201373626373626\\
3	0.351694139194139\\
4	0.454639804639805\\
5	0.551358363858364\\
6	0.623504273504274\\
7	0.696072446072446\\
8	0.732926332926333\\
9	0.78474765974766\\
10	0.808699633699634\\
11	0.862988400488401\\
12	0.906227106227106\\
13	0.916163901458019\\
14	0.924892771046617\\
15	0.948301698301698\\
16	0.949175824175824\\
17	0.984615384615385\\
18	1\\
19	1\\
20	1\\
21	1\\
22	1\\
23	1\\
};
\addlegendentry{PoP}

\addplot [color=mycolor3, line width=.6pt, mark size=1.5pt, mark=asterisk, mark options={solid, mycolor3}]
  table[row sep=crcr]{%
1	0\\
2	0.235393772893773\\
3	0.328250915750916\\
4	0.516391941391941\\
5	0.594912494912495\\
6	0.647476597476597\\
7	0.719459706959707\\
8	0.782885632885633\\
9	0.831283068783069\\
10	0.837032676318391\\
11	0.892195767195767\\
};
\addlegendentry{GC}

\addplot [color=mycolor4, line width=.6pt, mark size=1.5pt, mark=triangle, mark options={solid, mycolor4}]
  table[row sep=crcr]{%
1	0\\
2	0.143961131461131\\
3	0.229675417175417\\
4	0.323707773707774\\
5	0.398514448514449\\
6	0.477050264550265\\
7	0.532631257631258\\
8	0.586925111925112\\
9	0.644790394790395\\
10	0.70450752950753\\
11	0.741646316646317\\
12	0.761818274318274\\
13	0.791956654456655\\
14	0.827380952380953\\
15	0.866570004070004\\
16	0.877894790394791\\
17	0.915226902726903\\
18	0.928332315832316\\
19	0.939382376882377\\
20	0.941305453805454\\
21	0.95242673992674\\
22	0.963542938542939\\
23	0.974526862026862\\
24	0.981801994301994\\
25	0.983587708587708\\
26	0.98543956043956\\
27	0.989285714285714\\
28	0.989285714285714\\
29	0.989285714285714\\
30	0.989285714285714\\
31	0.989285714285714\\
32	0.989285714285714\\
33	0.989285714285714\\
34	0.989285714285714\\
35	0.989285714285714\\
36	0.992857142857143\\
37	0.992857142857143\\
38	0.992857142857143\\
39	0.992857142857143\\
40	0.992857142857143\\
41	0.992857142857143\\
42	0.992857142857143\\
43	0.992857142857143\\
44	0.992857142857143\\
45	0.992857142857143\\
46	0.992857142857143\\
47	0.992857142857143\\
48	0.992857142857143\\
49	0.992857142857143\\
50	1\\
51	1\\
52	1\\
53	1\\
54	1\\
55	1\\
};
\addlegendentry{GI}

\addplot [color=mycolor5, line width=.6pt, mark size=1.2pt, mark=square, mark options={solid, mycolor5}]
  table[row sep=crcr]{%
1	0\\
2	0.150254375254375\\
3	0.336747049247049\\
4	0.431827431827432\\
5	0.506435693935694\\
6	0.571494708994709\\
7	0.660719373219373\\
8	0.735617623117623\\
9	0.78970288970289\\
10	0.836655474155474\\
11	0.862454212454212\\
12	0.907346357346358\\
13	0.931781644281644\\
14	0.950178062678063\\
15	0.968432030932031\\
16	0.977696377696378\\
17	0.977696377696378\\
18	0.979619454619455\\
19	0.981471306471307\\
20	0.981471306471307\\
21	0.986894586894587\\
22	0.986894586894587\\
23	0.992592592592593\\
24	0.992592592592593\\
25	0.992592592592593\\
26	0.992592592592593\\
27	0.998148148148148\\
28	1\\
29	1\\
30	1\\
31	1\\
32	1\\
33	1\\
34	1\\
35	1\\
36	1\\
37	1\\
38	1\\
39	1\\
40	1\\
41	1\\
42	1\\
43	1\\
44	1\\
45	1\\
46	1\\
47	1\\
48	1\\
49	1\\
50	1\\
51	1\\
52	1\\
53	1\\
54	1\\
55	1\\
};
\addlegendentry{GD}

\addplot [color=mycolor6, line width=.6pt, mark size=1.5pt, mark=+, mark options={solid, mycolor6}]
  table[row sep=crcr]{%
1	0\\
2	0.0571428571428571\\
3	0.2\\
4	0.235714285714286\\
5	0.253571428571429\\
6	0.266071428571429\\
7	0.266071428571429\\
8	0.323214285714286\\
9	0.4\\
10	0.433928571428571\\
11	0.464285714285714\\
12	0.478571428571429\\
13	0.480357142857143\\
14	0.5375\\
15	0.614285714285714\\
16	0.632142857142857\\
17	0.667857142857143\\
18	0.685714285714286\\
19	0.694642857142857\\
20	0.7125\\
21	0.766071428571429\\
22	0.796428571428571\\
23	0.81203007518797\\
24	0.827067669172932\\
25	0.845982142857143\\
26	0.825\\
27	0.9\\
28	0.964285714285714\\
};
\addlegendentry{AF}

\addplot [color=mycolor7, line width=.6pt, mark size=1.5pt, mark=x, mark options={solid, mycolor7}]
  table[row sep=crcr]{%
1	0.201373626373626\\
2	0.246016483516484\\
3	0.36969881969882\\
4	0.500045787545788\\
5	0.544240944240944\\
6	0.665593203093203\\
7	0.701424501424501\\
8	0.729283679283679\\
9	0.791269841269841\\
10	0.876093813593814\\
11	0.89249592999593\\
12	0.962092434314657\\
13	0.973361823361823\\
14	0.987179487179487\\
};
\addlegendentry{APC}

\end{axis}
\end{tikzpicture}%

%% file: imgs/BICS2_it/AB2it.tex
%
%
\definecolor{mycolor1}{rgb}{0.00000,0.44700,0.74100}%
\definecolor{mycolor2}{rgb}{0.85000,0.32500,0.09800}%
\definecolor{mycolor3}{rgb}{0.92900,0.69400,0.12500}%
\definecolor{mycolor4}{rgb}{0.49400,0.18400,0.55600}%
\definecolor{mycolor5}{rgb}{0.46600,0.67400,0.18800}%
\definecolor{mycolor6}{rgb}{0.30100,0.74500,0.93300}%
\definecolor{mycolor7}{rgb}{0.63500,0.07800,0.18400}%
\pgfplotsset{
compat=1.11,
legend image code/.code={
\draw[mark repeat=2,mark phase=2]
plot coordinates {
(0cm,0cm)
(0.15cm,0cm)        
(0.3cm,0cm)         
};%
}
}
\begin{tikzpicture}

\begin{axis}[%
width=.85\columnwidth,
height=.55\columnwidth,
scale only axis,
xmin=0,
xmax=56,
xtick={1,5,10,15,20,25,30,35,40,45,50,55},
xticklabels={{1},{5},{},{15},{},{25},{},{35},{},{45},{},{55}},
xticklabel style={font=\footnotesize},
xlabel style={font=\color{white!15!black}\footnotesize,yshift=3pt},
xlabel={\# tested paths},
ymin=0,
ymax=1.05,
ytick={0,0.2,0.4,0.6,0.8,1},
yticklabels={{0},{.2},{.4},{.6},{.8},{1}},
yticklabel style={font=\footnotesize},
ylabel style={font=\color{white!15!black}\footnotesize,yshift=-5pt},
ylabel={$a_B$},
axis background/.style={fill=white},
legend style={at={(0.4,0.1)}, legend columns=2, anchor=south west, legend cell align=left, align=left, fill=none, draw=none, font=\scriptsize}
]

\addplot [color=mycolor1, line width=.6pt, mark size=1.5pt, mark=text,text mark={\large $\star$}, mark options={solid, fill=mycolor1, mycolor1}]
  table[row sep=crcr]{%
1	0\\
2	0\\
3	0\\
4	0\\
5	0\\
6	0\\
7	0.0131578947368421\\
8	0.0263157894736842\\
9	0.0263157894736842\\
10	0.144736842105263\\
11	0.337719298245614\\
12	0.592105263157895\\
13	0.863095238095238\\
14	1\\
15	1\\
};
\addlegendentry{FaCe}

\addplot [color=mycolor2, line width=.6pt, mark size=1.5pt, mark=o, mark options={solid, mycolor2}]
  table[row sep=crcr]{%
1	0\\
2	0\\
3	0\\
4	0\\
5	0\\
6	0\\
7	0\\
8	0\\
9	0\\
10	0.0666666666666667\\
11	0.188888888888889\\
12	0.333333333333333\\
13	0.538461538461538\\
14	0.633333333333333\\
15	0.614583333333333\\
16	0.5\\
17	0.5\\
18	0.8\\
19	1\\
20	1\\
21	1\\
22	1\\
23	1\\
};
\addlegendentry{PoP}

\addplot [color=mycolor3, line width=.6pt, mark size=1.5pt, mark=asterisk, mark options={solid, mycolor3}]
  table[row sep=crcr]{%
1	0\\
2	0\\
3	0\\
4	0\\
5	0\\
6	0\\
7	0\\
8	0.0277777777777778\\
9	0.305555555555556\\
10	0.488095238095238\\
11	0.5\\
};
\addlegendentry{GC}

\addplot [color=mycolor4, line width=.6pt, mark size=1.5pt, mark=triangle, mark options={solid, mycolor4}]
  table[row sep=crcr]{%
1	0\\
2	0\\
3	0\\
4	0\\
5	0.0131578947368421\\
6	0.0570175438596491\\
7	0.0921052631578947\\
8	0.105263157894737\\
9	0.140350877192982\\
10	0.140350877192982\\
11	0.197368421052632\\
12	0.320175438596491\\
13	0.429824561403509\\
14	0.521929824561403\\
15	0.614035087719298\\
16	0.684210526315789\\
17	0.758771929824561\\
18	0.798245614035088\\
19	0.824561403508772\\
20	0.824561403508772\\
21	0.824561403508772\\
22	0.824561403508772\\
23	0.87719298245614\\
24	0.87719298245614\\
25	0.929824561403509\\
26	0.947368421052632\\
27	0.947368421052632\\
28	0.947368421052632\\
29	0.947368421052632\\
30	0.947368421052632\\
31	0.947368421052632\\
32	0.947368421052632\\
33	0.947368421052632\\
34	0.947368421052632\\
35	0.947368421052632\\
36	0.973684210526316\\
37	1\\
38	1\\
39	1\\
40	1\\
41	1\\
42	1\\
43	1\\
44	1\\
45	1\\
46	1\\
47	1\\
48	1\\
49	1\\
50	1\\
51	1\\
52	1\\
53	1\\
54	1\\
55	1\\
};
\addlegendentry{GI}

\addplot [color=mycolor5, line width=.6pt, mark size=1.2pt, mark=square, mark options={solid, mycolor5}]
  table[row sep=crcr]{%
1	0\\
2	0\\
3	0\\
4	0\\
5	0\\
6	0\\
7	0.0175438596491228\\
8	0.043859649122807\\
9	0.0570175438596491\\
10	0.162280701754386\\
11	0.276315789473684\\
12	0.350877192982456\\
13	0.464912280701754\\
14	0.578947368421053\\
15	0.675438596491228\\
16	0.692982456140351\\
17	0.706140350877193\\
18	0.741228070175439\\
19	0.758771929824561\\
20	0.776315789473684\\
21	0.793859649122807\\
22	0.833333333333333\\
23	0.833333333333333\\
24	0.833333333333333\\
25	0.846491228070175\\
26	0.916666666666667\\
27	0.916666666666667\\
28	0.934210526315789\\
29	0.934210526315789\\
30	0.934210526315789\\
31	0.969298245614035\\
32	0.986842105263158\\
33	0.986842105263158\\
34	1\\
35	1\\
36	1\\
37	1\\
38	1\\
39	1\\
40	1\\
41	1\\
42	1\\
43	1\\
44	1\\
45	1\\
46	1\\
47	1\\
48	1\\
49	1\\
50	1\\
51	1\\
52	1\\
53	1\\
54	1\\
55	1\\
};
\addlegendentry{GD}

\addplot [color=mycolor6, line width=.6pt, mark size=1.5pt, mark=+, mark options={solid, mycolor6}]
  table[row sep=crcr]{%
1	0\\
2	0\\
3	0\\
4	0\\
5	0\\
6	0.25\\
7	0.25\\
8	0.25\\
9	0.25\\
10	0.2875\\
11	0.325\\
12	0.4875\\
13	0.5\\
14	0.5\\
15	0.5\\
16	0.5125\\
17	0.55\\
18	0.6875\\
19	0.725\\
20	0.75\\
21	0.7625\\
22	0.8\\
23	0.868421052631579\\
24	0.921052631578947\\
25	0.953125\\
26	1\\
27	1\\
28	1\\
};
\addlegendentry{AF}

\addplot [color=mycolor7, line width=.6pt, mark size=1.5pt, mark=x, mark options={solid, mycolor7}]
  table[row sep=crcr]{%
1	0\\
2	0\\
3	0\\
4	0\\
5	0\\
6	0\\
7	0\\
8	0.109649122807018\\
9	0.171052631578947\\
10	0.37719298245614\\
11	0.407894736842105\\
12	0.862745098039216\\
13	0.916666666666667\\
14	1\\
};
\addlegendentry{APC}

\end{axis}
\end{tikzpicture}%

%% file: imgs/BICS2_it/R12it.tex
%
%
\definecolor{mycolor1}{rgb}{0.00000,0.44700,0.74100}%
\definecolor{mycolor2}{rgb}{0.85000,0.32500,0.09800}%
\definecolor{mycolor3}{rgb}{0.92900,0.69400,0.12500}%
\definecolor{mycolor4}{rgb}{0.49400,0.18400,0.55600}%
\definecolor{mycolor5}{rgb}{0.46600,0.67400,0.18800}%
\definecolor{mycolor6}{rgb}{0.30100,0.74500,0.93300}%
\definecolor{mycolor7}{rgb}{0.63500,0.07800,0.18400}%
\pgfplotsset{
compat=1.11,
legend image code/.code={
\draw[mark repeat=2,mark phase=2]
plot coordinates {
(0cm,0cm)
(0.15cm,0cm)        
(0.3cm,0cm)         
};%
}
}
\begin{tikzpicture}

\begin{axis}[%
width=.85\columnwidth,
height=.55\columnwidth,
scale only axis,
xmin=0,
xmax=56,
xtick={1,5,10,15,20,25,30,35,40,45,50,55},
xticklabels={{1},{5},{},{15},{},{25},{},{35},{},{45},{},{55}},
xticklabel style={font=\footnotesize},
xlabel style={font=\color{white!15!black}\footnotesize,yshift=3pt},
xlabel={\# tested paths},
ymin=0,
ymax=1.05,
ytick={0,0.2,0.4,0.6,0.8,1},
yticklabels={{0},{.2},{.4},{.6},{.8},{1}},
yticklabel style={font=\footnotesize},
ylabel style={font=\color{white!15!black}\footnotesize,yshift=-5pt},
ylabel={$R_1$},
axis background/.style={fill=white},
legend style={at={(0.4,0.1)}, legend columns=2,, anchor=south west, legend cell align=left, align=left, fill=none, draw=none, font=\scriptsize}
]

\addplot [color=mycolor1, line width=.6pt, mark size=1.5pt, mark=text,text mark={\large $\star$}, mark options={solid, fill=mycolor1, mycolor1}]
  table[row sep=crcr]{%
1	0.1875\\
2	0.325\\
3	0.0625\\
4	0.0875\\
5	0.3125\\
6	0.3875\\
7	0.4875\\
8	0.5375\\
9	0.575\\
10	0.6\\
11	0.7\\
12	0.7625\\
13	0.783333333333333\\
14	0.863636363636364\\
15	0.916666666666667\\
};
\addlegendentry{FaCe}

\addplot [color=mycolor2, line width=.6pt, mark size=1.5pt, mark=o, mark options={solid, mycolor2}]
  table[row sep=crcr]{%
1	0.1875\\
2	0.325\\
3	0.2375\\
4	0.2125\\
5	0.2625\\
6	0.275\\
7	0.35\\
8	0.375\\
9	0.35\\
10	0.4625\\
11	0.575\\
12	0.65\\
13	0.661764705882353\\
14	0.711538461538462\\
15	0.795454545454545\\
16	0.75\\
17	0.9\\
18	0.9\\
19	1\\
20	1\\
21	1\\
22	1\\
23	1\\
};
\addlegendentry{PoP}

\addplot [color=mycolor3, line width=.6pt, mark size=1.5pt, mark=asterisk, mark options={solid, mycolor3}]
  table[row sep=crcr]{%
1	0.1875\\
2	0.1625\\
3	0.0625\\
4	0.0625\\
5	0.1875\\
6	0.325\\
7	0.3875\\
8	0.425\\
9	0.4875\\
10	0.517857142857143\\
11	0.75\\
};
\addlegendentry{GC}

\addplot [color=mycolor4, line width=.6pt, mark size=1.5pt, mark=triangle, mark options={solid, mycolor4}]
  table[row sep=crcr]{%
1	0.1875\\
2	0.1875\\
3	0.175\\
4	0.175\\
5	0.2125\\
6	0.3\\
7	0.3125\\
8	0.3875\\
9	0.4\\
10	0.4\\
11	0.4375\\
12	0.475\\
13	0.5375\\
14	0.6\\
15	0.65\\
16	0.6875\\
17	0.75\\
18	0.775\\
19	0.7875\\
20	0.7875\\
21	0.8375\\
22	0.8375\\
23	0.85\\
24	0.8625\\
25	0.8625\\
26	0.8875\\
27	0.8875\\
28	0.8875\\
29	0.8875\\
30	0.8875\\
31	0.8875\\
32	0.8875\\
33	0.8875\\
34	0.8875\\
35	0.8875\\
36	0.9\\
37	0.9125\\
38	0.9125\\
39	0.9125\\
40	0.9125\\
41	0.9125\\
42	0.9125\\
43	0.9125\\
44	0.9125\\
45	0.9125\\
46	0.9125\\
47	0.9125\\
48	0.9\\
49	0.9\\
50	0.9\\
51	0.9\\
52	0.9\\
53	0.9125\\
54	0.9125\\
55	0.9125\\
};
\addlegendentry{GI}

\addplot [color=mycolor5, line width=.6pt, mark size=1.2pt, mark=square, mark options={solid, mycolor5}]
  table[row sep=crcr]{%
1	0.1875\\
2	0.225\\
3	0.3125\\
4	0.3375\\
5	0.3375\\
6	0.375\\
7	0.3875\\
8	0.4375\\
9	0.525\\
10	0.5875\\
11	0.6125\\
12	0.65\\
13	0.7\\
14	0.725\\
15	0.8\\
16	0.8125\\
17	0.8125\\
18	0.8375\\
19	0.85\\
20	0.85\\
21	0.875\\
22	0.875\\
23	0.9\\
24	0.9\\
25	0.9\\
26	0.8875\\
27	0.9\\
28	0.9125\\
29	0.9125\\
30	0.9125\\
31	0.9125\\
32	0.9125\\
33	0.9125\\
34	0.9125\\
35	0.9125\\
36	0.9125\\
37	0.9125\\
38	0.9125\\
39	0.9125\\
40	0.9125\\
41	0.9125\\
42	0.9125\\
43	0.9125\\
44	0.9125\\
45	0.9125\\
46	0.9125\\
47	0.9125\\
48	0.9125\\
49	0.9125\\
50	0.9125\\
51	0.9125\\
52	0.9125\\
53	0.9125\\
54	0.9125\\
55	0.9125\\
};
\addlegendentry{GD}

\addplot [color=mycolor6, line width=.6pt, mark size=1.5pt, mark=+, mark options={solid, mycolor6}]
  table[row sep=crcr]{%
1	0.1875\\
2	0.125\\
3	0.2125\\
4	0.3125\\
5	0.425\\
6	0.3625\\
7	0.3625\\
8	0.4\\
9	0.3875\\
10	0.5\\
11	0.575\\
12	0.625\\
13	0.625\\
14	0.6375\\
15	0.675\\
16	0.725\\
17	0.7375\\
18	0.8125\\
19	0.8\\
20	0.8125\\
21	0.85\\
22	0.875\\
23	0.921052631578947\\
24	0.960526315789474\\
25	0.984375\\
26	1\\
27	1\\
28	1\\
};
\addlegendentry{AF}

\addplot [color=mycolor7, line width=.6pt, mark size=1.5pt, mark=x, mark options={solid, mycolor7}]
  table[row sep=crcr]{%
1	0.325\\
2	0.0625\\
3	0.0875\\
4	0.125\\
5	0.2875\\
6	0.325\\
7	0.275\\
8	0.35\\
9	0.4875\\
10	0.5875\\
11	0.5875\\
12	0.819444444444444\\
13	0.825\\
14	0.916666666666667\\
};
\addlegendentry{APC}

\end{axis}
\end{tikzpicture}%

%% file: imgs/BICS2_it/R22it.tex
%
%
\definecolor{mycolor1}{rgb}{0.00000,0.44700,0.74100}%
\definecolor{mycolor2}{rgb}{0.85000,0.32500,0.09800}%
\definecolor{mycolor3}{rgb}{0.92900,0.69400,0.12500}%
\definecolor{mycolor4}{rgb}{0.49400,0.18400,0.55600}%
\definecolor{mycolor5}{rgb}{0.46600,0.67400,0.18800}%
\definecolor{mycolor6}{rgb}{0.30100,0.74500,0.93300}%
\definecolor{mycolor7}{rgb}{0.63500,0.07800,0.18400}%
\pgfplotsset{
compat=1.11,
legend image code/.code={
\draw[mark repeat=2,mark phase=2]
plot coordinates {
(0cm,0cm)
(0.15cm,0cm)        
(0.3cm,0cm)         
};%
}
}
\begin{tikzpicture}

\begin{axis}[%
width=.85\columnwidth,
height=.55\columnwidth,
scale only axis,
xmin=0,
xmax=56,
xtick={1,5,10,15,20,25,30,35,40,45,50,55},
xticklabels={{1},{5},{},{15},{},{25},{},{35},{},{45},{},{55}},
xticklabel style={font=\footnotesize},
xlabel style={font=\color{white!15!black}\footnotesize,yshift=3pt},
xlabel={\# tested paths},
ymin=0,
ymax=1.05,
ytick={0,0.2,0.4,0.6,0.8,1},
yticklabels={{0},{.2},{.4},{.6},{.8},{1}},
yticklabel style={font=\footnotesize},
ylabel style={font=\color{white!15!black}\footnotesize,yshift=-5pt},
ylabel={$R_1$},
axis background/.style={fill=white},
legend style={at={(0.4,0.1)}, legend columns=2, anchor=south west, legend cell align=left, align=left, fill=none, draw=none, font=\scriptsize}
]

\addplot [color=mycolor1, line width=.6pt, mark size=1.5pt, mark=text,text mark={\large $\star$}, mark options={solid, fill=mycolor1, mycolor1}]
  table[row sep=crcr]{%
1	0.15625\\
2	0.166320166320166\\
3	0.186046511627907\\
4	0.212201591511936\\
5	0.298507462686567\\
6	0.346320346320346\\
7	0.398009950248756\\
8	0.451977401129944\\
9	0.526315789473684\\
10	0.601503759398496\\
11	0.64\\
12	0.727272727272727\\
13	0.8\\
14	0.862745098039216\\
15	0.923076923076923\\
};
\addlegendentry{FaCe}

\addplot [color=mycolor2, line width=.6pt, mark size=1.5pt, mark=o, mark options={solid, mycolor2}]
  table[row sep=crcr]{%
1	0.15625\\
2	0.166320166320166\\
3	0.201005025125628\\
4	0.224719101123595\\
5	0.25974025974026\\
6	0.296296296296296\\
7	0.336134453781513\\
8	0.368663594470046\\
9	0.43010752688172\\
10	0.457142857142857\\
11	0.555555555555556\\
12	0.634920634920635\\
13	0.647619047619048\\
14	0.712328767123288\\
15	0.8\\
16	0.8\\
17	0.909090909090909\\
18	0.909090909090909\\
19	1\\
20	1\\
21	1\\
22	1\\
23	1\\
};
\addlegendentry{PoP}

\addplot [color=mycolor3, line width=.6pt, mark size=1.5pt, mark=asterisk, mark options={solid, mycolor3}]
  table[row sep=crcr]{%
1	0.15625\\
2	0.198019801980198\\
3	0.208333333333333\\
4	0.256410256410256\\
5	0.282685512367491\\
6	0.303030303030303\\
7	0.355555555555556\\
8	0.41025641025641\\
9	0.465116279069767\\
10	0.482758620689655\\
11	0.727272727272727\\
};
\addlegendentry{GC}

\addplot [color=mycolor4, line width=.6pt, mark size=1.5pt, mark=triangle, mark options={solid, mycolor4}]
  table[row sep=crcr]{%
1	0.15625\\
2	0.173160173160173\\
3	0.190930787589499\\
4	0.21505376344086\\
5	0.238095238095238\\
6	0.258064516129032\\
7	0.284697508896797\\
8	0.313725490196078\\
9	0.347826086956522\\
10	0.392156862745098\\
11	0.427807486631016\\
12	0.441988950276243\\
13	0.473372781065089\\
14	0.509554140127389\\
15	0.571428571428571\\
16	0.601503759398496\\
17	0.683760683760684\\
18	0.727272727272727\\
19	0.747663551401869\\
20	0.747663551401869\\
21	0.792079207920792\\
22	0.808080808080808\\
23	0.824742268041237\\
24	0.860215053763441\\
25	0.869565217391304\\
26	0.898876404494382\\
27	0.898876404494382\\
28	0.898876404494382\\
29	0.898876404494382\\
30	0.898876404494382\\
31	0.898876404494382\\
32	0.898876404494382\\
33	0.898876404494382\\
34	0.898876404494382\\
35	0.898876404494382\\
36	0.909090909090909\\
37	0.919540229885058\\
38	0.919540229885058\\
39	0.919540229885058\\
40	0.919540229885058\\
41	0.919540229885058\\
42	0.919540229885058\\
43	0.919540229885058\\
44	0.919540229885058\\
45	0.919540229885058\\
46	0.919540229885058\\
47	0.919540229885058\\
48	0.909090909090909\\
49	0.909090909090909\\
50	0.909090909090909\\
51	0.909090909090909\\
52	0.909090909090909\\
53	0.919540229885058\\
54	0.919540229885058\\
55	0.919540229885058\\
};
\addlegendentry{GI}

\addplot [color=mycolor5, line width=.6pt, mark size=1.2pt, mark=square, mark options={solid, mycolor5}]
  table[row sep=crcr]{%
1	0.15625\\
2	0.175824175824176\\
3	0.191846522781775\\
4	0.227272727272727\\
5	0.25\\
6	0.272108843537415\\
7	0.321285140562249\\
8	0.363636363636364\\
9	0.421052631578947\\
10	0.481927710843373\\
11	0.547945205479452\\
12	0.606060606060606\\
13	0.650406504065041\\
14	0.720720720720721\\
15	0.8\\
16	0.824742268041237\\
17	0.824742268041237\\
18	0.833333333333333\\
19	0.851063829787234\\
20	0.851063829787234\\
21	0.869565217391304\\
22	0.869565217391304\\
23	0.898876404494382\\
24	0.898876404494382\\
25	0.898876404494382\\
26	0.888888888888889\\
27	0.909090909090909\\
28	0.919540229885058\\
29	0.919540229885058\\
30	0.919540229885058\\
31	0.919540229885058\\
32	0.919540229885058\\
33	0.919540229885058\\
34	0.919540229885058\\
35	0.919540229885058\\
36	0.919540229885058\\
37	0.919540229885058\\
38	0.919540229885058\\
39	0.919540229885058\\
40	0.919540229885058\\
41	0.919540229885058\\
42	0.919540229885058\\
43	0.919540229885058\\
44	0.919540229885058\\
45	0.919540229885058\\
46	0.919540229885058\\
47	0.919540229885058\\
48	0.919540229885058\\
49	0.919540229885058\\
50	0.919540229885058\\
51	0.919540229885058\\
52	0.919540229885058\\
53	0.919540229885058\\
54	0.919540229885058\\
55	0.919540229885058\\
};
\addlegendentry{GD}

\addplot [color=mycolor6, line width=.6pt, mark size=1.5pt, mark=+, mark options={solid, mycolor6}]
  table[row sep=crcr]{%
1	0.15625\\
2	0.143626570915619\\
3	0.175824175824176\\
4	0.196078431372549\\
5	0.191387559808612\\
6	0.187353629976581\\
7	0.187353629976581\\
8	0.202020202020202\\
9	0.229885057471264\\
10	0.282685512367491\\
11	0.299625468164794\\
12	0.323886639676113\\
13	0.32\\
14	0.321285140562249\\
15	0.37037037037037\\
16	0.404040404040404\\
17	0.446927374301676\\
18	0.467836257309941\\
19	0.493827160493827\\
20	0.50314465408805\\
21	0.592592592592593\\
22	0.695652173913043\\
23	0.844444444444445\\
24	0.938271604938272\\
25	0.984615384615385\\
26	1\\
27	1\\
28	1\\
};
\addlegendentry{AF}

\addplot [color=mycolor7, line width=.6pt, mark size=1.5pt, mark=x, mark options={solid, mycolor7}]
  table[row sep=crcr]{%
1	0.166320166320166\\
2	0.186046511627907\\
3	0.209424083769633\\
4	0.24024024024024\\
5	0.264900662251656\\
6	0.31496062992126\\
7	0.333333333333333\\
8	0.380952380952381\\
9	0.444444444444444\\
10	0.567375886524823\\
11	0.610687022900763\\
12	0.774193548387097\\
13	0.8\\
14	0.857142857142857\\
};
\addlegendentry{APC}

\end{axis}
\end{tikzpicture}%

%% file: imgs/BICS2/AW2.tex
%
%
\definecolor{mycolor1}{rgb}{0.00000,0.44700,0.74100}%
\definecolor{mycolor2}{rgb}{0.85000,0.32500,0.09800}%
\definecolor{mycolor3}{rgb}{0.92900,0.69400,0.12500}%
\definecolor{mycolor4}{rgb}{0.49400,0.18400,0.55600}%
\definecolor{mycolor5}{rgb}{0.46600,0.67400,0.18800}%
\definecolor{mycolor6}{rgb}{0.30100,0.74500,0.93300}%
\definecolor{mycolor7}{rgb}{0.63500,0.07800,0.18400}%
\pgfplotsset{
compat=1.11,
legend image code/.code={
\draw[mark repeat=2,mark phase=2]
plot coordinates {
(0cm,0cm)
(0.15cm,0cm)        
(0.3cm,0cm)         
};%
}
}
\begin{tikzpicture}

\begin{axis}[%
width=.85\columnwidth,
height=.55\columnwidth,
scale only axis,
xmin=0.8,
xmax=5.2,
xtick={1,2,3,4,5},
xticklabel style={font=\footnotesize},
xlabel style={font=\color{white!15!black}\footnotesize,yshift=3pt},
xlabel={\# failed nodes},
ymin=0,
ymax=1.05,
ytick={0,0.2,0.4,0.6,0.8,1},
yticklabel style={font=\footnotesize},
yticklabels={{0},{.2},{.4},{.6},{.8},{1}},
ylabel style={font=\color{white!15!black}\footnotesize,yshift=-5pt},
ylabel={$a_W$},
axis background/.style={fill=white},
legend style={at={(0.1,0.03)}, legend columns=3, anchor=south west, legend cell align=left, align=left, fill=none, draw=none, font=\scriptsize}
]

\addplot[area legend, draw=none, fill=mycolor1, fill opacity=0.2, forget plot]
table[row sep=crcr] {%
x	y\\
1	1\\
2	1\\
3	1\\
4	1\\
5	1\\
5	1\\
4	1\\
3	1\\
2	1\\
1	1\\
}--cycle;
\addplot [color=mycolor1, line width=.6pt, mark size=2.5pt, mark=text,text mark={\LARGE $\star$}, mark options={solid, fill=mycolor1, mycolor1}]
  table[row sep=crcr]{%
1	1\\
2	1\\
3	1\\
4	1\\
5	1\\
};
\addlegendentry{FaCe}

\addplot [color=mycolor1, line width=.6pt, mark size=2.5pt,  mark=text,text mark={\LARGE $\star$}, mark options={solid, fill=mycolor1, mycolor1}, forget plot]
 plot [error bars/.cd, y dir = both, y explicit]
 table[row sep=crcr, y error plus index=2, y error minus index=3]{%
1	1	0	0\\
2	1	0	0\\
3	1	0	0\\
4	1	0	0\\
5	1	0	0\\
};

\addplot[area legend, draw=none, fill=mycolor2, fill opacity=0.2, forget plot]
table[row sep=crcr] {%
x	y\\
1	1\\
2	1\\
3	1\\
4	1\\
5	1\\
5	1\\
4	1\\
3	1\\
2	1\\
1	1\\
}--cycle;
\addplot [color=mycolor2, line width=.6pt, mark size=2.5pt, mark=o, mark options={solid, mycolor2}]
  table[row sep=crcr]{%
1	1\\
2	1\\
3	1\\
4	1\\
5	1\\
};
\addlegendentry{PoP}

\addplot [color=mycolor2, line width=.6pt, mark size=2.5pt, mark=o, mark options={solid, mycolor2}, forget plot]
 plot [error bars/.cd, y dir = both, y explicit]
 table[row sep=crcr, y error plus index=2, y error minus index=3]{%
1	1	0	0\\
2	1	0	0\\
3	1	0	0\\
4	1	0	0\\
5	1	0	0\\
};

\addplot[area legend, draw=none, fill=mycolor3, fill opacity=0.2, forget plot]
table[row sep=crcr] {%
x	y\\
1	0.963486593310036\\
2	0.928807572104814\\
3	0.880762925992816\\
4	0.884501261246666\\
5	0.832932585102056\\
5	0.778751175581704\\
4	0.852553276807873\\
3	0.841612468727406\\
2	0.896502590274313\\
1	0.93800395174002\\
}--cycle;
\addplot [color=mycolor3, line width=.6pt, mark size=2.5pt, mark=asterisk, mark options={solid, mycolor3}]
  table[row sep=crcr]{%
1	0.950745272525028\\
2	0.912655081189564\\
3	0.861187697360111\\
4	0.868527269027269\\
5	0.80584188034188\\
};
\addlegendentry{GC}

\addplot [color=mycolor3, line width=.6pt, mark size=2.5pt, mark=asterisk, mark options={solid, mycolor3}, forget plot]
 plot [error bars/.cd, y dir = both, y explicit]
 table[row sep=crcr, y error plus index=2, y error minus index=3]{%
1	0.950745272525028	0.0127413207850081	0.0127413207850081\\
2	0.912655081189564	0.0161524909152505	0.0161524909152505\\
3	0.861187697360111	0.019575228632705	0.019575228632705\\
4	0.868527269027269	0.0159739922193964	0.0159739922193964\\
5	0.80584188034188	0.027090704760176	0.027090704760176\\
};

\addplot[area legend, draw=none, fill=mycolor4, fill opacity=0.2, forget plot]
table[row sep=crcr] {%
x	y\\
1	1\\
2	1\\
3	1\\
4	1\\
5	1\\
5	1\\
4	1\\
3	1\\
2	1\\
1	1\\
}--cycle;
\addplot [color=mycolor4, line width=.6pt, mark size=2.5pt, mark=triangle, mark options={solid, mycolor4}]
  table[row sep=crcr]{%
1	1\\
2	1\\
3	1\\
4	1\\
5	1\\
};
\addlegendentry{GI}

\addplot [color=mycolor4, line width=.6pt, mark size=2.5pt, mark=triangle, mark options={solid, mycolor4}, forget plot]
 plot [error bars/.cd, y dir = both, y explicit]
 table[row sep=crcr, y error plus index=2, y error minus index=3]{%
1	1	0	0\\
2	1	0	0\\
3	1	0	0\\
4	1	0	0\\
5	1	0	0\\
};

\addplot[area legend, draw=none, fill=mycolor5, fill opacity=0.2, forget plot]
table[row sep=crcr] {%
x	y\\
1	1\\
2	1\\
3	1\\
4	1\\
5	1\\
5	1\\
4	1\\
3	1\\
2	1\\
1	1\\
}--cycle;
\addplot [color=mycolor5, line width=.6pt, mark size=2pt, mark=square, mark options={solid, mycolor5}]
  table[row sep=crcr]{%
1	1\\
2	1\\
3	1\\
4	1\\
5	1\\
};
\addlegendentry{GD}

\addplot [color=mycolor5, line width=.6pt, mark size=2pt, mark=square, mark options={solid, mycolor5}, forget plot]
 plot [error bars/.cd, y dir = both, y explicit]
 table[row sep=crcr, y error plus index=2, y error minus index=3]{%
1	1	0	0\\
2	1	0	0\\
3	1	0	0\\
4	1	0	0\\
5	1	0	0\\
};

\addplot[area legend, draw=none, fill=mycolor6, fill opacity=0.2, forget plot]
table[row sep=crcr] {%
x	y\\
1	1\\
2	1\\
3	1\\
4	1\\
5	1\\
5	1\\
4	1\\
3	1\\
2	1\\
1	1\\
}--cycle;
\addplot [color=mycolor6, line width=.6pt, mark size=2.5pt, mark=+, mark options={solid, mycolor6}]
  table[row sep=crcr]{%
1	1\\
2	1\\
3	1\\
4	1\\
5	1\\
};
\addlegendentry{AF}

\addplot [color=mycolor6, line width=.6pt, mark size=2.5pt, mark=+, mark options={solid, mycolor6}, forget plot]
 plot [error bars/.cd, y dir = both, y explicit]
 table[row sep=crcr, y error plus index=2, y error minus index=3]{%
1	1	0	0\\
2	1	0	0\\
3	1	0	0\\
4	1	0	0\\
5	1	0	0\\
};
\addplot[area legend, draw=none, fill=mycolor7, fill opacity=0.2, forget plot]
table[row sep=crcr] {%
x	y\\
1	0.997803800070675\\
2	1\\
3	0.998867116206388\\
4	0.992874695974286\\
5	0.981981441066331\\
5	0.964839071754182\\
4	0.979894534794945\\
3	0.99383845408539\\
2	0.996666666666667\\
1	0.992518780574487\\
};--cycle;
\addplot [color=mycolor7, line width=.6pt, mark size=2.5pt, mark=x, mark options={solid, mycolor7}]
  table[row sep=crcr]{%
1	0.995161290322581\\
2	0.998333333333333\\
3	0.996352785145889\\
4	0.986384615384615\\
5	0.973410256410256\\
};
\addlegendentry{APC}

\addplot [color=mycolor7, line width=.6pt, mark size=2.5pt, mark=x, mark options={solid, mycolor7}, forget plot]
 plot [error bars/.cd, y dir = both, y explicit]
 table[row sep=crcr, y error plus index=2, y error minus index=3]{%
1	0.995161290322581	0.00264250974809395	0.00264250974809395\\
2	0.998333333333333	0.00166666666666667	0.00166666666666667\\
3	0.996352785145889	0.00251433106049896	0.00251433106049896\\
4	0.986384615384615	0.00649008058967047	0.00649008058967047\\
5	0.973410256410256	0.00857118465607446	0.00857118465607446\\
};
\end{axis}
\end{tikzpicture}%

%% file: imgs/BICS2/AB2.tex
%
%
\definecolor{mycolor1}{rgb}{0.00000,0.44700,0.74100}%
\definecolor{mycolor2}{rgb}{0.85000,0.32500,0.09800}%
\definecolor{mycolor3}{rgb}{0.92900,0.69400,0.12500}%
\definecolor{mycolor4}{rgb}{0.49400,0.18400,0.55600}%
\definecolor{mycolor5}{rgb}{0.46600,0.67400,0.18800}%
\definecolor{mycolor6}{rgb}{0.30100,0.74500,0.93300}%
\definecolor{mycolor7}{rgb}{0.63500,0.07800,0.18400}%
\pgfplotsset{
compat=1.11,
legend image code/.code={
\draw[mark repeat=2,mark phase=2]
plot coordinates {
(0cm,0cm)
(0.15cm,0cm)        
(0.3cm,0cm)         
};%
}
}
\begin{tikzpicture}

\begin{axis}[%
width=.85\columnwidth,
height=.55\columnwidth,
scale only axis,
xmin=0.8,
xmax=5.2,
xtick={1,2,3,4,5},
xticklabel style={font=\footnotesize},
xlabel style={font=\color{white!15!black}\footnotesize,yshift=3pt},
xlabel={\# failed nodes},
ymin=0,
ymax=1.05,
ytick={0,0.2,0.4,0.6,0.8,1},
yticklabels={{0},{.2},{.4},{.6},{.8},{1}},
yticklabel style={font=\footnotesize},
ylabel style={font=\color{white!15!black}\footnotesize,yshift=-5pt},
ylabel={$a_B$},
axis background/.style={fill=white},
legend style={at={(0.1,0.33)}, legend columns=3, anchor=south west, legend cell align=left, align=left, fill=none, draw=none, font=\scriptsize}
]

\addplot[area legend, draw=none, fill=mycolor1, fill opacity=0.2, forget plot]
table[row sep=crcr] {%
x	y\\
1	1\\
2	1\\
3	1\\
4	1\\
5	1\\
5	1\\
4	1\\
3	1\\
2	1\\
1	1\\
}--cycle;
\addplot [color=mycolor1, line width=.6pt, mark size=2.5pt, mark=text,text mark={\LARGE $\star$}, mark options={solid, fill=mycolor1, mycolor1}]
  table[row sep=crcr]{%
1	1\\
2	1\\
3	1\\
4	1\\
5	1\\
};
\addlegendentry{FaCe}

\addplot [color=mycolor1, line width=.6pt, mark size=2.5pt,  mark=text,text mark={\LARGE $\star$}, mark options={solid, fill=mycolor1, mycolor1}, forget plot]
 plot [error bars/.cd, y dir = both, y explicit]
 table[row sep=crcr, y error plus index=2, y error minus index=3]{%
1	1	0	0\\
2	1	0	0\\
3	1	0	0\\
4	1	0	0\\
5	1	0	0\\
};

\addplot[area legend, draw=none, fill=mycolor2, fill opacity=0.2, forget plot]
table[row sep=crcr] {%
x	y\\
1	1\\
2	1\\
3	1\\
4	1\\
5	1\\
5	1\\
4	1\\
3	1\\
2	1\\
1	1\\
}--cycle;
\addplot [color=mycolor2, line width=.6pt, mark size=2.5pt, mark=o, mark options={solid, mycolor2}]
  table[row sep=crcr]{%
1	1\\
2	1\\
3	1\\
4	1\\
5	1\\
};
\addlegendentry{PoP}

\addplot [color=mycolor2, line width=.6pt, mark size=2.5pt, mark=o, mark options={solid, mycolor2}, forget plot]
 plot [error bars/.cd, y dir = both, y explicit]
 table[row sep=crcr, y error plus index=2, y error minus index=3]{%
1	1	0	0\\
2	1	0	0\\
3	1	0	0\\
4	1	0	0\\
5	1	0	0\\
};

\addplot[area legend, draw=none, fill=mycolor3, fill opacity=0.2, forget plot]
table[row sep=crcr] {%
x	y\\
1	0.333063353756439\\
2	0.447860691039001\\
3	0.347660180287567\\
4	0.361073700278842\\
5	0.191556897050173\\
5	0.0917764362831607\\
4	0.235417527791333\\
3	0.213743328484363\\
2	0.236349835276789\\
1	0.137524881537678\\
}--cycle;
\addplot [color=mycolor3, line width=.6pt, mark size=2.5pt, mark=asterisk, mark options={solid, mycolor3}]
  table[row sep=crcr]{%
1	0.235294117647059\\
2	0.342105263157895\\
3	0.280701754385965\\
4	0.298245614035088\\
5	0.141666666666667\\
};
\addlegendentry{GC}

\addplot [color=mycolor3, line width=.6pt, mark size=2.5pt, mark=asterisk, mark options={solid, mycolor3}, forget plot]
 plot [error bars/.cd, y dir = both, y explicit]
 table[row sep=crcr, y error plus index=2, y error minus index=3]{%
1	0.235294117647059	0.0977692361093803	0.0977692361093803\\
2	0.342105263157895	0.105755427881106	0.105755427881106\\
3	0.280701754385965	0.0669584259016016	0.0669584259016016\\
4	0.298245614035088	0.0628280862437543	0.0628280862437543\\
5	0.141666666666667	0.0498902303835059	0.0498902303835059\\
};

\addplot[area legend, draw=none, fill=mycolor4, fill opacity=0.2, forget plot]
table[row sep=crcr] {%
x	y\\
1	1\\
2	1\\
3	1\\
4	1\\
5	1\\
5	1\\
4	1\\
3	1\\
2	1\\
1	1\\
}--cycle;
\addplot [color=mycolor4, line width=.6pt, mark size=2.5pt, mark=triangle, mark options={solid, mycolor4}]
  table[row sep=crcr]{%
1	1\\
2	1\\
3	1\\
4	1\\
5	1\\
};
\addlegendentry{GI}

\addplot [color=mycolor4, line width=.6pt, mark size=2.5pt, mark=triangle, mark options={solid, mycolor4}, forget plot]
 plot [error bars/.cd, y dir = both, y explicit]
 table[row sep=crcr, y error plus index=2, y error minus index=3]{%
1	1	0	0\\
2	1	0	0\\
3	1	0	0\\
4	1	0	0\\
5	1	0	0\\
};

\addplot[area legend, draw=none, fill=mycolor5, fill opacity=0.2, forget plot]
table[row sep=crcr] {%
x	y\\
1	1\\
2	1\\
3	1\\
4	1\\
5	1\\
5	1\\
4	1\\
3	1\\
2	1\\
1	1\\
}--cycle;
\addplot [color=mycolor5, line width=.6pt, mark size=2pt, mark=square, mark options={solid, mycolor5}]
  table[row sep=crcr]{%
1	1\\
2	1\\
3	1\\
4	1\\
5	1\\
};
\addlegendentry{GD}

\addplot [color=mycolor5, line width=.6pt, mark size=2pt, mark=square, mark options={solid, mycolor5}, forget plot]
 plot [error bars/.cd, y dir = both, y explicit]
 table[row sep=crcr, y error plus index=2, y error minus index=3]{%
1	1	0	0\\
2	1	0	0\\
3	1	0	0\\
4	1	0	0\\
5	1	0	0\\
};

\addplot[area legend, draw=none, fill=mycolor6, fill opacity=0.2, forget plot]
table[row sep=crcr] {%
x	y\\
1	1\\
2	1\\
3	1\\
4	1\\
5	1\\
5	1\\
4	1\\
3	1\\
2	1\\
1	1\\
}--cycle;
\addplot [color=mycolor6, line width=.6pt, mark size=2.5pt, mark=+, mark options={solid, mycolor6}]
  table[row sep=crcr]{%
1	1\\
2	1\\
3	1\\
4	1\\
5	1\\
};
\addlegendentry{AF}

\addplot [color=mycolor6, line width=.6pt, mark size=2.5pt, mark=+, mark options={solid, mycolor6}, forget plot]
 plot [error bars/.cd, y dir = both, y explicit]
 table[row sep=crcr, y error plus index=2, y error minus index=3]{%
1	1	0	0\\
2	1	0	0\\
3	1	0	0\\
4	1	0	0\\
5	1	0	0\\
};
\addplot[area legend, draw=none, fill=mycolor7, fill opacity=0.2, forget plot]
table[row sep=crcr] {%
x	y\\
1	1\\
2	1\\
3	1\\
4	1\\
5	1\\
5	1\\
4	1\\
3	1\\
2	1\\
1	1\\
};--cycle;
\addplot [color=mycolor7, line width=.6pt, mark size=2.5pt, mark=x, mark options={solid, mycolor7}]
  table[row sep=crcr]{%
1	1\\
2	1\\
3	1\\
4	1\\
5	1\\
};
\addlegendentry{APC}

\addplot [color=mycolor7, line width=.6pt, mark size=2.5pt, mark=x, mark options={solid, mycolor7}, forget plot]
 plot [error bars/.cd, y dir = both, y explicit]
 table[row sep=crcr, y error plus index=2, y error minus index=3]{%
1	1	0	0\\
2	1	0	0\\
3	1	0	0\\
4	1	0	0\\
5	1	0	0\\
};
\end{axis}
\end{tikzpicture}%

%% file: imgs/BICS2/R12.tex
%
%
\definecolor{mycolor1}{rgb}{0.00000,0.44700,0.74100}%
\definecolor{mycolor2}{rgb}{0.85000,0.32500,0.09800}%
\definecolor{mycolor3}{rgb}{0.92900,0.69400,0.12500}%
\definecolor{mycolor4}{rgb}{0.49400,0.18400,0.55600}%
\definecolor{mycolor5}{rgb}{0.46600,0.67400,0.18800}%
\definecolor{mycolor6}{rgb}{0.30100,0.74500,0.93300}%
\definecolor{mycolor7}{rgb}{0.63500,0.07800,0.18400}%
\pgfplotsset{
compat=1.11,
legend image code/.code={
\draw[mark repeat=2,mark phase=2]
plot coordinates {
(0cm,0cm)
(0.15cm,0cm)        
(0.3cm,0cm)         
};%
}
}
\begin{tikzpicture}

\begin{axis}[%
width=.85\columnwidth,
height=.55\columnwidth,
scale only axis,
xmin=0.8,
xmax=5.2,
xtick={1,2,3,4,5},
xticklabel style={font=\footnotesize},
xlabel style={font=\color{white!15!black}\footnotesize,yshift=3pt},
xlabel={\# failed nodes},
ymin=0,
ymax=1.05,
ytick={0,0.2,0.4,0.6,0.8,1},
yticklabels={{0},{.2},{.4},{.6},{.8},{1}},
yticklabel style={font=\footnotesize},
ylabel style={font=\color{white!15!black}\footnotesize,yshift=-7pt},
ylabel={$R_1$},
axis background/.style={fill=white},
legend style={at={(0.1,0.0)}, legend columns=3, anchor=south west, legend cell align=left, align=left, fill=none, draw=none, font=\scriptsize}
]

\addplot[area legend, draw=none, fill=mycolor1, fill opacity=0.2, forget plot]
table[row sep=crcr] {%
x	y\\
1	1\\
2	0.879712165490242\\
3	0.844590403603996\\
4	0.865958901095456\\
5	0.882036163775859\\
5	0.817963836224141\\
4	0.784041098904544\\
3	0.755409596396004\\
2	0.770287834509758\\
1	0.9\\
}--cycle;
\addplot [color=mycolor1, line width=.6pt, mark size=2.5pt, mark=text,text mark={\LARGE $\star$}, mark options={solid, fill=mycolor1, mycolor1}]
  table[row sep=crcr]{%
1	0.95\\
2	0.825\\
3	0.8\\
4	0.825\\
5	0.85\\
};
\addlegendentry{FaCe}

\addplot [color=mycolor1, line width=.6pt, mark size=2.5pt,  mark=text,text mark={\LARGE $\star$}, mark options={solid, fill=mycolor1, mycolor1}, forget plot]
 plot [error bars/.cd, y dir = both, y explicit]
 table[row sep=crcr, y error plus index=2, y error minus index=3]{%
1	0.95	0.05	0.05\\
2	0.825	0.0547121654902416	0.0547121654902416\\
3	0.8	0.0445904036039959	0.0445904036039959\\
4	0.825	0.0409589010954563	0.0409589010954563\\
5	0.85	0.0320361637758594	0.0320361637758594\\
};

\addplot[area legend, draw=none, fill=mycolor2, fill opacity=0.2, forget plot]
table[row sep=crcr] {%
x	y\\
1	1\\
2	0.879712165490242\\
3	0.844590403603996\\
4	0.865958901095456\\
5	0.87111946420862\\
5	0.80888053579138\\
4	0.784041098904544\\
3	0.755409596396004\\
2	0.770287834509758\\
1	0.9\\
}--cycle;
\addplot [color=mycolor2, line width=.6pt, mark size=2.5pt, mark=o, mark options={solid, mycolor2}]
  table[row sep=crcr]{%
1	0.95\\
2	0.825\\
3	0.8\\
4	0.825\\
5	0.84\\
};
\addlegendentry{PoP}

\addplot [color=mycolor2, line width=.6pt, mark size=2.5pt, mark=o, mark options={solid, mycolor2}, forget plot]
 plot [error bars/.cd, y dir = both, y explicit]
 table[row sep=crcr, y error plus index=2, y error minus index=3]{%
1	0.95	0.05	0.05\\
2	0.825	0.0547121654902416	0.0547121654902416\\
3	0.8	0.0445904036039959	0.0445904036039959\\
4	0.825	0.0409589010954563	0.0409589010954563\\
5	0.84	0.03111946420862	0.03111946420862\\
};

\addplot[area legend, draw=none, fill=mycolor3, fill opacity=0.2, forget plot]
table[row sep=crcr] {%
x	y\\
1	0.614707866935281\\
2	0.64064457728035\\
3	0.527844195698816\\
4	0.507932859659525\\
5	0.488534740022261\\
5	0.371465259977739\\
4	0.417067140340475\\
3	0.405489137634517\\
2	0.50935542271965\\
1	0.385292133064719\\
}--cycle;
\addplot [color=mycolor3, line width=.6pt, mark size=2.5pt, mark=asterisk, mark options={solid, mycolor3}]
  table[row sep=crcr]{%
1	0.5\\
2	0.575\\
3	0.466666666666667\\
4	0.4625\\
5	0.43\\
};
\addlegendentry{GC}

\addplot [color=mycolor3, line width=.6pt, mark size=2.5pt, mark=asterisk, mark options={solid, mycolor3}, forget plot]
 plot [error bars/.cd, y dir = both, y explicit]
 table[row sep=crcr, y error plus index=2, y error minus index=3]{%
1	0.5	0.114707866935281	0.114707866935281\\
2	0.575	0.0656445772803496	0.0656445772803496\\
3	0.466666666666667	0.0611775290321498	0.0611775290321498\\
4	0.4625	0.0454328596595251	0.0454328596595251\\
5	0.43	0.058534740022261	0.058534740022261\\
};

\addplot[area legend, draw=none, fill=mycolor4, fill opacity=0.2, forget plot]
table[row sep=crcr] {%
x	y\\
1	1\\
2	0.879712165490242\\
3	0.844590403603996\\
4	0.865958901095456\\
5	0.87111946420862\\
5	0.80888053579138\\
4	0.784041098904544\\
3	0.755409596396004\\
2	0.770287834509758\\
1	0.9\\
}--cycle;
\addplot [color=mycolor4, line width=.6pt, mark size=2.5pt, mark=triangle, mark options={solid, mycolor4}]
  table[row sep=crcr]{%
1	0.95\\
2	0.825\\
3	0.8\\
4	0.825\\
5	0.84\\
};
\addlegendentry{GI}

\addplot [color=mycolor4, line width=.6pt, mark size=2.5pt, mark=triangle, mark options={solid, mycolor4}, forget plot]
 plot [error bars/.cd, y dir = both, y explicit]
 table[row sep=crcr, y error plus index=2, y error minus index=3]{%
1	0.95	0.05	0.05\\
2	0.825	0.0547121654902416	0.0547121654902416\\
3	0.8	0.0445904036039959	0.0445904036039959\\
4	0.825	0.0409589010954563	0.0409589010954563\\
5	0.84	0.03111946420862	0.03111946420862\\
};

\addplot[area legend, draw=none, fill=mycolor5, fill opacity=0.2, forget plot]
table[row sep=crcr] {%
x	y\\
1	1\\
2	0.879712165490242\\
3	0.844590403603996\\
4	0.865958901095456\\
5	0.87111946420862\\
5	0.80888053579138\\
4	0.784041098904544\\
3	0.755409596396004\\
2	0.770287834509758\\
1	0.9\\
}--cycle;
\addplot [color=mycolor5, line width=.6pt, mark size=2pt, mark=square, mark options={solid, mycolor5}]
  table[row sep=crcr]{%
1	0.95\\
2	0.825\\
3	0.8\\
4	0.825\\
5	0.84\\
};
\addlegendentry{GD}

\addplot [color=mycolor5, line width=.6pt, mark size=2pt, mark=square, mark options={solid, mycolor5}, forget plot]
 plot [error bars/.cd, y dir = both, y explicit]
 table[row sep=crcr, y error plus index=2, y error minus index=3]{%
1	0.95	0.05	0.05\\
2	0.825	0.0547121654902416	0.0547121654902416\\
3	0.8	0.0445904036039959	0.0445904036039959\\
4	0.825	0.0409589010954563	0.0409589010954563\\
5	0.84	0.03111946420862	0.03111946420862\\
};

\addplot[area legend, draw=none, fill=mycolor6, fill opacity=0.2, forget plot]
table[row sep=crcr] {%
x	y\\
1	1\\
2	1\\
3	1\\
4	1\\
5	1\\
5	1\\
4	1\\
3	1\\
2	1\\
1	1\\
}--cycle;
\addplot [color=mycolor6, line width=.6pt, mark size=2.5pt, mark=+, mark options={solid, mycolor6}]
  table[row sep=crcr]{%
1	1\\
2	1\\
3	1\\
4	1\\
5	1\\
};
\addlegendentry{AF}

\addplot [color=mycolor6, line width=.6pt, mark size=2.5pt, mark=+, mark options={solid, mycolor6}, forget plot]
 plot [error bars/.cd, y dir = both, y explicit]
 table[row sep=crcr, y error plus index=2, y error minus index=3]{%
1	1	0	0\\
2	1	0	0\\
3	1	0	0\\
4	1	0	0\\
5	1	0	0\\
};
\addplot[area legend, draw=none, fill=mycolor7, fill opacity=0.2, forget plot]
table[row sep=crcr] {%
x	y\\
1	1\\
2	0.879712165490242\\
3	0.844590403603996\\
4	0.865958901095456\\
5	0.86\\
5	0.8\\
4	0.784041098904544\\
3	0.755409596396004\\
2	0.770287834509758\\
1	0.9\\
};--cycle;
\addplot [color=mycolor7, line width=.6pt, mark size=2.5pt, mark=x, mark options={solid, mycolor7}]
  table[row sep=crcr]{%
1	0.95\\
2	0.825\\
3	0.8\\
4	0.825\\
5	0.83\\
};
\addlegendentry{APC}

\addplot [color=mycolor7, line width=.6pt, mark size=2.5pt, mark=x, mark options={solid, mycolor7}, forget plot]
 plot [error bars/.cd, y dir = both, y explicit]
 table[row sep=crcr, y error plus index=2, y error minus index=3]{%
1	0.95	0.05	0.05\\
2	0.825	0.0547121654902416	0.0547121654902416\\
3	0.8	0.0445904036039959	0.0445904036039959\\
4	0.825	0.0409589010954563	0.0409589010954563\\
5	0.83	0.03	0.03\\
};

\addplot[area legend, draw=none, fill=black, fill opacity=0.08, forget plot]
table[row sep=crcr] {%
x	y\\
1	1\\
2	0.879712165490242\\
3	0.844590403603996\\
4	0.865958901095456\\
5	0.882036163775859\\
5	0.817963836224141\\
4	0.784041098904544\\
3	0.755409596396004\\
2	0.770287834509758\\
1	0.9\\
}--cycle;
\addplot [color=black, dashed, line width=0.8pt]
  table[row sep=crcr]{%
1	0.95\\
2	0.825\\
3	0.8\\
4	0.825\\
5	0.85\\
};
\addlegendentry{all}

\addplot [color=black, dashed, line width=.8pt, forget plot]
 plot [error bars/.cd, y dir = both, y explicit]
 table[row sep=crcr, y error plus index=2, y error minus index=3]{%
1	0.95	0.05	0.05\\
2	0.825	0.0547121654902416	0.0547121654902416\\
3	0.8	0.0445904036039959	0.0445904036039959\\
4	0.825	0.0409589010954563	0.0409589010954563\\
5	0.85	0.0320361637758594	0.0320361637758594\\
};
\end{axis}
\end{tikzpicture}%

%% file: imgs/BICS2/R22.tex
%
%
\definecolor{mycolor1}{rgb}{0.00000,0.44700,0.74100}%
\definecolor{mycolor2}{rgb}{0.85000,0.32500,0.09800}%
\definecolor{mycolor3}{rgb}{0.92900,0.69400,0.12500}%
\definecolor{mycolor4}{rgb}{0.49400,0.18400,0.55600}%
\definecolor{mycolor5}{rgb}{0.46600,0.67400,0.18800}%
\definecolor{mycolor6}{rgb}{0.30100,0.74500,0.93300}%
\definecolor{mycolor7}{rgb}{0.63500,0.07800,0.18400}%
\pgfplotsset{
compat=1.11,
legend image code/.code={
\draw[mark repeat=2,mark phase=2]
plot coordinates {
(0cm,0cm)
(0.15cm,0cm)        
(0.3cm,0cm)         
};%
}
}
\begin{tikzpicture}

\begin{axis}[%
width=.85\columnwidth,
height=.55\columnwidth,
scale only axis,
xmin=0.8,
xmax=5.2,
xtick={1,2,3,4,5},
xticklabel style={font=\footnotesize},
xlabel style={font=\color{white!15!black}\footnotesize,yshift=3pt},
xlabel={\# failed nodes},
ymin=0,
ymax=1.05,
ytick={0,0.2,0.4,0.6,0.8,1},
yticklabels={{0},{.2},{.4},{.6},{.8},{1}},
yticklabel style={font=\footnotesize},
ylabel style={font=\color{white!15!black}\footnotesize,yshift=-7pt},
ylabel={$R_2$},
axis background/.style={fill=white},
legend style={at={(0.1,0.0)}, legend columns=3, anchor=south west, legend cell align=left, align=left, fill=none, draw=none, font=\scriptsize}
]

\addplot[area legend, draw=none, fill=mycolor1, fill opacity=0.2, forget plot]
table[row sep=crcr] {%
x	y\\
1	1\\
2	0.920038969481215\\
3	0.895418715001545\\
4	0.888483025693972\\
5	0.873571717199528\\
5	0.811547330419519\\
4	0.824850307639362\\
3	0.819581284998455\\
2	0.829961030518785\\
1	0.95\\
}--cycle;
\addplot [color=mycolor1, line width=.6pt, mark size=2.5pt, mark=text,text mark={\LARGE $\star$}, mark options={solid, fill=mycolor1, mycolor1}]
  table[row sep=crcr]{%
1	0.975\\
2	0.875\\
3	0.8575\\
4	0.856666666666667\\
5	0.842559523809524\\
};
\addlegendentry{FaCe}

\addplot [color=mycolor1, line width=.6pt, mark size=2.5pt,  mark=text,text mark={\LARGE $\star$}, mark options={solid, fill=mycolor1, mycolor1}, forget plot]
 plot [error bars/.cd, y dir = both, y explicit]
 table[row sep=crcr, y error plus index=2, y error minus index=3]{%
1	0.975	0.025	0.025\\
2	0.875	0.0450389694812151	0.0450389694812151\\
3	0.8575	0.0379187150015454	0.0379187150015454\\
4	0.856666666666667	0.0318163590273052	0.0318163590273052\\
5	0.842559523809524	0.0310121933900047	0.0310121933900047\\
};

\addplot[area legend, draw=none, fill=mycolor2, fill opacity=0.2, forget plot]
table[row sep=crcr] {%
x	y\\
1	1\\
2	0.919808110326828\\
3	0.885971124072526\\
4	0.880038431051382\\
5	0.876679321110465\\
5	0.817368297937153\\
4	0.810437759424809\\
3	0.824028875927474\\
2	0.846858556339839\\
1	0.95\\
}--cycle;
\addplot [color=mycolor2, line width=.6pt, mark size=2.5pt, mark=o, mark options={solid, mycolor2}]
  table[row sep=crcr]{%
1	0.975\\
2	0.883333333333333\\
3	0.855\\
4	0.845238095238095\\
5	0.847023809523809\\
};
\addlegendentry{PoP}

\addplot [color=mycolor2, line width=.6pt, mark size=2.5pt, mark=o, mark options={solid, mycolor2}, forget plot]
 plot [error bars/.cd, y dir = both, y explicit]
 table[row sep=crcr, y error plus index=2, y error minus index=3]{%
1	0.975	0.025	0.025\\
2	0.883333333333333	0.0364747769934944	0.0364747769934944\\
3	0.855	0.0309711240725258	0.0309711240725258\\
4	0.845238095238095	0.0348003358132865	0.0348003358132865\\
5	0.847023809523809	0.0296555115866559	0.0296555115866559\\
};

\addplot[area legend, draw=none, fill=mycolor3, fill opacity=0.2, forget plot]
table[row sep=crcr] {%
x	y\\
1	0.783824941794373\\
2	0.679749011474595\\
3	0.597738703338766\\
4	0.597479901282913\\
5	0.568754619713295\\
5	0.449889236430561\\
4	0.543155019352007\\
3	0.529047010946949\\
2	0.572155750430167\\
1	0.649508391538961\\
}--cycle;
\addplot [color=mycolor3, line width=.6pt, mark size=2.5pt, mark=asterisk, mark options={solid, mycolor3}]
  table[row sep=crcr]{%
1	0.716666666666667\\
2	0.625952380952381\\
3	0.563392857142857\\
4	0.57031746031746\\
5	0.509321928071928\\
};
\addlegendentry{GC}

\addplot [color=mycolor3, line width=.6pt, mark size=2.5pt, mark=asterisk, mark options={solid, mycolor3}, forget plot]
 plot [error bars/.cd, y dir = both, y explicit]
 table[row sep=crcr, y error plus index=2, y error minus index=3]{%
1	0.716666666666667	0.067158275127706	0.067158275127706\\
2	0.625952380952381	0.0537966305222141	0.0537966305222141\\
3	0.563392857142857	0.0343458461959084	0.0343458461959084\\
4	0.57031746031746	0.0271624409654528	0.0271624409654528\\
5	0.509321928071928	0.0594326916413669	0.0594326916413669\\
};

\addplot[area legend, draw=none, fill=mycolor4, fill opacity=0.2, forget plot]
table[row sep=crcr] {%
x	y\\
1	1\\
2	0.919808110326828\\
3	0.885971124072526\\
4	0.885523127926238\\
5	0.881815590155505\\
5	0.824136790796876\\
4	0.818286395883286\\
3	0.824028875927474\\
2	0.846858556339839\\
1	0.95\\
}--cycle;
\addplot [color=mycolor4, line width=.6pt, mark size=2.5pt, mark=triangle, mark options={solid, mycolor4}]
  table[row sep=crcr]{%
1	0.975\\
2	0.883333333333333\\
3	0.855\\
4	0.851904761904762\\
5	0.852976190476191\\
};
\addlegendentry{GI}

\addplot [color=mycolor4, line width=.6pt, mark size=2.5pt, mark=triangle, mark options={solid, mycolor4}, forget plot]
 plot [error bars/.cd, y dir = both, y explicit]
 table[row sep=crcr, y error plus index=2, y error minus index=3]{%
1	0.975	0.025	0.025\\
2	0.883333333333333	0.0364747769934944	0.0364747769934944\\
3	0.855	0.0309711240725258	0.0309711240725258\\
4	0.851904761904762	0.033618366021476	0.033618366021476\\
5	0.852976190476191	0.0288393996793142	0.0288393996793142\\
};

\addplot[area legend, draw=none, fill=mycolor5, fill opacity=0.2, forget plot]
table[row sep=crcr] {%
x	y\\
1	1\\
2	0.919808110326828\\
3	0.885971124072526\\
4	0.885523127926238\\
5	0.881815590155505\\
5	0.824136790796876\\
4	0.818286395883286\\
3	0.824028875927474\\
2	0.846858556339839\\
1	0.95\\
}--cycle;
\addplot [color=mycolor5, line width=.6pt, mark size=2pt, mark=square, mark options={solid, mycolor5}]
  table[row sep=crcr]{%
1	0.975\\
2	0.883333333333333\\
3	0.855\\
4	0.851904761904762\\
5	0.852976190476191\\
};
\addlegendentry{GD}

\addplot [color=mycolor5, line width=.6pt, mark size=2pt, mark=square, mark options={solid, mycolor5}, forget plot]
 plot [error bars/.cd, y dir = both, y explicit]
 table[row sep=crcr, y error plus index=2, y error minus index=3]{%
1	0.975	0.025	0.025\\
2	0.883333333333333	0.0364747769934944	0.0364747769934944\\
3	0.855	0.0309711240725258	0.0309711240725258\\
4	0.851904761904762	0.033618366021476	0.033618366021476\\
5	0.852976190476191	0.0288393996793142	0.0288393996793142\\
};

\addplot[area legend, draw=none, fill=mycolor6, fill opacity=0.2, forget plot]
table[row sep=crcr] {%
x	y\\
1	1\\
2	1\\
3	1\\
4	1\\
5	1\\
5	1\\
4	1\\
3	1\\
2	1\\
1	1\\
}--cycle;
\addplot [color=mycolor6, line width=.6pt, mark size=2.5pt, mark=+, mark options={solid, mycolor6}]
  table[row sep=crcr]{%
1	1\\
2	1\\
3	1\\
4	1\\
5	1\\
};
\addlegendentry{AF}

\addplot [color=mycolor6, line width=.6pt, mark size=2.5pt, mark=+, mark options={solid, mycolor6}, forget plot]
 plot [error bars/.cd, y dir = both, y explicit]
 table[row sep=crcr, y error plus index=2, y error minus index=3]{%
1	1	0	0\\
2	1	0	0\\
3	1	0	0\\
4	1	0	0\\
5	1	0	0\\
};
\addplot[area legend, draw=none, fill=mycolor7, fill opacity=0.2, forget plot]
table[row sep=crcr] {%
x	y\\
1	1\\
2	0.919808110326828\\
3	0.885971124072526\\
4	0.865627882496836\\
5	0.839953196872682\\
5	0.763419819000334\\
4	0.785324498455545\\
3	0.824028875927474\\
2	0.846858556339839\\
1	0.95\\
};--cycle;
\addplot [color=mycolor7, line width=.6pt, mark size=2.5pt, mark=x, mark options={solid, mycolor7}]
  table[row sep=crcr]{%
1	0.975\\
2	0.883333333333333\\
3	0.855\\
4	0.82547619047619\\
5	0.801686507936508\\
};
\addlegendentry{APC}

\addplot [color=mycolor7, line width=.6pt, mark size=2.5pt, mark=x, mark options={solid, mycolor7}, forget plot]
 plot [error bars/.cd, y dir = both, y explicit]
 table[row sep=crcr, y error plus index=2, y error minus index=3]{%
1	0.975	0.025	0.025\\
2	0.883333333333333	0.0364747769934944	0.0364747769934944\\
3	0.855	0.0309711240725258	0.0309711240725258\\
4	0.82547619047619	0.0401516920206456	0.0401516920206456\\
5	0.801686507936508	0.0382666889361738	0.0382666889361738\\
};

\addplot[area legend, draw=none, fill=black, fill opacity=0.08, forget plot]
table[row sep=crcr] {%
x	y\\
1	1\\
2	0.919808110326828\\
3	0.885971124072526\\
4	0.890892433082694\\
5	0.885965713916079\\
5	0.824748571798206\\
4	0.826250424060163\\
3	0.824028875927474\\
2	0.846858556339839\\
1	0.95\\
}--cycle;
\addplot [color=black, dashed, line width=0.8pt]
  table[row sep=crcr]{%
1	0.975\\
2	0.883333333333333\\
3	0.855\\
4	0.858571428571429\\
5	0.855357142857143\\
};
\addlegendentry{all}

\addplot [color=black, dashed, line width=.8pt, forget plot]
 plot [error bars/.cd, y dir = both, y explicit]
 table[row sep=crcr, y error plus index=2, y error minus index=3]{%
1	0.975	0.025	0.025\\
2	0.883333333333333	0.0364747769934944	0.0364747769934944\\
3	0.855	0.0309711240725258	0.0309711240725258\\
4	0.858571428571429	0.0323210045112657	0.0323210045112657\\
5	0.855357142857143	0.0306085710589364	0.0306085710589364\\
};
\end{axis}
\end{tikzpicture}%

%% file: imgs/BICS2/ET2.tex
%
%
\definecolor{mycolor1}{rgb}{0.00000,0.44700,0.74100}%
\definecolor{mycolor2}{rgb}{0.85000,0.32500,0.09800}%
\definecolor{mycolor3}{rgb}{0.92900,0.69400,0.12500}%
\definecolor{mycolor4}{rgb}{0.49400,0.18400,0.55600}%
\definecolor{mycolor5}{rgb}{0.46600,0.67400,0.18800}%
\definecolor{mycolor6}{rgb}{0.30100,0.74500,0.93300}%
\definecolor{mycolor7}{rgb}{0.63500,0.07800,0.18400}%
\pgfplotsset{
compat=1.11,
legend image code/.code={
\draw[mark repeat=2,mark phase=2]
plot coordinates {
(0cm,0cm)
(0.15cm,0cm)        
(0.3cm,0cm)         
};%
}
}
\begin{tikzpicture}

\begin{axis}[%
width=.85\columnwidth,
height=.55\columnwidth,
scale only axis,
xmin=0.8,
xmax=5.2,
xtick={1,2,3,4,5},
xticklabel style={font=\footnotesize},
xlabel style={font=\color{white!15!black}\footnotesize,yshift=3pt},
xlabel={\# failed nodes},
ymin=0,
ymax=3.500009653552294,
ytick={0,1,2,3},
yticklabel style={font=\footnotesize},
ylabel style={font=\color{white!15!black}\footnotesize,yshift=-6pt},
ylabel={elapsed time (s)},
axis background/.style={fill=white},
legend style={at={(0.05,0.33)}, legend columns=3, anchor=south west, legend cell align=left, align=left, fill=none, draw=none, font=\scriptsize}
]

\addplot[area legend, draw=none, fill=mycolor1, fill opacity=0.2, forget plot]
table[row sep=crcr] {%
x	y\\
1	0.149651355564833\\
2	0.177605970024028\\
3	0.289267452024011\\
4	0.322049347993171\\
5	0.365789073015123\\
5	0.298273426984877\\
4	0.246700652006829\\
3	0.223232547975989\\
2	0.138019029975972\\
1	0.105036144435167\\
}--cycle;
\addplot [color=mycolor1, line width=.6pt, mark size=2.5pt, mark=text,text mark={\LARGE $\star$}, mark options={solid, fill=mycolor1, mycolor1}]
  table[row sep=crcr]{%
1	0.12734375\\
2	0.1578125\\
3	0.25625\\
4	0.284375\\
5	0.33203125\\
};
\addlegendentry{FaCe}

\addplot [color=mycolor1, line width=.6pt, mark size=2.5pt,  mark=text,text mark={\LARGE $\star$}, mark options={solid, fill=mycolor1, mycolor1}, forget plot]
 plot [error bars/.cd, y dir = both, y explicit]
 table[row sep=crcr, y error plus index=2, y error minus index=3]{%
1	0.12734375	0.0223076055648334	0.0223076055648334\\
2	0.1578125	0.0197934700240283	0.0197934700240283\\
3	0.25625	0.0330174520240114	0.0330174520240114\\
4	0.284375	0.0376743479931707	0.0376743479931707\\
5	0.33203125	0.0337578230151234	0.0337578230151234\\
};

\addplot[area legend, draw=none, fill=mycolor2, fill opacity=0.2, forget plot]
table[row sep=crcr] {%
x	y\\
1	0.06660573428621\\
2	0.230009425146034\\
3	0.958286800211447\\
4	1.60740763328961\\
5	3.00009653552294\\
5	1.89521596447706\\
4	0.92227986671039\\
3	0.421400699788553\\
2	0.123115574853966\\
1	0.03339426571379\\
}--cycle;
\addplot [color=mycolor2, line width=.6pt, mark size=2.5pt, mark=o, mark options={solid, mycolor2}]
  table[row sep=crcr]{%
1	0.05\\
2	0.1765625\\
3	0.68984375\\
4	1.26484375\\
5	2.44765625\\
};
\addlegendentry{PoP}

\addplot [color=mycolor2, line width=.6pt, mark size=2.5pt, mark=o, mark options={solid, mycolor2}, forget plot]
 plot [error bars/.cd, y dir = both, y explicit]
 table[row sep=crcr, y error plus index=2, y error minus index=3]{%
1	0.05	0.01660573428621	0.01660573428621\\
2	0.1765625	0.0534469251460342	0.0534469251460342\\
3	0.68984375	0.268443050211447	0.268443050211447\\
4	1.26484375	0.34256388328961	0.34256388328961\\
5	2.44765625	0.55244028552294	0.55244028552294\\
};

\addplot[area legend, draw=none, fill=mycolor3, fill opacity=0.2, forget plot]
table[row sep=crcr] {%
x	y\\
1	0.0244560285399714\\
2	0.0195517553902346\\
3	0.0197062434584265\\
4	0.0194260949981606\\
5	0.0226077772593672\\
5	0.0180172227406328\\
4	0.0149489050018394\\
3	0.0131062565415735\\
2	0.0116982446097654\\
1	0.0146064714600286\\
}--cycle;
\addplot [color=mycolor3, line width=.6pt, mark size=2.5pt, mark=asterisk, mark options={solid, mycolor3}]
  table[row sep=crcr]{%
1	0.01953125\\
2	0.015625\\
3	0.01640625\\
4	0.0171875\\
5	0.0203125\\
};
\addlegendentry{GC}

\addplot [color=mycolor3, line width=.6pt, mark size=2.5pt, mark=asterisk, mark options={solid, mycolor3}, forget plot]
 plot [error bars/.cd, y dir = both, y explicit]
 table[row sep=crcr, y error plus index=2, y error minus index=3]{%
1	0.01953125	0.00492477853997143	0.00492477853997143\\
2	0.015625	0.00392675539023464	0.00392675539023464\\
3	0.01640625	0.00329999345842653	0.00329999345842653\\
4	0.0171875	0.00223859499816056	0.00223859499816056\\
5	0.0203125	0.00229527725936725	0.00229527725936725\\
};

\addplot[area legend, draw=none, fill=mycolor4, fill opacity=0.2, forget plot]
table[row sep=crcr] {%
x	y\\
1	0.269711277059086\\
2	0.259144699348072\\
3	0.257447242521277\\
4	0.247739133347504\\
5	0.22979854345604\\
5	0.20770145654396\\
4	0.208510866652496\\
3	0.219115257478723\\
2	0.215855300651928\\
1	0.211538722940914\\
}--cycle;
\addplot [color=mycolor4, line width=.6pt, mark size=2.5pt, mark=triangle, mark options={solid, mycolor4}]
  table[row sep=crcr]{%
1	0.240625\\
2	0.2375\\
3	0.23828125\\
4	0.228125\\
5	0.21875\\
};
\addlegendentry{GI}

\addplot [color=mycolor4, line width=.6pt, mark size=2.5pt, mark=triangle, mark options={solid, mycolor4}, forget plot]
 plot [error bars/.cd, y dir = both, y explicit]
 table[row sep=crcr, y error plus index=2, y error minus index=3]{%
1	0.240625	0.0290862770590857	0.0290862770590857\\
2	0.2375	0.0216446993480718	0.0216446993480718\\
3	0.23828125	0.0191659925212773	0.0191659925212773\\
4	0.228125	0.0196141333475044	0.0196141333475044\\
5	0.21875	0.0110485434560398	0.0110485434560398\\
};

\addplot[area legend, draw=none, fill=mycolor5, fill opacity=0.2, forget plot]
table[row sep=crcr] {%
x	y\\
1	0.596851498148356\\
2	0.472609923098498\\
3	0.534916300286621\\
4	0.555636933242185\\
5	0.484178015429779\\
5	0.451759484570221\\
4	0.484988066757815\\
3	0.491646199713379\\
2	0.441452576901502\\
1	0.498461001851644\\
}--cycle;
\addplot [color=mycolor5, line width=.6pt, mark size=2pt, mark=square, mark options={solid, mycolor5}]
  table[row sep=crcr]{%
1	0.54765625\\
2	0.45703125\\
3	0.51328125\\
4	0.5203125\\
5	0.46796875\\
};
\addlegendentry{GD}

\addplot [color=mycolor5, line width=.6pt, mark size=2pt, mark=square, mark options={solid, mycolor5}, forget plot]
 plot [error bars/.cd, y dir = both, y explicit]
 table[row sep=crcr, y error plus index=2, y error minus index=3]{%
1	0.54765625	0.0491952481483556	0.0491952481483556\\
2	0.45703125	0.0155786730984981	0.0155786730984981\\
3	0.51328125	0.0216350502866207	0.0216350502866207\\
4	0.5203125	0.0353244332421849	0.0353244332421849\\
5	0.46796875	0.0162092654297789	0.0162092654297789\\
};

\addplot[area legend, draw=none, fill=mycolor6, fill opacity=0.2, forget plot]
table[row sep=crcr] {%
x	y\\
1	0.851001688492223\\
2	1.19198506511429\\
3	1.40278518433117\\
4	1.67780580281221\\
5	1.71959616036868\\
5	1.53196633963132\\
4	1.42531919718779\\
3	1.18002731566883\\
2	0.940827434885707\\
1	0.639623311507777\\
}--cycle;
\addplot [color=mycolor6, line width=.6pt, mark size=2.5pt, mark=+, mark options={solid, mycolor6}]
  table[row sep=crcr]{%
1	0.7453125\\
2	1.06640625\\
3	1.29140625\\
4	1.5515625\\
5	1.62578125\\
};
\addlegendentry{AF}

\addplot [color=mycolor6, line width=.6pt, mark size=2.5pt, mark=+, mark options={solid, mycolor6}, forget plot]
 plot [error bars/.cd, y dir = both, y explicit]
 table[row sep=crcr, y error plus index=2, y error minus index=3]{%
1	0.7453125	0.105689188492223	0.105689188492223\\
2	1.06640625	0.125578815114293	0.125578815114293\\
3	1.29140625	0.111378934331168	0.111378934331168\\
4	1.5515625	0.126243302812205	0.126243302812205\\
5	1.62578125	0.0938149103686805	0.0938149103686805\\
};
\addplot[area legend, draw=none, fill=mycolor7, fill opacity=0.2, forget plot]
table[row sep=crcr] {%
x	y\\
1	0.00888414759507351\\
2	0.02925618607642\\
3	0.15554855792375\\
4	0.78187632821906\\
5	1.99870127236706\\
5	0.940361227632938\\
4	0.41812367178094\\
3	0.0866389420762501\\
2	0.01449381392358\\
1	0.00361585240492649\\
};--cycle;
\addplot [color=mycolor7, line width=.6pt, mark size=2.5pt, mark=x, mark options={solid, mycolor7}]
  table[row sep=crcr]{%
1	0.00625\\
2	0.021875\\
3	0.12109375\\
4	0.6\\
5	1.46953125\\
};
\addlegendentry{APC}

\addplot [color=mycolor7, line width=.6pt, mark size=2.5pt, mark=x, mark options={solid, mycolor7}, forget plot]
 plot [error bars/.cd, y dir = both, y explicit]
 table[row sep=crcr, y error plus index=2, y error minus index=3]{%
1	0.00625	0.00263414759507351	0.00263414759507351\\
2	0.021875	0.00738118607642002	0.00738118607642002\\
3	0.12109375	0.0344548079237499	0.0344548079237499\\
4	0.6	0.18187632821906	0.18187632821906\\
5	1.46953125	0.529170022367062	0.529170022367062\\
};
\end{axis}
\end{tikzpicture}%

%% file: imgs/BICS2/NT2.tex
%
%
\definecolor{mycolor1}{rgb}{0.00000,0.44700,0.74100}%
\definecolor{mycolor2}{rgb}{0.85000,0.32500,0.09800}%
\definecolor{mycolor3}{rgb}{0.92900,0.69400,0.12500}%
\definecolor{mycolor4}{rgb}{0.49400,0.18400,0.55600}%
\definecolor{mycolor5}{rgb}{0.46600,0.67400,0.18800}%
\definecolor{mycolor6}{rgb}{0.30100,0.74500,0.93300}%
\definecolor{mycolor7}{rgb}{0.63500,0.07800,0.18400}%
\pgfplotsset{
compat=1.11,
legend image code/.code={
\draw[mark repeat=2,mark phase=2]
plot coordinates {
(0cm,0cm)
(0.15cm,0cm)        
(0.3cm,0cm)         
};%
}
}
\begin{tikzpicture}
\draw (0.2,2.6) node (1) [] {\footnotesize{$\cdot 10$}}; 
\begin{axis}[%
width=.85\columnwidth,
height=.55\columnwidth,
scale only axis,
xmin=0.8,
xmax=5.2,
xtick={1,2,3,4,5},
xticklabel style={font=\footnotesize},
xlabel style={font=\color{white!15!black}\footnotesize,yshift=3pt},
xlabel={\# failed nodes},
ymin=0,
ymax=60,
ytick={0,10,20,30,40,50,60},
yticklabels={{0},{1},{2},{3},{4},{5},{6}},
yticklabel style={font=\footnotesize},
ylabel style={font=\color{white!15!black}\footnotesize,yshift=-5pt},
ylabel={\# tests},
axis background/.style={fill=none},
legend style={at={(0.1,0.31)}, legend columns=3, anchor=south west, legend cell align=left, align=left, fill=none, draw=none, font=\scriptsize}
]

\addplot[area legend, draw=none, fill=mycolor1, fill opacity=0.2, forget plot]
table[row sep=crcr] {%
x	y\\
1	10\\
2	10.7835325870964\\
3	11.9523573072576\\
4	12.9535862677615\\
5	13.7532716343652\\
5	13.1467283656348\\
4	12.5464137322385\\
3	11.4476426927424\\
2	10.4164674129036\\
1	9.7\\
}--cycle;
\addplot [color=mycolor1, line width=.6pt, mark size=2.5pt, mark=text,text mark={\LARGE $\star$}, mark options={solid, fill=mycolor1, mycolor1}]
  table[row sep=crcr]{%
1	9.85\\
2	10.6\\
3	11.7\\
4	12.75\\
5	13.45\\
};
\addlegendentry{FaCe}

\addplot [color=mycolor1, line width=.6pt, mark size=2.5pt,  mark=text,text mark={\LARGE $\star$}, mark options={solid, fill=mycolor1, mycolor1}, forget plot]
 plot [error bars/.cd, y dir = both, y explicit]
 table[row sep=crcr, y error plus index=2, y error minus index=3]{%
1	9.85	0.15	0.15\\
2	10.6	0.183532587096449	0.183532587096449\\
3	11.7	0.252357307257618	0.252357307257618\\
4	12.75	0.203586267761489	0.203586267761489\\
5	13.45	0.303271634365178	0.303271634365178\\
};

\addplot[area legend, draw=none, fill=mycolor2, fill opacity=0.2, forget plot]
table[row sep=crcr] {%
x	y\\
1	11.2532716343652\\
2	13.6540624439178\\
3	15.4790085651735\\
4	16.6967530237148\\
5	17.8745298646859\\
5	16.8254701353141\\
4	15.4032469762852\\
3	14.1209914348265\\
2	12.6459375560822\\
1	10.6467283656348\\
}--cycle;
\addplot [color=mycolor2, line width=.6pt, mark size=2.5pt, mark=o, mark options={solid, mycolor2}]
  table[row sep=crcr]{%
1	10.95\\
2	13.15\\
3	14.8\\
4	16.05\\
5	17.35\\
};
\addlegendentry{PoP}

\addplot [color=mycolor2, line width=.6pt, mark size=2.5pt, mark=o, mark options={solid, mycolor2}, forget plot]
 plot [error bars/.cd, y dir = both, y explicit]
 table[row sep=crcr, y error plus index=2, y error minus index=3]{%
1	10.95	0.303271634365178	0.303271634365178\\
2	13.15	0.504062443917836	0.504062443917836\\
3	14.8	0.679008565173479	0.679008565173479\\
4	16.05	0.64675302371478	0.64675302371478\\
5	17.35	0.524529864685862	0.524529864685862\\
};

\addplot[area legend, draw=none, fill=mycolor3, fill opacity=0.2, forget plot]
table[row sep=crcr] {%
x	y\\
1	8.7123902973898\\
2	8.93191780219091\\
3	8.95942433098048\\
4	8.8277332747317\\
5	8.87301048307916\\
5	8.62698951692084\\
4	8.5722667252683\\
3	8.74057566901952\\
2	8.76808219780909\\
1	8.4876097026102\\
}--cycle;
\addplot [color=mycolor3, line width=.6pt, mark size=2.5pt, mark=asterisk, mark options={solid, mycolor3}]
  table[row sep=crcr]{%
1	8.6\\
2	8.85\\
3	8.85\\
4	8.7\\
5	8.75\\
};
\addlegendentry{GC}

\addplot [color=mycolor3, line width=.6pt, mark size=2.5pt, mark=asterisk, mark options={solid, mycolor3}, forget plot]
 plot [error bars/.cd, y dir = both, y explicit]
 table[row sep=crcr, y error plus index=2, y error minus index=3]{%
1	8.6	0.112390297389803	0.112390297389803\\
2	8.85	0.0819178021909125	0.0819178021909125\\
3	8.85	0.109424330980483	0.109424330980483\\
4	8.7	0.127733274731701	0.127733274731701\\
5	8.75	0.12301048307916	0.12301048307916\\
};

\addplot[area legend, draw=none, fill=mycolor4, fill opacity=0.2, forget plot]
table[row sep=crcr] {%
x	y\\
1	55\\
2	55\\
3	55\\
4	55\\
5	55\\
5	55\\
4	55\\
3	55\\
2	55\\
1	55\\
}--cycle;
\addplot [color=mycolor4, line width=.6pt, mark size=2.5pt, mark=triangle, mark options={solid, mycolor4}]
  table[row sep=crcr]{%
1	55\\
2	55\\
3	55\\
4	55\\
5	55\\
};
\addlegendentry{GI}

\addplot [color=mycolor4, line width=.6pt, mark size=2.5pt, mark=triangle, mark options={solid, mycolor4}, forget plot]
 plot [error bars/.cd, y dir = both, y explicit]
 table[row sep=crcr, y error plus index=2, y error minus index=3]{%
1	55	0	0\\
2	55	0	0\\
3	55	0	0\\
4	55	0	0\\
5	55	0	0\\
};

\addplot[area legend, draw=none, fill=mycolor5, fill opacity=0.2, forget plot]
table[row sep=crcr] {%
x	y\\
1	55\\
2	55\\
3	55\\
4	55\\
5	55\\
5	55\\
4	55\\
3	55\\
2	55\\
1	55\\
}--cycle;
\addplot [color=mycolor5, line width=.6pt, mark size=2pt, mark=square, mark options={solid, mycolor5}]
  table[row sep=crcr]{%
1	55\\
2	55\\
3	55\\
4	55\\
5	55\\
};
\addlegendentry{GD}

\addplot [color=mycolor5, line width=.6pt, mark size=2pt, mark=square, mark options={solid, mycolor5}, forget plot]
 plot [error bars/.cd, y dir = both, y explicit]
 table[row sep=crcr, y error plus index=2, y error minus index=3]{%
1	55	0	0\\
2	55	0	0\\
3	55	0	0\\
4	55	0	0\\
5	55	0	0\\
};

\addplot[area legend, draw=none, fill=mycolor6, fill opacity=0.2, forget plot]
table[row sep=crcr] {%
x	y\\
1	7.34933992677988\\
2	13.6957534065727\\
3	19.8846974386589\\
4	25.8590924232298\\
5	31.4445758416642\\
5	30.4554241583358\\
4	25.1409075767702\\
3	19.3153025613411\\
2	13.2042465934273\\
1	7.15066007322012\\
}--cycle;
\addplot [color=mycolor6, line width=.6pt, mark size=2.5pt, mark=+, mark options={solid, mycolor6}]
  table[row sep=crcr]{%
1	7.25\\
2	13.45\\
3	19.6\\
4	25.5\\
5	30.95\\
};
\addlegendentry{AF}

\addplot [color=mycolor6, line width=.6pt, mark size=2.5pt, mark=+, mark options={solid, mycolor6}, forget plot]
 plot [error bars/.cd, y dir = both, y explicit]
 table[row sep=crcr, y error plus index=2, y error minus index=3]{%
1	7.25	0.0993399267798783	0.0993399267798783\\
2	13.45	0.245753406572738	0.245753406572738\\
3	19.6	0.284697438658916	0.284697438658916\\
4	25.5	0.359092423229804	0.359092423229804\\
5	30.95	0.494575841664243	0.494575841664243\\
};
\addplot[area legend, draw=none, fill=mycolor7, fill opacity=0.2, forget plot]
table[row sep=crcr] {%
x	y\\
1	10.145095250022\\
2	11.145095250022\\
3	12.0586926725569\\
4	12.6957534065727\\
5	13.7026201370835\\
5	13.0973798629165\\
4	12.2042465934273\\
3	11.6413073274431\\
2	10.854904749978\\
1	9.854904749978\\
};--cycle;
\addplot [color=mycolor7, line width=.6pt, mark size=2.5pt, mark=x, mark options={solid, mycolor7}]
  table[row sep=crcr]{%
1	10\\
2	11\\
3	11.85\\
4	12.45\\
5	13.4\\
};
\addlegendentry{APC}

\addplot [color=mycolor7, line width=.6pt, mark size=2.5pt, mark=x, mark options={solid, mycolor7}, forget plot]
 plot [error bars/.cd, y dir = both, y explicit]
 table[row sep=crcr, y error plus index=2, y error minus index=3]{%
1	10	0.145095250022002	0.145095250022002\\
2	11	0.145095250022002	0.145095250022002\\
3	11.85	0.208692672556914	0.208692672556914\\
4	12.45	0.245753406572738	0.245753406572738\\
5	13.4	0.302620137083475	0.302620137083475\\
};
\end{axis}
\end{tikzpicture}%

%% file: imgs/MINNESOTA1/AW1MN.tex
%
%
\definecolor{mycolor1}{rgb}{0.00000,0.44700,0.74100}%
\definecolor{mycolor2}{rgb}{0.92900,0.69400,0.12500}%
\definecolor{mycolor3}{rgb}{0.49400,0.18400,0.55600}%
\definecolor{mycolor4}{rgb}{0.46600,0.67400,0.18800}%
\definecolor{mycolor5}{rgb}{0.30100,0.74500,0.93300}%
\definecolor{mycolor6}{rgb}{0.63500,0.07800,0.18400}%
\pgfplotsset{
compat=1.11,
legend image code/.code={
\draw[mark repeat=2,mark phase=2]
plot coordinates {
(0cm,0cm)
(0.15cm,0cm)        
(0.3cm,0cm)         
};%
}
}
\begin{tikzpicture}

\begin{axis}[%
width=.85\columnwidth,
height=.55\columnwidth,
scale only axis,
xmin=4,
xmax=46,
xtick={5,10,15,20,25,30,35,40,45},
xticklabel style={font=\footnotesize},
xlabel style={font=\color{white!15!black}\footnotesize,yshift=3pt},
xlabel={\# failed nodes},
ymin=0,
ymax=1.07,
ytick={0,0.2,0.4,0.6,0.8,1},
yticklabels={{0},{.2},{.4},{.6},{.8},{1}},
yticklabel style={font=\footnotesize},
ylabel style={font=\color{white!15!black}\footnotesize,yshift=-5pt},
ylabel={$a_W$},
axis background/.style={fill=white},
legend style={at={(0,0)}, legend columns=3, anchor=south west, legend cell align=left, align=left, fill=none, draw=none, font=\scriptsize}
]

\addplot[area legend, draw=none, fill=mycolor1, fill opacity=0.2, forget plot]
table[row sep=crcr] {%
x	y\\
5	1\\
10	1\\
15	1\\
20	1\\
25	1\\
30	1\\
35	1\\
40	1\\
45	1\\
45	1\\
40	1\\
35	1\\
30	1\\
25	1\\
20	1\\
15	1\\
10	1\\
5	1\\
}--cycle;
\addplot [color=mycolor1, line width=.6pt, mark size=2.5pt, mark=text,text mark={\LARGE $\star$}, mark options={solid, fill=mycolor1, mycolor1}]
  table[row sep=crcr]{%
5	1\\
10	1\\
15	1\\
20	1\\
25	1\\
30	1\\
35	1\\
40	1\\
45	1\\
};
\addlegendentry{FaCe}

\addplot [color=mycolor1, line width=.6pt, mark size=2.5pt, mark=text,text mark={\LARGE $\star$}, mark options={solid, fill=mycolor1, mycolor1}, forget plot]
 plot [error bars/.cd, y dir = both, y explicit]
 table[row sep=crcr, y error plus index=2, y error minus index=3]{%
5	1	0	0\\
10	1	0	0\\
15	1	0	0\\
20	1	0	0\\
25	1	0	0\\
30	1	0	0\\
35	1	0	0\\
40	1	0	0\\
45	1	0	0\\
};

\addplot[area legend, draw=none, fill=mycolor2, fill opacity=0.2, forget plot]
table[row sep=crcr] {%
x	y\\
5	0.973868475097671\\
10	0.923004435072424\\
15	0.897684405965968\\
20	0.846484385136813\\
25	0.81718592661508\\
30	0.781273849734906\\
35	0.760612556442786\\
40	0.69687665189696\\
45	0.66929547810027\\
45	0.64405504887965\\
40	0.664797340812745\\
35	0.708612440309305\\
30	0.754110651330396\\
25	0.795822856519781\\
20	0.816649483309623\\
15	0.882352422002685\\
10	0.913933650351168\\
5	0.968187445055731\\
}--cycle;
\addplot [color=mycolor2, line width=.6pt, mark size=2.5pt, mark=asterisk, mark options={solid, mycolor2}]
  table[row sep=crcr]{%
5	0.971027960076701\\
10	0.918469042711796\\
15	0.890018413984327\\
20	0.831566934223218\\
25	0.806504391567431\\
30	0.767692250532651\\
35	0.734612498376045\\
40	0.680836996354853\\
45	0.65667526348996\\
};
\addlegendentry{GC}

\addplot [color=mycolor2, line width=.6pt, mark size=2.5pt, mark=asterisk, mark options={solid, mycolor2}, forget plot]
 plot [error bars/.cd, y dir = both, y explicit]
 table[row sep=crcr, y error plus index=2, y error minus index=3]{%
5	0.971027960076701	0.00284051502097007	0.00284051502097007\\
10	0.918469042711796	0.00453539236062809	0.00453539236062809\\
15	0.890018413984327	0.00766599198164154	0.00766599198164154\\
20	0.831566934223218	0.014917450913595	0.014917450913595\\
25	0.806504391567431	0.0106815350476496	0.0106815350476496\\
30	0.767692250532651	0.0135815992022549	0.0135815992022549\\
35	0.734612498376045	0.0260000580667403	0.0260000580667403\\
40	0.680836996354853	0.0160396555421074	0.0160396555421074\\
45	0.65667526348996	0.0126202146103099	0.0126202146103099\\
};

\addplot[area legend, draw=none, fill=mycolor3, fill opacity=0.2, forget plot]
table[row sep=crcr] {%
x	y\\
5	0.622407281941556\\
10	0.664138557206244\\
15	0.700589141706152\\
20	0.701215990478132\\
25	0.732165073892255\\
30	0.747759741231442\\
35	0.78270426326326\\
40	0.752777002528261\\
45	0.738030396101973\\
45	0.698215125605986\\
40	0.725514257574762\\
35	0.7518093174093\\
30	0.713661099549645\\
25	0.714298737440147\\
20	0.686324717999919\\
15	0.681396666668141\\
10	0.63776140572551\\
5	0.601482556409393\\
}--cycle;
\addplot [color=mycolor3, line width=.6pt, mark size=2.5pt, mark=triangle, mark options={solid, mycolor3}]
  table[row sep=crcr]{%
5	0.611944919175475\\
10	0.650949981465877\\
15	0.690992904187146\\
20	0.693770354239026\\
25	0.723231905666201\\
30	0.730710420390543\\
35	0.76725679033628\\
40	0.739145630051511\\
45	0.71812276085398\\
};
\addlegendentry{GI}

\addplot [color=mycolor3, line width=.6pt, mark size=2.5pt, mark=triangle, mark options={solid, mycolor3}, forget plot]
 plot [error bars/.cd, y dir = both, y explicit]
 table[row sep=crcr, y error plus index=2, y error minus index=3]{%
5	0.611944919175475	0.0104623627660815	0.0104623627660815\\
10	0.650949981465877	0.0131885757403671	0.0131885757403671\\
15	0.690992904187146	0.00959623751900572	0.00959623751900572\\
20	0.693770354239026	0.0074456362391068	0.0074456362391068\\
25	0.723231905666201	0.00893316822605402	0.00893316822605402\\
30	0.730710420390543	0.0170493208408986	0.0170493208408986\\
35	0.76725679033628	0.01544747292698	0.01544747292698\\
40	0.739145630051511	0.0136313724767495	0.0136313724767495\\
45	0.71812276085398	0.0199076352479938	0.0199076352479938\\
};

\addplot[area legend, draw=none, fill=mycolor4, fill opacity=0.2, forget plot]
table[row sep=crcr] {%
x	y\\
5	0.785205171052891\\
10	0.820749378900369\\
15	0.80249712612065\\
20	0.780246573705837\\
25	0.760199684930134\\
30	0.755514337478655\\
35	0.780960858006276\\
40	0.772840494024298\\
45	0.737038524427794\\
45	0.705893223168226\\
40	0.758098444862937\\
35	0.773255312921738\\
30	0.745969465834834\\
25	0.726056511939237\\
20	0.755598442396675\\
15	0.793335475683139\\
10	0.815257000724302\\
5	0.775838190783784\\
}--cycle;
\addplot [color=mycolor4, line width=.6pt, mark size=2.5pt, mark=square, mark options={solid, mycolor4}]
  table[row sep=crcr]{%
5	0.780521680918338\\
10	0.818003189812335\\
15	0.797916300901895\\
20	0.767922508051256\\
25	0.743128098434686\\
30	0.750741901656745\\
35	0.777108085464007\\
40	0.765469469443617\\
45	0.72146587379801\\
};
\addlegendentry{GD}

\addplot [color=mycolor4, line width=.6pt, mark size=2.5pt, mark=square, mark options={solid, mycolor4}, forget plot]
 plot [error bars/.cd, y dir = both, y explicit]
 table[row sep=crcr, y error plus index=2, y error minus index=3]{%
5	0.780521680918338	0.00468349013455362	0.00468349013455362\\
10	0.818003189812335	0.00274618908803326	0.00274618908803326\\
15	0.797916300901895	0.00458082521875532	0.00458082521875532\\
20	0.767922508051256	0.012324065654581	0.012324065654581\\
25	0.743128098434686	0.0170715864954487	0.0170715864954487\\
30	0.750741901656745	0.00477243582191071	0.00477243582191071\\
35	0.777108085464007	0.00385277254226873	0.00385277254226873\\
40	0.765469469443617	0.00737102458068039	0.00737102458068039\\
45	0.72146587379801	0.015572650629784	0.015572650629784\\
};

\addplot[area legend, draw=none, fill=mycolor5, fill opacity=0.2, forget plot]
table[row sep=crcr] {%
x	y\\
5	0.998704101633161\\
10	0.988995855583816\\
15	0.72141071265115\\
20	0.630874843630153\\
25	0.504973761377788\\
30	0.466490323369915\\
35	0.43528915248685\\
40	0.391946037222722\\
45	0.376129061144522\\
45	0.332781829944586\\
40	0.363066187716153\\
35	0.417850944131507\\
30	0.416564569231517\\
25	0.454460200886363\\
20	0.540220727465418\\
15	0.688727536196777\\
10	0.980935807286343\\
5	0.994989592060533\\
}--cycle;
\addplot [color=mycolor5, line width=.6pt, mark size=2.5pt, mark=+, mark options={solid, mycolor5}]
  table[row sep=crcr]{%
5	0.996846846846847\\
10	0.98496583143508\\
15	0.705069124423963\\
20	0.585547785547786\\
25	0.479716981132076\\
30	0.441527446300716\\
35	0.426570048309179\\
40	0.377506112469438\\
45	0.354455445544554\\
};
\addlegendentry{AF}

\addplot [color=mycolor5, line width=.6pt, mark size=2.5pt, mark=+, mark options={solid, mycolor5}, forget plot]
 plot [error bars/.cd, y dir = both, y explicit]
 table[row sep=crcr, y error plus index=2, y error minus index=3]{%
5	0.996846846846847	0.00185725478631426	0.00185725478631426\\
10	0.98496583143508	0.0040300241487367	0.0040300241487367\\
15	0.705069124423963	0.0163415882271863	0.0163415882271863\\
20	0.585547785547786	0.0453270580823677	0.0453270580823677\\
25	0.479716981132076	0.0252567802457124	0.0252567802457124\\
30	0.441527446300716	0.0249628770691988	0.0249628770691988\\
35	0.426570048309179	0.00871910417767152	0.00871910417767152\\
40	0.377506112469438	0.0144399247532845	0.0144399247532845\\
45	0.354455445544554	0.0216736155999681	0.0216736155999681\\
};

\addplot[area legend, draw=none, fill=mycolor6, fill opacity=0.2, forget plot]
table[row sep=crcr] {%
x	y\\
5	0.977971148929524\\
10	0.946138184056253\\
15	0.907815588418551\\
20	0.852899940079066\\
25	0.787636082859221\\
30	0.76261039424745\\
35	0.720884266948725\\
40	0.671661990134628\\
45	0.646871150597793\\
45	0.627393170121028\\
40	0.662232409642751\\
35	0.686606431814963\\
30	0.732913939590645\\
25	0.765201809004041\\
20	0.821315018564585\\
15	0.885934789397762\\
10	0.933881175393351\\
5	0.961754749790147\\
}--cycle;
\addplot [color=mycolor6, line width=.6pt, mark size=2.5pt, mark=x, mark options={solid, mycolor6}]
  table[row sep=crcr]{%
5	0.969862949359835\\
10	0.940009679724802\\
15	0.896875188908157\\
20	0.837107479321825\\
25	0.776418945931631\\
30	0.747762166919048\\
35	0.703745349381844\\
40	0.66694719988869\\
45	0.637132160359411\\
};
\addlegendentry{APC}

\addplot [color=mycolor6, line width=.6pt, mark size=2.5pt, mark=x, mark options={solid, mycolor6}, forget plot]
 plot [error bars/.cd, y dir = both, y explicit]
 table[row sep=crcr, y error plus index=2, y error minus index=3]{%
5	0.969862949359835	0.00810819956968828	0.00810819956968828\\
10	0.940009679724802	0.00612850433145123	0.00612850433145123\\
15	0.896875188908157	0.0109403995103948	0.0109403995103948\\
20	0.837107479321825	0.0157924607572407	0.0157924607572407\\
25	0.776418945931631	0.0112171369275902	0.0112171369275902\\
30	0.747762166919048	0.0148482273284026	0.0148482273284026\\
35	0.703745349381844	0.0171389175668811	0.0171389175668811\\
40	0.66694719988869	0.00471479024593845	0.00471479024593845\\
45	0.637132160359411	0.00973899023838242	0.00973899023838242\\
};
\end{axis}
\end{tikzpicture}%

%% file: imgs/MINNESOTA1/AB1MN.tex
%
%
\definecolor{mycolor1}{rgb}{0.00000,0.44700,0.74100}%
\definecolor{mycolor2}{rgb}{0.92900,0.69400,0.12500}%
\definecolor{mycolor3}{rgb}{0.49400,0.18400,0.55600}%
\definecolor{mycolor4}{rgb}{0.46600,0.67400,0.18800}%
\definecolor{mycolor5}{rgb}{0.30100,0.74500,0.93300}%
\definecolor{mycolor6}{rgb}{0.63500,0.07800,0.18400}%
\pgfplotsset{
compat=1.11,
legend image code/.code={
\draw[mark repeat=2,mark phase=2]
plot coordinates {
(0cm,0cm)
(0.15cm,0cm)        
(0.3cm,0cm)         
};%
}
}
\begin{tikzpicture}

\begin{axis}[%
width=.85\columnwidth,
height=.55\columnwidth,
scale only axis,
xmin=4,
xmax=46,
xtick={5,10,15,20,25,30,35,40,45},
xticklabel style={font=\footnotesize},
xlabel style={font=\color{white!15!black}\footnotesize,yshift=3pt},
xlabel={\# failed nodes},
ymin=0,
ymax=1.07,
ytick={0,0.2,0.4,0.6,0.8,1},
yticklabels={{0},{.2},{.4},{.6},{.8},{1}},
yticklabel style={font=\footnotesize},
ylabel style={font=\color{white!15!black}\footnotesize,yshift=-5pt},
ylabel={$a_B$},
axis background/.style={fill=white},
legend style={at={(0.1,0.55)}, legend columns=3, anchor=south west, legend cell align=left, align=left, fill=none, draw=none, font=\scriptsize}
]

\addplot[area legend, draw=none, fill=mycolor1, fill opacity=0.2, forget plot]
table[row sep=crcr] {%
x	y\\
5	1\\
10	1\\
15	1\\
20	1\\
25	1\\
30	1\\
35	1\\
40	1\\
45	1\\
45	1\\
40	1\\
35	1\\
30	1\\
25	1\\
20	1\\
15	1\\
10	1\\
5	1\\
}--cycle;
\addplot [color=mycolor1, line width=.6pt, mark size=2.5pt, mark=text,text mark={\LARGE $\star$}, mark options={solid, fill=mycolor1, mycolor1}]
  table[row sep=crcr]{%
5	1\\
10	1\\
15	1\\
20	1\\
25	1\\
30	1\\
35	1\\
40	1\\
45	1\\
};
\addlegendentry{FaCe}

\addplot [color=mycolor1, line width=.6pt, mark size=2.5pt, mark=text,text mark={\LARGE $\star$}, mark options={solid, fill=mycolor1, mycolor1}, forget plot]
 plot [error bars/.cd, y dir = both, y explicit]
 table[row sep=crcr, y error plus index=2, y error minus index=3]{%
5	1	0	0\\
10	1	0	0\\
15	1	0	0\\
20	1	0	0\\
25	1	0	0\\
30	1	0	0\\
35	1	0	0\\
40	1	0	0\\
45	1	0	0\\
};

\addplot[area legend, draw=none, fill=mycolor2, fill opacity=0.2, forget plot]
table[row sep=crcr] {%
x	y\\
5	0.24120226591666\\
10	0.377534305418231\\
15	0.201809404463352\\
20	0.0768542943042952\\
25	0.063689453417432\\
30	0.10885876908021\\
35	0.0341421356237309\\
40	0.0802584284067835\\
45	0.0620766102249654\\
45	0.0106506625023074\\
40	0.0288324806841255\\
35	0.00585786437626905\\
30	0.0755856753642344\\
25	0.0251994354714569\\
20	0.030288562838562\\
15	0.125809643155695\\
10	0.235799027915102\\
5	0.0921310674166737\\
}--cycle;
\addplot [color=mycolor2, line width=.6pt, mark size=2.5pt, mark=asterisk, mark options={solid, mycolor2}]
  table[row sep=crcr]{%
5	0.166666666666667\\
10	0.306666666666667\\
15	0.163809523809524\\
20	0.0535714285714286\\
25	0.0444444444444444\\
30	0.0922222222222222\\
35	0.02\\
40	0.0545454545454545\\
45	0.0363636363636364\\
};
\addlegendentry{GC}

\addplot [color=mycolor2, line width=.6pt, mark size=2.5pt, mark=asterisk, mark options={solid, mycolor2}, forget plot]
 plot [error bars/.cd, y dir = both, y explicit]
 table[row sep=crcr, y error plus index=2, y error minus index=3]{%
5	0.166666666666667	0.074535599249993	0.074535599249993\\
10	0.306666666666667	0.0708676387515643	0.0708676387515643\\
15	0.163809523809524	0.0379998806538286	0.0379998806538286\\
20	0.0535714285714286	0.0232828657328666	0.0232828657328666\\
25	0.0444444444444444	0.0192450089729875	0.0192450089729875\\
30	0.0922222222222222	0.0166365468579878	0.0166365468579878\\
35	0.02	0.0141421356237309	0.0141421356237309\\
40	0.0545454545454545	0.025712973861329	0.025712973861329\\
45	0.0363636363636364	0.025712973861329	0.025712973861329\\
};

\addplot[area legend, draw=none, fill=mycolor3, fill opacity=0.2, forget plot]
table[row sep=crcr] {%
x	y\\
5	0.462432778206914\\
10	0.440237792317455\\
15	0.601569117263222\\
20	0.408106059429445\\
25	0.465485511143483\\
30	0.439380649042436\\
35	0.401982005981707\\
40	0.501488549718934\\
45	0.49203412834873\\
45	0.299957213642612\\
40	0.425784177553793\\
35	0.286906882907181\\
30	0.325381255719469\\
25	0.34721290155493\\
20	0.308560607237222\\
15	0.396526120832016\\
10	0.253095541015879\\
5	0.204233888459753\\
}--cycle;
\addplot [color=mycolor3, line width=.6pt, mark size=2.5pt, mark=triangle, mark options={solid, mycolor3}]
  table[row sep=crcr]{%
5	0.333333333333333\\
10	0.346666666666667\\
15	0.499047619047619\\
20	0.358333333333333\\
25	0.406349206349206\\
30	0.382380952380952\\
35	0.344444444444444\\
40	0.463636363636364\\
45	0.395995670995671\\
};
\addlegendentry{GI}

\addplot [color=mycolor3, line width=.6pt, mark size=2.5pt, mark=triangle, mark options={solid, mycolor3}, forget plot]
 plot [error bars/.cd, y dir = both, y explicit]
 table[row sep=crcr, y error plus index=2, y error minus index=3]{%
5	0.333333333333333	0.129099444873581	0.129099444873581\\
10	0.346666666666667	0.093571125650788	0.093571125650788\\
15	0.499047619047619	0.102521498215603	0.102521498215603\\
20	0.358333333333333	0.0497727260961116	0.0497727260961116\\
25	0.406349206349206	0.0591363047942767	0.0591363047942767\\
30	0.382380952380952	0.0569996966614835	0.0569996966614835\\
35	0.344444444444444	0.057537561537263	0.057537561537263\\
40	0.463636363636364	0.0378521860825706	0.0378521860825706\\
45	0.395995670995671	0.0960384573530595	0.0960384573530595\\
};

\addplot[area legend, draw=none, fill=mycolor4, fill opacity=0.2, forget plot]
table[row sep=crcr] {%
x	y\\
5	0.573205080756888\\
10	0.678320104986824\\
15	0.547768918697798\\
20	0.317087891752952\\
25	0.188855805740577\\
30	0.231016348301187\\
35	0.34099554987341\\
40	0.356460642542435\\
45	0.121926111162296\\
45	0.0443076550714703\\
40	0.247435461353669\\
35	0.272337783459923\\
30	0.13612650884167\\
25	0.112731495846724\\
20	0.168626393961334\\
15	0.408421557492678\\
10	0.468346561679843\\
5	0.226794919243112\\
}--cycle;
\addplot [color=mycolor4, line width=.6pt, mark size=2.5pt, mark=square, mark options={solid, mycolor4}]
  table[row sep=crcr]{%
5	0.4\\
10	0.573333333333333\\
15	0.478095238095238\\
20	0.242857142857143\\
25	0.150793650793651\\
30	0.183571428571429\\
35	0.306666666666667\\
40	0.301948051948052\\
45	0.0831168831168831\\
};
\addlegendentry{GD}

\addplot [color=mycolor4, line width=.6pt, mark size=2.5pt, mark=square, mark options={solid, mycolor4}, forget plot]
 plot [error bars/.cd, y dir = both, y explicit]
 table[row sep=crcr, y error plus index=2, y error minus index=3]{%
5	0.4	0.173205080756888	0.173205080756888\\
10	0.573333333333333	0.104986771653491	0.104986771653491\\
15	0.478095238095238	0.0696736806025595	0.0696736806025595\\
20	0.242857142857143	0.074230748895809	0.074230748895809\\
25	0.150793650793651	0.0380621549469263	0.0380621549469263\\
30	0.183571428571429	0.0474449197297589	0.0474449197297589\\
35	0.306666666666667	0.0343288832067433	0.0343288832067433\\
40	0.301948051948052	0.0545125905943833	0.0545125905943833\\
45	0.0831168831168831	0.0388092280454128	0.0388092280454128\\
};

\addplot[area legend, draw=none, fill=mycolor5, fill opacity=0.2, forget plot]
table[row sep=crcr] {%
x	y\\
5	1\\
10	1\\
15	0.784037008503093\\
20	0.653979157616564\\
25	0.544492422502471\\
30	0.487748517734456\\
35	0.463531622792857\\
40	0.416726039399559\\
45	0.391384679168039\\
45	0.373059765276405\\
40	0.393273960600442\\
35	0.450754091492858\\
30	0.445584815598877\\
25	0.511507577497529\\
20	0.606020842383436\\
15	0.735962991496907\\
10	1\\
5	1\\
}--cycle;
\addplot [color=mycolor5, line width=.6pt, mark size=2.5pt, mark=+, mark options={solid, mycolor5}]
  table[row sep=crcr]{%
5	1\\
10	1\\
15	0.76\\
20	0.63\\
25	0.528\\
30	0.466666666666667\\
35	0.457142857142857\\
40	0.405\\
45	0.382222222222222\\
};
\addlegendentry{AF}

\addplot [color=mycolor5, line width=.6pt, mark size=2.5pt, mark=+, mark options={solid, mycolor5}, forget plot]
 plot [error bars/.cd, y dir = both, y explicit]
 table[row sep=crcr, y error plus index=2, y error minus index=3]{%
5	1	0	0\\
10	1	0	0\\
15	0.76	0.0240370085030933	0.0240370085030933\\
20	0.63	0.0239791576165636	0.0239791576165636\\
25	0.528	0.0164924225024706	0.0164924225024706\\
30	0.466666666666667	0.0210818510677892	0.0210818510677892\\
35	0.457142857142857	0.0063887656499994	0.0063887656499994\\
40	0.405	0.0117260393995586	0.0117260393995586\\
45	0.382222222222222	0.00916245694581702	0.00916245694581702\\
};

\addplot[area legend, draw=none, fill=mycolor6, fill opacity=0.2, forget plot]
table[row sep=crcr] {%
x	y\\
5	0.404103520853922\\
10	0.336587256544677\\
15	0.191963370778639\\
20	0.148256778645765\\
25	0.063689453417432\\
30	0.0646850196850295\\
35	0.0605471361138563\\
40	0.0678146425197751\\
45	0.0753757814363994\\
45	0.0276545215939036\\
40	0.0256918509867184\\
35	0.0238973083305882\\
30	0.0253149803149705\\
25	0.0251994354714569\\
20	0.073171792782806\\
15	0.126131867316599\\
10	0.196746076788657\\
5	0.129229812479411\\
}--cycle;
\addplot [color=mycolor6, line width=.6pt, mark size=2.5pt, mark=x, mark options={solid, mycolor6}]
  table[row sep=crcr]{%
5	0.266666666666667\\
10	0.266666666666667\\
15	0.159047619047619\\
20	0.110714285714286\\
25	0.0444444444444444\\
30	0.045\\
35	0.0422222222222222\\
40	0.0467532467532468\\
45	0.0515151515151515\\
};
\addlegendentry{APC}

\addplot [color=mycolor6, line width=.6pt, mark size=2.5pt, mark=x, mark options={solid, mycolor6}, forget plot]
 plot [error bars/.cd, y dir = both, y explicit]
 table[row sep=crcr, y error plus index=2, y error minus index=3]{%
5	0.266666666666667	0.137436854187255	0.137436854187255\\
10	0.266666666666667	0.0699205898780101	0.0699205898780101\\
15	0.159047619047619	0.0329157517310199	0.0329157517310199\\
20	0.110714285714286	0.0375424929314797	0.0375424929314797\\
25	0.0444444444444444	0.0192450089729875	0.0192450089729875\\
30	0.045	0.0196850196850295	0.0196850196850295\\
35	0.0422222222222222	0.018324913891634	0.018324913891634\\
40	0.0467532467532468	0.0210613957665284	0.0210613957665284\\
45	0.0515151515151515	0.0238606299212479	0.0238606299212479\\
};
\end{axis}
\end{tikzpicture}%

%% file: imgs/MINNESOTA1/R11MN.tex
%
%
\definecolor{mycolor1}{rgb}{0.00000,0.44700,0.74100}%
\definecolor{mycolor2}{rgb}{0.92900,0.69400,0.12500}%
\definecolor{mycolor3}{rgb}{0.49400,0.18400,0.55600}%
\definecolor{mycolor4}{rgb}{0.46600,0.67400,0.18800}%
\definecolor{mycolor5}{rgb}{0.30100,0.74500,0.93300}%
\definecolor{mycolor6}{rgb}{0.63500,0.07800,0.18400}%
\pgfplotsset{
compat=1.11,
legend image code/.code={
\draw[mark repeat=2,mark phase=2]
plot coordinates {
(0cm,0cm)
(0.15cm,0cm)        
(0.3cm,0cm)         
};%
}
}
\begin{tikzpicture}

\begin{axis}[%
width=.85\columnwidth,
height=.55\columnwidth,
scale only axis,
xmin=4,
xmax=46,
xtick={5,10,15,20,25,30,35,40,45},
xticklabel style={font=\footnotesize},
xlabel style={font=\color{white!15!black}\footnotesize,yshift=3pt},
xlabel={\# failed nodes},
ymin=0,
ymax=1.07,
ytick={0,0.2,0.4,0.6,0.8,1},
yticklabels={{0},{.2},{.4},{.6},{.8},{1}},
yticklabel style={font=\footnotesize},
ylabel style={font=\color{white!15!black}\footnotesize,yshift=-7pt},
ylabel={$R_1$},
axis background/.style={fill=white},
legend style={at={(0.15,0.5)}, legend columns=3, anchor=south west, legend cell align=right, align=left, fill=none, draw=none, font=\scriptsize}
]

\addplot[area legend, draw=none, fill=mycolor1, fill opacity=0.2, forget plot]
table[row sep=crcr] {%
x	y\\
5	0.714641016151378\\
10	0.69605551275464\\
15	0.571547005383793\\
20	0.561180339887499\\
25	0.590422205101856\\
30	0.610653841409022\\
35	0.59059486837857\\
40	0.557776083947861\\
45	0.520706269029503\\
45	0.457071508748275\\
40	0.512223916052139\\
35	0.552262274478573\\
30	0.576012825257645\\
25	0.561577794898144\\
20	0.538819660112501\\
15	0.548452994616208\\
10	0.62394448724536\\
5	0.645358983848622\\
}--cycle;
\addplot [color=mycolor1, line width=.6pt, mark size=2.5pt, mark=text,text mark={\LARGE $\star$}, mark options={solid, fill=mycolor1, mycolor1}]
  table[row sep=crcr]{%
5	0.68\\
10	0.66\\
15	0.56\\
20	0.55\\
25	0.576\\
30	0.593333333333333\\
35	0.571428571428571\\
40	0.535\\
45	0.488888888888889\\
};
\addlegendentry{FaCe}

\addplot [color=mycolor1, line width=.6pt, mark size=2.5pt, mark=text,text mark={\LARGE $\star$}, mark options={solid, fill=mycolor1, mycolor1}, forget plot]
 plot [error bars/.cd, y dir = both, y explicit]
 table[row sep=crcr, y error plus index=2, y error minus index=3]{%
5	0.68	0.0346410161513776	0.0346410161513776\\
10	0.66	0.0360555127546399	0.0360555127546399\\
15	0.56	0.0115470053837925	0.0115470053837925\\
20	0.55	0.0111803398874989	0.0111803398874989\\
25	0.576	0.014422205101856	0.014422205101856\\
30	0.593333333333333	0.0173205080756888	0.0173205080756888\\
35	0.571428571428571	0.0191662969499982	0.0191662969499982\\
40	0.535	0.0227760839478607	0.0227760839478607\\
45	0.488888888888889	0.0318173801406141	0.0318173801406141\\
};

\addplot[area legend, draw=none, fill=mycolor2, fill opacity=0.2, forget plot]
table[row sep=crcr] {%
x	y\\
5	0.444721359549996\\
10	0.444721359549996\\
15	0.384037008503093\\
20	0.329580398915498\\
25	0.281941125496954\\
30	0.277600366297602\\
35	0.284984464704394\\
40	0.313831406486677\\
45	0.269511355309512\\
45	0.237155311357154\\
40	0.256168593513323\\
35	0.252158392438463\\
30	0.229066300369065\\
25	0.214058874503046\\
20	0.270419601084502\\
15	0.335962991496907\\
10	0.355278640450004\\
5	0.355278640450004\\
}--cycle;
\addplot [color=mycolor2, line width=.6pt, mark size=2.5pt, mark=asterisk, mark options={solid, mycolor2}]
  table[row sep=crcr]{%
5	0.4\\
10	0.4\\
15	0.36\\
20	0.3\\
25	0.248\\
30	0.253333333333333\\
35	0.268571428571429\\
40	0.285\\
45	0.253333333333333\\
};
\addlegendentry{GC}

\addplot [color=mycolor2, line width=.6pt, mark size=2.5pt, mark=asterisk, mark options={solid, mycolor2}, forget plot]
 plot [error bars/.cd, y dir = both, y explicit]
 table[row sep=crcr, y error plus index=2, y error minus index=3]{%
5	0.4	0.0447213595499958	0.0447213595499958\\
10	0.4	0.0447213595499958	0.0447213595499958\\
15	0.36	0.0240370085030933	0.0240370085030933\\
20	0.3	0.0295803989154981	0.0295803989154981\\
25	0.248	0.0339411254969543	0.0339411254969543\\
30	0.253333333333333	0.0242670329642684	0.0242670329642684\\
35	0.268571428571429	0.0164130361329658	0.0164130361329658\\
40	0.285	0.028831406486677	0.028831406486677\\
45	0.253333333333333	0.0161780219761789	0.0161780219761789\\
};

\addplot[area legend, draw=none, fill=mycolor3, fill opacity=0.2, forget plot]
table[row sep=crcr] {%
x	y\\
5	0.444721359549996\\
10	0.507958315233127\\
15	0.471972210155418\\
20	0.369364916731037\\
25	0.453120439557122\\
30	0.415485837703549\\
35	0.411665035800852\\
40	0.423874586088177\\
45	0.383431119434039\\
45	0.34545776945485\\
40	0.386125413911823\\
35	0.376906392770577\\
30	0.397847495629785\\
25	0.394879560442878\\
20	0.330635083268963\\
15	0.408027789844582\\
10	0.412041684766873\\
5	0.355278640450004\\
}--cycle;
\addplot [color=mycolor3, line width=.6pt, mark size=2.5pt, mark=triangle, mark options={solid, mycolor3}]
  table[row sep=crcr]{%
5	0.4\\
10	0.46\\
15	0.44\\
20	0.35\\
25	0.424\\
30	0.406666666666667\\
35	0.394285714285714\\
40	0.405\\
45	0.364444444444444\\
};
\addlegendentry{GI}

\addplot [color=mycolor3, line width=.6pt, mark size=2.5pt, mark=triangle, mark options={solid, mycolor3}, forget plot]
 plot [error bars/.cd, y dir = both, y explicit]
 table[row sep=crcr, y error plus index=2, y error minus index=3]{%
5	0.4	0.0447213595499958	0.0447213595499958\\
10	0.46	0.0479583152331272	0.0479583152331272\\
15	0.44	0.0319722101554181	0.0319722101554181\\
20	0.35	0.0193649167310371	0.0193649167310371\\
25	0.424	0.0291204395571221	0.0291204395571221\\
30	0.406666666666667	0.00881917103688197	0.00881917103688197\\
35	0.394285714285714	0.0173793215151378	0.0173793215151378\\
40	0.405	0.0188745860881769	0.0188745860881769\\
45	0.364444444444444	0.0189866749895945	0.0189866749895945\\
};

\addplot[area legend, draw=none, fill=mycolor4, fill opacity=0.2, forget plot]
table[row sep=crcr] {%
x	y\\
5	0.59211102550928\\
10	0.507958315233127\\
15	0.440820704170784\\
20	0.420413812651491\\
25	0.390627416997969\\
30	0.334142135623731\\
35	0.356130718031613\\
40	0.34193741096848\\
45	0.36462836117844\\
45	0.337593861043782\\
40	0.29806258903152\\
35	0.341012139111244\\
30	0.305857864376269\\
25	0.34537258300203\\
20	0.359586187348509\\
15	0.385845962495882\\
10	0.412041684766873\\
5	0.44788897449072\\
}--cycle;
\addplot [color=mycolor4, line width=.6pt, mark size=2.5pt, mark=square, mark options={solid, mycolor4}]
  table[row sep=crcr]{%
5	0.52\\
10	0.46\\
15	0.413333333333333\\
20	0.39\\
25	0.368\\
30	0.32\\
35	0.348571428571429\\
40	0.32\\
45	0.351111111111111\\
};
\addlegendentry{GD}

\addplot [color=mycolor4, line width=.6pt, mark size=2.5pt, mark=square, mark options={solid, mycolor4}, forget plot]
 plot [error bars/.cd, y dir = both, y explicit]
 table[row sep=crcr, y error plus index=2, y error minus index=3]{%
5	0.52	0.0721110255092798	0.0721110255092798\\
10	0.46	0.0479583152331272	0.0479583152331272\\
15	0.413333333333333	0.0274873708374511	0.0274873708374511\\
20	0.39	0.0304138126514911	0.0304138126514911\\
25	0.368	0.0226274169979695	0.0226274169979695\\
30	0.32	0.0141421356237309	0.0141421356237309\\
35	0.348571428571429	0.00755928946018455	0.00755928946018455\\
40	0.32	0.0219374109684803	0.0219374109684803\\
45	0.351111111111111	0.0135172500673294	0.0135172500673294\\
};

\addplot[area legend, draw=none, fill=mycolor5, fill opacity=0.2, forget plot]
table[row sep=crcr] {%
x	y\\
5	1\\
10	1\\
15	0.830971675407097\\
20	0.669364916731037\\
25	0.572784609690826\\
30	0.526942541767661\\
35	0.532355872343819\\
40	0.492677669529664\\
45	0.472645368355163\\
45	0.451799076089281\\
40	0.457322330470336\\
35	0.496215556227609\\
30	0.486390791565673\\
25	0.531215390309173\\
20	0.630635083268963\\
15	0.79569499125957\\
10	1\\
5	1\\
}--cycle;
\addplot [color=mycolor5, line width=.6pt, mark size=2.5pt, mark=+, mark options={solid, mycolor5}]
  table[row sep=crcr]{%
5	1\\
10	1\\
15	0.813333333333333\\
20	0.65\\
25	0.552\\
30	0.506666666666667\\
35	0.514285714285714\\
40	0.475\\
45	0.462222222222222\\
};
\addlegendentry{AF}

\addplot [color=mycolor5, line width=.6pt, mark size=2.5pt, mark=+, mark options={solid, mycolor5}, forget plot]
 plot [error bars/.cd, y dir = both, y explicit]
 table[row sep=crcr, y error plus index=2, y error minus index=3]{%
5	1	0	0\\
10	1	0	0\\
15	0.813333333333333	0.0176383420737639	0.0176383420737639\\
20	0.65	0.0193649167310371	0.0193649167310371\\
25	0.552	0.0207846096908265	0.0207846096908265\\
30	0.506666666666667	0.0202758751009941	0.0202758751009941\\
35	0.514285714285714	0.018070158058105	0.018070158058105\\
40	0.475	0.0176776695296637	0.0176776695296637\\
45	0.462222222222222	0.0104231461329409	0.0104231461329409\\
};

\addplot[area legend, draw=none, fill=mycolor6, fill opacity=0.2, forget plot]
table[row sep=crcr] {%
x	y\\
5	0.376568542494924\\
10	0.407958315233127\\
15	0.372773865220664\\
20	0.249370039370059\\
25	0.246978250586152\\
30	0.26721471334323\\
35	0.23587839578372\\
40	0.2125\\
45	0.22932478316157\\
45	0.206230772393985\\
40	0.1875\\
35	0.198407318501994\\
30	0.226118619990103\\
25	0.201021749413848\\
20	0.170629960629941\\
15	0.293892801446003\\
10	0.312041684766873\\
5	0.263431457505076\\
}--cycle;
\addplot [color=mycolor6, line width=.6pt, mark size=2.5pt, mark=x, mark options={solid, mycolor6}]
  table[row sep=crcr]{%
5	0.32\\
10	0.36\\
15	0.333333333333333\\
20	0.21\\
25	0.224\\
30	0.246666666666667\\
35	0.217142857142857\\
40	0.2\\
45	0.217777777777778\\
};
\addlegendentry{APC}

\addplot [color=mycolor6, line width=.6pt, mark size=2.5pt, mark=x, mark options={solid, mycolor6}, forget plot]
 plot [error bars/.cd, y dir = both, y explicit]
 table[row sep=crcr, y error plus index=2, y error minus index=3]{%
5	0.32	0.0565685424949238	0.0565685424949238\\
10	0.36	0.0479583152331272	0.0479583152331272\\
15	0.333333333333333	0.0394405318873308	0.0394405318873308\\
20	0.21	0.0393700393700591	0.0393700393700591\\
25	0.224	0.0229782505861521	0.0229782505861521\\
30	0.246666666666667	0.0205480466765633	0.0205480466765633\\
35	0.217142857142857	0.0187355386408629	0.0187355386408629\\
40	0.2	0.0125	0.0125\\
45	0.217777777777778	0.0115470053837925	0.0115470053837925\\
};
\addplot [color=mycolor6, line width=.6pt, mark size=2.5pt, mark=x, mark options={solid, mycolor6}, forget plot]
 plot [error bars/.cd, y dir = both, y explicit]
 table[row sep=crcr, y error plus index=2, y error minus index=3]{%
5	0.32	0.0565685424949238	0.0565685424949238\\
10	0.36	0.0479583152331272	0.0479583152331272\\
15	0.333333333333333	0.0394405318873308	0.0394405318873308\\
20	0.21	0.0393700393700591	0.0393700393700591\\
25	0.224	0.0229782505861521	0.0229782505861521\\
30	0.246666666666667	0.0205480466765633	0.0205480466765633\\
35	0.217142857142857	0.0187355386408629	0.0187355386408629\\
40	0.2	0.0125	0.0125\\
45	0.217777777777778	0.0115470053837925	0.0115470053837925\\
};
\addplot[area legend, draw=none, fill=black, fill opacity=0.08, forget plot]
table[row sep=crcr] {%
x	y\\
5	0.714641016151378\\
10	0.69605551275464\\
15	0.571547005383793\\
20	0.561180339887499\\
25	0.590422205101856\\
30	0.610653841409022\\
35	0.59059486837857\\
40	0.557776083947861\\
45	0.520706269029503\\
45	0.457071508748275\\
40	0.512223916052139\\
35	0.552262274478573\\
30	0.576012825257645\\
25	0.561577794898144\\
20	0.538819660112501\\
15	0.548452994616208\\
10	0.62394448724536\\
5	0.645358983848622\\
}--cycle;
\addplot [color=black, dashed, line width=.6pt]
  table[row sep=crcr]{%
5	0.68\\
10	0.66\\
15	0.56\\
20	0.55\\
25	0.576\\
30	0.593333333333333\\
35	0.571428571428571\\
40	0.535\\
45	0.488888888888889\\
};
\addlegendentry{all}

\addplot [color=black, dashed, line width=.6pt, forget plot]
 plot [error bars/.cd, y dir = both, y explicit]
 table[row sep=crcr, y error plus index=2, y error minus index=3]{%
5	0.68	0.0346410161513776	0.0346410161513776\\
10	0.66	0.0360555127546399	0.0360555127546399\\
15	0.56	0.0115470053837925	0.0115470053837925\\
20	0.55	0.0111803398874989	0.0111803398874989\\
25	0.576	0.014422205101856	0.014422205101856\\
30	0.593333333333333	0.0173205080756888	0.0173205080756888\\
35	0.571428571428571	0.0191662969499982	0.0191662969499982\\
40	0.535	0.0227760839478607	0.0227760839478607\\
45	0.488888888888889	0.0318173801406141	0.0318173801406141\\
};
\end{axis}
\end{tikzpicture}%

%% file: imgs/MINNESOTA1/R21MN.tex
\definecolor{mycolor1}{rgb}{0.00000,0.44700,0.74100}%
\definecolor{mycolor2}{rgb}{0.92900,0.69400,0.12500}%
\definecolor{mycolor3}{rgb}{0.49400,0.18400,0.55600}%
\definecolor{mycolor4}{rgb}{0.46600,0.67400,0.18800}%
\definecolor{mycolor5}{rgb}{0.30100,0.74500,0.93300}%
\definecolor{mycolor6}{rgb}{0.63500,0.07800,0.18400}%
\pgfplotsset{
compat=1.11,
legend image code/.code={
\draw[mark repeat=2,mark phase=2]
plot coordinates {
(0cm,0cm)
(0.15cm,0cm)        
(0.3cm,0cm)         
};%
}
}
\begin{tikzpicture}

\begin{axis}[%
width=.85\columnwidth,
height=.55\columnwidth,
scale only axis,
xmin=4,
xmax=46,
xtick={5,10,15,20,25,30,35,40,45},
xticklabel style={font=\footnotesize},
xlabel style={font=\color{white!15!black}\footnotesize,yshift=3pt},
xlabel={\# failed nodes},
ymin=0,
ymax=1.07,
ytick={0,0.2,0.4,0.6,0.8,1},
yticklabels={{0},{.2},{.4},{.6},{.8},{1}},
yticklabel style={font=\footnotesize},
ylabel style={font=\color{white!15!black}\footnotesize,yshift=-7pt},
ylabel={$R_2$},
axis background/.style={fill=white},
legend style={at={(0.15,0.4)}, legend columns=3, anchor=south west, legend cell align=left, align=left, fill=none, draw=none, font=\scriptsize}
]

\addplot[area legend, draw=none, fill=mycolor1, fill opacity=0.2, forget plot]
table[row sep=crcr] {%
x	y\\
5	0.41261584532173\\
10	0.299998734259163\\
15	0.257151579923651\\
20	0.258557260119854\\
25	0.273392368043676\\
30	0.272823859138674\\
35	0.282444631026368\\
40	0.27964395820779\\
45	0.272545647857007\\
45	0.245634291654107\\
40	0.256719537691901\\
35	0.260843828153785\\
30	0.251778312694801\\
25	0.24267090156929\\
20	0.231863743875258\\
15	0.238162947668802\\
10	0.23718353875606\\
5	0.29636448594121\\
}--cycle;
\addplot [color=mycolor1, line width=.6pt, mark size=2.5pt, mark=text,text mark={\LARGE $\star$}, mark options={solid, fill=mycolor1, mycolor1}]
  table[row sep=crcr]{%
5	0.35449016563147\\
10	0.268591136507612\\
15	0.247657263796226\\
20	0.245210501997556\\
25	0.258031634806483\\
30	0.262301085916737\\
35	0.271644229590077\\
40	0.268181747949845\\
45	0.259089969755557\\
};
\addlegendentry{FaCe}

\addplot [color=mycolor1, line width=.6pt, mark size=2.5pt, mark=text,text mark={\LARGE $\star$}, mark options={solid, fill=mycolor1, mycolor1}, forget plot]
 plot [error bars/.cd, y dir = both, y explicit]
 table[row sep=crcr, y error plus index=2, y error minus index=3]{%
5	0.35449016563147	0.0581256796902601	0.0581256796902601\\
10	0.268591136507612	0.0314075977515516	0.0314075977515516\\
15	0.247657263796226	0.00949431612742442	0.00949431612742442\\
20	0.245210501997556	0.0133467581222979	0.0133467581222979\\
25	0.258031634806483	0.015360733237193	0.015360733237193\\
30	0.262301085916737	0.0105227732219365	0.0105227732219365\\
35	0.271644229590077	0.0108004014362912	0.0108004014362912\\
40	0.268181747949845	0.0114622102579442	0.0114622102579442\\
45	0.259089969755557	0.01345567810145	0.01345567810145\\
};

\addplot[area legend, draw=none, fill=mycolor2, fill opacity=0.2, forget plot]
table[row sep=crcr] {%
x	y\\
5	0.196867639286788\\
10	0.172934039401805\\
15	0.16810674976305\\
20	0.150315450102326\\
25	0.155775407060324\\
30	0.157149993030951\\
35	0.165181815030788\\
40	0.167590969452723\\
45	0.178989730925197\\
45	0.164626554870646\\
40	0.157272879476461\\
35	0.152259806868565\\
30	0.151891842583959\\
25	0.148564628807168\\
20	0.139867791979571\\
15	0.15739925912057\\
10	0.139920862926782\\
5	0.169799027379879\\
}--cycle;
\addplot [color=mycolor2, line width=.6pt, mark size=2.5pt, mark=asterisk, mark options={solid, mycolor2}]
  table[row sep=crcr]{%
5	0.183333333333333\\
10	0.156427451164293\\
15	0.16275300444181\\
20	0.145091621040948\\
25	0.152170017933746\\
30	0.154520917807455\\
35	0.158720810949676\\
40	0.162431924464592\\
45	0.171808142897921\\
};
\addlegendentry{GC}

\addplot [color=mycolor2, line width=.6pt, mark size=2.5pt, mark=asterisk, mark options={solid, mycolor2}, forget plot]
 plot [error bars/.cd, y dir = both, y explicit]
 table[row sep=crcr, y error plus index=2, y error minus index=3]{%
5	0.183333333333333	0.0135343059534548	0.0135343059534548\\
10	0.156427451164293	0.0165065882375115	0.0165065882375115\\
15	0.16275300444181	0.00535374532124017	0.00535374532124017\\
20	0.145091621040948	0.00522382906137768	0.00522382906137768\\
25	0.152170017933746	0.00360538912657761	0.00360538912657761\\
30	0.154520917807455	0.0026290752234957	0.0026290752234957\\
35	0.158720810949676	0.00646100408111181	0.00646100408111181\\
40	0.162431924464592	0.00515904498813106	0.00515904498813106\\
45	0.171808142897921	0.0071815880272758	0.0071815880272758\\
};

\addplot[area legend, draw=none, fill=mycolor3, fill opacity=0.2, forget plot]
table[row sep=crcr] {%
x	y\\
5	0.0433760792674507\\
10	0.0592239655431879\\
15	0.0950381422605029\\
20	0.11542641891412\\
25	0.133754809940766\\
30	0.148873399487349\\
35	0.179222935945029\\
40	0.177014175964332\\
45	0.175701831722734\\
45	0.170112686499644\\
40	0.169426418122687\\
35	0.169282931771406\\
30	0.143712056245323\\
25	0.126578935960695\\
20	0.108199331194767\\
15	0.0855397403124441\\
10	0.0566752856361991\\
5	0.0351272699499424\\
}--cycle;
\addplot [color=mycolor3, line width=.6pt, mark size=2.5pt, mark=triangle, mark options={solid, mycolor3}]
  table[row sep=crcr]{%
5	0.0392516746086966\\
10	0.0579496255896935\\
15	0.0902889412864735\\
20	0.111812875054443\\
25	0.130166872950731\\
30	0.146292727866336\\
35	0.174252933858217\\
40	0.173220297043509\\
45	0.172907259111189\\
};
\addlegendentry{GI}

\addplot [color=mycolor3, line width=.6pt, mark size=2.5pt, mark=triangle, mark options={solid, mycolor3}, forget plot]
 plot [error bars/.cd, y dir = both, y explicit]
 table[row sep=crcr, y error plus index=2, y error minus index=3]{%
5	0.0392516746086966	0.00412440465875415	0.00412440465875415\\
10	0.0579496255896935	0.00127433995349441	0.00127433995349441\\
15	0.0902889412864735	0.0047492009740294	0.0047492009740294\\
20	0.111812875054443	0.00361354385967643	0.00361354385967643\\
25	0.130166872950731	0.00358793699003561	0.00358793699003561\\
30	0.146292727866336	0.00258067162101294	0.00258067162101294\\
35	0.174252933858217	0.00497000208681139	0.00497000208681139\\
40	0.173220297043509	0.00379387892082256	0.00379387892082256\\
45	0.172907259111189	0.0027945726115447	0.0027945726115447\\
};

\addplot[area legend, draw=none, fill=mycolor4, fill opacity=0.2, forget plot]
table[row sep=crcr] {%
x	y\\
5	0.167987164507799\\
10	0.110524620664865\\
15	0.129270455792382\\
20	0.128491577025274\\
25	0.144196358670383\\
30	0.1540546520409\\
35	0.183537205783077\\
40	0.182505804100556\\
45	0.183174939883478\\
45	0.176373657137032\\
40	0.17400434413229\\
35	0.173348665373108\\
30	0.149512143799467\\
25	0.131638485167596\\
20	0.118583089360926\\
15	0.112559834944603\\
10	0.0929192364762037\\
5	0.125473534613278\\
}--cycle;
\addplot [color=mycolor4, line width=.6pt, mark size=2.5pt, mark=square, mark options={solid, mycolor4}]
  table[row sep=crcr]{%
5	0.146730349560538\\
10	0.101721928570535\\
15	0.120915145368492\\
20	0.1235373331931\\
25	0.137917421918989\\
30	0.151783397920184\\
35	0.178442935578093\\
40	0.178255074116423\\
45	0.179774298510255\\
};
\addlegendentry{GD}

\addplot [color=mycolor4, line width=.6pt, mark size=2.5pt, mark=square, mark options={solid, mycolor4}, forget plot]
 plot [error bars/.cd, y dir = both, y explicit]
 table[row sep=crcr, y error plus index=2, y error minus index=3]{%
5	0.146730349560538	0.0212568149472605	0.0212568149472605\\
10	0.101721928570535	0.00880269209433083	0.00880269209433083\\
15	0.120915145368492	0.00835531042388935	0.00835531042388935\\
20	0.1235373331931	0.00495424383217392	0.00495424383217392\\
25	0.137917421918989	0.0062789367513935	0.0062789367513935\\
30	0.151783397920184	0.00227125412071643	0.00227125412071643\\
35	0.178442935578093	0.00509427020498454	0.00509427020498454\\
40	0.178255074116423	0.00425072998413324	0.00425072998413324\\
45	0.179774298510255	0.00340064137322296	0.00340064137322296\\
};

\addplot[area legend, draw=none, fill=mycolor5, fill opacity=0.2, forget plot]
table[row sep=crcr] {%
x	y\\
5	1\\
10	1\\
15	0.188924384950107\\
20	0.131873870536581\\
25	0.120382785945246\\
30	0.128868096692887\\
35	0.143648016640687\\
40	0.155757491794604\\
45	0.159275498497493\\
45	0.148789119037323\\
40	0.139852441563973\\
35	0.132979375909822\\
30	0.117543718405723\\
25	0.110973859436817\\
20	0.113286858098002\\
15	0.138668281743688\\
10	1\\
5	1\\
}--cycle;
\addplot [color=mycolor5, line width=.6pt, mark size=2.5pt, mark=+, mark options={solid, mycolor5}]
  table[row sep=crcr]{%
5	1\\
10	1\\
15	0.163796333346898\\
20	0.122580364317292\\
25	0.115678322691031\\
30	0.123205907549305\\
35	0.138313696275254\\
40	0.147804966679288\\
45	0.154032308767408\\
};
\addlegendentry{AF}

\addplot [color=mycolor5, line width=.6pt, mark size=2.5pt, mark=+, mark options={solid, mycolor5}, forget plot]
 plot [error bars/.cd, y dir = both, y explicit]
 table[row sep=crcr, y error plus index=2, y error minus index=3]{%
5	1	0	0\\
10	1	0	0\\
15	0.163796333346898	0.0251280516032098	0.0251280516032098\\
20	0.122580364317292	0.0092935062192894	0.0092935062192894\\
25	0.115678322691031	0.00470446325421465	0.00470446325421465\\
30	0.123205907549305	0.00566218914358212	0.00566218914358212\\
35	0.138313696275254	0.00533432036543256	0.00533432036543256\\
40	0.147804966679288	0.00795252511531552	0.00795252511531552\\
45	0.154032308767408	0.0052431897300852	0.0052431897300852\\
};

\addplot[area legend, draw=none, fill=mycolor6, fill opacity=0.2, forget plot]
table[row sep=crcr] {%
x	y\\
5	0.253869456143882\\
10	0.207589376246797\\
15	0.176521861037473\\
20	0.156172416178812\\
25	0.14739384897913\\
30	0.161727453146705\\
35	0.161449106687644\\
40	0.163524504005546\\
45	0.171050484014421\\
45	0.16125452172155\\
40	0.155841922630711\\
35	0.152778301705326\\
30	0.148274387990916\\
25	0.13944720374185\\
20	0.139889683240238\\
15	0.161979059493299\\
10	0.173063759324371\\
5	0.18709693041074\\
}--cycle;
\addplot [color=mycolor6, line width=.6pt, mark size=2.5pt, mark=x, mark options={solid, mycolor6}]
  table[row sep=crcr]{%
5	0.220483193277311\\
10	0.190326567785584\\
15	0.169250460265386\\
20	0.148031049709525\\
25	0.14342052636049\\
30	0.15500092056881\\
35	0.157113704196485\\
40	0.159683213318129\\
45	0.166152502867985\\
};
\addlegendentry{APC}

\addplot [color=mycolor6, line width=.6pt, mark size=2.5pt, mark=x, mark options={solid, mycolor6}, forget plot]
 plot [error bars/.cd, y dir = both, y explicit]
 table[row sep=crcr, y error plus index=2, y error minus index=3]{%
5	0.220483193277311	0.0333862628665706	0.0333862628665706\\
10	0.190326567785584	0.0172628084612129	0.0172628084612129\\
15	0.169250460265386	0.00727140077208708	0.00727140077208708\\
20	0.148031049709525	0.00814136646928677	0.00814136646928677\\
25	0.14342052636049	0.00397332261864002	0.00397332261864002\\
30	0.15500092056881	0.00672653257789458	0.00672653257789458\\
35	0.157113704196485	0.00433540249115891	0.00433540249115891\\
40	0.159683213318129	0.00384129068741718	0.00384129068741718\\
45	0.166152502867985	0.0048979811464355	0.0048979811464355\\
};

\addplot[area legend, draw=none, fill=black, fill opacity=0.08, forget plot]
table[row sep=crcr] {%
x	y\\
5	0.41261584532173\\
10	0.299998734259163\\
15	0.257151579923651\\
20	0.258557260119854\\
25	0.273392368043676\\
30	0.272823859138674\\
35	0.282444631026368\\
40	0.27964395820779\\
45	0.272545647857007\\
45	0.245634291654107\\
40	0.256719537691901\\
35	0.260843828153785\\
30	0.251778312694801\\
25	0.24267090156929\\
20	0.231863743875258\\
15	0.238162947668802\\
10	0.23718353875606\\
5	0.29636448594121\\
}--cycle;
\addplot [color=black, dashed, line width=.6pt]
  table[row sep=crcr]{%
5	0.35449016563147\\
10	0.268591136507612\\
15	0.247657263796226\\
20	0.245210501997556\\
25	0.258031634806483\\
30	0.262301085916737\\
35	0.271644229590077\\
40	0.268181747949845\\
45	0.259089969755557\\
};
\addlegendentry{all}

\addplot [color=black, dashed, line width=.6pt, forget plot]
 plot [error bars/.cd, y dir = both, y explicit]
 table[row sep=crcr, y error plus index=2, y error minus index=3]{%
5	0.35449016563147	0.0581256796902601	0.0581256796902601\\
10	0.268591136507612	0.0314075977515516	0.0314075977515516\\
15	0.247657263796226	0.00949431612742442	0.00949431612742442\\
20	0.245210501997556	0.0133467581222979	0.0133467581222979\\
25	0.258031634806483	0.015360733237193	0.015360733237193\\
30	0.262301085916737	0.0105227732219365	0.0105227732219365\\
35	0.271644229590077	0.0108004014362912	0.0108004014362912\\
40	0.268181747949845	0.0114622102579442	0.0114622102579442\\
45	0.259089969755557	0.01345567810145	0.01345567810145\\
};
\end{axis}
\end{tikzpicture}%

%% file: imgs/MINNESOTA2/AW2MN.tex
%
%
\definecolor{mycolor1}{rgb}{0.00000,0.44700,0.74100}%
\definecolor{mycolor2}{rgb}{0.92900,0.69400,0.12500}%
\definecolor{mycolor3}{rgb}{0.49400,0.18400,0.55600}%
\definecolor{mycolor4}{rgb}{0.46600,0.67400,0.18800}%
\definecolor{mycolor5}{rgb}{0.30100,0.74500,0.93300}%
\definecolor{mycolor6}{rgb}{0.63500,0.07800,0.18400}%
\pgfplotsset{
compat=1.11,
legend image code/.code={
\draw[mark repeat=2,mark phase=2]
plot coordinates {
(0cm,0cm)
(0.15cm,0cm)        
(0.3cm,0cm)         
};%
}
}
\begin{tikzpicture}

\begin{axis}[%
width=.85\columnwidth,
height=.55\columnwidth,
scale only axis,
xmin=4,
xmax=46,
xtick={5,10,15,20,25,30,35,40,45},
xticklabel style={font=\footnotesize},
xlabel style={font=\color{white!15!black}\footnotesize,yshift=3pt},
xlabel={\# failed nodes},
ymin=0,
ymax=1.07,
ytick={0,.2,.4,.6,.8,1},
yticklabels={{0},{.2},{.4},{.6},{.8},{1}},
yticklabel style={font=\footnotesize},
ylabel style={font=\color{white!15!black}\footnotesize,yshift=-5pt},
ylabel={$a_W$},
axis background/.style={fill=white},
legend style={at={(0.05,0.1)}, legend columns=3, anchor=south west, legend cell align=left, align=left, fill=none, draw=none, font=\scriptsize}
]

\addplot[area legend, draw=none, fill=mycolor1, fill opacity=0.2, forget plot]
table[row sep=crcr] {%
x	y\\
5	1\\
10	1\\
15	1\\
20	1\\
25	1\\
30	1\\
35	1\\
40	1\\
45	1\\
45	1\\
40	1\\
35	1\\
30	1\\
25	1\\
20	1\\
15	1\\
10	1\\
5	1\\
}--cycle;
\addplot [color=mycolor1, line width=.6pt, mark size=2.5pt, mark=text,text mark={\LARGE $\star$}, mark options={solid, fill=mycolor1, mycolor1}]
  table[row sep=crcr]{%
5	1\\
10	1\\
15	1\\
20	1\\
25	1\\
30	1\\
35	1\\
40	1\\
45	1\\
};
\addlegendentry{FaCe}

\addplot [color=mycolor1, line width=.6pt, mark size=2.5pt, mark=text,text mark={\LARGE $\star$}, mark options={solid, fill=mycolor1, mycolor1}, forget plot]
 plot [error bars/.cd, y dir = both, y explicit]
 table[row sep=crcr, y error plus index=2, y error minus index=3]{%
5	1	0	0\\
10	1	0	0\\
15	1	0	0\\
20	1	0	0\\
25	1	0	0\\
30	1	0	0\\
35	1	0	0\\
40	1	0	0\\
45	1	0	0\\
};

\addplot[area legend, draw=none, fill=mycolor2, fill opacity=0.2, forget plot]
table[row sep=crcr] {%
x	y\\
5	0.964686578045619\\
10	0.941534759357559\\
15	0.921719642772959\\
20	0.877955820407077\\
25	0.789804314182656\\
30	0.761916184000021\\
35	0.726674637393434\\
40	0.64617296383899\\
45	0.646532826436449\\
45	0.601336212326943\\
40	0.601631557448463\\
35	0.681431514955088\\
30	0.736793587974264\\
25	0.761999292715343\\
20	0.848457537840861\\
15	0.90189741699515\\
10	0.918689219692317\\
5	0.948760172880746\\
}--cycle;
\addplot [color=mycolor2, line width=.6pt, mark size=2.5pt, mark=asterisk, mark options={solid, mycolor2}]
  table[row sep=crcr]{%
5	0.956723375463183\\
10	0.930111989524938\\
15	0.911808529884054\\
20	0.863206679123969\\
25	0.775901803448999\\
30	0.749354885987143\\
35	0.704053076174261\\
40	0.623902260643727\\
45	0.623934519381696\\
};
\addlegendentry{GC}

\addplot [color=mycolor2, line width=.6pt, mark size=2.5pt, mark=asterisk, mark options={solid, mycolor2}, forget plot]
 plot [error bars/.cd, y dir = both, y explicit]
 table[row sep=crcr, y error plus index=2, y error minus index=3]{%
5	0.956723375463183	0.00796320258243663	0.00796320258243663\\
10	0.930111989524938	0.0114227698326209	0.0114227698326209\\
15	0.911808529884054	0.00991111288890451	0.00991111288890451\\
20	0.863206679123969	0.014749141283108	0.014749141283108\\
25	0.775901803448999	0.0139025107336566	0.0139025107336566\\
30	0.749354885987143	0.0125612980128785	0.0125612980128785\\
35	0.704053076174261	0.0226215612191729	0.0226215612191729\\
40	0.623902260643727	0.0222707031952634	0.0222707031952634\\
45	0.623934519381696	0.0225983070547531	0.0225983070547531\\
};

\addplot[area legend, draw=none, fill=mycolor3, fill opacity=0.2, forget plot]
table[row sep=crcr] {%
x	y\\
5	1\\
10	1\\
15	1\\
20	1\\
25	1\\
30	1\\
35	1\\
40	1\\
45	1\\
45	1\\
40	1\\
35	1\\
30	1\\
25	1\\
20	1\\
15	1\\
10	1\\
5	1\\
}--cycle;
\addplot [color=mycolor3, line width=.6pt, mark size=2.5pt, mark=triangle, mark options={solid, mycolor3}]
  table[row sep=crcr]{%
5	1\\
10	1\\
15	1\\
20	1\\
25	1\\
30	1\\
35	1\\
40	1\\
45	1\\
};
\addlegendentry{GI}

\addplot [color=mycolor3, line width=.6pt, mark size=2.5pt, mark=triangle, mark options={solid, mycolor3}, forget plot]
 plot [error bars/.cd, y dir = both, y explicit]
 table[row sep=crcr, y error plus index=2, y error minus index=3]{%
5	1	0	0\\
10	1	0	0\\
15	1	0	0\\
20	1	0	0\\
25	1	0	0\\
30	1	0	0\\
35	1	0	0\\
40	1	0	0\\
45	1	0	0\\
};

\addplot[area legend, draw=none, fill=mycolor4, fill opacity=0.2, forget plot]
table[row sep=crcr] {%
x	y\\
5	1\\
10	1\\
15	1\\
20	1\\
25	1\\
30	1\\
35	1\\
40	1\\
45	1\\
45	1\\
40	1\\
35	1\\
30	1\\
25	1\\
20	1\\
15	1\\
10	1\\
5	1\\
}--cycle;
\addplot [color=mycolor4, line width=.6pt, mark size=2.5pt, mark=square, mark options={solid, mycolor4}]
  table[row sep=crcr]{%
5	1\\
10	1\\
15	1\\
20	1\\
25	1\\
30	1\\
35	1\\
40	1\\
45	1\\
};
\addlegendentry{GD}

\addplot [color=mycolor4, line width=.6pt, mark size=2.5pt, mark=square, mark options={solid, mycolor4}, forget plot]
 plot [error bars/.cd, y dir = both, y explicit]
 table[row sep=crcr, y error plus index=2, y error minus index=3]{%
5	1	0	0\\
10	1	0	0\\
15	1	0	0\\
20	1	0	0\\
25	1	0	0\\
30	1	0	0\\
35	1	0	0\\
40	1	0	0\\
45	1	0	0\\
};

\addplot[area legend, draw=none, fill=mycolor5, fill opacity=0.2, forget plot]
table[row sep=crcr] {%
x	y\\
5	0.999489200632335\\
10	0.993631342728654\\
15	0.98669267540351\\
20	0.988078241165595\\
25	0.984859550428011\\
30	0.971614588350678\\
35	0.965349045579373\\
40	0.960071169741884\\
45	0.950814763773298\\
45	0.930373355038583\\
40	0.941151324145646\\
35	0.950592983406135\\
30	0.96060498205505\\
25	0.976461204288971\\
20	0.984882131794778\\
15	0.982892578052711\\
10	0.9881454226472\\
5	0.998708997565863\\
}--cycle;
\addplot [color=mycolor5, line width=.6pt, mark size=2.5pt, mark=+, mark options={solid, mycolor5}]
  table[row sep=crcr]{%
5	0.999099099099099\\
10	0.990888382687927\\
15	0.984792626728111\\
20	0.986480186480187\\
25	0.980660377358491\\
30	0.966109785202864\\
35	0.957971014492754\\
40	0.950611246943765\\
45	0.940594059405941\\
};
\addlegendentry{AF}

\addplot [color=mycolor5, line width=.6pt, mark size=2.5pt, mark=+, mark options={solid, mycolor5}, forget plot]
 plot [error bars/.cd, y dir = both, y explicit]
 table[row sep=crcr, y error plus index=2, y error minus index=3]{%
5	0.999099099099099	0.00039010153323623	0.00039010153323623\\
10	0.990888382687927	0.00274296004072717	0.00274296004072717\\
15	0.984792626728111	0.00190004867539985	0.00190004867539985\\
20	0.986480186480187	0.00159805468540817	0.00159805468540817\\
25	0.980660377358491	0.00419917306952002	0.00419917306952002\\
30	0.966109785202864	0.00550480314781422	0.00550480314781422\\
35	0.957971014492754	0.00737803108661896	0.00737803108661896\\
40	0.950611246943765	0.00945992279811888	0.00945992279811888\\
45	0.940594059405941	0.0102207043673577	0.0102207043673577\\
};

\addplot[area legend, draw=none, fill=mycolor6, fill opacity=0.2, forget plot]
table[row sep=crcr] {%
x	y\\
5	0.999724982893133\\
10	0.999422870504801\\
15	0.991765704355614\\
20	0.988221832051246\\
25	0.979962569560492\\
30	0.972993448229007\\
35	0.959709192592314\\
40	0.963161577743882\\
45	0.964365151255606\\
45	0.948725627958734\\
40	0.94340682552604\\
35	0.952191721235659\\
30	0.964514135153754\\
25	0.97379961942792\\
20	0.981141917406098\\
15	0.983601198715229\\
10	0.996636242795692\\
5	0.998397082834567\\
}--cycle;
\addplot [color=mycolor6, line width=.6pt, mark size=2.5pt, mark=x, mark options={solid, mycolor6}]
  table[row sep=crcr]{%
5	0.99906103286385\\
10	0.998029556650246\\
15	0.987683451535421\\
20	0.984681874728672\\
25	0.976881094494206\\
30	0.96875379169138\\
35	0.955950456913987\\
40	0.953284201634961\\
45	0.95654538960717\\
};
\addlegendentry{APC}

\addplot [color=mycolor6, line width=.6pt, mark size=2.5pt, mark=x, mark options={solid, mycolor6}, forget plot]
 plot [error bars/.cd, y dir = both, y explicit]
 table[row sep=crcr, y error plus index=2, y error minus index=3]{%
5	0.99906103286385	0.000663950029283137	0.000663950029283137\\
10	0.998029556650246	0.00139331385455477	0.00139331385455477\\
15	0.987683451535421	0.00408225282019286	0.00408225282019286\\
20	0.984681874728672	0.00353995732257372	0.00353995732257372\\
25	0.976881094494206	0.00308147506628578	0.00308147506628578\\
30	0.96875379169138	0.00423965653762678	0.00423965653762678\\
35	0.955950456913987	0.00375873567832771	0.00375873567832771\\
40	0.953284201634961	0.00987737610892124	0.00987737610892124\\
45	0.95654538960717	0.00781976164843571	0.00781976164843571\\
};
\end{axis}
\end{tikzpicture}%

%% file: imgs/MINNESOTA2/AB2MN.tex
%
%
\definecolor{mycolor1}{rgb}{0.00000,0.44700,0.74100}%
\definecolor{mycolor2}{rgb}{0.92900,0.69400,0.12500}%
\definecolor{mycolor3}{rgb}{0.49400,0.18400,0.55600}%
\definecolor{mycolor4}{rgb}{0.46600,0.67400,0.18800}%
\definecolor{mycolor5}{rgb}{0.30100,0.74500,0.93300}%
\definecolor{mycolor6}{rgb}{0.63500,0.07800,0.18400}%
\pgfplotsset{
compat=1.11,
legend image code/.code={
\draw[mark repeat=2,mark phase=2]
plot coordinates {
(0cm,0cm)
(0.15cm,0cm)        
(0.3cm,0cm)         
};%
}
}
\begin{tikzpicture}

\begin{axis}[%
width=.85\columnwidth,
height=.55\columnwidth,
scale only axis,
xmin=4,
xmax=46,
xtick={5,10,15,20,25,30,35,40,45},
xticklabel style={font=\footnotesize},
xlabel style={font=\color{white!15!black}\footnotesize,yshift=3pt},
xlabel={\# failed nodes},
ymin=0,
ymax=1.07,
ytick={0,.2,.4,.6,.8,1},
yticklabels={{0},{.2},{.4},{.6},{.8},{1}},
yticklabel style={font=\footnotesize},
ylabel style={font=\color{white!15!black}\footnotesize,yshift=-5pt},
ylabel={$a_B$},
axis background/.style={fill=white},
legend style={at={(0.05,0.4)}, legend columns=3, anchor=south west, legend cell align=left, align=left, fill=none, draw=none, font=\scriptsize}
]

\addplot[area legend, draw=none, fill=mycolor1, fill opacity=0.2, forget plot]
table[row sep=crcr] {%
x	y\\
5	1\\
10	1\\
15	1\\
20	1\\
25	1\\
30	1\\
35	1\\
40	1\\
45	1\\
45	1\\
40	1\\
35	1\\
30	1\\
25	1\\
20	1\\
15	1\\
10	1\\
5	1\\
}--cycle;
\addplot [color=mycolor1, line width=.6pt, mark size=2.5pt, mark=text,text mark={\LARGE $\star$}, mark options={solid, fill=mycolor1, mycolor1}]
  table[row sep=crcr]{%
5	1\\
10	1\\
15	1\\
20	1\\
25	1\\
30	1\\
35	1\\
40	1\\
45	1\\
};
\addlegendentry{FaCe}

\addplot [color=mycolor1, line width=.6pt, mark size=2.5pt, mark=text,text mark={\LARGE $\star$}, mark options={solid, fill=mycolor1, mycolor1}, forget plot]
 plot [error bars/.cd, y dir = both, y explicit]
 table[row sep=crcr, y error plus index=2, y error minus index=3]{%
5	1	0	0\\
10	1	0	0\\
15	1	0	0\\
20	1	0	0\\
25	1	0	0\\
30	1	0	0\\
35	1	0	0\\
40	1	0	0\\
45	1	0	0\\
};

\addplot[area legend, draw=none, fill=mycolor2, fill opacity=0.2, forget plot]
table[row sep=crcr] {%
x	y\\
5	0.204056941504209\\
10	0.275515860353926\\
15	0.193701445263814\\
20	0.131852731421388\\
25	0.0569035593728849\\
30	0.107138347472157\\
35	0.0677289209169908\\
40	0\\
45	0.0487744794624728\\
45	0.00836837768038436\\
40	0\\
35	0.0267155235274536\\
30	0.0373060969722874\\
25	0.00976310729378175\\
20	0.0490996495309934\\
15	0.132965221402852\\
10	0.151150806312741\\
5	0.0459430584957905\\
}--cycle;
\addplot [color=mycolor2, line width=.6pt, mark size=2.5pt, mark=asterisk, mark options={solid, mycolor2}]
  table[row sep=crcr]{%
5	0.125\\
10	0.213333333333333\\
15	0.163333333333333\\
20	0.0904761904761905\\
25	0.0333333333333333\\
30	0.0722222222222222\\
35	0.0472222222222222\\
40	0\\
45	0.0285714285714286\\
};
\addlegendentry{GC}

\addplot [color=mycolor2, line width=.6pt, mark size=2.5pt, mark=asterisk, mark options={solid, mycolor2}, forget plot]
 plot [error bars/.cd, y dir = both, y explicit]
 table[row sep=crcr, y error plus index=2, y error minus index=3]{%
5	0.125	0.0790569415042095	0.0790569415042095\\
10	0.213333333333333	0.0621825270205921	0.0621825270205921\\
15	0.163333333333333	0.030368111930481	0.030368111930481\\
20	0.0904761904761905	0.0413765409451971	0.0413765409451971\\
25	0.0333333333333333	0.0235702260395516	0.0235702260395516\\
30	0.0722222222222222	0.0349161252499348	0.0349161252499348\\
35	0.0472222222222222	0.0205066986947686	0.0205066986947686\\
40	0	0	0\\
45	0.0285714285714286	0.0202030508910442	0.0202030508910442\\
};

\addplot[area legend, draw=none, fill=mycolor3, fill opacity=0.2, forget plot]
table[row sep=crcr] {%
x	y\\
5	1\\
10	1\\
15	1\\
20	1\\
25	1\\
30	1\\
35	1\\
40	1\\
45	1\\
45	1\\
40	1\\
35	1\\
30	1\\
25	1\\
20	1\\
15	1\\
10	1\\
5	1\\
}--cycle;
\addplot [color=mycolor3, line width=.6pt, mark size=2.5pt, mark=triangle, mark options={solid, mycolor3}]
  table[row sep=crcr]{%
5	1\\
10	1\\
15	1\\
20	1\\
25	1\\
30	1\\
35	1\\
40	1\\
45	1\\
};
\addlegendentry{GI}

\addplot [color=mycolor3, line width=.6pt, mark size=2.5pt, mark=triangle, mark options={solid, mycolor3}, forget plot]
 plot [error bars/.cd, y dir = both, y explicit]
 table[row sep=crcr, y error plus index=2, y error minus index=3]{%
5	1	0	0\\
10	1	0	0\\
15	1	0	0\\
20	1	0	0\\
25	1	0	0\\
30	1	0	0\\
35	1	0	0\\
40	1	0	0\\
45	1	0	0\\
};

\addplot[area legend, draw=none, fill=mycolor4, fill opacity=0.2, forget plot]
table[row sep=crcr] {%
x	y\\
5	1\\
10	1\\
15	1\\
20	1\\
25	1\\
30	1\\
35	1\\
40	1\\
45	1\\
45	1\\
40	1\\
35	1\\
30	1\\
25	1\\
20	1\\
15	1\\
10	1\\
5	1\\
}--cycle;
\addplot [color=mycolor4, line width=.6pt, mark size=2.5pt, mark=square, mark options={solid, mycolor4}]
  table[row sep=crcr]{%
5	1\\
10	1\\
15	1\\
20	1\\
25	1\\
30	1\\
35	1\\
40	1\\
45	1\\
};
\addlegendentry{GD}

\addplot [color=mycolor4, line width=.6pt, mark size=2.5pt, mark=square, mark options={solid, mycolor4}, forget plot]
 plot [error bars/.cd, y dir = both, y explicit]
 table[row sep=crcr, y error plus index=2, y error minus index=3]{%
5	1	0	0\\
10	1	0	0\\
15	1	0	0\\
20	1	0	0\\
25	1	0	0\\
30	1	0	0\\
35	1	0	0\\
40	1	0	0\\
45	1	0	0\\
};

\addplot[area legend, draw=none, fill=mycolor5, fill opacity=0.2, forget plot]
table[row sep=crcr] {%
x	y\\
5	1\\
10	1\\
15	1\\
20	1\\
25	1\\
30	1\\
35	1\\
40	1\\
45	1\\
45	1\\
40	1\\
35	1\\
30	1\\
25	1\\
20	1\\
15	1\\
10	1\\
5	1\\
}--cycle;
\addplot [color=mycolor5, line width=.6pt, mark size=2.5pt, mark=+, mark options={solid, mycolor5}]
  table[row sep=crcr]{%
5	1\\
10	1\\
15	1\\
20	1\\
25	1\\
30	1\\
35	1\\
40	1\\
45	1\\
};
\addlegendentry{AF}

\addplot [color=mycolor5, line width=.6pt, mark size=2.5pt, mark=+, mark options={solid, mycolor5}, forget plot]
 plot [error bars/.cd, y dir = both, y explicit]
 table[row sep=crcr, y error plus index=2, y error minus index=3]{%
5	1	0	0\\
10	1	0	0\\
15	1	0	0\\
20	1	0	0\\
25	1	0	0\\
30	1	0	0\\
35	1	0	0\\
40	1	0	0\\
45	1	0	0\\
};

\addplot[area legend, draw=none, fill=mycolor6, fill opacity=0.2, forget plot]
table[row sep=crcr] {%
x	y\\
5	1\\
10	1\\
15	1\\
20	0.991631622319616\\
25	0.964368350311177\\
30	0.943580759995809\\
35	0.932544406316838\\
40	0.948658348185149\\
45	0.948205501742037\\
45	0.904651641115105\\
40	0.904501824975024\\
35	0.86586829209586\\
30	0.89530812889308\\
25	0.905631649688823\\
20	0.951225520537527\\
15	1\\
10	1\\
5	1\\
}--cycle;
\addplot [color=mycolor6, line width=.6pt, mark size=2.5pt, mark=x, mark options={solid, mycolor6}]
  table[row sep=crcr]{%
5	1\\
10	1\\
15	1\\
20	0.971428571428572\\
25	0.935\\
30	0.919444444444445\\
35	0.899206349206349\\
40	0.926580086580087\\
45	0.926428571428571\\
};
\addlegendentry{APC}

\addplot [color=mycolor6, line width=.6pt, mark size=2.5pt, mark=x, mark options={solid, mycolor6}, forget plot]
 plot [error bars/.cd, y dir = both, y explicit]
 table[row sep=crcr, y error plus index=2, y error minus index=3]{%
5	1	0	0\\
10	1	0	0\\
15	1	0	0\\
20	0.971428571428572	0.0202030508910442	0.0202030508910442\\
25	0.935	0.0293683503111768	0.0293683503111768\\
30	0.919444444444445	0.024136315551365	0.024136315551365\\
35	0.899206349206349	0.0333380571104892	0.0333380571104892\\
40	0.926580086580087	0.0220782616050626	0.0220782616050626\\
45	0.926428571428571	0.0217769303134659	0.0217769303134659\\
};
\end{axis}
\end{tikzpicture}%

%% file: imgs/MINNESOTA2/R12MN.tex
%
%
\definecolor{mycolor1}{rgb}{0.00000,0.44700,0.74100}%
\definecolor{mycolor2}{rgb}{0.92900,0.69400,0.12500}%
\definecolor{mycolor3}{rgb}{0.49400,0.18400,0.55600}%
\definecolor{mycolor4}{rgb}{0.46600,0.67400,0.18800}%
\definecolor{mycolor5}{rgb}{0.30100,0.74500,0.93300}%
\definecolor{mycolor6}{rgb}{0.63500,0.07800,0.18400}%
\pgfplotsset{
compat=1.11,
legend image code/.code={
\draw[mark repeat=2,mark phase=2]
plot coordinates {
(0cm,0cm)
(0.15cm,0cm)        
(0.3cm,0cm)         
};%
}
}
\begin{tikzpicture}

\begin{axis}[%
width=.85\columnwidth,
height=.55\columnwidth,
scale only axis,
xmin=4,
xmax=46,
xtick={5,10,15,20,25,30,35,40,45},
xticklabel style={font=\footnotesize},
xlabel style={font=\color{white!15!black}\footnotesize,yshift=3pt},
xlabel={\# failed nodes},
ymin=0,
ymax=1.07,
ytick={0,0.2,0.4,0.6,0.8,1},
yticklabels={{0},{.2},{.4},{.6},{.8},{1}},
yticklabel style={font=\footnotesize},
ylabel style={font=\color{white!15!black}\footnotesize,yshift=-6pt},
ylabel={$R_1$},
axis background/.style={fill=white},
legend style={at={(0.05,0.55)}, legend columns=3, anchor=south west, legend cell align=left, align=left, fill=none, draw=none, font=\scriptsize}
]

\addplot[area legend, draw=none, fill=mycolor1, fill opacity=0.2, forget plot]
table[row sep=crcr] {%
x	y\\
5	0.692915026221292\\
10	0.646457513110646\\
15	0.571547005383792\\
20	0.575\\
25	0.55062741699797\\
30	0.535916547957974\\
35	0.495373804341848\\
40	0.488831406486677\\
45	0.491511866416664\\
45	0.441821466916669\\
40	0.431168593513323\\
35	0.441769052801009\\
30	0.47741678537536\\
25	0.50537258300203\\
20	0.525\\
15	0.548452994616207\\
10	0.593542486889354\\
5	0.587084973778708\\
}--cycle;
\addplot [color=mycolor1, line width=.6pt, mark size=2.5pt, mark=text,text mark={\LARGE $\star$}, mark options={solid, fill=mycolor1, mycolor1}]
  table[row sep=crcr]{%
5	0.64\\
10	0.62\\
15	0.56\\
20	0.55\\
25	0.528\\
30	0.506666666666667\\
35	0.468571428571429\\
40	0.46\\
45	0.466666666666667\\
};
\addlegendentry{FaCe}

\addplot [color=mycolor1, line width=.6pt, mark size=2.5pt, mark=text,text mark={\LARGE $\star$}, mark options={solid, fill=mycolor1, mycolor1}, forget plot]
 plot [error bars/.cd, y dir = both, y explicit]
 table[row sep=crcr, y error plus index=2, y error minus index=3]{%
5	0.64	0.0529150262212918	0.0529150262212918\\
10	0.62	0.0264575131106459	0.0264575131106459\\
15	0.56	0.0115470053837925	0.0115470053837925\\
20	0.55	0.025	0.025\\
25	0.528	0.0226274169979695	0.0226274169979695\\
30	0.506666666666667	0.0292498812913071	0.0292498812913071\\
35	0.468571428571429	0.0268023757704196	0.0268023757704196\\
40	0.46	0.028831406486677	0.028831406486677\\
45	0.466666666666667	0.0248451997499977	0.0248451997499977\\
};

\addplot[area legend, draw=none, fill=mycolor2, fill opacity=0.2, forget plot]
table[row sep=crcr] {%
x	y\\
5	0.412915026221292\\
10	0.431622776601684\\
15	0.348240453183332\\
20	0.322403703492039\\
25	0.288944271909999\\
30	0.267475468957064\\
35	0.227444432215611\\
40	0.264330127018922\\
45	0.253051074102683\\
45	0.235837814786206\\
40	0.255669872981078\\
35	0.206841282070103\\
30	0.239191197709602\\
25	0.271055728090001\\
20	0.257596296507961\\
15	0.318426213483335\\
10	0.368377223398316\\
5	0.307084973778708\\
}--cycle;
\addplot [color=mycolor2, line width=.6pt, mark size=2.5pt, mark=asterisk, mark options={solid, mycolor2}]
  table[row sep=crcr]{%
5	0.36\\
10	0.4\\
15	0.333333333333333\\
20	0.29\\
25	0.28\\
30	0.253333333333333\\
35	0.217142857142857\\
40	0.26\\
45	0.244444444444444\\
};
\addlegendentry{GC}

\addplot [color=mycolor2, line width=.6pt, mark size=2.5pt, mark=asterisk, mark options={solid, mycolor2}, forget plot]
 plot [error bars/.cd, y dir = both, y explicit]
 table[row sep=crcr, y error plus index=2, y error minus index=3]{%
5	0.36	0.0529150262212918	0.0529150262212918\\
10	0.4	0.0316227766016838	0.0316227766016838\\
15	0.333333333333333	0.0149071198499986	0.0149071198499986\\
20	0.29	0.0324037034920393	0.0324037034920393\\
25	0.28	0.00894427190999916	0.00894427190999916\\
30	0.253333333333333	0.0141421356237309	0.0141421356237309\\
35	0.217142857142857	0.0103015750727543	0.0103015750727543\\
40	0.26	0.0043301270189222	0.0043301270189222\\
45	0.244444444444444	0.0086066296582387	0.0086066296582387\\
};

\addplot[area legend, draw=none, fill=mycolor3, fill opacity=0.2, forget plot]
table[row sep=crcr] {%
x	y\\
5	0.692915026221292\\
10	0.646457513110646\\
15	0.571547005383792\\
20	0.575\\
25	0.55062741699797\\
30	0.535916547957974\\
35	0.495373804341848\\
40	0.488831406486677\\
45	0.491511866416664\\
45	0.441821466916669\\
40	0.431168593513323\\
35	0.441769052801009\\
30	0.47741678537536\\
25	0.50537258300203\\
20	0.525\\
15	0.548452994616207\\
10	0.593542486889354\\
5	0.587084973778708\\
}--cycle;
\addplot [color=mycolor3, line width=.6pt, mark size=2.5pt, mark=triangle, mark options={solid, mycolor3}]
  table[row sep=crcr]{%
5	0.64\\
10	0.62\\
15	0.56\\
20	0.55\\
25	0.528\\
30	0.506666666666667\\
35	0.468571428571429\\
40	0.46\\
45	0.466666666666667\\
};
\addlegendentry{GI}

\addplot [color=mycolor3, line width=.6pt, mark size=2.5pt, mark=triangle, mark options={solid, mycolor3}, forget plot]
 plot [error bars/.cd, y dir = both, y explicit]
 table[row sep=crcr, y error plus index=2, y error minus index=3]{%
5	0.64	0.0529150262212918	0.0529150262212918\\
10	0.62	0.0264575131106459	0.0264575131106459\\
15	0.56	0.0115470053837925	0.0115470053837925\\
20	0.55	0.025	0.025\\
25	0.528	0.0226274169979695	0.0226274169979695\\
30	0.506666666666667	0.0292498812913071	0.0292498812913071\\
35	0.468571428571429	0.0268023757704196	0.0268023757704196\\
40	0.46	0.028831406486677	0.028831406486677\\
45	0.466666666666667	0.0248451997499977	0.0248451997499977\\
};

\addplot[area legend, draw=none, fill=mycolor4, fill opacity=0.2, forget plot]
table[row sep=crcr] {%
x	y\\
5	0.692915026221292\\
10	0.646457513110646\\
15	0.571547005383792\\
20	0.575\\
25	0.55062741699797\\
30	0.535916547957974\\
35	0.495373804341848\\
40	0.488831406486677\\
45	0.491511866416664\\
45	0.441821466916669\\
40	0.431168593513323\\
35	0.441769052801009\\
30	0.47741678537536\\
25	0.50537258300203\\
20	0.525\\
15	0.548452994616207\\
10	0.593542486889354\\
5	0.587084973778708\\
}--cycle;
\addplot [color=mycolor4, line width=.6pt, mark size=2.5pt, mark=square, mark options={solid, mycolor4}]
  table[row sep=crcr]{%
5	0.64\\
10	0.62\\
15	0.56\\
20	0.55\\
25	0.528\\
30	0.506666666666667\\
35	0.468571428571429\\
40	0.46\\
45	0.466666666666667\\
};
\addlegendentry{GD}

\addplot [color=mycolor4, line width=.6pt, mark size=2.5pt, mark=square, mark options={solid, mycolor4}, forget plot]
 plot [error bars/.cd, y dir = both, y explicit]
 table[row sep=crcr, y error plus index=2, y error minus index=3]{%
5	0.64	0.0529150262212918	0.0529150262212918\\
10	0.62	0.0264575131106459	0.0264575131106459\\
15	0.56	0.0115470053837925	0.0115470053837925\\
20	0.55	0.025	0.025\\
25	0.528	0.0226274169979695	0.0226274169979695\\
30	0.506666666666667	0.0292498812913071	0.0292498812913071\\
35	0.468571428571429	0.0268023757704196	0.0268023757704196\\
40	0.46	0.028831406486677	0.028831406486677\\
45	0.466666666666667	0.0248451997499977	0.0248451997499977\\
};

\addplot[area legend, draw=none, fill=mycolor5, fill opacity=0.2, forget plot]
table[row sep=crcr] {%
x	y\\
5	1\\
10	1\\
15	1\\
20	1\\
25	1\\
30	1\\
35	1\\
40	1\\
45	1\\
45	1\\
40	1\\
35	1\\
30	1\\
25	1\\
20	1\\
15	1\\
10	1\\
5	1\\
}--cycle;
\addplot [color=mycolor5, line width=.6pt, mark size=2.5pt, mark=+, mark options={solid, mycolor5}]
  table[row sep=crcr]{%
5	1\\
10	1\\
15	1\\
20	1\\
25	1\\
30	1\\
35	1\\
40	1\\
45	1\\
};
\addlegendentry{AF}

\addplot [color=mycolor5, line width=.6pt, mark size=2.5pt, mark=+, mark options={solid, mycolor5}, forget plot]
 plot [error bars/.cd, y dir = both, y explicit]
 table[row sep=crcr, y error plus index=2, y error minus index=3]{%
5	1	0	0\\
10	1	0	0\\
15	1	0	0\\
20	1	0	0\\
25	1	0	0\\
30	1	0	0\\
35	1	0	0\\
40	1	0	0\\
45	1	0	0\\
};

\addplot[area legend, draw=none, fill=mycolor6, fill opacity=0.2, forget plot]
table[row sep=crcr] {%
x	y\\
5	0.692915026221292\\
10	0.646457513110646\\
15	0.571547005383792\\
20	0.575\\
25	0.55062741699797\\
30	0.535916547957974\\
35	0.495373804341848\\
40	0.488831406486677\\
45	0.491511866416664\\
45	0.441821466916669\\
40	0.431168593513323\\
35	0.441769052801009\\
30	0.47741678537536\\
25	0.50537258300203\\
20	0.525\\
15	0.548452994616207\\
10	0.593542486889354\\
5	0.587084973778708\\
}--cycle;
\addplot [color=mycolor6, line width=.6pt, mark size=2.5pt, mark=x, mark options={solid, mycolor6}]
  table[row sep=crcr]{%
5	0.64\\
10	0.62\\
15	0.56\\
20	0.55\\
25	0.528\\
30	0.506666666666667\\
35	0.468571428571429\\
40	0.46\\
45	0.466666666666667\\
};
\addlegendentry{APC}

\addplot [color=mycolor6, line width=.6pt, mark size=2.5pt, mark=x, mark options={solid, mycolor6}, forget plot]
 plot [error bars/.cd, y dir = both, y explicit]
 table[row sep=crcr, y error plus index=2, y error minus index=3]{%
5	0.64	0.0529150262212918	0.0529150262212918\\
10	0.62	0.0264575131106459	0.0264575131106459\\
15	0.56	0.0115470053837925	0.0115470053837925\\
20	0.55	0.025	0.025\\
25	0.528	0.0226274169979695	0.0226274169979695\\
30	0.506666666666667	0.0292498812913071	0.0292498812913071\\
35	0.468571428571429	0.0268023757704196	0.0268023757704196\\
40	0.46	0.028831406486677	0.028831406486677\\
45	0.466666666666667	0.0248451997499977	0.0248451997499977\\
};
\addplot[area legend, draw=none, fill=black, fill opacity=0.08, forget plot]
table[row sep=crcr] {%
5	0.64\\
10	0.62\\
15	0.56\\
20	0.55\\
25	0.528\\
30	0.506666666666667\\
35	0.468571428571429\\
40	0.46\\
45	0.466666666666667\\
}--cycle;
\addlegendentry{all}

\addplot [color=black, dashed, line width=.6pt, forget plot]
 plot [error bars/.cd, y dir = both, y explicit]
 table[row sep=crcr, y error plus index=2, y error minus index=3]{%
5	0.64	0.0529150262212918	0.0529150262212918\\
10	0.62	0.0264575131106459	0.0264575131106459\\
15	0.56	0.0115470053837925	0.0115470053837925\\
20	0.55	0.025	0.025\\
25	0.528	0.0226274169979695	0.0226274169979695\\
30	0.506666666666667	0.0292498812913071	0.0292498812913071\\
35	0.468571428571429	0.0268023757704196	0.0268023757704196\\
40	0.46	0.028831406486677	0.028831406486677\\
45	0.466666666666667	0.0248451997499977	0.0248451997499977\\
};
\end{axis}
\end{tikzpicture}%

%% file: imgs/MINNESOTA2/R22MN.tex
\definecolor{mycolor1}{rgb}{0.00000,0.44700,0.74100}%
\definecolor{mycolor2}{rgb}{0.92900,0.69400,0.12500}%
\definecolor{mycolor3}{rgb}{0.49400,0.18400,0.55600}%
\definecolor{mycolor4}{rgb}{0.46600,0.67400,0.18800}%
\definecolor{mycolor5}{rgb}{0.30100,0.74500,0.93300}%
\definecolor{mycolor6}{rgb}{0.63500,0.07800,0.18400}%
\pgfplotsset{
compat=1.11,
legend image code/.code={
\draw[mark repeat=2,mark phase=2]
plot coordinates {
(0cm,0cm)
(0.15cm,0cm)        
(0.3cm,0cm)         
};%
}
}
\begin{tikzpicture}

\begin{axis}[%
width=.85\columnwidth,
height=.55\columnwidth,
scale only axis,
xmin=4,
xmax=46,
xtick={5,10,15,20,25,30,35,40,45},
xticklabel style={font=\footnotesize},
xlabel style={font=\color{white!15!black}\footnotesize,yshift=5pt},
xlabel={\# failed nodes},
ymin=0,
ymax=1.07,
ytick={0,0.2,0.4,0.6,0.8,1},
yticklabels={{0},{.2},{.4},{.6},{.8},{1}},
yticklabel style={font=\footnotesize},
ylabel style={font=\color{white!15!black}\footnotesize,yshift=-5pt},
ylabel={$R_2$},
axis background/.style={fill=white},
legend style={at={(0.05,0.4)}, legend columns=3, anchor=south west, legend cell align=left, align=left, fill=none, draw=none, font=\scriptsize}
]

\addplot[area legend, draw=none, fill=mycolor1, fill opacity=0.2, forget plot]
table[row sep=crcr] {%
x	y\\
5	0.425597058590669\\
10	0.308374700193678\\
15	0.256336317197321\\
20	0.262430607635593\\
25	0.242898187764617\\
30	0.252814309990973\\
35	0.249484228093142\\
40	0.260505618097133\\
45	0.271588010878401\\
45	0.249384352461232\\
40	0.237030605006675\\
35	0.226466194073715\\
30	0.228983340515657\\
25	0.220120587955098\\
20	0.227171115328285\\
15	0.222878944533724\\
10	0.248952654586987\\
5	0.359785042967903\\
}--cycle;
\addplot [color=mycolor1, line width=.6pt, mark size=2.5pt, mark=text,text mark={\LARGE $\star$}, mark options={solid, fill=mycolor1, mycolor1}]
  table[row sep=crcr]{%
5	0.35449016563147\\
10	0.268591136507612\\
15	0.247657263796226\\
20	0.245210501997556\\
25	0.258031634806483\\
30	0.262301085916737\\
35	0.271644229590077\\
40	0.268181747949845\\
45	0.259089969755557\\
};
\addlegendentry{FaCe}

\addplot [color=mycolor1, line width=.6pt, mark size=2.5pt, mark=text,text mark={\LARGE $\star$}, mark options={solid, fill=mycolor1, mycolor1}, forget plot]
 plot [error bars/.cd, y dir = both, y explicit]
 table[row sep=crcr, y error plus index=2, y error minus index=3]{%
5	0.392691050779286	0.0329060078113833	0.0329060078113833\\
10	0.278663677390333	0.0297110228033454	0.0297110228033454\\
15	0.239607630865523	0.0167286863317986	0.0167286863317986\\
20	0.244800861481939	0.0176297461536543	0.0176297461536543\\
25	0.231509387859857	0.0113887999047597	0.0113887999047597\\
30	0.240898825253315	0.0119154847376583	0.0119154847376583\\
35	0.237975211083428	0.0115090170097134	0.0115090170097134\\
40	0.248768111551904	0.0117375065452288	0.0117375065452288\\
45	0.260486181669816	0.0111018292085845	0.0111018292085845\\
};

\addplot[area legend, draw=none, fill=mycolor2, fill opacity=0.2, forget plot]
table[row sep=crcr] {%
x	y\\
5	0.243972397216764\\
10	0.183676868978555\\
15	0.16733035552809\\
20	0.169350711061485\\
25	0.142204253067225\\
30	0.157356490408127\\
35	0.155910441832684\\
40	0.157263523920396\\
45	0.172024582242081\\
45	0.160962829265967\\
40	0.148796950329971\\
35	0.14459524491977\\
30	0.143505316571287\\
25	0.131641591932145\\
20	0.151718302080805\\
15	0.150980461867563\\
10	0.161400440552222\\
5	0.202747343311026\\
}--cycle;
\addplot [color=mycolor2, line width=.6pt, mark size=2.5pt, mark=asterisk, mark options={solid, mycolor2}]
  table[row sep=crcr]{%
5	0.223359870263895\\
10	0.172538654765389\\
15	0.159155408697827\\
20	0.160534506571145\\
25	0.136922922499685\\
30	0.150430903489707\\
35	0.150252843376227\\
40	0.153030237125184\\
45	0.166493705754024\\
};
\addlegendentry{GC}

\addplot [color=mycolor2, line width=.6pt, mark size=2.5pt, mark=asterisk, mark options={solid, mycolor2}, forget plot]
 plot [error bars/.cd, y dir = both, y explicit]
 table[row sep=crcr, y error plus index=2, y error minus index=3]{%
5	0.223359870263895	0.0206125269528686	0.0206125269528686\\
10	0.172538654765389	0.0111382142131664	0.0111382142131664\\
15	0.159155408697827	0.00817494683026339	0.00817494683026339\\
20	0.160534506571145	0.00881620449033991	0.00881620449033991\\
25	0.136922922499685	0.00528133056754015	0.00528133056754015\\
30	0.150430903489707	0.00692558691842006	0.00692558691842006\\
35	0.150252843376227	0.00565759845645703	0.00565759845645703\\
40	0.153030237125184	0.00423328679521223	0.00423328679521223\\
45	0.166493705754024	0.00553087648805734	0.00553087648805734\\
};

\addplot[area legend, draw=none, fill=mycolor3, fill opacity=0.2, forget plot]
table[row sep=crcr] {%
x	y\\
5	0.425597058590669\\
10	0.308374700193678\\
15	0.256336317197321\\
20	0.262430607635593\\
25	0.242898187764617\\
30	0.252814309990973\\
35	0.249484228093142\\
40	0.260505618097133\\
45	0.271588010878401\\
45	0.249384352461232\\
40	0.237030605006675\\
35	0.226466194073715\\
30	0.228983340515657\\
25	0.220120587955098\\
20	0.227171115328285\\
15	0.222878944533724\\
10	0.248952654586987\\
5	0.359785042967903\\
}--cycle;
\addplot [color=mycolor3, line width=.6pt, mark size=2.5pt, mark=triangle, mark options={solid, mycolor3}]
  table[row sep=crcr]{%
5	0.392691050779286\\
10	0.278663677390333\\
15	0.239607630865523\\
20	0.244800861481939\\
25	0.231509387859857\\
30	0.240898825253315\\
35	0.237975211083428\\
40	0.248768111551904\\
45	0.260486181669816\\
};
\addlegendentry{GI}

\addplot [color=mycolor3, line width=.6pt, mark size=2.5pt, mark=triangle, mark options={solid, mycolor3}, forget plot]
 plot [error bars/.cd, y dir = both, y explicit]
 table[row sep=crcr, y error plus index=2, y error minus index=3]{%
5	0.392691050779286	0.0329060078113833	0.0329060078113833\\
10	0.278663677390333	0.0297110228033454	0.0297110228033454\\
15	0.239607630865523	0.0167286863317986	0.0167286863317986\\
20	0.244800861481939	0.0176297461536543	0.0176297461536543\\
25	0.231509387859857	0.0113887999047597	0.0113887999047597\\
30	0.240898825253315	0.0119154847376583	0.0119154847376583\\
35	0.237975211083428	0.0115090170097134	0.0115090170097134\\
40	0.248768111551904	0.0117375065452288	0.0117375065452288\\
45	0.260486181669816	0.0111018292085845	0.0111018292085845\\
};

\addplot[area legend, draw=none, fill=mycolor4, fill opacity=0.2, forget plot]
table[row sep=crcr] {%
x	y\\
5	0.425597058590669\\
10	0.308374700193678\\
15	0.256336317197321\\
20	0.262430607635593\\
25	0.242898187764617\\
30	0.252814309990973\\
35	0.249484228093142\\
40	0.260505618097133\\
45	0.271588010878401\\
45	0.249384352461232\\
40	0.237030605006675\\
35	0.226466194073715\\
30	0.228983340515657\\
25	0.220120587955098\\
20	0.227171115328285\\
15	0.222878944533724\\
10	0.248952654586987\\
5	0.359785042967903\\
}--cycle;
\addplot [color=mycolor4, line width=.6pt, mark size=2.5pt, mark=square, mark options={solid, mycolor4}]
  table[row sep=crcr]{%
5	0.392691050779286\\
10	0.278663677390333\\
15	0.239607630865523\\
20	0.244800861481939\\
25	0.231509387859857\\
30	0.240898825253315\\
35	0.237975211083428\\
40	0.248768111551904\\
45	0.260486181669816\\
};
\addlegendentry{GD}

\addplot [color=mycolor4, line width=.6pt, mark size=2.5pt, mark=square, mark options={solid, mycolor4}, forget plot]
 plot [error bars/.cd, y dir = both, y explicit]
 table[row sep=crcr, y error plus index=2, y error minus index=3]{%
5	0.392691050779286	0.0329060078113833	0.0329060078113833\\
10	0.278663677390333	0.0297110228033454	0.0297110228033454\\
15	0.239607630865523	0.0167286863317986	0.0167286863317986\\
20	0.244800861481939	0.0176297461536543	0.0176297461536543\\
25	0.231509387859857	0.0113887999047597	0.0113887999047597\\
30	0.240898825253315	0.0119154847376583	0.0119154847376583\\
35	0.237975211083428	0.0115090170097134	0.0115090170097134\\
40	0.248768111551904	0.0117375065452288	0.0117375065452288\\
45	0.260486181669816	0.0111018292085845	0.0111018292085845\\
};

\addplot[area legend, draw=none, fill=mycolor5, fill opacity=0.2, forget plot]
table[row sep=crcr] {%
x	y\\
5	1\\
10	1\\
15	1\\
20	1\\
25	1\\
30	1\\
35	1\\
40	1\\
45	1\\
45	1\\
40	1\\
35	1\\
30	1\\
25	1\\
20	1\\
15	1\\
10	1\\
5	1\\
}--cycle;
\addplot [color=mycolor5, line width=.6pt, mark size=2.5pt, mark=+, mark options={solid, mycolor5}]
  table[row sep=crcr]{%
5	1\\
10	1\\
15	1\\
20	1\\
25	1\\
30	1\\
35	1\\
40	1\\
45	1\\
};
\addlegendentry{AF}

\addplot [color=mycolor5, line width=.6pt, mark size=2.5pt, mark=+, mark options={solid, mycolor5}, forget plot]
 plot [error bars/.cd, y dir = both, y explicit]
 table[row sep=crcr, y error plus index=2, y error minus index=3]{%
5	1	0	0\\
10	1	0	0\\
15	1	0	0\\
20	1	0	0\\
25	1	0	0\\
30	1	0	0\\
35	1	0	0\\
40	1	0	0\\
45	1	0	0\\
};

\addplot[area legend, draw=none, fill=mycolor6, fill opacity=0.2, forget plot]
table[row sep=crcr] {%
x	y\\
5	0.425597058590669\\
10	0.308374700193678\\
15	0.256336317197321\\
20	0.262430607635593\\
25	0.242898187764617\\
30	0.252814309990973\\
35	0.249484228093142\\
40	0.260505618097133\\
45	0.271588010878401\\
45	0.249384352461232\\
40	0.237030605006675\\
35	0.226466194073715\\
30	0.228983340515657\\
25	0.220120587955098\\
20	0.227171115328285\\
15	0.222878944533724\\
10	0.248952654586987\\
5	0.359785042967903\\
}--cycle;
\addplot [color=mycolor6, line width=.6pt, mark size=2.5pt, mark=x, mark options={solid, mycolor6}]
  table[row sep=crcr]{%
5	0.35449016563147\\
10	0.268591136507612\\
15	0.247657263796226\\
20	0.245210501997556\\
25	0.258031634806483\\
30	0.262301085916737\\
35	0.271644229590077\\
40	0.268181747949845\\
45	0.259089969755557\\
};
\addlegendentry{APC}

\addplot [color=mycolor6, line width=.6pt, mark size=2.5pt, mark=x, mark options={solid, mycolor6}, forget plot]
 plot [error bars/.cd, y dir = both, y explicit]
 table[row sep=crcr, y error plus index=2, y error minus index=3]{%
5	0.392691050779286	0.0329060078113833	0.0329060078113833\\
10	0.278663677390333	0.0297110228033454	0.0297110228033454\\
15	0.239607630865523	0.0167286863317986	0.0167286863317986\\
20	0.244800861481939	0.0176297461536543	0.0176297461536543\\
25	0.231509387859857	0.0113887999047597	0.0113887999047597\\
30	0.240898825253315	0.0119154847376583	0.0119154847376583\\
35	0.237975211083428	0.0115090170097134	0.0115090170097134\\
40	0.248768111551904	0.0117375065452288	0.0117375065452288\\
45	0.260486181669816	0.0111018292085845	0.0111018292085845\\
};

\addplot[area legend, draw=none, fill=black, fill opacity=0.08, forget plot]
table[row sep=crcr] {%
x	y\\
5	0.425597058590669\\
10	0.308374700193678\\
15	0.256336317197321\\
20	0.262430607635593\\
25	0.242898187764617\\
30	0.252814309990973\\
35	0.249484228093142\\
40	0.260505618097133\\
45	0.271588010878401\\
45	0.249384352461232\\
40	0.237030605006675\\
35	0.226466194073715\\
30	0.228983340515657\\
25	0.220120587955098\\
20	0.227171115328285\\
15	0.222878944533724\\
10	0.248952654586987\\
5	0.359785042967903\\
}--cycle;
\addplot [color=black, dashed, line width=.6pt]
  table[row sep=crcr]{%
5	0.392691050779286\\
10	0.278663677390333\\
15	0.239607630865523\\
20	0.244800861481939\\
25	0.231509387859857\\
30	0.240898825253315\\
35	0.237975211083428\\
40	0.248768111551904\\
45	0.260486181669816\\
};
\addlegendentry{all}

\addplot [color=black, dashed, line width=.6pt, forget plot]
 plot [error bars/.cd, y dir = both, y explicit]
 table[row sep=crcr, y error plus index=2, y error minus index=3]{%
5	0.392691050779286	0.0329060078113833	0.0329060078113833\\
10	0.278663677390333	0.0297110228033454	0.0297110228033454\\
15	0.239607630865523	0.0167286863317986	0.0167286863317986\\
20	0.244800861481939	0.0176297461536543	0.0176297461536543\\
25	0.231509387859857	0.0113887999047597	0.0113887999047597\\
30	0.240898825253315	0.0119154847376583	0.0119154847376583\\
35	0.237975211083428	0.0115090170097134	0.0115090170097134\\
40	0.248768111551904	0.0117375065452288	0.0117375065452288\\
45	0.260486181669816	0.0111018292085845	0.0111018292085845\\
};
\end{axis}
\end{tikzpicture}%

%% file: imgs/MINNESOTA2/NT2MN.tex
\definecolor{mycolor1}{rgb}{0.00000,0.44700,0.74100}%
\definecolor{mycolor2}{rgb}{0.92900,0.69400,0.12500}%
\definecolor{mycolor3}{rgb}{0.49400,0.18400,0.55600}%
\definecolor{mycolor4}{rgb}{0.46600,0.67400,0.18800}%
\definecolor{mycolor5}{rgb}{0.30100,0.74500,0.93300}%
\definecolor{mycolor6}{rgb}{0.63500,0.07800,0.18400}%
\pgfplotsset{
compat=1.11,
legend image code/.code={
\draw[mark repeat=2,mark phase=2]
plot coordinates {
(0cm,0cm)
(0.15cm,0cm)        
(0.3cm,0cm)         
};%
}
}
\begin{tikzpicture}
\begin{axis}[%
width=.85\columnwidth,
height=.55\columnwidth,
scale only axis,
xmin=4,
xmax=46,
xtick={5,10,15,20,25,30,35,40,45},
xticklabel style={font=\footnotesize},
xlabel style={font=\color{white!15!black}\footnotesize, yshift=3pt},
xlabel={\# failed nodes},
ymin=0,
ymax=900,
ytick={200,400,600,866},
yticklabels={{200},{400},{600},{$|\mathcal{M}|$}},
yticklabel style={font=\scriptsize},
ylabel style={font=\color{white!15!black}\footnotesize,yshift=-10pt},
ylabel={\# tests},
axis background/.style={fill=white},
legend style={at={(0.05,0.55)}, legend columns=3, anchor=south west, legend cell align=left, align=left, fill=none, draw=none, font=\scriptsize}
]

\addplot[area legend, draw=none, fill=mycolor1, fill opacity=0.2, forget plot]
table[row sep=crcr] {%
x	y\\
5	105.185640646055\\
10	113.97113092008\\
15	121.980449381476\\
20	131.223534186399\\
25	136.729468812791\\
30	144.608314132003\\
35	151.07576076891\\
40	163.613303834664\\
45	173.778661543544\\
45	163.021338456456\\
40	157.586696165336\\
35	145.32423923109\\
30	140.991685867997\\
25	131.270531187209\\
20	123.976465813601\\
15	118.419550618524\\
10	111.22886907992\\
5	102.414359353945\\
}--cycle;
\addplot [color=mycolor1, line width=0.6pt, mark size=2.5pt, mark=text,text mark={\LARGE $\star$}, mark options={solid, fill=mycolor1, mycolor1}]
  table[row sep=crcr]{%
5	103.8\\
10	112.6\\
15	120.2\\
20	127.6\\
25	134\\
30	142.8\\
35	148.2\\
40	160.6\\
45	168.4\\
};
\addlegendentry{FaCe}

\addplot [color=mycolor1, line width=0.6pt, mark size=2.5pt,mark=text,text mark={\LARGE $\star$}, mark options={solid, fill=mycolor1, mycolor1}, forget plot]
 plot [error bars/.cd, y dir = both, y explicit]
 table[row sep=crcr, y error plus index=2, y error minus index=3]{%
5	103.8	1.3856406460551	1.3856406460551\\
10	112.6	1.37113092008021	1.37113092008021\\
15	120.2	1.78044938147649	1.78044938147649\\
20	127.6	3.62353418639869	3.62353418639869\\
25	134	2.72946881279124	2.72946881279124\\
30	142.8	1.80831413200251	1.80831413200251\\
35	148.2	2.87576076890968	2.87576076890968\\
40	160.6	3.01330383466387	3.01330383466387\\
45	168.4	5.37866154354408	5.37866154354408\\
};

\addplot[area legend, draw=none, fill=mycolor2, fill opacity=0.2, forget plot]
table[row sep=crcr] {%
x	y\\
5	97.6690415759823\\
10	96.3656854249492\\
15	97.0645751311065\\
20	98.3656854249492\\
25	98.0242640687119\\
30	97.6828427124746\\
35	97.6828427124746\\
40	96.6690415759823\\
45	97.2123105625618\\
45	96.3876894374382\\
40	95.7309584240177\\
35	97.1171572875254\\
30	97.1171572875254\\
25	97.1757359312881\\
20	97.2343145750508\\
15	96.5354248688935\\
10	95.2343145750508\\
5	96.7309584240177\\
}--cycle;
\addplot [color=mycolor2, line width=0.6pt, mark size=2.5pt, mark=asterisk, mark options={solid, mycolor2}]
  table[row sep=crcr]{%
5	97.2\\
10	95.8\\
15	96.8\\
20	97.8\\
25	97.6\\
30	97.4\\
35	97.4\\
40	96.2\\
45	96.8\\
};
\addlegendentry{GC}

\addplot [color=mycolor2, line width=0.6pt, mark size=2.5pt, mark=asterisk, mark options={solid, mycolor2}, forget plot]
 plot [error bars/.cd, y dir = both, y explicit]
 table[row sep=crcr, y error plus index=2, y error minus index=3]{%
5	97.2	0.469041575982343	0.469041575982343\\
10	95.8	0.565685424949238	0.565685424949238\\
15	96.8	0.264575131106459	0.264575131106459\\
20	97.8	0.565685424949238	0.565685424949238\\
25	97.6	0.424264068711929	0.424264068711929\\
30	97.4	0.282842712474619	0.282842712474619\\
35	97.4	0.282842712474619	0.282842712474619\\
40	96.2	0.469041575982343	0.469041575982343\\
45	96.8	0.412310562561766	0.412310562561766\\
};

\addplot[area legend, draw=none, fill=mycolor3, fill opacity=0.2, forget plot]
table[row sep=crcr] {%
x	y\\
5	866\\
10	866\\
15	866\\
20	866\\
25	866\\
30	866\\
35	866\\
40	866\\
45	866\\
45	866\\
40	866\\
35	866\\
30	866\\
25	866\\
20	866\\
15	866\\
10	866\\
5	866\\
}--cycle;
\addplot [color=mycolor3, line width=0.6pt, mark size=2.5pt, mark=triangle, mark options={solid, mycolor3}]
  table[row sep=crcr]{%
5	866\\
10	866\\
15	866\\
20	866\\
25	866\\
30	866\\
35	866\\
40	866\\
45	866\\
};
\addlegendentry{GI}

\addplot [color=mycolor3, line width=0.6pt, mark size=2.5pt, mark=triangle, mark options={solid, mycolor3}, forget plot]
 plot [error bars/.cd, y dir = both, y explicit]
 table[row sep=crcr, y error plus index=2, y error minus index=3]{%
5	866	0	0\\
10	866	0	0\\
15	866	0	0\\
20	866	0	0\\
25	866	0	0\\
30	866	0	0\\
35	866	0	0\\
40	866	0	0\\
45	866	0	0\\
};

\addplot[area legend, draw=none, fill=mycolor4, fill opacity=0.2, forget plot]
table[row sep=crcr] {%
x	y\\
5	866\\
10	866\\
15	866\\
20	866\\
25	866\\
30	866\\
35	866\\
40	866\\
45	866\\
45	866\\
40	866\\
35	866\\
30	866\\
25	866\\
20	866\\
15	866\\
10	866\\
5	866\\
}--cycle;
\addplot [color=mycolor4, line width=0.6pt, mark size=2.5pt, mark=square, mark options={solid, mycolor4}]
  table[row sep=crcr]{%
5	866\\
10	866\\
15	866\\
20	866\\
25	866\\
30	866\\
35	866\\
40	866\\
45	866\\
};
\addlegendentry{GD}

\addplot [color=mycolor4, line width=0.6pt, mark size=2.5pt, mark=square, mark options={solid, mycolor4}, forget plot]
 plot [error bars/.cd, y dir = both, y explicit]
 table[row sep=crcr, y error plus index=2, y error minus index=3]{%
5	866	0	0\\
10	866	0	0\\
15	866	0	0\\
20	866	0	0\\
25	866	0	0\\
30	866	0	0\\
35	866	0	0\\
40	866	0	0\\
45	866	0	0\\
};

\addplot[area legend, draw=none, fill=mycolor5, fill opacity=0.2, forget plot]
table[row sep=crcr] {%
x	y\\
5	53.4082762530298\\
10	106.048528137424\\
15	158.538083151965\\
20	205.890454496037\\
25	255.937814719698\\
30	302.949358868962\\
35	353.955386467836\\
40	402.044799302252\\
45	444.572404813208\\
45	436.627595186792\\
40	391.955200697748\\
35	348.844613532164\\
30	299.050641131038\\
25	252.462185280302\\
20	201.709545503963\\
15	156.661916848035\\
10	104.351471862576\\
5	52.1917237469702\\
}--cycle;
\addplot [color=mycolor5, line width=0.6pt, mark size=2.5pt, mark=+, mark options={solid, mycolor5}]
  table[row sep=crcr]{%
5	52.8\\
10	105.2\\
15	157.6\\
20	203.8\\
25	254.2\\
30	301\\
35	351.4\\
40	397\\
45	440.6\\
};
\addlegendentry{AF}

\addplot [color=mycolor5, line width=0.6pt, mark size=2.5pt, mark=+, mark options={solid, mycolor5}, forget plot]
 plot [error bars/.cd, y dir = both, y explicit]
 table[row sep=crcr, y error plus index=2, y error minus index=3]{%
5	52.8	0.608276253029822	0.608276253029822\\
10	105.2	0.848528137423857	0.848528137423857\\
15	157.6	0.938083151964686	0.938083151964686\\
20	203.8	2.09045449603669	2.09045449603669\\
25	254.2	1.73781471969828	1.73781471969828\\
30	301	1.94935886896179	1.94935886896179\\
35	351.4	2.55538646783613	2.55538646783613\\
40	397	5.04479930225177	5.04479930225177\\
45	440.6	3.97240481320824	3.97240481320824\\
};

\addplot[area legend, draw=none, fill=mycolor6, fill opacity=0.2, forget plot]
table[row sep=crcr] {%
x	y\\
5	115.266397831977\\
10	145.215158056547\\
15	174.013222438758\\
20	237.479570225768\\
25	302.115894296472\\
30	372.152263870422\\
35	431.268157561222\\
40	500.169580224005\\
45	567.681132292187\\
45	469.118867707813\\
40	417.030419775995\\
35	389.531842438778\\
30	320.247736129578\\
25	254.284105703528\\
20	199.320429774232\\
15	160.386777561242\\
10	133.984841943453\\
5	111.133602168023\\
}--cycle;
\addplot [color=mycolor6, line width=0.6pt, mark size=2.5pt, mark=x, mark options={solid, mycolor6}]
  table[row sep=crcr]{%
5	113.2\\
10	139.6\\
15	167.2\\
20	218.4\\
25	278.2\\
30	346.2\\
35	410.4\\
40	458.6\\
45	518.4\\
};
\addlegendentry{APC}

\addplot [color=mycolor6, line width=0.6pt, mark size=2.5pt, mark=x, mark options={solid, mycolor6}, forget plot]
 plot [error bars/.cd, y dir = both, y explicit]
 table[row sep=crcr, y error plus index=2, y error minus index=3]{%
5	113.2	2.06639783197718	2.06639783197718\\
10	139.6	5.61515805654658	5.61515805654658\\
15	167.2	6.81322243875833	6.81322243875833\\
20	218.4	19.0795702257677	19.0795702257677\\
25	278.2	23.9158942964715	23.9158942964715\\
30	346.2	25.9522638704218	25.9522638704218\\
35	410.4	20.8681575612223	20.8681575612223\\
40	458.6	41.5695802240052	41.5695802240052\\
45	518.4	49.2811322921866	49.2811322921866\\
};
\end{axis}
\end{tikzpicture}%

%% file: imgs/MINNESOTA/ABcen.tex
%
%
\definecolor{mycolor1}{rgb}{0.00000,0.60000,1.00000}%
\definecolor{mycolor2}{rgb}{0.00000,1.00000,1.00000}%
\pgfplotsset{
compat=1.11,
legend image code/.code={
\draw[mark repeat=2,mark phase=2]
plot coordinates {
(0cm,0cm)
(0.15cm,0cm)        
(0.3cm,0cm)         
};%
}
}

\begin{tikzpicture}

\begin{axis}[%
width=.85\columnwidth,
height=.55\columnwidth,
scale only axis,
xmin=0,
xmax=200,
xtick={0,40,80,120,160,200},
xticklabel style={font=\footnotesize},
xlabel style={font=\color{white!15!black}\footnotesize,yshift=3pt},
xlabel={\# tested paths},
ymin=0,
ymax=1,
ytick={0,0.2,0.4,0.6,0.8,1},
yticklabels={{0},{.2},{.4},{.6},{.8},{1}},
yticklabel style={font=\footnotesize},
ylabel style={font=\color{white!15!black}\footnotesize,yshift=-5pt},
ylabel={$a_B$},
axis background/.style={fill=white},
legend style={at={(0.05,0.4)}, legend columns=1, anchor=south west, legend cell align=left, align=left, fill=none, draw=none, font=\scriptsize}
]
\addplot [color=blue, line width=0.6pt]
  table[row sep=crcr]{%
1	0\\
2	0\\
3	0\\
4	0\\
5	0\\
6	0\\
7	0\\
8	0\\
9	0\\
10	0\\
11	0\\
12	0\\
13	0\\
14	0\\
15	0\\
16	0\\
17	0\\
18	0\\
19	0\\
20	0\\
21	0\\
22	0\\
23	0\\
24	0\\
25	0\\
26	0\\
27	0\\
28	0\\
29	0\\
30	0\\
31	0\\
32	0\\
33	0\\
34	0\\
35	0\\
36	0\\
37	0\\
38	0\\
39	0\\
40	0\\
41	0\\
42	0\\
43	0\\
44	0\\
45	0\\
46	0\\
47	0\\
48	0\\
49	0\\
50	0\\
51	0\\
52	0\\
53	0\\
54	0\\
55	0\\
56	0\\
57	0\\
58	0\\
59	0\\
60	0\\
61	0\\
62	0\\
63	0\\
64	0\\
65	0\\
66	0\\
67	0\\
68	0\\
69	0\\
70	0\\
71	0\\
72	0\\
73	0\\
74	0\\
75	0\\
76	0\\
77	0\\
78	0\\
79	0\\
80	0.00833333333333333\\
81	0.0183333333333333\\
82	0.0183333333333333\\
83	0.0183333333333333\\
84	0.0183333333333333\\
85	0.0283333333333333\\
86	0.0283333333333333\\
87	0.0283333333333333\\
88	0.0283333333333333\\
89	0.0283333333333333\\
90	0.0283333333333333\\
91	0.0283333333333333\\
92	0.0283333333333333\\
93	0.0283333333333333\\
94	0.0283333333333333\\
95	0.0283333333333333\\
96	0.0283333333333333\\
97	0.046025641025641\\
98	0.066025641025641\\
99	0.086025641025641\\
100	0.0937179487179487\\
101	0.0937179487179487\\
102	0.0937179487179487\\
103	0.0937179487179487\\
104	0.0937179487179487\\
105	0.102051282051282\\
106	0.102051282051282\\
107	0.102051282051282\\
108	0.102051282051282\\
109	0.102051282051282\\
110	0.10974358974359\\
111	0.118076923076923\\
112	0.128076923076923\\
113	0.16474358974359\\
114	0.16474358974359\\
115	0.16474358974359\\
116	0.16474358974359\\
117	0.181410256410256\\
118	0.221410256410256\\
119	0.231410256410256\\
120	0.242521367521368\\
121	0.252521367521368\\
122	0.276807081807082\\
123	0.310140415140415\\
124	0.34045177045177\\
125	0.358785103785104\\
126	0.383070818070818\\
127	0.390763125763126\\
128	0.409854034854035\\
129	0.417546342546342\\
130	0.417546342546342\\
131	0.434213009213009\\
132	0.452546342546343\\
133	0.472546342546343\\
134	0.516832056832057\\
135	0.527943167943168\\
136	0.548145188145188\\
137	0.575837495837496\\
138	0.565793465793466\\
139	0.592777592777593\\
140	0.603888703888704\\
141	0.640654407321074\\
142	0.620988733488733\\
143	0.641020784770785\\
144	0.683969502719503\\
145	0.663714063714064\\
146	0.667709142709143\\
147	0.681598031598032\\
148	0.634584304584305\\
149	0.64996891996892\\
150	0.707575757575758\\
151	0.727575757575758\\
152	0.709974747474748\\
153	0.730808080808081\\
154	0.783585858585859\\
155	0.808585858585859\\
156	0.861363636363636\\
157	0.815151515151515\\
158	0.815151515151515\\
159	0.815151515151515\\
160	0.815151515151515\\
161	0.848484848484849\\
162	0.848484848484849\\
163	0.848484848484849\\
164	0.772727272727273\\
165	0.772727272727273\\
166	0.772727272727273\\
167	0.636363636363636\\
168	0.636363636363636\\
169	0.727272727272727\\
170	0.818181818181818\\
171	0.909090909090909\\
172	0.909090909090909\\
173	0.909090909090909\\
174	0.909090909090909\\
175	0.909090909090909\\
176	0.909090909090909\\
177	0.909090909090909\\
178	1\\
179	1\\
180	1\\
181	1\\
182	1\\
183	1\\
};
\addlegendentry{$c=0.05$}

\addplot [color=mycolor1, line width=0.6pt]
  table[row sep=crcr]{%
1	0\\
2	0\\
3	0\\
4	0\\
5	0\\
6	0\\
7	0\\
8	0\\
9	0\\
10	0\\
11	0\\
12	0\\
13	0\\
14	0\\
15	0\\
16	0\\
17	0\\
18	0\\
19	0\\
20	0\\
21	0\\
22	0\\
23	0\\
24	0\\
25	0\\
26	0\\
27	0\\
28	0\\
29	0\\
30	0\\
31	0\\
32	0\\
33	0\\
34	0\\
35	0\\
36	0\\
37	0\\
38	0\\
39	0\\
40	0\\
41	0\\
42	0\\
43	0\\
44	0\\
45	0\\
46	0\\
47	0\\
48	0\\
49	0\\
50	0\\
51	0\\
52	0\\
53	0\\
54	0\\
55	0\\
56	0\\
57	0\\
58	0\\
59	0\\
60	0\\
61	0\\
62	0\\
63	0\\
64	0\\
65	0\\
66	0\\
67	0\\
68	0\\
69	0\\
70	0\\
71	0\\
72	0\\
73	0\\
74	0\\
75	0\\
76	0\\
77	0.00909090909090909\\
78	0.00909090909090909\\
79	0.00909090909090909\\
80	0.00909090909090909\\
81	0.00909090909090909\\
82	0.00909090909090909\\
83	0.00909090909090909\\
84	0.00909090909090909\\
85	0.0167832167832168\\
86	0.0167832167832168\\
87	0.0167832167832168\\
88	0.0167832167832168\\
89	0.0167832167832168\\
90	0.0167832167832168\\
91	0.0310689310689311\\
92	0.0310689310689311\\
93	0.0510689310689311\\
94	0.0510689310689311\\
95	0.0610689310689311\\
96	0.0710689310689311\\
97	0.0810689310689311\\
98	0.0810689310689311\\
99	0.0810689310689311\\
100	0.0810689310689311\\
101	0.0894022644022644\\
102	0.0894022644022644\\
103	0.0977355977355977\\
104	0.0977355977355977\\
105	0.0977355977355977\\
106	0.0977355977355977\\
107	0.0977355977355977\\
108	0.107735597735598\\
109	0.107735597735598\\
110	0.107735597735598\\
111	0.107735597735598\\
112	0.107735597735598\\
113	0.158687978687979\\
114	0.168687978687979\\
115	0.17979908979909\\
116	0.197491397491397\\
117	0.222491397491397\\
118	0.230824730824731\\
119	0.248517038517039\\
120	0.28962814962815\\
121	0.28962814962815\\
122	0.297961482961483\\
123	0.327052392052392\\
124	0.353719058719059\\
125	0.4047446997447\\
126	0.4147446997447\\
127	0.4147446997447\\
128	0.4247446997447\\
129	0.445855810855811\\
130	0.473548118548119\\
131	0.493548118548119\\
132	0.516167166167166\\
133	0.543859473859474\\
134	0.559885114885115\\
135	0.604170829170829\\
136	0.612504162504162\\
137	0.631595071595071\\
138	0.70539849039849\\
139	0.714489399489399\\
140	0.722794489461156\\
141	0.698560467310467\\
142	0.655497676926248\\
143	0.669783391211963\\
144	0.646086321086321\\
145	0.763608613608614\\
146	0.804084804084804\\
147	0.798468198468198\\
148	0.827039627039627\\
149	0.842424242424242\\
150	0.737373737373737\\
151	0.774410774410775\\
152	0.661616161616162\\
153	0.661616161616162\\
154	0.661616161616162\\
155	0.661616161616162\\
156	0.772727272727273\\
157	0.772727272727273\\
158	0.545454545454546\\
159	0.545454545454546\\
160	0.545454545454546\\
161	0.545454545454546\\
162	0.545454545454546\\
163	0.545454545454546\\
164	0.636363636363636\\
165	0.636363636363636\\
166	0.727272727272727\\
167	0.727272727272727\\
168	0.727272727272727\\
169	0.818181818181818\\
170	0.909090909090909\\
171	0.909090909090909\\
172	0.909090909090909\\
173	0.909090909090909\\
174	1\\
175	1\\
176	1\\
177	1\\
178	1\\
179	1\\
180	1\\
181	1\\
};
\addlegendentry{$c=0.08$}

\addplot [color=mycolor2, line width=0.6pt]
  table[row sep=crcr]{%
1	0\\
2	0\\
3	0\\
4	0\\
5	0\\
6	0\\
7	0\\
8	0\\
9	0\\
10	0\\
11	0\\
12	0\\
13	0\\
14	0\\
15	0\\
16	0\\
17	0\\
18	0\\
19	0\\
20	0\\
21	0\\
22	0\\
23	0\\
24	0\\
25	0\\
26	0\\
27	0\\
28	0\\
29	0\\
30	0\\
31	0\\
32	0\\
33	0\\
34	0\\
35	0\\
36	0\\
37	0\\
38	0\\
39	0\\
40	0\\
41	0\\
42	0\\
43	0\\
44	0\\
45	0\\
46	0.00833333333333333\\
47	0.00833333333333333\\
48	0.00833333333333333\\
49	0.00833333333333333\\
50	0.00833333333333333\\
51	0.00833333333333333\\
52	0.00833333333333333\\
53	0.00833333333333333\\
54	0.00833333333333333\\
55	0.00833333333333333\\
56	0.00833333333333333\\
57	0.00833333333333333\\
58	0.00833333333333333\\
59	0.00833333333333333\\
60	0.00833333333333333\\
61	0.00833333333333333\\
62	0.00833333333333333\\
63	0.00833333333333333\\
64	0.00833333333333333\\
65	0.0183333333333333\\
66	0.0183333333333333\\
67	0.0183333333333333\\
68	0.0183333333333333\\
69	0.0183333333333333\\
70	0.0183333333333333\\
71	0.0183333333333333\\
72	0.0183333333333333\\
73	0.0183333333333333\\
74	0.0183333333333333\\
75	0.0183333333333333\\
76	0.0183333333333333\\
77	0.0183333333333333\\
78	0.0183333333333333\\
79	0.0183333333333333\\
80	0.0183333333333333\\
81	0.0183333333333333\\
82	0.0383333333333333\\
83	0.0383333333333333\\
84	0.0383333333333333\\
85	0.0383333333333333\\
86	0.0383333333333333\\
87	0.0483333333333333\\
88	0.0677777777777778\\
89	0.0677777777777778\\
90	0.0677777777777778\\
91	0.0677777777777778\\
92	0.0677777777777778\\
93	0.0677777777777778\\
94	0.0677777777777778\\
95	0.0927777777777778\\
96	0.100470085470085\\
97	0.117136752136752\\
98	0.117136752136752\\
99	0.126227661227661\\
100	0.126227661227661\\
101	0.142894327894328\\
102	0.142894327894328\\
103	0.142894327894328\\
104	0.151227661227661\\
105	0.151227661227661\\
106	0.151227661227661\\
107	0.151227661227661\\
108	0.151227661227661\\
109	0.159560994560995\\
110	0.159560994560995\\
111	0.159560994560995\\
112	0.169560994560995\\
113	0.183846708846709\\
114	0.203846708846709\\
115	0.203846708846709\\
116	0.222180042180042\\
117	0.222180042180042\\
118	0.22987234987235\\
119	0.22987234987235\\
120	0.277009102009102\\
121	0.291294816294816\\
122	0.305580530580531\\
123	0.343913863913864\\
124	0.381606171606172\\
125	0.391606171606172\\
126	0.413584193584194\\
127	0.443584193584194\\
128	0.443584193584194\\
129	0.461917526917527\\
130	0.471917526917527\\
131	0.533895548895549\\
132	0.55992118992119\\
133	0.603254523254523\\
134	0.590653173986507\\
135	0.62998297998298\\
136	0.650875667542334\\
137	0.661986778653445\\
138	0.728999395666062\\
139	0.761003194336528\\
140	0.753243978243978\\
141	0.786722999222999\\
142	0.795937395937396\\
143	0.736534576534577\\
144	0.736534576534577\\
145	0.772898212898213\\
146	0.760353535353535\\
147	0.717508417508418\\
148	0.717508417508418\\
149	0.717508417508418\\
150	0.781144781144781\\
151	0.811447811447811\\
152	0.841750841750842\\
153	0.841750841750842\\
154	0.808080808080808\\
155	0.909090909090909\\
156	0.818181818181818\\
157	0.818181818181818\\
158	0.818181818181818\\
159	0.818181818181818\\
160	0.818181818181818\\
161	0.818181818181818\\
162	0.818181818181818\\
163	0.909090909090909\\
164	0.909090909090909\\
165	0.909090909090909\\
166	0.909090909090909\\
167	0.909090909090909\\
168	0.909090909090909\\
169	0.909090909090909\\
170	0.909090909090909\\
171	0.909090909090909\\
172	0.909090909090909\\
173	1\\
174	1\\
175	1\\
176	1\\
177	1\\
178	1\\
};
\addlegendentry{$c=0.1$}

\end{axis}
\end{tikzpicture}%

%% file: imgs/dynamicBICS/precision.tex
%
%
\definecolor{mycolor1}{rgb}{0.00000,0.44700,0.74100}%
\definecolor{mycolor2}{rgb}{0.85000,0.32500,0.09800}%
\pgfplotsset{
compat=1.11,
legend image code/.code={
\draw[mark repeat=2,mark phase=2]
plot coordinates {
(0cm,0cm)
(0.15cm,0cm)        
(0.3cm,0cm)         
};%
}
}
\begin{tikzpicture}

\begin{axis}[%
width=.85\columnwidth,
height=.5\columnwidth,
scale only axis,
xmin=1,
xmax=188,
xtick={1,24.375,47.75,71.125,94.5,117.875,141.25,164.625,188},
xticklabel style={font=\footnotesize},
xticklabel style={rotate=45},
xlabel style={font=\color{white!15!black}\footnotesize,yshift=5pt},
xticklabels={{ 13},{ 37},{ 61},{ 85},{109},{133},{157},{181},{200}},
xlabel={steps},
ymin=0,
ymax=1,
ytick={  0, 0.2, 0.4, 0.6, 0.8,   1},
yticklabels={{0},{.2},{.4},{.6},{.8},{1}},
yticklabel style={font=\footnotesize},
ylabel style={font=\color{white!15!black}\footnotesize,yshift=-5pt},
ylabel={precision},
axis background/.style={fill=white},
legend style={at={(0.1,0.03)}, legend columns=1, anchor=south west, legend cell align=left, align=left, fill=none, draw=none, font=\scriptsize}
]
\addplot [color=mycolor1, line width=1pt]
  table[row sep=crcr]{%
1	1\\
2	1\\
3	0.998\\
4	0.998063172043011\\
5	0.997354838709678\\
6	0.99545685020393\\
7	0.993608870967742\\
8	0.990967741935484\\
9	0.989655913978495\\
10	0.986732526881721\\
11	0.983416367980884\\
12	0.982139976958526\\
13	0.979987604540024\\
14	0.981445073450218\\
15	0.978416742133604\\
16	0.975732473731645\\
17	0.977591422354399\\
18	0.975409848309529\\
19	0.975318705830194\\
20	0.977070087492433\\
21	0.976544711149272\\
22	0.977920734487001\\
23	0.977233145431528\\
24	0.977849841469825\\
25	0.975071651316093\\
26	0.975449995653833\\
27	0.972156833603168\\
28	0.971495220835071\\
29	0.952199449611063\\
30	0.971633402190886\\
31	0.949124644057413\\
32	0.947153457887712\\
33	0.946365449140688\\
34	0.939695507739012\\
35	0.955857894578994\\
36	0.93608793644005\\
37	0.956072785582065\\
38	0.931176944823857\\
39	0.929337452193575\\
40	0.925106054623639\\
41	0.919206736428074\\
42	0.923230345340093\\
43	0.926068670153598\\
44	0.925396737980789\\
45	0.923826315858439\\
46	0.924581786407408\\
47	0.920319804414178\\
48	0.899211786480189\\
49	0.91330075294172\\
50	0.911878653403011\\
51	0.910460110174298\\
52	0.919047310824295\\
53	0.919562260359005\\
54	0.912823487160483\\
55	0.914805831144965\\
56	0.899805781631044\\
57	0.921713512650792\\
58	0.920331393345783\\
59	0.91968179385667\\
60	0.907878756525605\\
61	0.891915558367196\\
62	0.880236531982161\\
63	0.907855291594665\\
64	0.900912543602932\\
65	0.902277327627064\\
66	0.909717127418164\\
67	0.911695660516252\\
68	0.906668423297323\\
69	0.907703329965116\\
70	0.906257201602175\\
71	0.907032952359433\\
72	0.905476213189892\\
73	0.905875566949048\\
74	0.908335170464835\\
75	0.925920935498963\\
76	0.931108913268813\\
77	0.931792530019074\\
78	0.933782891518268\\
79	0.929717441946609\\
80	0.931986214711424\\
81	0.936561560802066\\
82	0.935468906683537\\
83	0.921277142012638\\
84	0.930782848601468\\
85	0.938754946441553\\
86	0.934596118463561\\
87	0.910143040003511\\
88	0.929288850373499\\
89	0.929771714920614\\
90	0.92722986061976\\
91	0.929935063262346\\
92	0.935189316930914\\
93	0.933198029535183\\
94	0.936136602700123\\
95	0.934392986536252\\
96	0.930310799036405\\
97	0.91157573990909\\
98	0.924325156553122\\
99	0.926496562888592\\
100	0.932954954062399\\
101	0.928151682346804\\
102	0.935613435621596\\
103	0.932155967485528\\
104	0.919138661532748\\
105	0.917230519121046\\
106	0.934686041885762\\
107	0.939841700879297\\
108	0.941231057232852\\
109	0.93153375866456\\
110	0.935784830866887\\
111	0.925544193539368\\
112	0.912006258803047\\
113	0.900459930208222\\
114	0.914002348747874\\
115	0.916569313356015\\
116	0.912514247800406\\
117	0.916440628254066\\
118	0.900598176605641\\
119	0.8956591973212\\
120	0.896104879460731\\
121	0.886184811766636\\
122	0.884449701537091\\
123	0.901729431907065\\
124	0.896436030974138\\
125	0.893167778028504\\
126	0.892113200947715\\
127	0.890408436115371\\
128	0.891274555634027\\
129	0.889177912895447\\
130	0.889320947130037\\
131	0.89394268806332\\
132	0.898673609631367\\
133	0.896444323550848\\
134	0.887994565722034\\
135	0.887641064106193\\
136	0.878745170494659\\
137	0.86594912371936\\
138	0.841174857682079\\
139	0.841324332153483\\
140	0.855968470889178\\
141	0.870313109440621\\
142	0.880458474201662\\
143	0.904809266923949\\
144	0.906620758332464\\
145	0.889555325730284\\
146	0.888460967871196\\
147	0.904579985578942\\
148	0.902132560173401\\
149	0.903318841662439\\
150	0.898860077087321\\
151	0.898862093483926\\
152	0.900070199094876\\
153	0.897392037918443\\
154	0.90068532615564\\
155	0.894336613483579\\
156	0.884051387743836\\
157	0.89217168076526\\
158	0.882380228473834\\
159	0.882388071802401\\
160	0.877753052344144\\
161	0.896199926661615\\
162	0.897216058012357\\
163	0.877536991330127\\
164	0.860257346429539\\
165	0.851933631555483\\
166	0.859552925920197\\
167	0.862365251860005\\
168	0.855311698803932\\
169	0.86445792918852\\
170	0.850848272763491\\
171	0.845423614011395\\
172	0.884187713389194\\
173	0.887942479821079\\
174	0.881017025043631\\
175	0.884823112668364\\
176	0.891426391477677\\
177	0.875577680790019\\
178	0.880128117533886\\
179	0.892517035804695\\
180	0.895752357332683\\
181	0.88352196737499\\
182	0.896137567183517\\
183	0.898358039215904\\
184	0.894817120803143\\
185	0.885181020691269\\
186	0.885861122463121\\
187	0.883452656812799\\
188	0.887352659814104\\
};
\addlegendentry{DFaCe}

\addplot [color=mycolor2, line width=1pt]
  table[row sep=crcr]{%
1	1\\
2	1\\
3	0.998\\
4	0.998040183537264\\
5	0.996662058371736\\
6	0.995926267281106\\
7	0.995446428571429\\
8	0.994124423963134\\
9	0.992860215053764\\
10	0.988545698924731\\
11	0.986080367074527\\
12	0.986604262672811\\
13	0.98265787677078\\
14	0.981152161011032\\
15	0.976287248894145\\
16	0.975999596316837\\
17	0.977654195597898\\
18	0.977143812354723\\
19	0.976888928881792\\
20	0.977908202662347\\
21	0.978571749794451\\
22	0.979034903449059\\
23	0.980298128036703\\
24	0.980737664667634\\
25	0.984003945620636\\
26	0.983447231047913\\
27	0.979711252681391\\
28	0.977225932604122\\
29	0.979108832693402\\
30	0.979777160479079\\
31	0.975061528204706\\
32	0.972358321466507\\
33	0.971556467376639\\
34	0.964176793687917\\
35	0.961622859139893\\
36	0.961718964486161\\
37	0.959943142773408\\
38	0.956576429022144\\
39	0.955107832004439\\
40	0.952410211915038\\
41	0.952200091473228\\
42	0.956743076687803\\
43	0.96189639346255\\
44	0.957904732632461\\
45	0.962625136513262\\
46	0.962975545129402\\
47	0.960568624934078\\
48	0.95516042956652\\
49	0.956412844470107\\
50	0.961135011409376\\
51	0.964294199553793\\
52	0.968979216641925\\
53	0.969058672399788\\
54	0.969211438617001\\
55	0.968667621895264\\
56	0.966214049984128\\
57	0.961477276825335\\
58	0.961631570645809\\
59	0.962200446084906\\
60	0.961478268577712\\
61	0.963864937368691\\
62	0.960666449519736\\
63	0.954602718176439\\
64	0.956782439630159\\
65	0.959416268618911\\
66	0.959516157999501\\
67	0.965683955412404\\
68	0.962569156562567\\
69	0.961426616430236\\
70	0.958055493283135\\
71	0.960561970109773\\
72	0.960878833639425\\
73	0.962343351117777\\
74	0.964218061023574\\
75	0.965312273736037\\
76	0.961841329668972\\
77	0.966024043560362\\
78	0.967428130444732\\
79	0.965353467157476\\
80	0.968300033196112\\
81	0.97153830493818\\
82	0.970712235704394\\
83	0.966169749629338\\
84	0.96365477160936\\
85	0.962319078531848\\
86	0.960213067522917\\
87	0.96127238672784\\
88	0.959522990097221\\
89	0.9606172810417\\
90	0.964821748244\\
91	0.965321829743887\\
92	0.962057227387429\\
93	0.960594648174992\\
94	0.962292368847184\\
95	0.95959378572546\\
96	0.961599581162813\\
97	0.963256980654784\\
98	0.961106172400439\\
99	0.963279835104066\\
100	0.964832622542584\\
101	0.962824043995059\\
102	0.964722872444592\\
103	0.963700695908646\\
104	0.968096610748252\\
105	0.966666196581316\\
106	0.966201766121038\\
107	0.967156435479137\\
108	0.966341357800461\\
109	0.96500073813597\\
110	0.97174519518816\\
111	0.968730816206966\\
112	0.969424522521519\\
113	0.966979020218513\\
114	0.960483711427204\\
115	0.959630285889491\\
116	0.960645030936188\\
117	0.959707414157136\\
118	0.952650207631818\\
119	0.949913472512884\\
120	0.953962921570079\\
121	0.952883859881324\\
122	0.949608700803353\\
123	0.952853928611876\\
124	0.956138330185356\\
125	0.954634905507709\\
126	0.955173583864606\\
127	0.951464870715121\\
128	0.950591409504689\\
129	0.952506651761505\\
130	0.95396267536896\\
131	0.9519252044962\\
132	0.961145282806017\\
133	0.962131757765573\\
134	0.963473603764585\\
135	0.959541124437987\\
136	0.958367460645552\\
137	0.957765795407751\\
138	0.960177079376302\\
139	0.963892681478084\\
140	0.961555152200619\\
141	0.9613230047586\\
142	0.960066893938951\\
143	0.958995842413682\\
144	0.958141819303137\\
145	0.960565171418346\\
146	0.95807644322363\\
147	0.960191514807793\\
148	0.958112212366383\\
149	0.965089169692006\\
150	0.962180938898584\\
151	0.965908460492665\\
152	0.966746817157218\\
153	0.967709057261148\\
154	0.962899216019508\\
155	0.961209159186571\\
156	0.959025228111656\\
157	0.961508913713279\\
158	0.963905543966135\\
159	0.960438369915011\\
160	0.961426493588261\\
161	0.960754333945476\\
162	0.958102841732083\\
163	0.956498137286713\\
164	0.962131795799354\\
165	0.962685873035979\\
166	0.958770009025292\\
167	0.960785284996094\\
168	0.965918355221252\\
169	0.9687905281534\\
170	0.968941397598898\\
171	0.974089989476391\\
172	0.97338536748946\\
173	0.971627462511582\\
174	0.973150162959282\\
175	0.970532598834105\\
176	0.971748502066467\\
177	0.97043789412602\\
178	0.975335965945599\\
179	0.97686311121211\\
180	0.975141386576592\\
181	0.975738346311405\\
182	0.97316288864937\\
183	0.975388030479799\\
184	0.974197641819722\\
185	0.975161687784746\\
186	0.973694538600738\\
187	0.974225201237241\\
188	0.972362572544664\\
};
\addlegendentry{DPoP}

\addplot[area legend, draw=none, fill=mycolor1, fill opacity=0.2, forget plot]
table[row sep=crcr] {%
x	y\\
1	1\\
2	1\\
3	1.00599659791607\\
4	1.00581004924228\\
5	1.00641706367757\\
6	1.00847384661848\\
7	1.01076061653661\\
8	1.01262182511423\\
9	1.01266345683818\\
10	1.01147420286839\\
11	1.01160586348887\\
12	1.00961476059554\\
13	1.00839813116388\\
14	1.00649316596142\\
15	1.00660293677367\\
16	1.00225243601413\\
17	1.00325750553389\\
18	1.00579689583819\\
19	1.00537445748238\\
20	1.00727464437504\\
21	1.00959646733588\\
22	1.00997442487872\\
23	1.01072091122119\\
24	1.00949153697413\\
25	1.0102220710847\\
26	1.00940173036304\\
27	1.0107697728589\\
28	1.0121355529914\\
29	1.09611966953064\\
30	1.01142591674541\\
31	1.09132223039162\\
32	1.09100340940929\\
33	1.08962989977121\\
34	1.08387098559504\\
35	1.00270950608379\\
36	1.0801029283536\\
37	1.0083570119694\\
38	1.07975463828998\\
39	1.07261670889663\\
40	1.07070987880449\\
41	1.06488236286765\\
42	1.0693489593759\\
43	1.07269669743103\\
44	1.07232562941934\\
45	1.07124875863189\\
46	1.07216811185697\\
47	1.06977928792199\\
48	1.09616704490814\\
49	1.064289149018\\
50	1.06069602479321\\
51	1.06034629850369\\
52	1.06820703443917\\
53	1.06954539621269\\
54	1.06291169884597\\
55	1.0660917968082\\
56	1.06186224116622\\
57	1.01517434289005\\
58	1.02153658821624\\
59	1.01080665045579\\
60	1.06591488841984\\
61	1.09366865665678\\
62	1.10332336226543\\
63	1.06383859647537\\
64	1.06175125795778\\
65	1.06491168217364\\
66	1.06357887594058\\
67	1.06626165273751\\
68	1.06302336232623\\
69	1.06740078798699\\
70	1.06312831255799\\
71	1.06498246099113\\
72	1.06308288859336\\
73	1.062285601554\\
74	1.0670957103849\\
75	1.02325988695605\\
76	1.02190125208866\\
77	1.02036743462941\\
78	1.01201488055435\\
79	1.01432095730733\\
80	1.0177513679077\\
81	1.01950905456283\\
82	1.01587032029197\\
83	1.06656556360007\\
84	1.00277032021065\\
85	1.01216311516597\\
86	1.00936956584524\\
87	1.07206494284836\\
88	1.01094088055352\\
89	1.01901712887523\\
90	1.00991991109583\\
91	1.00323464852316\\
92	0.996788792997926\\
93	0.991911614096142\\
94	0.994450757027086\\
95	0.991433844032279\\
96	1.01465263747413\\
97	1.05525669318562\\
98	0.989846579826211\\
99	0.992777924742799\\
100	0.994694549575521\\
101	1.00314469972128\\
102	0.994728913221263\\
103	0.994845921201311\\
104	1.06037639705414\\
105	1.06259500218389\\
106	0.990257832161705\\
107	0.992172001170073\\
108	1.00138625649086\\
109	0.995966175476054\\
110	0.995083891227062\\
111	1.00110906492675\\
112	1.03381530496958\\
113	1.04897325093381\\
114	0.989586606372979\\
115	0.994373502332863\\
116	0.993703206061432\\
117	0.987366654319494\\
118	1.04622461399096\\
119	1.04299290930531\\
120	1.04957283399557\\
121	1.07895421488957\\
122	1.07755568141394\\
123	1.04755845821803\\
124	1.04543188255676\\
125	1.0444136519838\\
126	1.03848579056858\\
127	1.03894371224169\\
128	1.04023844445406\\
129	1.00022656099542\\
130	0.999798889597729\\
131	0.99349562986539\\
132	1.00124899841195\\
133	1.00165266298208\\
134	0.999325319617285\\
135	1.02229048231158\\
136	0.986753861780802\\
137	1.02143110230471\\
138	1.06500782460768\\
139	1.08960797114836\\
140	1.08538865011599\\
141	1.06477559660355\\
142	1.0339837899783\\
143	0.984269111084515\\
144	0.97979876256514\\
145	1.03622912734173\\
146	1.03832184830655\\
147	0.997210171588219\\
148	0.996655680689167\\
149	0.990140069101761\\
150	0.981691391210878\\
151	1.00659248859452\\
152	1.00841934682464\\
153	1.00313535925914\\
154	1.01096378174673\\
155	1.00454837593303\\
156	1.0392068678397\\
157	0.999113960936441\\
158	1.03815281237821\\
159	1.03675982774592\\
160	1.04109756376431\\
161	1.00021099531435\\
162	1.00044520852652\\
163	1.00241416995722\\
164	1.05055031575711\\
165	1.0485377137893\\
166	1.04800131829058\\
167	1.05179148987014\\
168	1.05897764891182\\
169	1.0568605384267\\
170	1.07947378502845\\
171	1.0831215821382\\
172	1.05865220385895\\
173	1.05518204101567\\
174	1.06088493814323\\
175	1.06077867155791\\
176	1.06059154093204\\
177	1.08561325792021\\
178	1.08595820302128\\
179	1.05987427724159\\
180	1.06104047231026\\
181	1.0662232665762\\
182	1.05920487327984\\
183	1.06095921471533\\
184	1.06129625250473\\
185	1.0918438835225\\
186	1.09407970198669\\
187	1.09491533554693\\
188	1.0958446429038\\
188	0.678860676724406\\
187	0.671989978078668\\
186	0.677642542939557\\
185	0.678518157860034\\
184	0.728337989101552\\
183	0.735756863716478\\
182	0.733070261087194\\
181	0.70082066817378\\
180	0.73046424235511\\
179	0.725159794367804\\
178	0.674298032046496\\
177	0.665542103659831\\
176	0.722261242023311\\
175	0.708867553778823\\
174	0.701149111944029\\
173	0.720702918626487\\
172	0.709723222919433\\
171	0.60772564588459\\
170	0.622222760498536\\
169	0.672055319950335\\
168	0.651645748696047\\
167	0.672939013849874\\
166	0.671104533549819\\
165	0.655329549321661\\
164	0.669964377101972\\
163	0.75265981270303\\
162	0.793986907498196\\
161	0.792188858008879\\
160	0.71440854092398\\
159	0.72801631585888\\
158	0.726607644569461\\
157	0.78522940059408\\
156	0.728895907647974\\
155	0.784124851034131\\
154	0.790406870564548\\
153	0.791648716577744\\
152	0.791721051365114\\
151	0.791131698373335\\
150	0.816028762963763\\
149	0.816497614223116\\
148	0.807609439657635\\
147	0.811949799569666\\
146	0.738600087435845\\
145	0.742881524118838\\
144	0.833442754099788\\
143	0.825349422763383\\
142	0.726933158425019\\
141	0.675850622277694\\
140	0.626548291662365\\
139	0.593040693158604\\
138	0.617341890756479\\
137	0.710467145134015\\
136	0.770736479208516\\
135	0.752991645900802\\
134	0.776663811826783\\
133	0.791235984119612\\
132	0.796098220850788\\
131	0.794389746261249\\
130	0.778843004662344\\
129	0.778129264795472\\
128	0.742310666813997\\
127	0.741873159989053\\
126	0.745740611326846\\
125	0.741921904073211\\
124	0.747440179391521\\
123	0.755900405596099\\
122	0.69134372166024\\
121	0.693415408643701\\
120	0.74263692492589\\
119	0.748325485337085\\
118	0.754971739220327\\
117	0.845514602188639\\
116	0.83132528953938\\
115	0.838765124379167\\
114	0.838418091122769\\
113	0.751946609482638\\
112	0.790197212636514\\
111	0.849979322151985\\
110	0.876485770506712\\
109	0.867101341853066\\
108	0.881075857974846\\
107	0.887511400588521\\
106	0.87911425160982\\
105	0.771866036058198\\
104	0.777900926011362\\
103	0.869466013769745\\
102	0.87649795802193\\
101	0.853158664972324\\
100	0.871215358549278\\
99	0.860215201034385\\
98	0.858803733280032\\
97	0.76789478663256\\
96	0.845968960598682\\
95	0.877352129040226\\
94	0.87782244837316\\
93	0.874484444974225\\
92	0.873589840863903\\
91	0.856635478001533\\
90	0.844539810143694\\
89	0.840526300966001\\
88	0.84763682019348\\
87	0.748221137158661\\
86	0.859822671081885\\
85	0.865346777717131\\
84	0.858795376992281\\
83	0.775988720425206\\
82	0.855067493075106\\
81	0.853614067041298\\
80	0.846221061515149\\
79	0.845113926585884\\
78	0.855550902482188\\
77	0.843217625408734\\
76	0.840316574448966\\
75	0.828581984041872\\
74	0.749574630544765\\
73	0.749465532344096\\
72	0.74786953778642\\
71	0.749083443727736\\
70	0.749386090646361\\
69	0.748005871943241\\
68	0.750313484268416\\
67	0.757129668294998\\
66	0.755855378895748\\
65	0.739642973080493\\
64	0.740073829248081\\
63	0.751871986713964\\
62	0.657149701698892\\
61	0.690162460077611\\
60	0.749842624631374\\
59	0.828556937257548\\
58	0.819126198475325\\
57	0.828252682411538\\
56	0.73774932209587\\
55	0.763519865481733\\
54	0.762735275475001\\
53	0.769579124505325\\
52	0.769887587209423\\
51	0.760573921844907\\
50	0.763061282012815\\
49	0.762312356865437\\
48	0.702256528052238\\
47	0.770860320906367\\
46	0.776995460957847\\
45	0.776403873084987\\
44	0.778467846542234\\
43	0.779440642876169\\
42	0.777111731304287\\
41	0.773531109988496\\
40	0.779502230442792\\
39	0.786058195490517\\
38	0.782599251357735\\
37	0.903788559194728\\
36	0.792072944526495\\
35	0.909006283074199\\
34	0.795520029882981\\
33	0.803100998510168\\
32	0.803303506366137\\
31	0.806927057723204\\
30	0.931840887636359\\
29	0.808279229691481\\
28	0.93085488867874\\
27	0.933543894347433\\
26	0.941498260944626\\
25	0.939921231547487\\
24	0.946208145965518\\
23	0.943745379641864\\
22	0.945867044095277\\
21	0.943492954962664\\
20	0.946865530609828\\
19	0.945262954178004\\
18	0.945022800780865\\
17	0.951925339174909\\
16	0.949212511449158\\
15	0.95023054749354\\
14	0.956396980939012\\
13	0.951577077916166\\
12	0.954665193321515\\
11	0.955226872472896\\
10	0.961990850895051\\
9	0.96664837111881\\
8	0.969313658756736\\
7	0.976457125398875\\
6	0.982439853789379\\
5	0.988292613741782\\
4	0.990316294843739\\
3	0.99000340208393\\
2	1\\
1	1\\
}--cycle;

\addplot[area legend, draw=none, fill=mycolor2, fill opacity=0.2, forget plot]
table[row sep=crcr] {%
x	y\\
1	1\\
2	1\\
3	1.00599659791607\\
4	1.00588323823168\\
5	1.00870193312185\\
6	1.00920982468109\\
7	1.00895991592218\\
8	1.00837811505623\\
9	1.00936131457193\\
10	1.01063790876771\\
11	1.01096898589061\\
12	1.01074154652855\\
13	1.01096777425225\\
14	1.00903445625598\\
15	1.01426047290014\\
16	1.00614538752189\\
17	1.00853640862664\\
18	1.00973333536267\\
19	1.00769032694333\\
20	1.00588103034832\\
21	1.0074284163745\\
22	1.01136238892551\\
23	1.01260340456758\\
24	1.00808051890199\\
25	1.00856095124442\\
26	1.00961519526886\\
27	1.00579946803413\\
28	1.00551460013413\\
29	1.00545707299011\\
30	1.00534532058137\\
31	1.00272433191453\\
32	1.00179105643057\\
33	1.00239915045939\\
34	1.00003902566374\\
35	0.999379807118802\\
36	1.00107426348592\\
37	1.00096316839154\\
38	0.99646910870513\\
39	0.998180816837389\\
40	0.999136003832121\\
41	0.99229887042697\\
42	0.995030158754915\\
43	1.00369300683876\\
44	0.999804846635031\\
45	1.00279061008202\\
46	1.00436030578839\\
47	1.00287697000421\\
48	1.0078201183297\\
49	1.00289284284232\\
50	1.00270918334595\\
51	1.00487509394078\\
52	1.00494247500473\\
53	1.00630389551276\\
54	1.00501214662886\\
55	1.00433101708953\\
56	1.00546441339506\\
57	1.00782184322421\\
58	1.0066813313052\\
59	1.0039746769923\\
60	1.00121626435615\\
61	1.00319447494192\\
62	1.00564853336238\\
63	1.0008743934587\\
64	1.00084166912818\\
65	1.00179931674019\\
66	1.0008158755622\\
67	1.00925144360395\\
68	1.01291795697286\\
69	1.01119394463399\\
70	1.01060796850627\\
71	1.00850215937261\\
72	1.00884409774979\\
73	1.00529574817312\\
74	1.00440139350042\\
75	1.00587665995111\\
76	1.0054903431534\\
77	1.0059057659489\\
78	1.01117664378725\\
79	1.00885720140745\\
80	1.00801061852408\\
81	1.00923615928194\\
82	1.01132431104829\\
83	1.00596994128866\\
84	1.00516213701233\\
85	1.00265885911176\\
86	1.00026623988964\\
87	1.00706090854567\\
88	1.00500627285214\\
89	1.0021345541769\\
90	1.00647294181528\\
91	1.00268129691323\\
92	0.999713274488563\\
93	1.00024937177379\\
94	0.999284933562968\\
95	0.999897195997313\\
96	1.00406403160185\\
97	1.00471845588493\\
98	1.0017895104418\\
99	1.0119723193044\\
100	1.00817694964241\\
101	1.00218426498196\\
102	0.998306250262289\\
103	1.00030133874126\\
104	1.00483635216581\\
105	1.00890807346052\\
106	1.00852854369933\\
107	1.00786693208228\\
108	1.00930018845332\\
109	1.00236267967531\\
110	1.00561243617679\\
111	1.00233102914663\\
112	1.00143912108255\\
113	0.999775812241154\\
114	0.992708047534662\\
115	0.998666099447207\\
116	1.01011183922127\\
117	1.00908425568497\\
118	1.00359869694404\\
119	1.00540802274594\\
120	1.01258312965812\\
121	1.00786398014838\\
122	1.00803425425032\\
123	1.01091119524077\\
124	1.00501861559495\\
125	1.00299305224993\\
126	1.00029028273907\\
127	1.00091570185024\\
128	0.998094156827135\\
129	1.00098440567086\\
130	0.999617149393896\\
131	1.00296228968675\\
132	1.00641886466053\\
133	1.01082253722431\\
134	1.00546203686776\\
135	1.00704105834987\\
136	1.00881847580085\\
137	1.0081441880759\\
138	1.0085871549722\\
139	1.01149494482444\\
140	1.01233045077116\\
141	1.01268500258492\\
142	1.01327052401007\\
143	1.00166750927447\\
144	0.998774726295504\\
145	1.00321824889254\\
146	1.00515456519133\\
147	1.00699382477993\\
148	1.00847134909916\\
149	1.00171642930414\\
150	1.00126442635014\\
151	1.00772712811879\\
152	1.01090995946218\\
153	1.00322378664598\\
154	1.00234567057929\\
155	1.00550282362474\\
156	1.00312466116957\\
157	1.00474229234834\\
158	1.00185127442772\\
159	0.99715403557327\\
160	0.998739515402046\\
161	1.00075366511522\\
162	0.997412460258072\\
163	0.998073776296617\\
164	1.00571297824773\\
165	1.00710999670746\\
166	1.00772948890123\\
167	1.00634377377322\\
168	1.02178059863968\\
169	1.02167692865748\\
170	1.02228320928893\\
171	1.02018598337794\\
172	1.01438723975267\\
173	1.01596253228678\\
174	1.01626918406158\\
175	1.01239672464534\\
176	1.01250013214616\\
177	1.02511915707335\\
178	1.02176795004384\\
179	1.01664399019059\\
180	1.01414713669086\\
181	1.01731394824662\\
182	1.01906097632461\\
183	1.01893236888056\\
184	1.01842801235142\\
185	1.01501223353152\\
186	1.0144593283965\\
187	1.01616221258849\\
188	1.0146977505977\\
188	0.93002739449163\\
187	0.932288189885989\\
186	0.932929748804975\\
185	0.935311142037975\\
184	0.929967271288019\\
183	0.931843692079039\\
182	0.927264800974133\\
181	0.934162744376191\\
180	0.936135636462323\\
179	0.93708223223363\\
178	0.928903981847354\\
177	0.915756631178694\\
176	0.930996871986775\\
175	0.928668473022872\\
174	0.930031141856988\\
173	0.92729239273638\\
172	0.93238349522625\\
171	0.927993995574845\\
170	0.915599585908863\\
169	0.915904127649319\\
168	0.910056111802825\\
167	0.915226796218965\\
166	0.909810529149358\\
165	0.918261749364497\\
164	0.918550613350979\\
163	0.914922498276808\\
162	0.918793223206094\\
161	0.920755002775728\\
160	0.924113471774477\\
159	0.923722704256752\\
158	0.925959813504549\\
157	0.918275535078216\\
156	0.914925795053746\\
155	0.916915494748399\\
154	0.923452761459726\\
153	0.932194327876314\\
152	0.922583674852254\\
151	0.924089792866536\\
150	0.923097451447025\\
149	0.928461910079869\\
148	0.907753075633608\\
147	0.913389204835655\\
146	0.910998321255932\\
145	0.917912093944155\\
144	0.91750891231077\\
143	0.916324175552895\\
142	0.906863263867829\\
141	0.909961006932284\\
140	0.910779853630075\\
139	0.916290418131728\\
138	0.911767003780405\\
137	0.907387402739606\\
136	0.907916445490254\\
135	0.912041190526106\\
134	0.92148517066141\\
133	0.913440978306836\\
132	0.915871700951499\\
131	0.900888119305651\\
130	0.908308201344024\\
129	0.904028897852146\\
128	0.903088662182243\\
127	0.902014039580004\\
126	0.910056884990143\\
125	0.906276758765487\\
124	0.907258044775766\\
123	0.894796661982982\\
122	0.891183147356389\\
121	0.897903739614265\\
120	0.895342713482039\\
119	0.894418922279823\\
118	0.901701718319597\\
117	0.910330572629306\\
116	0.911178222651103\\
115	0.920594472331774\\
114	0.928259375319746\\
113	0.934182228195873\\
112	0.937409923960489\\
111	0.935130603267298\\
110	0.937877954199533\\
109	0.927638796596629\\
108	0.923382527147602\\
107	0.926445938875996\\
106	0.923874988542746\\
105	0.924424319702115\\
104	0.931356869330694\\
103	0.927100053076035\\
102	0.931139494626895\\
101	0.923463823008162\\
100	0.921488295442753\\
99	0.914587350903731\\
98	0.920422834359075\\
97	0.921795505424641\\
96	0.919135130723774\\
95	0.919290375453606\\
94	0.925299804131401\\
93	0.920939924576194\\
92	0.924401180286296\\
91	0.927962362574546\\
90	0.923170554672719\\
89	0.919100007906498\\
88	0.914039707342304\\
87	0.915483864910009\\
86	0.920159895156197\\
85	0.92197929795193\\
84	0.922147406206387\\
83	0.926369557970019\\
82	0.930100160360502\\
81	0.933840450594421\\
80	0.928589447868148\\
79	0.921849732907506\\
78	0.923679617102219\\
77	0.926142321171824\\
76	0.918192316184544\\
75	0.92474788752096\\
74	0.92403472854673\\
73	0.91939095406243\\
72	0.91291356952906\\
71	0.912621780846934\\
70	0.905503018060004\\
69	0.911659288226481\\
68	0.912220356152275\\
67	0.922116467220856\\
66	0.918216440436803\\
65	0.917033220497631\\
64	0.912723210132134\\
63	0.908331042894182\\
62	0.915684365677093\\
61	0.924535399795463\\
60	0.92174027279927\\
59	0.920426215177512\\
58	0.916581809986413\\
57	0.915132710426459\\
56	0.926963686573192\\
55	0.933004226701\\
54	0.93341073060514\\
53	0.931813449286816\\
52	0.933015958279116\\
51	0.923713305166806\\
50	0.919560839472801\\
49	0.909932846097899\\
48	0.902500740803337\\
47	0.91826027986395\\
46	0.921590784470416\\
45	0.922459662944507\\
44	0.916004618629891\\
43	0.920099780086344\\
42	0.91845599462069\\
41	0.912101312519486\\
40	0.905684419997954\\
39	0.912034847171489\\
38	0.916683749339159\\
37	0.918923117155277\\
36	0.922363665486399\\
35	0.923865911160983\\
34	0.928314561712092\\
33	0.940713784293886\\
32	0.942925586502447\\
31	0.947398724494882\\
30	0.954209000376783\\
29	0.952760592396694\\
28	0.948937265074116\\
27	0.953623037328655\\
26	0.957279266826968\\
25	0.959446939996848\\
24	0.953394810433274\\
23	0.94799285150583\\
22	0.94670741797261\\
21	0.949715083214402\\
20	0.949935374976377\\
19	0.946087530820258\\
18	0.944554289346774\\
17	0.946771982569158\\
16	0.945853805111785\\
15	0.938314024888145\\
14	0.953269865766081\\
13	0.954347979289306\\
12	0.962466978817076\\
11	0.961191748258449\\
10	0.966453489081751\\
9	0.976359115535597\\
8	0.979870732870033\\
7	0.981932941220674\\
6	0.98264270988112\\
5	0.98462218362162\\
4	0.99019712884285\\
3	0.99000340208393\\
2	1\\
1	1\\
}--cycle;
\end{axis}
\end{tikzpicture}%

%% file: imgs/dynamicBICS/recall.tex
%
%
\definecolor{mycolor1}{rgb}{0.00000,0.44700,0.74100}%
\definecolor{mycolor2}{rgb}{0.85000,0.32500,0.09800}%
\pgfplotsset{
compat=1.11,
legend image code/.code={
\draw[mark repeat=2,mark phase=2]
plot coordinates {
(0cm,0cm)
(0.15cm,0cm)        
(0.3cm,0cm)         
};%
}
}
\begin{tikzpicture}

\begin{axis}[%
width=.85\columnwidth,
height=.5\columnwidth,
scale only axis,
xmin=1,
xmax=188,
xtick={1,24.375,47.75,71.125,94.5,117.875,141.25,164.625,188},
xticklabel style={font=\footnotesize},
xticklabel style={rotate=45},
xlabel style={font=\color{white!15!black}\footnotesize,yshift=5pt},
xticklabels={{ 13},{ 37},{ 61},{ 85},{109},{133},{157},{181},{200}},
xlabel={steps},
ymin=0,
ymax=1,
ytick={  0, 0.2, 0.4, 0.6, 0.8,   1},
yticklabels={{0},{.2},{.4},{.6},{.8},{1}},
yticklabel style={font=\footnotesize},
ylabel style={font=\color{white!15!black}\footnotesize,yshift=-5pt},
ylabel={recall},
axis background/.style={fill=white},
legend style={at={(0.1,0.03)}, legend columns=1, anchor=south west, legend cell align=left, align=left, fill=none, draw=none, font=\scriptsize}
]
\addplot [color=mycolor1, line width=1pt]
  table[row sep=crcr]{%
1	0.99\\
2	0.9025\\
3	0.935625\\
4	0.9875\\
5	0.93625\\
6	0.94625\\
7	0.985625\\
8	0.956875\\
9	0.9575\\
10	0.975\\
11	0.970625\\
12	0.966875\\
13	0.959375\\
14	0.948125\\
15	0.85875\\
16	0.8625\\
17	0.88625\\
18	0.850625\\
19	0.87375\\
20	0.86375\\
21	0.823125\\
22	0.823125\\
23	0.81\\
24	0.808125\\
25	0.80125\\
26	0.810625\\
27	0.81\\
28	0.740625\\
29	0.731875\\
30	0.746875\\
31	0.718125\\
32	0.733125\\
33	0.728125\\
34	0.725\\
35	0.746875\\
36	0.739375\\
37	0.7675\\
38	0.754375\\
39	0.7425\\
40	0.744375\\
41	0.713125\\
42	0.7225\\
43	0.72\\
44	0.715\\
45	0.735\\
46	0.721875\\
47	0.720625\\
48	0.705\\
49	0.703125\\
50	0.701875\\
51	0.699375\\
52	0.7125\\
53	0.70375\\
54	0.663125\\
55	0.68125\\
56	0.663125\\
57	0.670625\\
58	0.689375\\
59	0.680625\\
60	0.691875\\
61	0.669375\\
62	0.68875\\
63	0.711875\\
64	0.69625\\
65	0.708125\\
66	0.69875\\
67	0.69875\\
68	0.673125\\
69	0.70625\\
70	0.6775\\
71	0.70625\\
72	0.711875\\
73	0.715\\
74	0.7275\\
75	0.738125\\
76	0.75375\\
77	0.76875\\
78	0.75875\\
79	0.7375\\
80	0.735\\
81	0.74\\
82	0.738125\\
83	0.730625\\
84	0.74875\\
85	0.776875\\
86	0.77\\
87	0.75875\\
88	0.784375\\
89	0.775\\
90	0.768125\\
91	0.77875\\
92	0.799375\\
93	0.770625\\
94	0.78\\
95	0.774375\\
96	0.781875\\
97	0.749375\\
98	0.749375\\
99	0.75375\\
100	0.7725\\
101	0.779375\\
102	0.7825\\
103	0.78625\\
104	0.775625\\
105	0.769375\\
106	0.7525\\
107	0.77375\\
108	0.793125\\
109	0.7775\\
110	0.7725\\
111	0.749375\\
112	0.74\\
113	0.739375\\
114	0.739375\\
115	0.74\\
116	0.74375\\
117	0.7525\\
118	0.753125\\
119	0.738125\\
120	0.74375\\
121	0.746875\\
122	0.749375\\
123	0.75625\\
124	0.755\\
125	0.74625\\
126	0.724375\\
127	0.720625\\
128	0.713125\\
129	0.686875\\
130	0.699375\\
131	0.701875\\
132	0.715625\\
133	0.7\\
134	0.686875\\
135	0.704375\\
136	0.690625\\
137	0.69375\\
138	0.688125\\
139	0.6925\\
140	0.70625\\
141	0.7125\\
142	0.7175\\
143	0.73875\\
144	0.7375\\
145	0.73\\
146	0.718125\\
147	0.720625\\
148	0.7175\\
149	0.71375\\
150	0.713125\\
151	0.72875\\
152	0.724375\\
153	0.72375\\
154	0.731875\\
155	0.709375\\
156	0.703125\\
157	0.709375\\
158	0.719375\\
159	0.708125\\
160	0.721875\\
161	0.738125\\
162	0.7375\\
163	0.721875\\
164	0.70625\\
165	0.704375\\
166	0.70625\\
167	0.715\\
168	0.72\\
169	0.720625\\
170	0.73\\
171	0.715625\\
172	0.74625\\
173	0.7375\\
174	0.738125\\
175	0.7375\\
176	0.745\\
177	0.738125\\
178	0.745625\\
179	0.74625\\
180	0.756875\\
181	0.758125\\
182	0.76125\\
183	0.773125\\
184	0.75875\\
185	0.76875\\
186	0.761875\\
187	0.765625\\
188	0.77\\
};
\addlegendentry{DFaCe}

\addplot [color=mycolor2, line width=1pt]
  table[row sep=crcr]{%
1	0.988125\\
2	0.905625\\
3	0.935625\\
4	0.99\\
5	0.93125\\
6	0.946875\\
7	0.98625\\
8	0.955625\\
9	0.9575\\
10	0.96875\\
11	0.974375\\
12	0.966875\\
13	0.96375\\
14	0.961875\\
15	0.85375\\
16	0.86125\\
17	0.895625\\
18	0.85625\\
19	0.888125\\
20	0.896875\\
21	0.8875\\
22	0.886875\\
23	0.89625\\
24	0.901875\\
25	0.91125\\
26	0.90875\\
27	0.906875\\
28	0.845\\
29	0.843125\\
30	0.868125\\
31	0.845\\
32	0.84875\\
33	0.868125\\
34	0.856875\\
35	0.856875\\
36	0.845625\\
37	0.84125\\
38	0.844375\\
39	0.855\\
40	0.851875\\
41	0.8375\\
42	0.820625\\
43	0.833125\\
44	0.811875\\
45	0.834375\\
46	0.839375\\
47	0.82625\\
48	0.831875\\
49	0.843125\\
50	0.863125\\
51	0.8775\\
52	0.874375\\
53	0.873125\\
54	0.840625\\
55	0.864375\\
56	0.851875\\
57	0.854375\\
58	0.870625\\
59	0.8575\\
60	0.854375\\
61	0.86\\
62	0.845\\
63	0.8475\\
64	0.85625\\
65	0.860625\\
66	0.860625\\
67	0.840625\\
68	0.844375\\
69	0.846875\\
70	0.83875\\
71	0.84375\\
72	0.846875\\
73	0.839375\\
74	0.848125\\
75	0.8575\\
76	0.854375\\
77	0.869375\\
78	0.856875\\
79	0.865\\
80	0.8625\\
81	0.863125\\
82	0.866875\\
83	0.855625\\
84	0.855\\
85	0.869375\\
86	0.855\\
87	0.83375\\
88	0.855625\\
89	0.84875\\
90	0.871875\\
91	0.87125\\
92	0.858125\\
93	0.845\\
94	0.865\\
95	0.868125\\
96	0.853125\\
97	0.864375\\
98	0.858125\\
99	0.8575\\
100	0.855\\
101	0.87\\
102	0.876875\\
103	0.87875\\
104	0.89125\\
105	0.8775\\
106	0.883125\\
107	0.880625\\
108	0.87375\\
109	0.8675\\
110	0.88\\
111	0.879375\\
112	0.888125\\
113	0.878125\\
114	0.860625\\
115	0.865625\\
116	0.865625\\
117	0.870625\\
118	0.8725\\
119	0.85125\\
120	0.850625\\
121	0.86125\\
122	0.859375\\
123	0.863125\\
124	0.8725\\
125	0.861875\\
126	0.854375\\
127	0.845\\
128	0.84625\\
129	0.845625\\
130	0.854375\\
131	0.841875\\
132	0.85125\\
133	0.855625\\
134	0.850625\\
135	0.85125\\
136	0.859375\\
137	0.863125\\
138	0.8725\\
139	0.870625\\
140	0.87375\\
141	0.87125\\
142	0.86\\
143	0.865625\\
144	0.876875\\
145	0.859375\\
146	0.8625\\
147	0.85125\\
148	0.850625\\
149	0.864375\\
150	0.86125\\
151	0.876875\\
152	0.880625\\
153	0.874375\\
154	0.874375\\
155	0.855\\
156	0.860625\\
157	0.86125\\
158	0.861875\\
159	0.853125\\
160	0.865\\
161	0.848125\\
162	0.861875\\
163	0.863125\\
164	0.869375\\
165	0.86125\\
166	0.8625\\
167	0.87125\\
168	0.859375\\
169	0.871875\\
170	0.870625\\
171	0.87625\\
172	0.87625\\
173	0.879375\\
174	0.8775\\
175	0.876875\\
176	0.87375\\
177	0.88375\\
178	0.881875\\
179	0.884375\\
180	0.89625\\
181	0.883125\\
182	0.885625\\
183	0.890625\\
184	0.8875\\
185	0.881875\\
186	0.886875\\
187	0.876875\\
188	0.88375\\
};
\addlegendentry{DPoP}

\addplot[area legend, draw=none, fill=mycolor1, fill opacity=0.2, forget plot]
table[row sep=crcr] {%
x	y\\
1	1.00472537723435\\
2	0.912758147583104\\
3	0.943121810546315\\
4	1.00746489265625\\
5	0.954062444056482\\
6	0.970734480715718\\
7	1.01329637003091\\
8	0.979541968347832\\
9	0.980919423301079\\
10	1.00461272134246\\
11	1.01075890374225\\
12	1.00497081937801\\
13	1.00457244806097\\
14	1.00755535818914\\
15	0.936392643579677\\
16	0.963317765987823\\
17	1.01849537529046\\
18	0.992823537640662\\
19	1.03421111010816\\
20	1.05553355279442\\
21	1.03418800593541\\
22	1.0310485979309\\
23	1.04553835425792\\
24	1.04628617018669\\
25	1.05776274483991\\
26	1.05672095176372\\
27	1.05360769001132\\
28	0.987516548786598\\
29	0.997121954867208\\
30	1.01460190286056\\
31	0.985646366922727\\
32	1.01463777406946\\
33	1.00607832640241\\
34	0.985769319044826\\
35	0.997133953895964\\
36	0.999578909688819\\
37	0.983652582162486\\
38	0.985486203520853\\
39	0.985511374345368\\
40	0.972245192472265\\
41	0.917674070069572\\
42	0.933943200507409\\
43	0.946666166466035\\
44	0.940315113807761\\
45	0.955360326441462\\
46	0.961474507735602\\
47	0.963951915655981\\
48	0.974944249724241\\
49	0.956549410347946\\
50	0.960511688933038\\
51	0.960670371267687\\
52	0.967316485586256\\
53	0.971773757704196\\
54	0.933152664162888\\
55	0.944074750527411\\
56	0.948580379344896\\
57	0.924046264641405\\
58	0.939168441452847\\
59	0.922792646487102\\
60	0.93393728247486\\
61	0.932130736699254\\
62	0.962637475963754\\
63	0.961179677185019\\
64	0.945934111653595\\
65	0.94533027497189\\
66	0.934535297744904\\
67	0.930614530953498\\
68	0.915615038981196\\
69	0.960675122299554\\
70	0.925534364390032\\
71	0.947899439334364\\
72	0.962455480421695\\
73	0.951113028536966\\
74	0.965496259188575\\
75	0.954713261749226\\
76	0.964910242373378\\
77	0.967598727243929\\
78	0.967097788614154\\
79	0.955108171552749\\
80	0.960633320695688\\
81	0.959381377458321\\
82	0.957819571438139\\
83	0.976111196305476\\
84	0.949291359162378\\
85	0.971469727011325\\
86	0.968924884551626\\
87	0.987697792819094\\
88	0.971262818651201\\
89	0.97722770808761\\
90	0.966882500500238\\
91	0.967398692234091\\
92	0.971593464797904\\
93	0.95998756020973\\
94	0.949872964239728\\
95	0.955966646691028\\
96	0.970486712756063\\
97	0.956615009544785\\
98	0.956903312858162\\
99	0.968002229526656\\
100	0.957866260441872\\
101	0.973600675137591\\
102	0.97708960608398\\
103	0.997572518044465\\
104	0.987637684028045\\
105	0.983184925661145\\
106	0.950428526318498\\
107	0.964888172588335\\
108	0.986783429644694\\
109	0.987239725734112\\
110	0.973941710564377\\
111	0.967119591828151\\
112	0.968197937040166\\
113	0.965746422767801\\
114	0.946048768427738\\
115	0.945981194653833\\
116	0.964450126684076\\
117	0.967791538218694\\
118	0.967340948464466\\
119	0.964168674949541\\
120	0.980988040278329\\
121	0.986308090276569\\
122	0.988744821305947\\
123	0.963442881890577\\
124	0.976003335924587\\
125	0.980583754874259\\
126	0.949606067575103\\
127	0.956465232856214\\
128	0.956556732300488\\
129	0.931352427122524\\
130	0.94216730358183\\
131	0.936766862060491\\
132	0.953555095258233\\
133	0.950318674444218\\
134	0.944916120803468\\
135	0.971800992289532\\
136	0.939926479495352\\
137	0.974112829787635\\
138	0.979399715382967\\
139	0.974306422378692\\
140	0.97992200221418\\
141	0.976459741055259\\
142	0.965495792837915\\
143	0.956702262606999\\
144	0.955291266386615\\
145	0.963662607803889\\
146	0.936144004566209\\
147	0.934390178778405\\
148	0.945218833975011\\
149	0.916996137348139\\
150	0.91254319872994\\
151	0.933079728025811\\
152	0.940852812939104\\
153	0.942396112357342\\
154	0.956875885769232\\
155	0.957474440286366\\
156	0.960063938241548\\
157	0.968014771206982\\
158	0.971635981284406\\
159	0.953750795164107\\
160	0.960474261902908\\
161	0.974764990981306\\
162	0.955748331456791\\
163	0.963462575757124\\
164	0.981302175013381\\
165	0.978135104882925\\
166	0.974032727758372\\
167	0.994482826860979\\
168	1.00375162950175\\
169	0.993670728817556\\
170	1.02149837224315\\
171	1.01921284094289\\
172	1.01880113263012\\
173	0.995116750825956\\
174	1.01258890812317\\
175	1.00108195391296\\
176	1.00171467326467\\
177	1.0316761430687\\
178	1.02705280660701\\
179	1.00935062481019\\
180	1.0191904416802\\
181	1.03192878164269\\
182	1.01471451468417\\
183	1.02287056567941\\
184	1.02564111515174\\
185	1.04300397416799\\
186	1.04094749465042\\
187	1.05818051213158\\
188	1.05232927837646\\
188	0.487670721623542\\
187	0.473069487868419\\
186	0.482802505349581\\
185	0.494496025832006\\
184	0.491858884848261\\
183	0.523379434320586\\
182	0.50778548531583\\
181	0.484321218357312\\
180	0.494559558319802\\
179	0.483149375189814\\
178	0.46419719339299\\
177	0.444573856931303\\
176	0.488285326735327\\
175	0.473918046087045\\
174	0.463661091876828\\
173	0.479883249174045\\
172	0.473698867369884\\
171	0.412037159057107\\
170	0.438501627756846\\
169	0.447579271182444\\
168	0.436248370498251\\
167	0.435517173139021\\
166	0.438467272241628\\
165	0.430614895117075\\
164	0.431197824986619\\
163	0.480287424242876\\
162	0.519251668543209\\
161	0.501485009018694\\
160	0.483275738097093\\
159	0.462499204835893\\
158	0.467114018715594\\
157	0.450735228793018\\
156	0.446186061758452\\
155	0.461275559713634\\
154	0.506874114230768\\
153	0.505103887642658\\
152	0.507897187060895\\
151	0.524420271974189\\
150	0.51370680127006\\
149	0.510503862651861\\
148	0.489781166024989\\
147	0.506859821221595\\
146	0.500105995433791\\
145	0.496337392196111\\
144	0.519708733613385\\
143	0.520797737393001\\
142	0.469504207162085\\
141	0.448540258944741\\
140	0.43257799778582\\
139	0.410693577621308\\
138	0.396850284617033\\
137	0.413387170212365\\
136	0.441323520504648\\
135	0.436949007710468\\
134	0.428833879196532\\
133	0.449681325555782\\
132	0.477694904741767\\
131	0.466983137939509\\
130	0.45658269641817\\
129	0.442397572877476\\
128	0.469693267699512\\
127	0.484784767143785\\
126	0.499143932424897\\
125	0.511916245125741\\
124	0.533996664075413\\
123	0.549057118109423\\
122	0.510005178694053\\
121	0.507441909723431\\
120	0.506511959721671\\
119	0.512081325050459\\
118	0.538909051535534\\
117	0.537208461781306\\
116	0.523049873315924\\
115	0.534018805346167\\
114	0.532701231572262\\
113	0.513003577232199\\
112	0.511802062959834\\
111	0.531630408171849\\
110	0.571058289435623\\
109	0.567760274265888\\
108	0.599466570355306\\
107	0.582611827411665\\
106	0.554571473681502\\
105	0.555565074338855\\
104	0.563612315971955\\
103	0.574927481955535\\
102	0.58791039391602\\
101	0.585149324862409\\
100	0.587133739558128\\
99	0.539497770473344\\
98	0.541846687141838\\
97	0.542134990455215\\
96	0.593263287243937\\
95	0.592783353308972\\
94	0.610127035760272\\
93	0.58126243979027\\
92	0.627156535202096\\
91	0.590101307765909\\
90	0.569367499499762\\
89	0.572772291912391\\
88	0.597487181348799\\
87	0.529802207180906\\
86	0.571075115448375\\
85	0.582280272988675\\
84	0.548208640837622\\
83	0.485138803694524\\
82	0.518430428561861\\
81	0.520618622541679\\
80	0.509366679304312\\
79	0.519891828447251\\
78	0.550402211385846\\
77	0.569901272756071\\
76	0.542589757626622\\
75	0.521536738250775\\
74	0.489503740811426\\
73	0.478886971463034\\
72	0.461294519578305\\
71	0.464600560665636\\
70	0.429465635609968\\
69	0.451824877700446\\
68	0.430634961018804\\
67	0.466885469046502\\
66	0.462964702255096\\
65	0.47091972502811\\
64	0.446565888346405\\
63	0.462570322814981\\
62	0.414862524036246\\
61	0.406619263300747\\
60	0.44981271752514\\
59	0.438457353512898\\
58	0.439581558547152\\
57	0.417203735358595\\
56	0.377669620655103\\
55	0.418425249472589\\
54	0.393097335837112\\
53	0.435726242295804\\
52	0.457683514413744\\
51	0.438079628732313\\
50	0.443238311066962\\
49	0.449700589652054\\
48	0.435055750275758\\
47	0.477298084344019\\
46	0.482275492264398\\
45	0.514639673558538\\
44	0.489684886192239\\
43	0.493333833533965\\
42	0.511056799492591\\
41	0.508575929930428\\
40	0.516504807527735\\
39	0.499488625654632\\
38	0.523263796479147\\
37	0.551347417837514\\
36	0.479171090311181\\
35	0.496616046104036\\
34	0.464230680955174\\
33	0.450171673597593\\
32	0.451612225930535\\
31	0.450603633077273\\
30	0.479148097139443\\
29	0.466628045132792\\
28	0.493733451213402\\
27	0.566392309988677\\
26	0.564529048236283\\
25	0.544737255160087\\
24	0.569963829813309\\
23	0.574461645742078\\
22	0.6152014020691\\
21	0.612061994064593\\
20	0.671966447205583\\
19	0.713288889891841\\
18	0.708426462359338\\
17	0.754004624709541\\
16	0.761682234012177\\
15	0.781107356420323\\
14	0.888694641810857\\
13	0.914177551939025\\
12	0.928779180621994\\
11	0.930491096257754\\
10	0.945387278657541\\
9	0.934080576698921\\
8	0.934208031652168\\
7	0.957953629969086\\
6	0.921765519284282\\
5	0.918437555943518\\
4	0.967535107343752\\
3	0.928128189453685\\
2	0.892241852416896\\
1	0.975274622765651\\
}--cycle;

\addplot[area legend, draw=none, fill=mycolor2, fill opacity=0.2, forget plot]
table[row sep=crcr] {%
x	y\\
1	1.00344732348369\\
2	0.910044417382416\\
3	0.943121810546315\\
4	1.01226779270239\\
5	0.954872779563077\\
6	0.977965152404429\\
7	1.01791812116244\\
8	0.980949740363719\\
9	0.980919423301079\\
10	1.00031726701726\\
11	1.00759720527703\\
12	0.998700065929682\\
13	1.00719830378784\\
14	1.02186609734122\\
15	0.973279530549858\\
16	0.970813430980247\\
17	1.02769695250224\\
18	0.983305798846051\\
19	1.01920938287482\\
20	1.02014911849539\\
21	0.988712357578166\\
22	0.993811184598716\\
23	1.01792779047089\\
24	1.00957629764913\\
25	1.0142889638012\\
26	1.01341631867657\\
27	1.00146941145528\\
28	0.94401459642963\\
29	0.96915550985085\\
30	0.980761326130986\\
31	0.947184369686187\\
32	0.968099320082192\\
33	0.977530061466505\\
34	0.954904452701414\\
35	0.961016365975019\\
36	0.965648714174682\\
37	0.969218576739187\\
38	0.965074293764736\\
39	0.967896127986241\\
40	0.975723329827335\\
41	0.930930578287752\\
42	0.913372617412601\\
43	0.941674178288658\\
44	0.921316488537324\\
45	0.951854440776257\\
46	0.950177456250863\\
47	0.936719198773011\\
48	0.96155308700779\\
49	0.950172949036201\\
50	0.964315695027775\\
51	0.965399960530278\\
52	0.95608240558573\\
53	0.963830233659725\\
54	0.941096227097562\\
55	0.940296738202129\\
56	0.971632740685442\\
57	0.952510114193418\\
58	0.959240790301821\\
59	0.97273738808323\\
60	0.950042049244891\\
61	0.950667873391329\\
62	0.954708856357476\\
63	0.94275576472994\\
64	0.944863541255711\\
65	0.951286278804624\\
66	0.946080804456491\\
67	0.939092611607646\\
68	0.95290214381148\\
69	0.973183520989242\\
70	0.962525894013235\\
71	0.965027479528626\\
72	0.968848893681377\\
73	0.970902585470866\\
74	0.969404122843774\\
75	0.969046412048153\\
76	0.955115640548704\\
77	0.966094551054182\\
78	0.969298797888315\\
79	0.971478595626213\\
80	0.959077266307071\\
81	0.961519720620155\\
82	0.985797739561002\\
83	0.986526809769844\\
84	0.969821567429376\\
85	0.96671076237825\\
86	0.962469739452365\\
87	0.982945333201074\\
88	0.972020495411295\\
89	0.963675663640744\\
90	0.983794206577872\\
91	0.977203216957643\\
92	0.96493437685753\\
93	0.957754813463478\\
94	0.953936818320992\\
95	0.961404003942638\\
96	0.968550745302213\\
97	0.977597242284629\\
98	0.976947145109266\\
99	1.00341591442208\\
100	0.980121114794214\\
101	0.974193023876312\\
102	0.970818253198027\\
103	0.972103754447198\\
104	0.981034322071261\\
105	0.994909865559127\\
106	0.976087251637919\\
107	0.97505071113563\\
108	0.974361966769082\\
109	0.974001053937489\\
110	0.978287310514447\\
111	0.969551361359057\\
112	0.975399216714109\\
113	0.961703773334148\\
114	0.966120142672784\\
115	0.971127699090147\\
116	0.965897667166144\\
117	0.96553911598527\\
118	0.974660962431341\\
119	0.983012241424871\\
120	0.995462005943979\\
121	0.99609212234752\\
122	0.988492454142675\\
123	0.990006565880171\\
124	0.979240315241989\\
125	0.984961437294766\\
126	0.967773129798673\\
127	0.945809858401946\\
128	0.95916024971098\\
129	0.956858302461676\\
130	0.943125628770409\\
131	0.953466070271262\\
132	0.960156613711951\\
133	0.966045477087378\\
134	0.963261326130985\\
135	0.951980748401217\\
136	0.976684672709562\\
137	0.980033042638963\\
138	0.98072371481787\\
139	0.980524435129744\\
140	0.971345453013855\\
141	0.975878229109216\\
142	0.965877950046846\\
143	0.956074998166763\\
144	0.959533031234424\\
145	0.970579631313225\\
146	0.969456683488491\\
147	0.971795634512412\\
148	0.960575200345986\\
149	0.960250148005633\\
150	0.955352229473253\\
151	0.968672265115324\\
152	0.990284796853352\\
153	0.990358826832601\\
154	0.989323205394732\\
155	0.965216334116693\\
156	0.97690400395258\\
157	0.983045663607\\
158	0.972691844778827\\
159	0.972955942245264\\
160	0.978016106401355\\
161	0.97136031135306\\
162	0.968475188215364\\
163	0.969874650503658\\
164	0.984357876387073\\
165	0.989112629571022\\
166	0.98492240357859\\
167	0.98777384357911\\
168	0.979372143354604\\
169	1.00099245414267\\
170	1.0023040242737\\
171	0.99586620172153\\
172	0.983012718498662\\
173	0.995104234594358\\
174	0.991464392040063\\
175	0.983721688874618\\
176	0.977866484511046\\
177	1.01618813534582\\
178	1.000037820584\\
179	0.988163642299667\\
180	1.00403083866659\\
181	0.992442586972219\\
182	1.00024678544666\\
183	1.01095385689235\\
184	0.999907842071555\\
185	0.988885707193634\\
186	0.999085867116245\\
187	0.986119638055502\\
188	0.990153700353792\\
188	0.777346299646208\\
187	0.767630361944498\\
186	0.774664132883755\\
185	0.774864292806366\\
184	0.775092157928445\\
183	0.770296143107647\\
182	0.771003214553343\\
181	0.773807413027781\\
180	0.78846916133341\\
179	0.780586357700333\\
178	0.763712179415997\\
177	0.751311864654181\\
176	0.769633515488955\\
175	0.770028311125382\\
174	0.763535607959937\\
173	0.763645765405642\\
172	0.769487281501338\\
171	0.75663379827847\\
170	0.738945975726301\\
169	0.742757545857325\\
168	0.739377856645396\\
167	0.75472615642089\\
166	0.74007759642141\\
165	0.733387370428978\\
164	0.754392123612927\\
163	0.756375349496342\\
162	0.755274811784636\\
161	0.72488968864694\\
160	0.751983893598645\\
159	0.733294057754736\\
158	0.751058155221173\\
157	0.739454336393\\
156	0.74434599604742\\
155	0.744783665883307\\
154	0.759426794605268\\
153	0.758391173167399\\
152	0.770965203146648\\
151	0.785077734884676\\
150	0.767147770526747\\
149	0.768499851994367\\
148	0.740674799654014\\
147	0.730704365487588\\
146	0.75554331651151\\
145	0.748170368686775\\
144	0.794216968765576\\
143	0.775175001833237\\
142	0.754122049953154\\
141	0.766621770890784\\
140	0.776154546986145\\
139	0.760725564870256\\
138	0.76427628518213\\
137	0.746216957361038\\
136	0.742065327290438\\
135	0.750519251598782\\
134	0.737988673869014\\
133	0.745204522912622\\
132	0.742343386288049\\
131	0.730283929728738\\
130	0.765624371229591\\
129	0.734391697538324\\
128	0.73333975028902\\
127	0.744190141598054\\
126	0.740976870201327\\
125	0.738788562705234\\
124	0.765759684758011\\
123	0.736243434119829\\
122	0.730257545857325\\
121	0.726407877652479\\
120	0.705787994056021\\
119	0.719487758575128\\
118	0.770339037568659\\
117	0.77571088401473\\
116	0.765352332833856\\
115	0.760122300909853\\
114	0.755129857327216\\
113	0.794546226665853\\
112	0.800850783285891\\
111	0.789198638640943\\
110	0.781712689485553\\
109	0.760998946062511\\
108	0.773138033230918\\
107	0.78619928886437\\
106	0.790162748362081\\
105	0.760090134440873\\
104	0.801465677928739\\
103	0.785396245552802\\
102	0.782931746801973\\
101	0.765806976123688\\
100	0.729878885205786\\
99	0.711584085577917\\
98	0.739302854890735\\
97	0.751152757715371\\
96	0.737699254697787\\
95	0.774845996057362\\
94	0.776063181679008\\
93	0.732245186536522\\
92	0.75131562314247\\
91	0.765296783042356\\
90	0.759955793422128\\
89	0.733824336359256\\
88	0.739229504588705\\
87	0.684554666798926\\
86	0.747530260547635\\
85	0.77203923762175\\
84	0.740178432570624\\
83	0.724723190230156\\
82	0.747952260438998\\
81	0.764730279379845\\
80	0.765922733692929\\
79	0.758521404373787\\
78	0.744451202111685\\
77	0.772655448945818\\
76	0.753634359451296\\
75	0.745953587951847\\
74	0.726845877156227\\
73	0.707847414529134\\
72	0.724901106318623\\
71	0.722472520471374\\
70	0.714974105986765\\
69	0.720566479010758\\
68	0.73584785618852\\
67	0.742157388392354\\
66	0.775169195543508\\
65	0.769963721195376\\
64	0.767636458744289\\
63	0.75224423527006\\
62	0.735291143642524\\
61	0.769332126608671\\
60	0.758707950755109\\
59	0.74226261191677\\
58	0.782009209698179\\
57	0.756239885806582\\
56	0.732117259314558\\
55	0.788453261797871\\
54	0.740153772902438\\
53	0.782419766340275\\
52	0.79266759441427\\
51	0.789600039469722\\
50	0.761934304972225\\
49	0.736077050963799\\
48	0.70219691299221\\
47	0.715780801226989\\
46	0.728572543749137\\
45	0.716895559223743\\
44	0.702433511462676\\
43	0.724575821711342\\
42	0.727877382587399\\
41	0.744069421712248\\
40	0.728026670172665\\
39	0.742103872013759\\
38	0.723675706235264\\
37	0.713281423260814\\
36	0.725601285825318\\
35	0.752733634024981\\
34	0.758845547298586\\
33	0.758719938533495\\
32	0.729400679917808\\
31	0.742815630313813\\
30	0.755488673869015\\
29	0.71709449014915\\
28	0.74598540357037\\
27	0.812280588544717\\
26	0.804083681323426\\
25	0.808211036198803\\
24	0.794173702350872\\
23	0.774572209529111\\
22	0.779938815401284\\
21	0.786287642421834\\
20	0.773600881504606\\
19	0.757040617125176\\
18	0.729194201153949\\
17	0.763553047497756\\
16	0.751686569019753\\
15	0.734220469450142\\
14	0.901883902658777\\
13	0.920301696212156\\
12	0.935049934070319\\
11	0.941152794722966\\
10	0.937182732982743\\
9	0.934080576698921\\
8	0.930300259636281\\
7	0.954581878837557\\
6	0.915784847595571\\
5	0.907627220436923\\
4	0.967732207297607\\
3	0.928128189453685\\
2	0.901205582617584\\
1	0.972802676516312\\
}--cycle;
\end{axis}
\end{tikzpicture}%

%% file: Appendix.tex
\newcommand{\frazione}{\frac{|\FPaths^{(m, \TPaths)}|}{|\hat{m}^{(\TPaths)|}}}
\onecolumn
\appendix

\section*{The minimum value of $\mathcal{U}(a|\Obst)$ s.t. $P(Z|\Obst)\in(0,1)$ is $(|\hat{m}^{(\TPaths)}|+|\Fonem)|P_{min}$}
\label{app:appA}
In Section~\ref{sec:optApp} we discuss what is the minimum value of $\mathcal{U}(a|\Obst)$ subject to $P(Z|\Obst)\in(0,1)$. First of all, notice that when $|\hat{m}^{(\TPaths)}|> 1$, the second component of $\mathcal{U}(a|\Obst)$, that is $\lfloor \frac{1}{|\hat{m}^{(\TPaths)}|} \rfloor P(\Bar{Z}|\Obst)$, is equal to 0. We can observe from Equation~\ref{eq:DeltaMin} that $\Delta_{min}$ decreases exponentially with $|\hat{m}|$, and therefore, $\exists n_0\in \mathbb{N}$ such that $\forall\,|\hat{m}^{(\TPaths)}| \ge n_0$, $\mathcal{U}(a|\Obst)$ decreases with $|\hat{m}^{(\TPaths)}|$. Such $n_0$ is $-\left [\frac{1}{\ln(P_{min})}\right ]$, that is 0 for all $p\in (0,1)$. Therefore the decreasing trend of $\Delta_{min}$ holds for all possible values of $|\hat{m}^{(\TPaths)}|$. For this reason, for our analysis we legitimately consider the second component of $\mathcal{U}(a|\Obst)$ to be zero. In Section~\ref{sec:optApp}, we claim that the minimum non-zero value of $P(Z|\Obst)$ is $P_{min}=\prod\limits_{v\in\hat{m}^{(\TPaths)}}1-\frac{p}{1+(1-p)(p^{\partial_v}-1)}$. As a matter of fact, the probability of failure of a path is the product of the conditional probability of failure of its nodes, $P(\bar{S}_v|\Obst)$. Such probability depends on the number of failing paths traversing each node $v$, on their lengths and on their intersections. In particular, it is easy to see that the shorter the failing paths traversing the node and the least the cardinality of their intersections, the more $P(\bar{S}_v|\Obst)$ grows. Therefore, the limit situation that we are seeking for occurs indeed when each node of $m^{(\TPaths)}$ is traversed by a large number of failing paths of length 2, i.e., paths passing through $v$ and another node that is not in $\hat{m}^{(\TPaths)}$. When such condition holds though, then the set of all such 2-length paths is the set $\Fonem$, as if $m$ works, it would not only be possible to classify all of its nodes as working, but also all of the other nodes of such 2-length paths as failed. Therefore, in this situation it holds that $\mathcal{U}(a|\Obst) = (|\hat{m}^{(\TPaths)}+|\Fonem|)P_{min}:=\Delta_{min}$. We wonder if it is possible to get a smaller value of such $\mathcal{U}(a|\Obst)$  in a situation where $P_{min}$ is sacrificed  for a slightly higher value of $P(Z|\Obst)$, but where at the same time $|\Fonem| = 0$. The next-most smaller value of $P(a|\Obst)$ such that $\FPaths_1^{m,(\TPaths)} = \emptyset$ results when all nodes $v$ of $\hat{m}$ are traversed by paths of length 3, that only intersect in $v$ and such that the two remaining nodes are not in $\hat{m}^{(\TPaths)}$. In this situation, if $m$ works, all nodes it traverses would be identified as working, but none of the nodes of the original 3-length paths that are not in $\hat{m}^{(\TPaths)}$ would be uniquely identified, hence $\Fonem=\emptyset$. We call $P_3$ such probability. It holds that: 
\begin{align*}
P_3 &= \prod\limits_{v\in \hat{m}^{(\TPaths)}}1-\frac{p}{1-\sum\limits_{i = 1}^{\partial_v}(-1)^{i+1}(1-p)^{2i+1}\binom{\partial_v}{i}}\\
&=\prod\limits_{v\in \hat{m}^{(\TPaths)}}1-\frac{p}{1-(1-p)\sum\limits_{i = 1}^{\partial_v}(-1)^{i+1}(1-p)^{2i}\binom{\partial_v}{i}}\\
&=\prod\limits_{v\in \hat{m}^{(\TPaths)}}1-\frac{p}{1+(1-p)\sum\limits_{i = 1}^{\partial_v}(-1)^{i}(1-p)^{2i}\binom{\partial_v}{i}}\\
&=\prod\limits_{v\in \hat{m}^{(\TPaths)}}1-\frac{p}{1+(1-p)\sum\limits_{i = 1}^{\partial_v}(-1)^{i}[(1-p)^2]^i\binom{\partial_v}{i}}\\
&=\prod\limits_{v\in \hat{m}^{(\TPaths)}}1-\frac{p}{1+(1-p)\sum\limits_{i = 1}^{\partial_v}[-(1-p)^2]^i\binom{\partial_v}{i}}\\
\text{\footnotesize{[Newton's binomial]}}\qquad&=\prod\limits_{v\in \hat{m}^{(\TPaths)}}1-\frac{p}{1+(1-p)[(1-(1-p)^2)^{\partial_v}-1]}\\
&=\prod\limits_{v\in \hat{m}^{(\TPaths)}}1-\frac{p}{1+(1-p)[(1-1+2p-p^2)^{\partial_v}-1]}\\
&=\prod\limits_{v\in \hat{m}^{(\TPaths)}}1-\frac{p}{1+(1-p)[(2p-p^2)^{\partial_v}-1]}.
\end{align*}
Hence the resulting conditional expected marginal benefit is $\mathcal{U}(a|\Obst) = |\hat{m}^{(\TPaths)}|P_3=:\Delta_3$. \\
We see now that $\Delta_{min}< \Delta_3$ for growing values of $|\FPaths^{(m, \TPaths)}|$. Asymptotically speaking, we can assume without loss of generality that $\partial_v = \partial_w$ $\forall v,w\in\hat{m}^{(\TPaths)}$. Since $\sum_{v}\partial_v=|\FPaths^{(m, \TPaths)}|$, it results that $\partial_v = \frac{|\FPaths^{(m, \TPaths)}|}{|\hat{m}^{(\TPaths)}|}$. We show that $\frac{\Delta_{min}}{\Delta_3}<1$ for growing values of $|\FPaths^{(m, \TPaths)}|$. 
\begingroup
\allowdisplaybreaks
\begin{align*}
    \frac{\Delta_{min}}{\Delta_3}  =& \frac{|\hat{m}^{(\TPaths)}|+|\FPaths^{(m, \TPaths)}|\left [ 1-\frac{p}{1+(1-p)(p^{\partial_v}-1)}\right]^{|\hat{m}^{(\TPaths)}|}}{|\hat{m}^{(\TPaths)}| \left [1- \frac{p}{1+(1-p)[(2p-p^2)^{\partial_v}-1]}\right ]^{|\hat{m}^{(\TPaths)}|} }\\
    =& \frac{|\hat{m}^{(\TPaths)}|+|\FPaths^{(m, \TPaths)}|}{|\hat{m}^{(\TPaths)}|}\left [ \frac{1+(1-p)(p^{\partial_v}-1)-p}{1+(1-p)(p^{\partial_v}-1)}\cdot \frac{1+(1-p)[(2p-p^2)^{\partial_v}-1]}{1+(1-p)[(2p-p^2)^{\partial_v}-1]-p}\right]^{|\hat{m}^{(\TPaths)}|}\\
    =&\frac{|\hat{m}^{(\TPaths)}|+|\FPaths^{(m, \TPaths)}|}{|\hat{m}^{(\TPaths)}|} \left [ \frac{p^{\partial_v}(1-p)(2-p)^{\partial_v}+p}{(2-p)^{\partial_v} [p^{\partial_v}(1-p)+p]}\right]^{|\hat{m}^{(\TPaths)}|}\le 1\\
    \intertext{this is true $\forall|\hat{m}^{(\TPaths)}|$ if and only if $|\FPaths^{(m, \TPaths)}|$ is sufficiently large and}
     & \frac{p^{\partial_v}(1-p)(2-p)^{\partial_v}+p}{(2-p)^{\partial_v} [p^{\partial_v}(1-p)+p]} \le 1\\
  \intertext{which is always true for $p\in(0,1)$, as it holds if and only if}
    & p^{\partial_v}(1-p)(2-p)^{\partial_v}+p\leq (2-p)^{\partial_v} [p^{\partial_v}(1-p)+p]\\
    \iff & p\leq (2-p)^{\partial_v}p\\
    \iff &(2-p)^{\partial_v} \geq 1 \qquad \text{\footnotesize{since $p\in(0,1)$}}\\
    \iff &\partial_v = \frazione\geq 0.
\end{align*}
\endgroup
Therefore $ \frac{\Delta_{min}}{\Delta_3}\rightarrow 0$ for growing values of $|\Fonem|$ and it is $<1$ for $|\Fonem|$ sufficiently large and $\forall |\hat{m}^{\TPaths}|$ and $\forall p\in(0,1)$. Hence, it holds that the minimum value of $\mathcal{U}(a|\Obst)$ subject to $P(Z|\Obst)\neq 0, 1$ is indeed $\Delta{min} = (|\hat{m}^{(\TPaths)}|+|\Fonem|) \prod\limits_{v\in\hat{m}} 1 - \frac{p}{1+(1-p)(p^{\partial_{v}}-1)}$.